\pdfoutput=1
\newcommand*{\ATLASLATEXPATH}{latex/}
\documentclass[cernpreprint,texlive=2016,txfonts,UKenglish]{latex/atlasdoc}

\usepackage[subfigure,biblatex=true]{\ATLASLATEXPATH atlaspackage}
\usepackage{latex/atlasphysics} 
\usepackage{\ATLASLATEXPATH atlasbiblatex}
\usepackage{\ATLASLATEXPATH atlascontribute}
\usepackage{\ATLASLATEXPATH atlasphysics}
\usepackage{multicol}
\usepackage{multirow}
\graphicspath{{logos/}{figures/}{figures_final/}}
\usepackage{MaterialPaper-defs}
\AtlasTitle{Study of the material of the ATLAS inner detector for Run 2 of the LHC}

\author{The ATLAS Collaboration}

\AtlasRefCode{PERF-2015-07}

\AtlasJournal{JINST}

\AtlasAbstract{
The ATLAS inner detector comprises three different sub-detectors: the pixel detector, the silicon strip tracker, and the transition-radiation drift-tube tracker. The Insertable $B$-Layer, a new innermost pixel layer, was installed during the shutdown period in 2014, together with modifications to the layout of the cables and support structures of the existing pixel detector. The material in the inner detector is studied with several methods, using a low-luminosity $\sqrt{s}=13$ TeV $pp$ collision sample corresponding to around $2.0\,\inb$ collected in 2015 with the ATLAS experiment at the LHC. In this paper, the material within the innermost barrel region is studied using reconstructed hadronic interaction and photon conversion vertices. For the forward rapidity region, the material is probed by a measurement of the efficiency with which single tracks reconstructed from pixel detector hits alone can be extended with hits on the track in the strip layers. The results of these studies have been taken into account in an improved description of the material in the ATLAS inner detector simulation, resulting in a reduction in the uncertainties associated with the charged-particle reconstruction efficiency determined from simulation.
}

\PreprintIdNumber{CERN-EP-2017-081}
\hypersetup{pdftitle={ATLAS document},pdfauthor={The ATLAS Collaboration}}

\begin{document}

\maketitle

\newcommand{\PV}{\mathrm{pv}}
\newcommand{\SV}{\mathrm{sv}}
\newcommand{\electron}{\ensuremath{e}\xspace}
\newcommand{\lambdaBaryon}{\ensuremath{\Lambda}\xspace}
\newcommand{\millimeter}{\ensuremath{\mathrm{mm}}\xspace}
\newcommand{\micrometer}{\ensuremath{\mathrm{\mu m}}\xspace}
\newcommand{\xAOD}{\texttt{xAOD}\xspace}
\newcommand{\DxAOD}{\texttt{DxAOD}\xspace}
\newcommand{\AOD}{\texttt{AOD}\xspace}
\newcommand{\ESD}{\texttt{ESD}\xspace}
\newcommand{\EPOS}{\texttt{EPOS}\xspace}
\newcommand{\PythiaEight}{\texttt{Py8}\xspace}
\newcommand{\PythiaEightOriginal}{\texttt{Py8(Orig)}\xspace}
\newcommand{\PythiaEightOriginalAlt}{\texttt{Py8(Orig\_Alt)}\xspace}
\newcommand{\SctExtEff}{\ensuremath{{\cal E}_{\mathrm{ext}}}\xspace}
\newcommand{\NormSctExtEff}{\ensuremath{\varDelta{\cal E}_{\mathrm{ext~norm}}^{\mathrm{mod-nom}}}\xspace}

\newcommand{\etagamma}{{\eta^{\gamma}}}
\newcommand{\ptgamma}{{\pT^{\gamma}}}

\section{Introduction}
\label{sec:intro}

Data recorded by tracking detectors are used to reconstruct the trajectories of charged particles and determine their momenta. The location of particle interactions with the material of the detector can be identified by reconstructing interaction vertices. Obtaining an accurate description of this material is essential to understand the performance of the detector. For the ATLAS detector~\cite{PERF-2007-01}, nuclear interactions of~\emph{primary particles} with the material are the largest source of \emph{secondary particles};\footnote{In this paper, \emph{primary particles} refer to particles which are promptly produced in the $pp$ collision, while \emph{secondary particles} refer to those which are produced in the decays of primary particles or through their interaction with material.} hence the uncertainty in the track reconstruction efficiency is directly coupled to the accuracy of knowing the amount and type of material~\cite{STDM-2010-01,STDM-2010-06}. For electromagnetic calorimeters, the knowledge of the material situated between the collision point and the calorimeter is essential to calibrate the energy of reconstructed electrons, unconverted and converted photons~\cite{PERF-2013-05}. The precision of track reconstruction parameters is also sensitive to the amount of material of the tracking detector. The precise knowledge of it is important for the performance of the reconstruction of the high-level objects based on track reconstruction like identification of $b$-hadrons ($b$-tagging). Furthermore, searches for new physics performed by reconstructing the decay vertex of long-lived particles require a precise description of the material to define fiducial decay volumes with minimal background~\cite{SUSY-2014-02}. The accuracy of the description of the material structure is thus an essential foundation for physics analysis with the ATLAS detector. In fact, it plays a central role together with other key ingredients required for particle reconstruction, e.g.~the magnetic field description, understanding of the processes occurring inside semiconductor sensors or gases, and the alignment of the components.

The description of the material including the geometrical layout and atomic composition, hereafter referred to as the \emph{geometry model}, is based on engineering design drawings of the detector, together with supporting measurements of the masses, dimensions and compositions of detector components. During construction of the ATLAS inner detector (ID)~\cite{PERF-2007-01}, detailed measurements of the mass of detector components were undertaken, and the corresponding masses in the geometry model were adapted to agree with the measurement as accurately as possible~\cite{PERF-2007-01}. The amount of material in the Run~1\footnote{Run~1 refers to the period of data-taking in 2008--2012, while Run~2 refers to the period since 2015.} as-built ID~\cite{PERF-2007-01} is generally known to an accuracy of about 4--5\%. However, obtaining a satisfactory geometry model is challenging because of the complexity of the detector design and the need to thoroughly validate the description.

Several \emph{in situ} methods using collision data have been developed to estimate the amount of material within the tracking detectors of high-energy physics experiments~\cite{STDM-2010-01,ATLAS-CONF-2010-019,ATLAS-CONF-2010-037}. Reconstructing photon conversion vertices is a traditional method to measure the material of tracking detectors~\cite{ATLAS-CONF-2010-007}, taking advantage of precise theoretical understanding of electromagnetic interaction processes. The reconstruction of hadronic interaction vertices instead of photon conversions is a complementary approach~\cite{PERF-2011-08,PERF-2015-06} -- it is sensitive to the material through nuclear interactions, and offers much better resolution in the radial position of the vertex compared to the case of photon conversion. However, the description of hadronic interactions is complex and only phenomenologically modelled in the simulation. Another complementary approach which is applicable to the all tracking acceptance is to measure the nuclear interaction rate of charged hadrons through hadronic interactions, referred to as the \emph{track-extension efficiency} method. The precision of each measurement varies depending on the detector region.  All of these approaches are used together to measure a large part of the inner detector's volume and cross-check individual measurements. Using the hadronic interaction approach, ATLAS has performed measurements of the inner detector's material in Run~1 of the LHC~\cite{PERF-2011-08,PERF-2015-06}. The measurements were performed by comparing observables sensitive to the material in data and Monte Carlo (MC) simulation.

The ATLAS inner detector system is immersed in a \SI{2}{\tesla} axial magnetic field, and provides measurements for charged-particle trajectory reconstruction with full coverage in $\phi$ in the range $|\eta| < 2.5$.\footnote{\AtlasCoordFootnote} It consists of a silicon pixel detector (pixel), a silicon micro-strip detector (SCT) and a transition-radiation straw-tube tracker (TRT). During the LHC shutdown period in 2013--2014, between Run~1 and Run~2, the inner detector was upgraded with the installation of a new pixel-detector layer together with a new, thinner beam pipe, referred to as the insertable $B$-layer (IBL)~\cite{Capeans:1291633}. In addition, the pixel detector was extracted and renovated. This involved replacement of pixel service panels (cables, cooling pipes and support structures) located in the forward $\eta$ region of the pixel detector. These changes motivated the material re-evaluation and creation of a new ID geometry model.

The characteristics of a material in terms of interaction with high-energy particles are quantified by the radiation length, $X_{0}$, and nuclear interaction length, $\lambda_{I}$. In this  paper, the unit of $\mathrm{mm}$ is used to quantify these properties.\footnote{There is another common convention of using $\mathrm{g/cm^{2}}$ in the literature~\cite{Agashe:2014kda}.} The radiation length $X_{0}$ is the mean path length over which a high-energy ($E\gg 2m_{e}$) electron loses all but $1/\mathrm{e}$ of its energy due to bremsstrahlung. Similarly, $\lambda_{I}$ is the mean path length to reduce the flux of relativistic primary hadrons to a fraction $1/\mathrm{e}$. The amount of material associated with electromagnetic interactions along a particular trajectory $C$ is represented by a dimensionless number $N_{X_{0}}^{[C]}$, frequently referred to as the \emph{number of radiation lengths} in the literature.\footnote{The bracket ``$[C]$'' indicates that the value is defined with respect to the trajectory $C$, but this can be omitted if the specified trajectory is clear.} This is calculated as a line integral:
\begin{eqnarray*}
\label{eq:radiation_length}
N_{X_{0}}^{[C]} = \int\limits_C \mathrm{d}s\,\frac{1}{X_{0}(s)}~,
\end{eqnarray*}
where $X_{0}(s)$ is the local radiation length of the material at the position $s$ along the trajectory $C$. Similarly, the amount of material associated with nuclear interactions is represented by a dimensionless number denoted by $N_{\lambda_{I}}^{[C]}$:
\begin{eqnarray*}
N_{\lambda_{I}}^{[C]} = {\displaystyle \int\limits_C \mathrm{d}s\,\frac{1}{\lambda_{I}(s)}}~.
\end{eqnarray*}

This paper presents studies of the ATLAS Run~2 ID material using hadronic interactions, photon conversions and the track-extension efficiency measurement. An additional study of the transverse impact parameter resolution of tracks is also presented. The paper is organised as follows. Section~\ref{sec:detector} provides an overview of the ATLAS detector and further details of the inner detector. Data and MC simulation samples together with the various geometry model versions used in this paper are introduced in Section~\ref{sec:datasim}. The methodology of the measurements presented in this paper is described in Section~\ref{sec:methods}. Event reconstruction and data selection are presented in Section~\ref{sec:trackvertexreco}. A qualitative overview of the comparison of data to MC simulation is discussed in Section~\ref{sec:results_qualitative}. Analysis methods and systematic uncertainties of each of the individual measurements are described in Section~\ref{sec:analysis}. The results of the measurements are presented and discussed in Section~\ref{sec:results}. Finally, conclusions of this work are presented in Section~\ref{sec:conclusion}.

\section{ATLAS detector}
\label{sec:detector}

The ATLAS detector at the LHC~\cite{PERF-2007-01} covers nearly the entire solid angle around the collision point. It consists of an inner detector (ID) surrounded by a thin superconducting solenoid, electromagnetic and hadronic calorimeters and a muon spectrometer incorporating three large superconducting toroidal magnet systems. Only the inner detector and trigger system are used for the measurements presented in this paper.

\begin{figure}[t!]
\begin{center}
\includegraphics[width=1.0\textwidth]{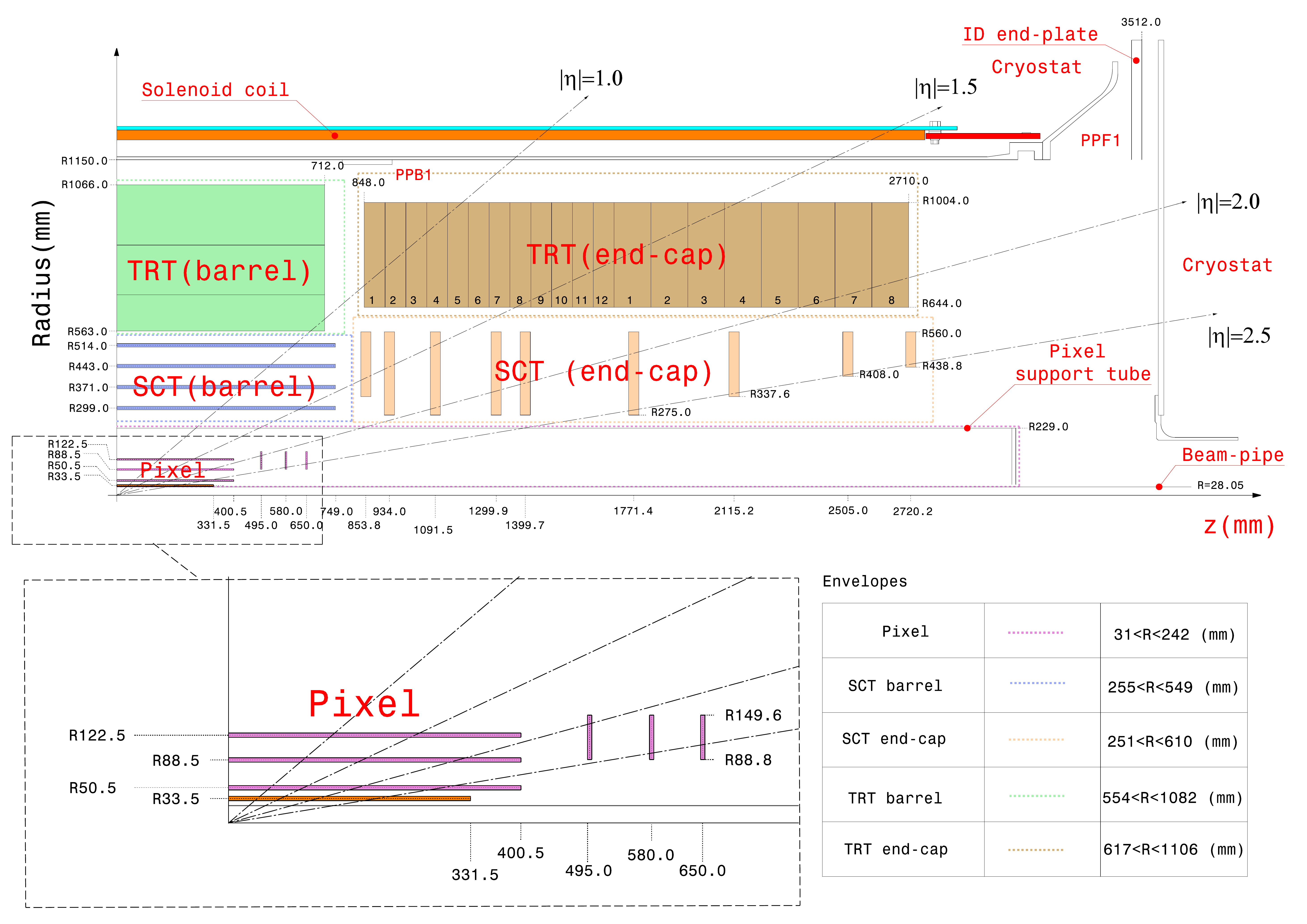}
\caption{The $r$--$z$ cross-section view of the layout of a quadrant of the ATLAS inner detector for Run~2. The top panel shows the whole inner detector, whereas the bottom-left panel shows a magnified view of the pixel detector region. Compared to Run~1, the IBL (shown in red in the bottom-left panel) and its services, together with the new beam pipe, were added. }
\label{fig:NewID}
\end{center}
\end{figure}

The pixel detector (including the IBL) spans the radial region (measured from the interaction point) of 33--150 mm, while the SCT and TRT detectors span the radial regions 299--560 mm and 563--1066 mm, respectively. The ID is designed such that its material content has a minimal effect on the particles traversing its volume. Figure~\ref{fig:NewID} shows the layout of the ID in Run~2.

The innermost pixel layer, the IBL, consists of 14 staves which cover the region $|\eta|< 3.03$ with over 12 million silicon pixels with a typical size of $50~\micrometer~(r$--$\phi)\times250~\micrometer~(z)$ each~\cite{Capeans:1291633}. The addition of the IBL improves the track reconstruction performance; for example, both the transverse and longitudinal impact parameter resolution improve by more than 40\% in the best case of tracks with transverse momentum ($\pt$) around $0.5~{\GeV}$~\cite{ATL-PHYS-PUB-2015-018}. Here, the transverse impact parameter, $d_{0}$, is defined as the shortest distance between a track and the beam line in the transverse plane. The longitudinal impact parameter, $z_{0}$, is defined as the distance in $z$ between the primary vertex and the point on the track used to evaluate $d_{0}$.  The average amount of material introduced by the IBL staves corresponds to approximately $N_{X_{0}}=1.5\%$, for particles produced perpendicular to the beam line, originating from $r = 0$. The IBL staves are placed between the inner positioning tube (IPT) at $r=29.0~\millimeter$ and the inner support tube (IST) at $r=42.5~\millimeter$. The IPT and IST are made from carbon fibre and resin. The thickness of the IPT varies from 0.325~mm at $|z|<311$ mm, to 0.455~mm at the outermost edge.

To minimise the distance of the IBL from the beam line, a new, thinner beam pipe was installed. The new beam pipe mainly consists of a 0.8-mm-thick beryllium pipe with an inner radius of 23.5 mm and an outer radius ranging from 24.3 mm ($|z|<30$~mm) to 28.2~mm ($|z|>311$~mm), wrapped with polyimide tapes and aerogel thermal insulators. There is no thermal insulator in the central part of the new beam pipe at $|z|<311$ mm, in order to reduce material thickness. The material composition of the new beam pipe was measured using X-rays as well as by mass measurements to a precision of 1\% before installation.

The pre-existing pixel detector consists of three barrel layers (referred to as PIX1, PIX2, PIX3 from inner to outer) and two end-caps with three disks each. It hosts 1744 pixel-sensor modules, and each module contains 46\,080 pixels with a typical size of $50~\micrometer~(r$--$\phi)\times400~\micrometer~(z)$ each. The detector contains over 80 million pixels in total. The radii of the three barrel layers are 50.5~mm, 88.5~mm and 122.5~mm, respectively. The barrel and end-cap layers of the pixel detector are supported by an octagonal prism structure referred to as the pixel support frame (PSF) with a radius of $r\simeq 200~\millimeter$. It is inserted inside the pixel support tube (PST), which has a radius of 229 mm. During the LHC shutdown period in 2013--2014, the optical--electrical signal conversion boards, which were previously placed on the old service panels at $(r,~z)\simeq (174,~1070)~\millimeter$, were moved to a region referred to as the ID end-plate located outside the ID acceptance and in front of the end-cap calorimeters. This change reduced the amount of material within the pixel service panels.

The SCT consists of 4088 silicon micro-strip modules, arranged in four barrel layers (referred to as SCT1, SCT2, SCT3, SCT4 from inner to outer) and two end-caps with nine wheels each. Each module is composed of two layers of silicon micro-strip detector sensors glued back to back with a relative stereo angle of $40\,\mathrm{mrad}$. The SCT barrel layers are enclosed by the inner and outer thermal enclosures, referred to as the SCT-ITE and SCT-OTE respectively, which are located at $r\simeq 255~\millimeter$ and $r\simeq 550~\millimeter$. The TRT is the outermost of the ID sub-detectors and consists of more than 350~000 gas-filled straw tubes. The structures of the SCT and TRT are unchanged since Run~1.

The ATLAS trigger system consists of a level-1 hardware stage and a high-level trigger software stage~\cite{Aaboud:2016leb}. The level-1 decision used in the measurements presented in this paper are provided by the minimum-bias trigger scintillators (MBTS), which were replaced between Run~1 and Run~2. The MBTS are mounted at each end of the detector in front of the liquid-argon end-cap calorimeter cryostats at $z = \pm3.56\,\mathrm{m}$ and segmented into two rings in pseudorapidity ($2.07 < |\eta| < 2.76$ and $2.76 < |\eta| < 3.86$). The inner ring is segmented into eight azimuthal sectors while the outer ring is segmented into four azimuthal sectors, giving a total of twelve sectors per side.

\section{Data and simulation samples}
\label{sec:datasim}

\subsection{Data sample}
\label{sec:data}

The $pp$ collision data sample used to perform the measurements described in this paper was collected in June 2015 at a centre-of-mass energy of $\rts=13~\TeV$ by the ATLAS detector at the LHC. During this running period, the LHC was operating in a special configuration with a low instantaneous luminosity. The average number of collisions per bunch crossing was approximately 0.005. The data were collected with triggers which required one or more counters above threshold on either side of the MBTS detectors. Events are retained for analysis if they were collected under stable LHC beam conditions and the detector components were operating normally. Approximately 130 million events passing the trigger condition are used in this study, corresponding to an integrated luminosity of around $2.0\,\inb$.

\subsection{Monte Carlo simulation}
\label{sec:mc}

The \textsc{Pythia~8}~\cite{Sjostrand:2007gs} (version 8.185), and \textsc{Epos}~\cite{Porteboeuf:2010um} (version LHCv3400) MC event generators are used to simulate minimum-bias inelastic $pp$ collisions. The \textsc{Pythia~8} model of inclusive $pp$ interactions splits the total inelastic cross-section into non-diffractive (ND) processes, dominated by $t$-channel gluon exchange, and diffractive processes involving a colour-singlet exchange. The \textsc{A2} set of tuned parameters for \textsc{Pythia~8}~\cite{ATL-PHYS-PUB-2011-014} was used in conjunction with the \textsc{MSTW2008lo} parton distribution functions~\cite{Martin:2009iq}. These samples were produced for the ND component, since there is little contribution from the diffractive components after full selection. \textsc{Epos} models inclusive $pp$ interactions with a parton-based Gribov--Regge~\cite{Drescher:2000ha} theory, which is an effective field theory inspired by quantum chromodynamics describing hard and soft scattering simultaneously. The \textsc{LHC} set of tuned parameters~\cite{Pierog:2013ria} of the \textsc{Epos} MC event generator was used. Both \textsc{Pythia~8} and \textsc{Epos}, tuned and set up as described above, are found to provide reasonable descriptions of the charged-particle multiplicity distributions measured in $pp$ collisions at $\rts = 13~\TeV$~\cite{STDM-2015-02}.

The modelling of the interactions of particles with material in the \textsc{Geant4} simulation~\cite{Agostinelli:2002hh}, is referred to as a \emph{physics list}. The analysis presented in this paper uses the \texttt{FTFP\_BERT} physics list. For hadronic interactions, this model employs the Fritiof model~\cite{FTF1,FTF2} for particles of kinetic energy larger than $4~\GeV$, and the Bertini-style cascade for hadrons below $5~\GeV$~\cite{BERT}.
In the energy region where these two models overlap, one model is randomly selected to simulate a given interaction, with a probability weight which changes linearly as a function of kinetic energy.

\subsection{Simulated descriptions of the inner detector }
\label{sec:geo}

Simulated $pp$ collision events generated by \textsc{Pythia~8} and \textsc{Epos} were processed through the ATLAS detector simulation~\cite{Aad:2010ah}, based on \textsc{Geant4}, and are reconstructed by the same software as used to process the data. The ATLAS detector is described within \textsc{Geant4} by a collection of geometry models, each describing the sub-detectors that constitute the full detector. The geometry model for the inner detector describes both the active elements of the detectors (e.g. the silicon pixel sensors) and the passive material (e.g. support structures and cables). The measurements presented in this paper make use of several alternative ID geometry models, summarised below:

\begin{description}
\item \emph{Original} - This ID geometry model represents the nominal geometry model used to generate MC simulation samples produced in 2015. The studies presented in this paper identified a number of missing components in the simulated description of the IBL.
\item \emph{Updated} - A modified version of the \emph{original} geometry model which was created for this study. In this model, several additional components are added to the simulated description of the IBL reflecting the observations which are described in Section~\ref{sec:results_qualitative}. These additional components include flex buses and a number of surface-mounted devices on the front-end of the modules. Small modifications to the positioning of each IBL stave and the material densities of the IBL support structures were also made.
\end{description}

\begin{figure}[htbp]
\begin{center}
\subfigure[$r < 600$ \millimeter]{
  \label{fig:GeoModelMap:matmap_unzoom}
  \includegraphics[width=1.0\textwidth]{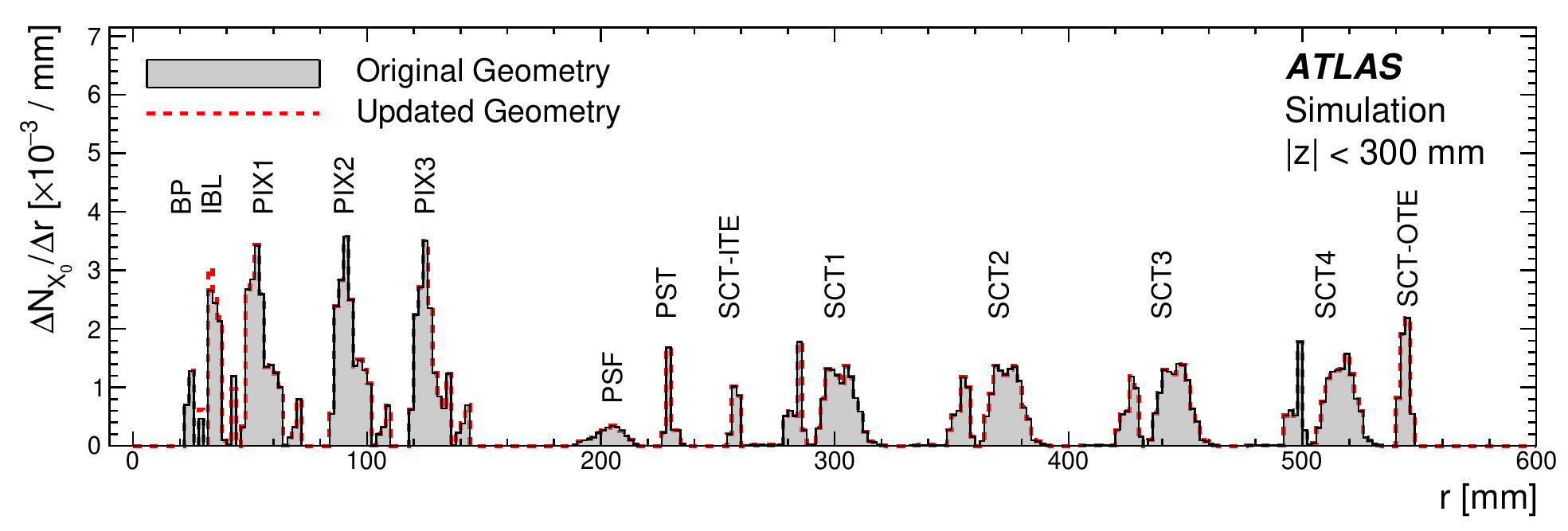}
}
\subfigure[$20~\millimeter< r < 75$ \millimeter]{
  \label{fig:GeoModelMap:matmap_zoom}
  \includegraphics[width=1.0\textwidth]{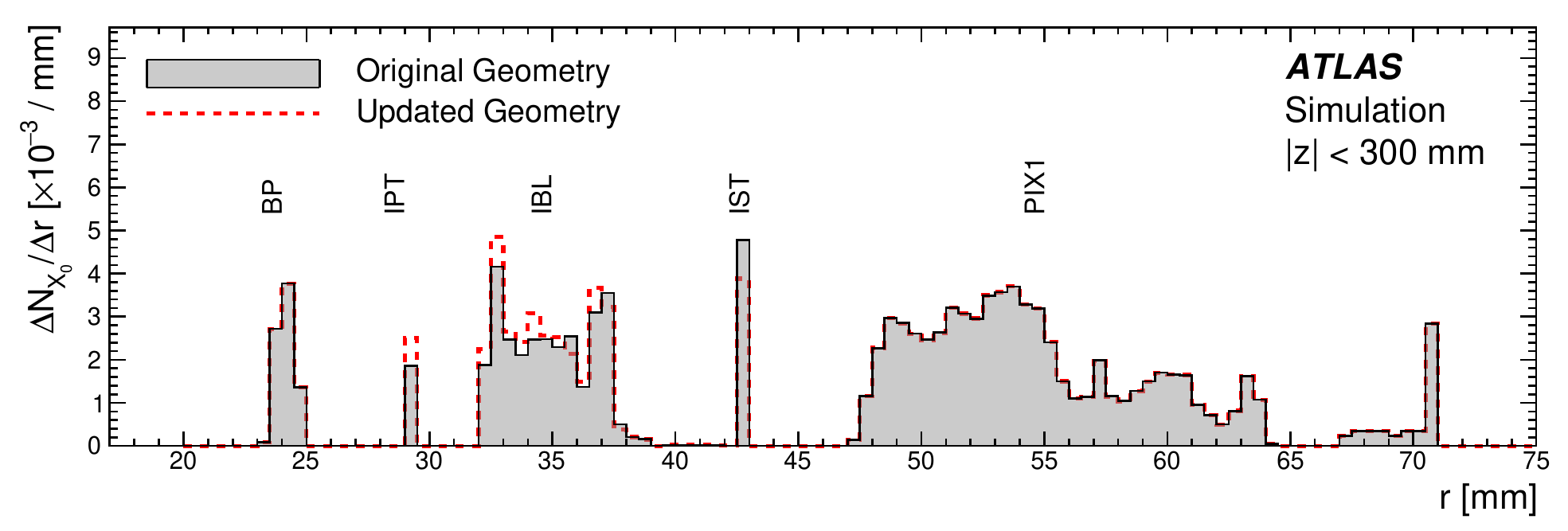}
}
\caption{The differential number of radiation lengths as a function of the radius, $\varDelta N_{X_{0}}/\varDelta r$, averaged over $|z|<300$~\millimeter \subref{fig:GeoModelMap:matmap_unzoom} for $r < 600$ \millimeter and \subref{fig:GeoModelMap:matmap_zoom} for $20~\millimeter< r < 75$ \millimeter for the \emph{original} geometry and the \emph{updated} geometry. The simulated material is sampled for each $z$-position along a straight radial path (perpendicular to the beam line).}
\label{fig:GeoModelMap:radial}
\end{center}
\end{figure}

\begin{figure}[htbp]
\begin{center}
\subfigure[Beam Pipe]{
  \label{fig:GeoModelMap:matmap_z_bp}
  \includegraphics[width=0.47\textwidth]{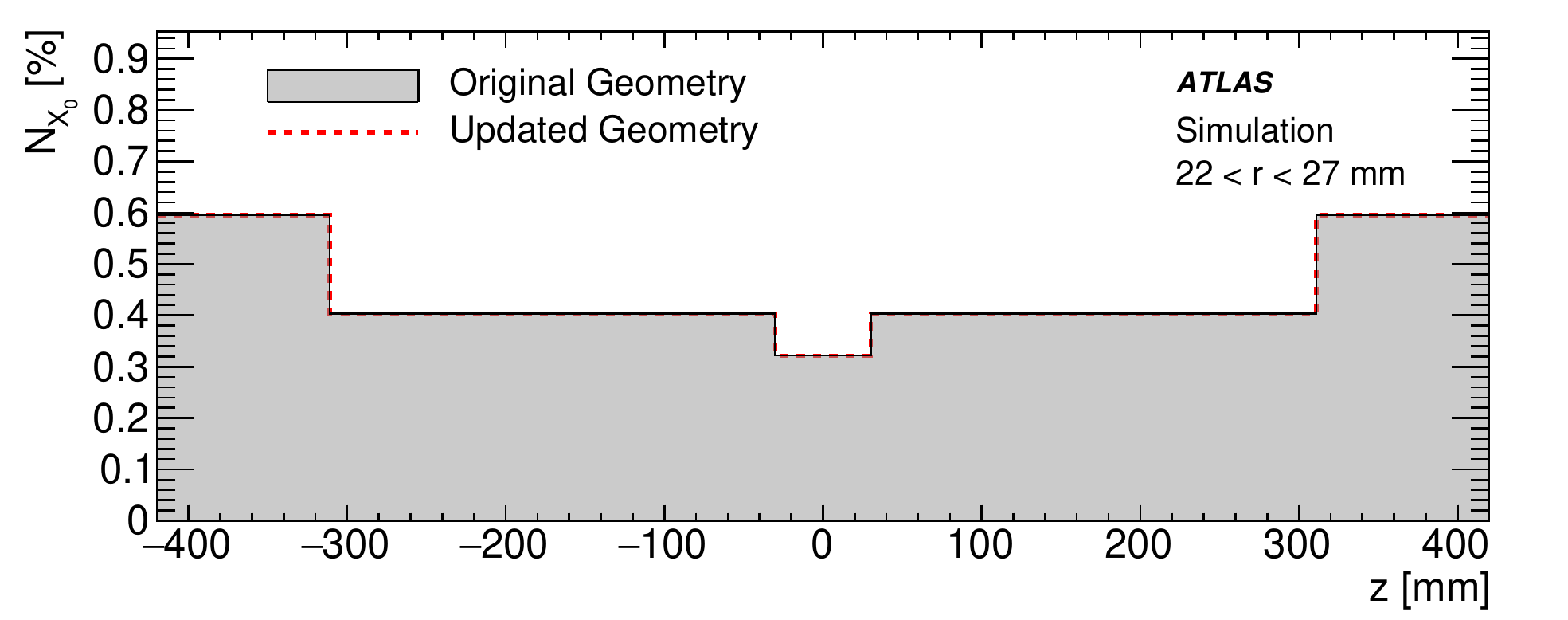}
}
\subfigure[IPT]{
  \label{fig:GeoModelMap:matmap_z_ipt}
  \includegraphics[width=0.47\textwidth]{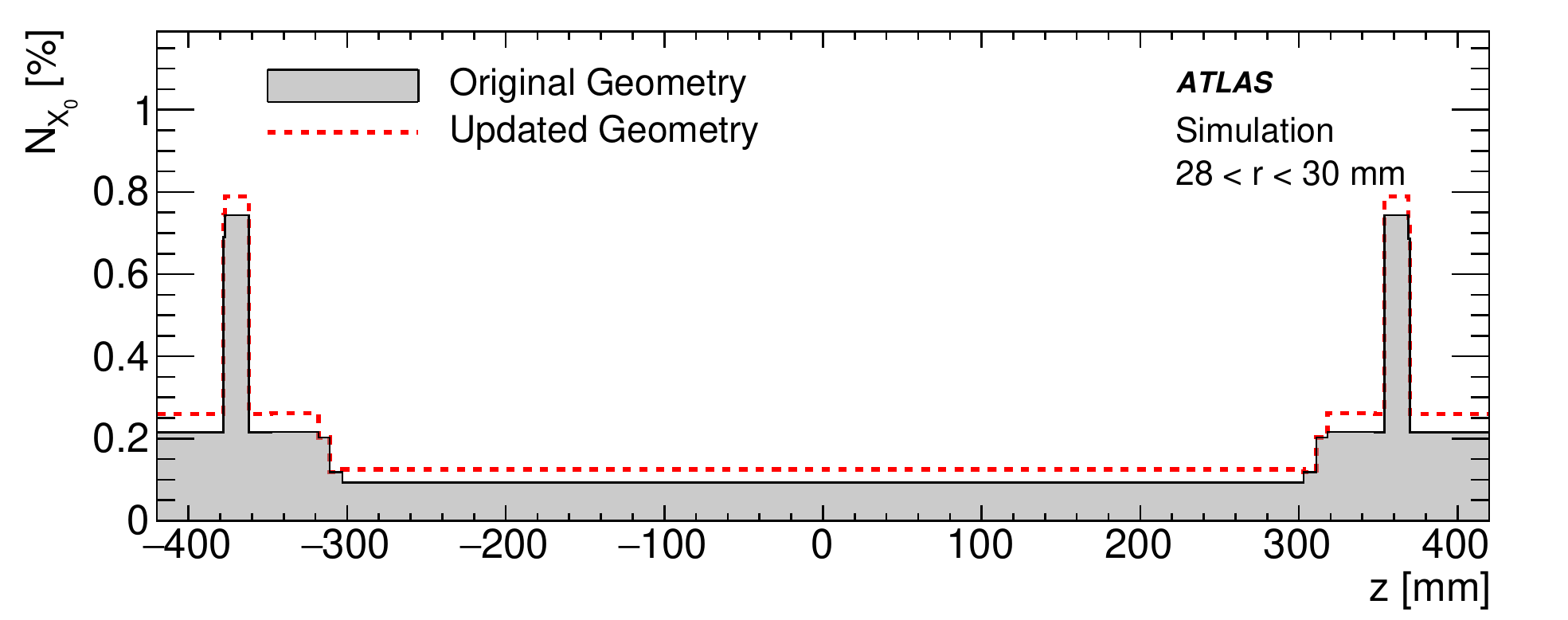}
}
\subfigure[IBL]{
  \label{fig:GeoModelMap:matmap_z_ipt}
  \includegraphics[width=0.47\textwidth]{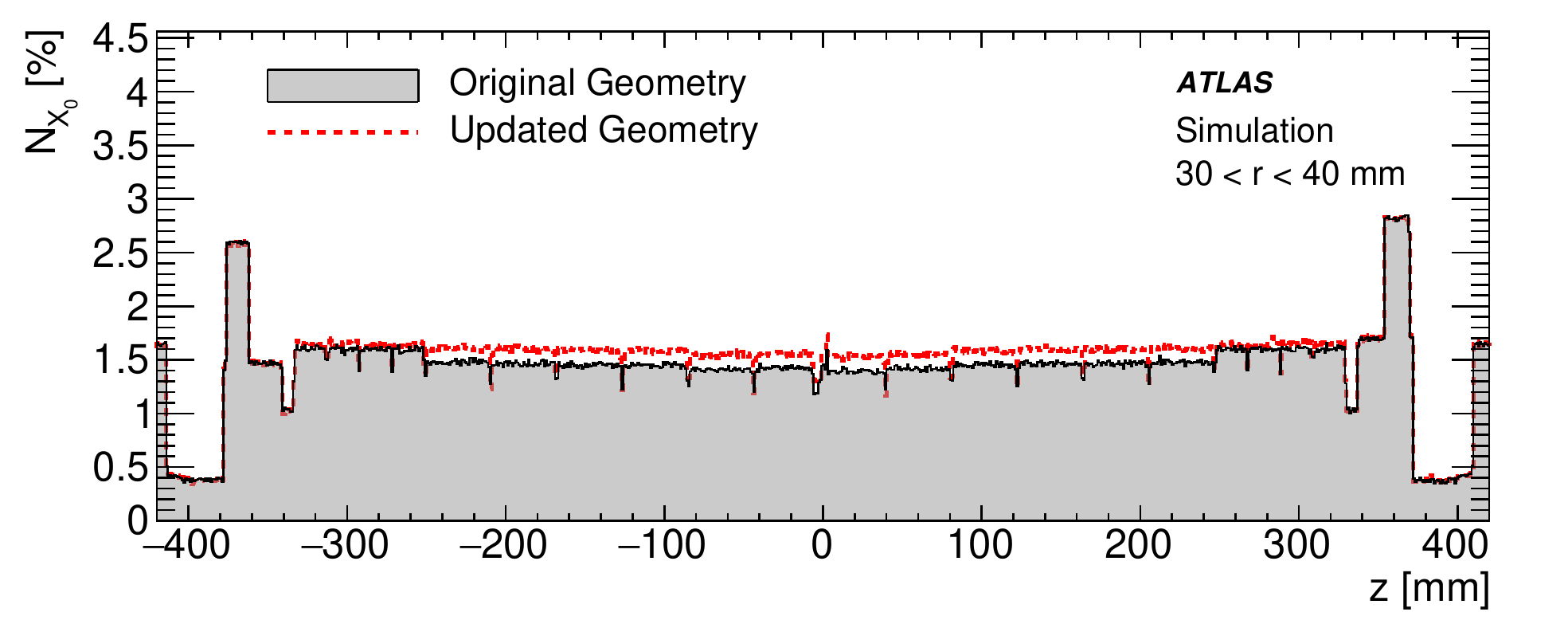}
}
\subfigure[IST]{
  \label{fig:GeoModelMap:matmap_z_ist}
  \includegraphics[width=0.47\textwidth]{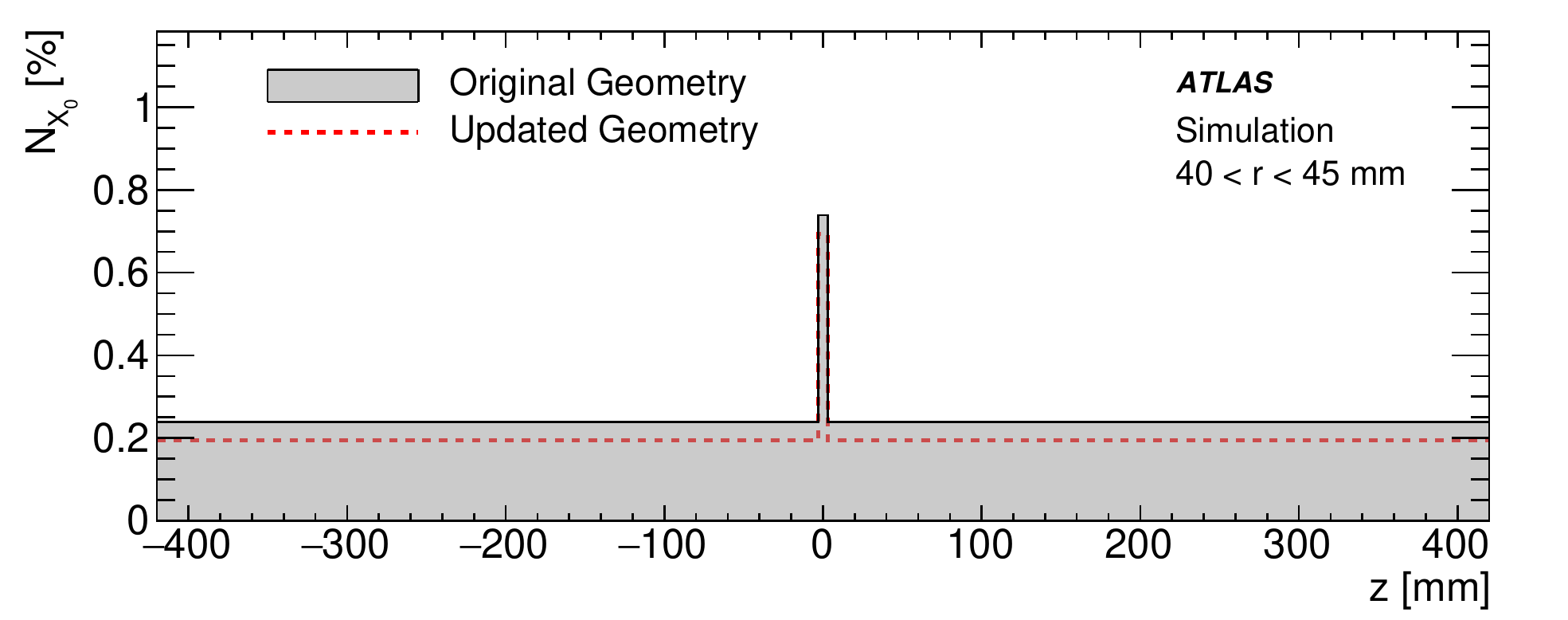}
}
\caption{Number of radiation lengths as a function of the $z$-coordinate for different radial sections for the \emph{original} geometry and the \emph{updated} geometry.}
\label{fig:GeoModelMap:z}
\end{center}
\end{figure}

\begin{figure}[tbp]
\begin{center}
\includegraphics[width=1.0\textwidth]{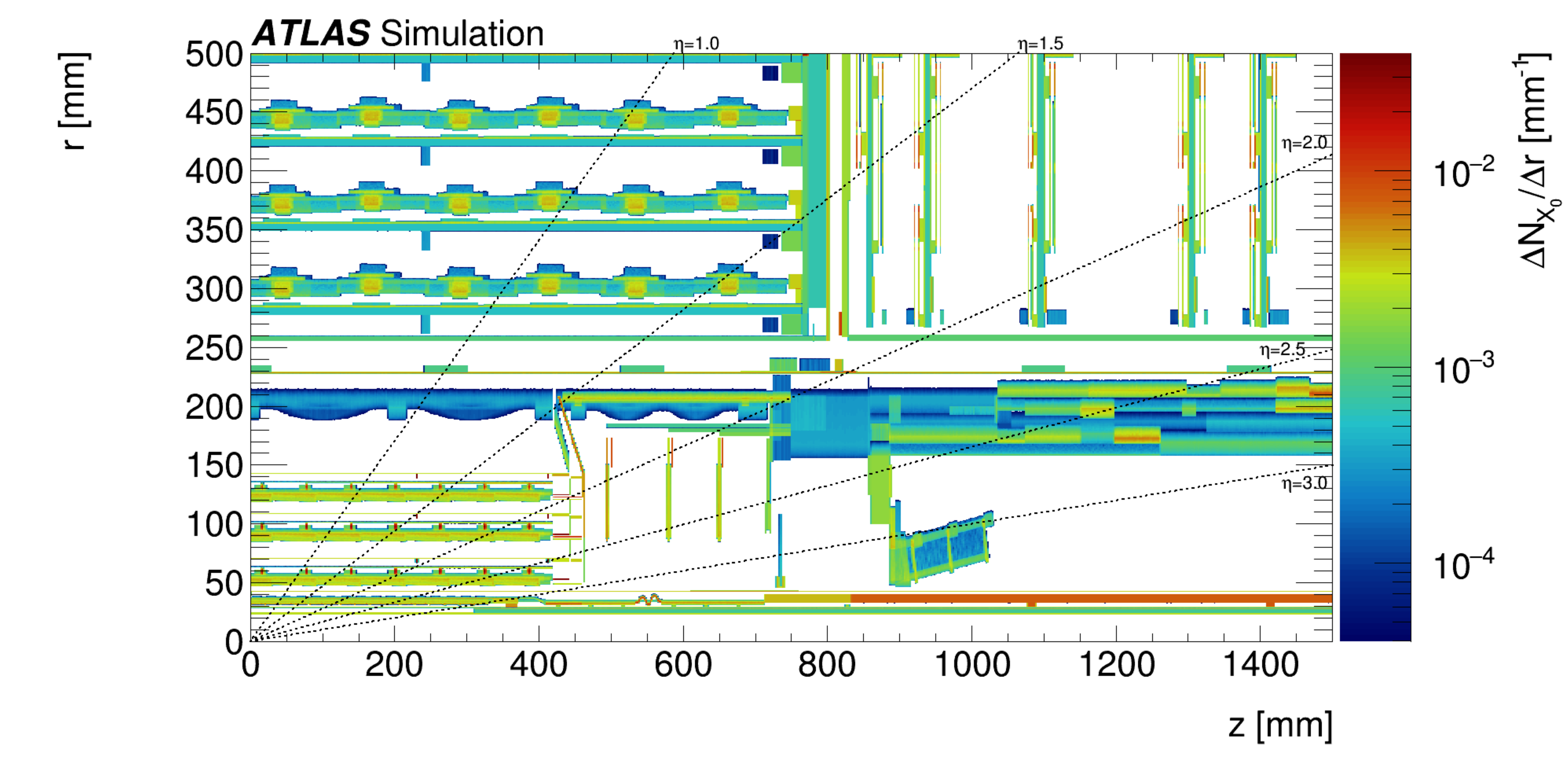}
\caption{The $r$--$z$ distribution of the differential number of radiation lengths, $\varDelta N_{X_{0}}/\varDelta r$, for the \emph{updated} geometry model of a quadrant of the inner detector barrel region of the pixel detector and the SCT. The simulated material is sampled for each $z$-position along a straight radial path (perpendicular to the beam line).
}
\label{fig:rzmap}
\end{center}

\begin{center}
\includegraphics[width=0.7\textwidth]{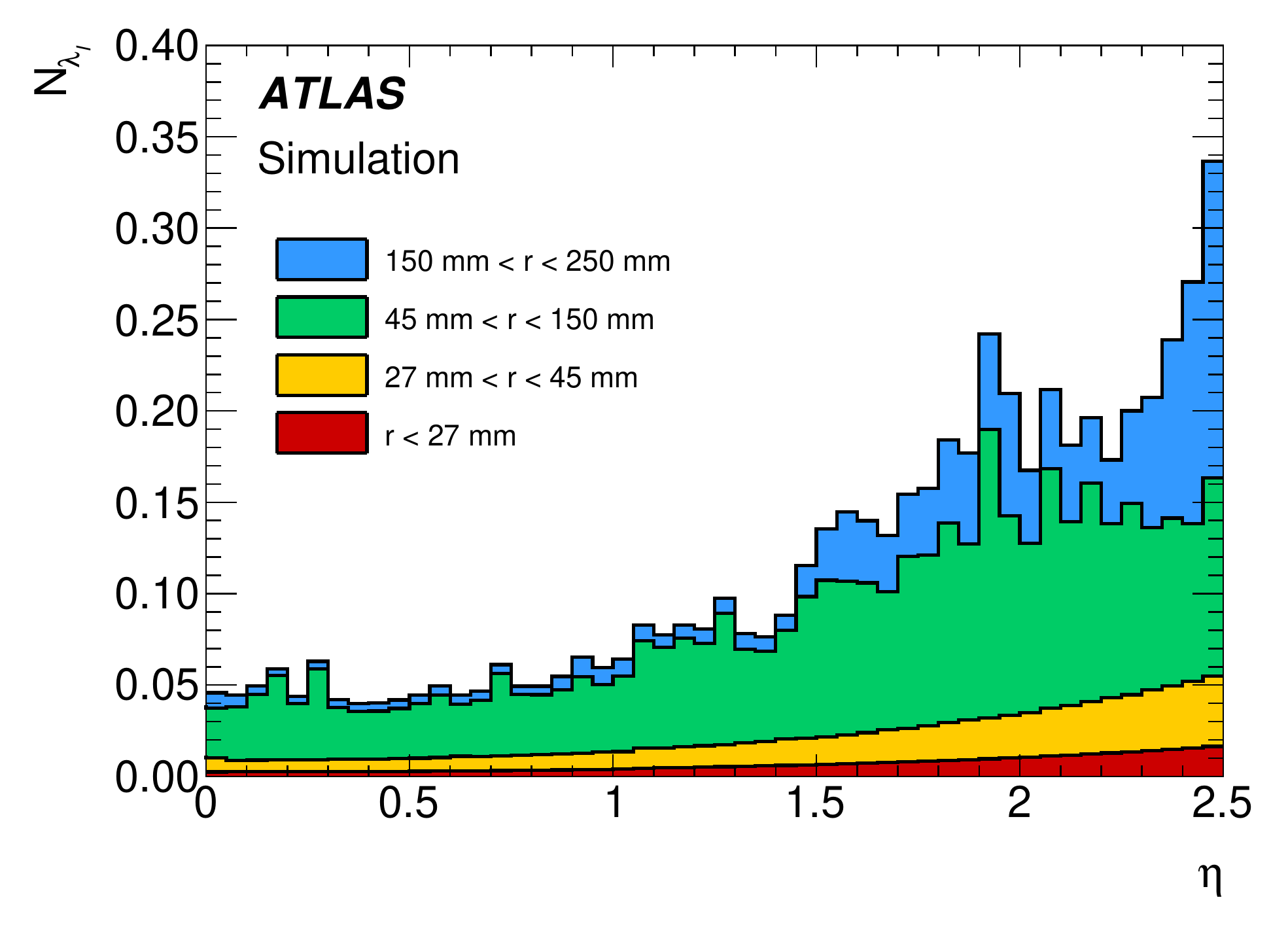}
\caption{The amount of material associated with nuclear interactions, $N_{\lambda_{I}} = \int\!{\mathrm d}s\,\lambda_{I}^{-1}$, averaged over $\phi$,
as a function of $\eta$ in the positive $\eta$ range integrated up to $r=250~\millimeter$ for the \emph{updated} geometry model.
The simulated material is sampled from $z=0$ along a straight path with fixed $\phi$.
The material within the inner detector is shown separately for the regions $r<27~\millimeter$, $27~\millimeter<r<45~\millimeter$, $45~\millimeter<r<150~\millimeter$ and $150~\millimeter<r<250~\millimeter$,
corresponding approximately to the beam pipe, IBL, pixel barrel and pixel service region, respectively. The statistical uncertainty in each bin is negligible.}
\label{figures:matmap_etascan}
\end{center}
\end{figure}

Figures~\ref{fig:GeoModelMap:radial} and \ref{fig:GeoModelMap:z} show respectively the radial and $z$-distributions of the number of radiation lengths for both the \emph{original} and \emph{updated} geometry models. For the \emph{updated} geometry, Figure~\ref{fig:rzmap} shows the distribution of the number of radiation lengths in the $r$--$z$ view, and Figure \ref{figures:matmap_etascan} shows $N_{\lambda_{I}}$ as a function of $\eta$.

Based on the \emph{original} and \emph{updated} geometry models, collections of \emph{distorted} geometry models were created, in which the density of a variety of components is artificially scaled by a known amount.
Furthermore, three modified geometry models were created in which a ring of passive material was added to the
\emph{original} geometry model. The rings were positioned at different $r$--$z$ coordinates and orientations in the region between the
pixel and SCT detectors. This was done in order to test the sensitivity of the track-extension efficiency method to the material location.
These distorted models are used to calibrate the material measurement methods and assess the systematic uncertainties associated with the measurements.
Table~\ref{tbl:mc_lists} summarises the collection of MC samples used in this paper. \textsc{Pythia~8} is used as the nominal event generator for all of the studies except for the hadronic interaction study, which uses \textsc{Epos} as the nominal event generator since it is found to provide a better description of events with decays in flight than \textsc{Pythia~8}.

\begin{table}[t!]
\caption{List of MC samples used in the analyses, with the base geometry model, presence of an additional distortion, the event generator used and the number of generated events.}
\small
\centering
\begin{tabular}{lllc}
\hline
\hline
Base geometry & Distortion & Event generator & Number of generated events\\
\hline
\multirow{6}{*}{\emph{updated}} & nominal     & \textsc{Pythia~8} (A2) ND & $2\times10^{7}$ \\
& nominal & \textsc{Epos} (LHC)           & $2\times10^{7}$\\
& IBL $+10\%$ & \textsc{Pythia~8} (A2) ND & $2\times10^{7}$\\
& IBL $-10\%$ & \textsc{Pythia~8} (A2) ND & $2\times10^{7}$\\
& IST $+10\%$ & \textsc{Pythia~8} (A2) ND & $2\times10^{7}$\\
& IST $-10\%$ & \textsc{Pythia~8} (A2) ND & $2\times10^{7}$\\
\hline
\multirow{6}{*}{\emph{original}} & nominal & \textsc{Pythia~8} (A2) ND  & $2\times10^{7}$\\
&  nominal &\textsc{Epos} (LHC)       & $2\times10^{7}$\\
& pixel service $+50\%$ & \textsc{Pythia~8} (A2) ND & $5\times10^{6}$\\
& pixel service $+50\%$ & \textsc{Epos} (LHC) & $5\times10^{6}$      \\
& ring layout 1 & \textsc{Pythia~8} (A2) ND & $5\times10^{6}$\\
& ring layout 2 & \textsc{Pythia~8} (A2) ND & $5\times10^{6}$\\
& ring layout 3 & \textsc{Pythia~8} (A2) ND & $5\times10^{6}$\\
\hline
\hline
\end{tabular}
\label{tbl:mc_lists}
\end{table}%

\section{Overview of analysis methods}
\label{sec:methods}

In this paper, the data are compared to the MC simulations which use the ATLAS Run~2 geometry models for various observables. In this section, the methods are described. Further details of the analysis techniques are provided in Section~\ref{sec:analysis}.

\subsection{Reconstruction of hadronic interaction and photon conversion vertices}
\label{sec:hadint}
The hadronic interaction and photon conversion analyses aim to identify and reconstruct the interaction vertices of hadronic interactions and photon conversions to probe the accuracy of the ID material content within the detector simulation. Vertices corresponding to $pp$ interaction positions are referred to as \emph{primary vertices}, while other vertices corresponding to in-flight decays, photon conversions and hadronic interactions are collectively referred to as \emph{secondary vertices}. In this paper, only secondary vertices with a distance from the beam line of more than 10~mm are considered. The properties of hadronic-interaction
and photon-conversion candidates are compared between the data and the MC simulation.

Photon conversions are well-understood electromagnetic processes which exhibit a high reconstruction purity. The vertex radial position resolution is around 2 mm, limited by the collinearity of the electron--positron pair. In contrast, hadronic interactions are complex phenomena which are difficult to model in simulation. Their reconstruction suffers from backgrounds associated with hadron decays and combinatoric fake vertices. However, the radial position resolution is far better than for photon conversions. Resolutions of ${\cal O}(0.1)~\millimeter$ can be achieved due to large opening angles between the daughter particles. This facilitates a detailed radiography of the material, including minute components, e.g.~the capacitors mounted on the surfaces of the pixel modules, allowing their location to be determined precisely.

Qualitative comparisons of the distributions of reconstructed photon conversion and hadronic interaction vertices in data and simulation samples can identify absent or inaccurately positioned components within the ID geometry model.
Such comparisons are effectively able to probe the central barrel region of $|z|<400~\millimeter$ in the radial range from the beam pipe up to the first layer of the SCT at $r\simeq300~\millimeter$, and are suitable for probing the barrel structures including the IBL and the new beam pipe. Further details can be found in Section~\ref{sec:analysis}.

\subsection{Track-extension efficiency}
\label{sec:sctext}

In an attempt to have as few particle interactions with material as possible, most of the pixel services, which reside between the pixel and SCT detectors, are located together in the forward region (approximately $1.0<|\eta|<2.5$). This region of the inner detector is more challenging to model within the simulation than the central barrel region, due to the complexity of the structure and the amount of material, as shown in Figures~\ref{fig:rzmap} and \ref{figures:matmap_etascan}.

The pixel service region is also expected to exhibit a relatively high rate of hadronic interactions due to the high density of material and the longer path length of the trajectories of hadrons produced at high pseudorapidity. If a charged hadron undergoes a hadronic interaction while traversing the region between the pixel and SCT detectors, it will typically only leave signals in the pixel detector. A track associated with the particle's trajectory can be reconstructed from pixel detector hits alone (referred to as a \emph{tracklet} hereafter) or from hits in all ID sub-detectors (referred to as a \emph{combined track} hereafter). Hits in the SCT and TRT detectors associated with the secondary particles produced in the hadronic interaction are unlikely to be associated to the tracklet. The rate of hadronic interactions can be related to the so-called \emph{track-extension efficiency}, denoted by \SctExtEff and defined as:

\begin{equation*}
\SctExtEff \equiv \frac{n_{\mathrm{tracklet}}^{\mathrm{matched}}}{n_{\mathrm{tracklet}}}\,,
\end{equation*}

where $n_{\mathrm{tracklet}}$ is the number of tracklets satisfying a given set of selection criteria and $n_{\mathrm{tracklet}}^{\mathrm{matched}}$ is the number of those tracklets that are matched to a combined track.
The efficiency \SctExtEff is related to the amount of material crossed by a particle and is therefore dependent on the kinematics and origin of the particle. For particles with sufficiently high $\pt$ , when averaging over $\phi$ and restricting the $z$-position of the primary vertex ($z_{\mathrm{vtx}}$) to a sufficiently narrow range, the particle trajectory $C$ can be approximately described as a function of $\eta$ alone.
\subsection{Track impact parameter resolution}
\label{sec:trackd0}

The resolution of the track transverse impact parameter $d_{0}$, denoted $\sigma_{d_{0}}$, depends on the track momentum. The resolution in the high-$\pt$ limit is largely determined by the intrinsic position resolution of detector sensors and the accuracy of the alignment of each detector component. At sufficiently low $\pt$, the effects of multiple scattering dominate the resolution. At low $\pt$, $\sigma_{d_{0}}$ is sensitive to the amount of the material in the ID, particularly that closest to the collision point. By measuring $\sigma_{d_{0}}$, it is possible to cross-check the accuracy of the geometry model using an independent observable which does not rely on vertex reconstruction and is insensitive to radial positions. Since this method is used as a cross-check of the validity of the vertex-based methods, results are presented without a full assessment of systematic uncertainties.

\section{Reconstruction and data selection}
\label{sec:trackvertexreco}

\subsection{Track reconstruction}
Charged-particle tracks are reconstructed with a special configuration (for low-luminosity running) of the ATLAS track reconstruction algorithm~\cite{ATL-SOFT-PUB-2007-005} optimised for Run~2~\cite{Salzburger:2015} down to $\pt$ of $100~\MeV$. The \emph{inside-out} tracking refers to the track reconstruction algorithm seeded from pixel and SCT hits and extended to the TRT. In this algorithm, candidates are rejected from the set of reconstructed tracks if the absolute value of the transverse (longitudinal) impact parameter, $d_{0}$ ($z_{0}$) is greater than 10 \millimeter (250 \millimeter). Tracks originating from in-flight decay vertices (e.g.~\Kshort decays) inside the inner detector's volume or from photon conversions may not have a sufficient number of pixel and SCT hits to satisfy the inside-out track finding. A second tracking algorithm, referred to as the \emph{outside-in} approach, complements this by finding track seeds in the TRT and extending them back to match hits in the pixels and SCT which are not already associated with tracks reconstructed with the inside-out approach. No $d_{0}$ and $z_{0}$ requirements are applied for the outside-in tracking. Tracklets are reconstructed using the inside-out approach down to $\pt$ of $50~\MeV$. Not all tracks reconstructed in the ID correspond directly to a charged particle traversing the detector. Coincidental arrangements of unrelated hits can give rise to so-called \emph{fake} tracks. Fake tracks (and similarly fake tracklets) are identified in the MC simulation as those with a small fraction of hits originating
from a single simulated charged particle \cite{ATL-PHYS-PUB-2015-051}.

\subsection{Vertex reconstruction}
\label{sec:vertex_reconstruction}

The \emph{primary} vertices, i.e.~positions of the inelastic proton-proton interactions, are reconstructed by the \emph{iterative vertex finding} algorithm~\cite{PERF-2015-01}. At least two charged-particle tracks with $\pt$ greater than $100~\MeV$~are required to form a primary vertex. The \emph{hard-scatter} primary vertex is defined as the primary vertex with the highest sum of the $\pt^{\,2}$ of the associated tracks. Other primary vertices are referred to as \emph{pile-up} vertices.

Secondary vertices are reconstructed by the \emph{inclusive secondary-vertex finding} algorithm, which is designed to find vertices from a predefined set of input tracks within an event~\cite{PERF-2011-08}.

The configuration of the secondary-vertex reconstruction differs between the hadronic-interaction and photon-conversion analyses, reflecting the different topologies associated with the interactions. These differences are summarised in Table~\ref{tbl:selection_summary}. For hadronic interactions, tracks are required to have at least one SCT hit, $|d_{0}|> 5~\millimeter$, a track-fitting $\chi^{2}$ divided by the number of degrees of freedom ($N_{\mathrm{dof}}$) less than 5, and satisfy certain quality criteria. The requirement on $d_{0}$ is imposed in order to efficiently reject combinatorial fake vertices. The fitted vertex must have a vertex-fitting $\chi^{2}/N_{\mathrm{dof}}$ of less than 10. In addition, a geometrical compatibility criterion is applied. The tracks associated with the reconstructed vertex are required to have no hits in any detector layer inside the \emph{vertex radius}, defined as the distance of the vertex in the $x$--$y$ plane from the origin of the ATLAS coordinate system, and are required to have hits in the closest outer layer beyond the vertex radius.

The reconstruction of photon conversion vertices begins with the identification of pairs of charged-particle tracks, as described in Ref.~\cite{PERF-2013-04}. Collinear track pairs with oppositely signed charges compatible with the photon conversion topology, including the requirement on the minimum distance of approach between the two track helices, and the distance between the first hits on the two tracks, are selected. The vertices of the track pairs are fitted while constraining the opening angle between the two tracks (in both the transverse and longitudinal planes) at the vertex to be zero. Finally, only vertices built from two tracks, both of which are associated with silicon detector hits, are retained for further analysis.

In MC simulation, the truth matching of the secondary vertices is defined as follows, using ``truth'' information from the generator-level event record. If the vertex has any fake tracks, it is classified as a \emph{fake} vertex. If all of the truth particles linked from the reconstructed tracks originate from a single truth vertex, the reconstructed vertex is classified as \emph {truth-matched}. Otherwise the vertex is classified as fake. In the case of a truth-matched vertex, it is further classified into \emph {in-flight decay}, \emph {photon conversion} or \emph {inelastic} vertex. A truth vertex is labelled as a photon conversion if the truth particle identifier of the parent particle of the vertex is a photon. A truth vertex is labelled as an in-flight decay if the difference between the energy sum of outgoing truth particles and incoming particle is less than 100~\MeV. If the incoming particle is a hadron and it is not labelled as an in-flight decay, then such a truth vertex is considered to be an inelastic interaction, i.e.~a hadronic interaction vertex.

\subsection{Data selection}

\begin{table}[t!]
\caption{Summary of selection criteria for different methods of hadronic interaction vertex reconstruction, photon conversion vertex reconstruction, track-extension efficiency and transverse impact parameter studies.}
\centering
\scriptsize
\renewcommand\arraystretch{1.06}
\begin{tabular}{p{3mm}lp{3mm}l}
\hline
\hline
\multicolumn{4}{l}{Notation:}\\
\multicolumn{4}{l}{$N_{\mathrm{Si}}$: number of hits on the track within the pixel and SCT layers;}\\
\multicolumn{4}{l}{$N_{\mathrm{pix}}$: number of hits on the track within the pixel layers;}\\
\multicolumn{4}{l}{$N_{\mathrm{SCT}}$: number of hits on the track within the SCT layers;}\\
\multicolumn{4}{l}{$N_{\mathrm{Si}}^{\mathrm{sh}}$: number of hits on the track within the pixel and SCT layers that are shared with other tracks;}\\
\multicolumn{4}{l}{$N_{\mathrm{Si}}^{\mathrm{hole}}$: number of sensors crossed by the track within the pixel and SCT detectors where expected hits are missing.}\\
\multicolumn{4}{l}{$N_{\mathrm{pix}}^{\mathrm{hole}}$: number of sensors crossed by the track within the pixel detector where expected hits are missing.}\\
\hline
\hline
\multicolumn{4}{c}{Hadronic Interactions}\\
\hline
\multicolumn{4}{l}{\textbf{Requirements applied to tracks associated with primary vertices: the \emph{loose-primary} requirement}} \\
 & \multicolumn{3}{l}{$\pt>400~\MeV$ and $|\eta|<2.5$;}\\
 & \multicolumn{3}{l}{$N_{\mathrm{Si}}\geq 7$; $N_{\mathrm{Si}}^{\mathrm{sh}}\leq1$; $N_{\mathrm{Si}}^{\mathrm{hole}}\leq 2$; $N_{\mathrm{pix}}^{\mathrm{hole}}\leq1$; either $(N_{\mathrm{Si}}\geq 7~\mathrm{and}~N_{\mathrm{Si}}^{\mathrm{sh}} = 0)$ or $N_{\mathrm{Si}} \geq 10$.}\\
\multicolumn{4}{l}{\textbf{Requirement on primary vertices}}\\
 & \multicolumn{3}{l}{at least five tracks satisfying the loose-primary selection criteria are associated with the primary vertex;}\\
 & \multicolumn{3}{l}{pile-up veto.}\\
\multicolumn{4}{l}{\textbf{Acceptance}}\\
 & \multicolumn{3}{l}{$|d_{0}|>5~\millimeter$ and at least one SCT hit, $\chi^{2}/N_{\mathrm{dof}}<5.0$ for tracks associated with secondary vertices;}\\
 & \multicolumn{3}{l}{hit pattern recognition for combinatorial fake rejection: see Section~\ref{sec:vertex_reconstruction} for details;}\\
 & \multicolumn{3}{l}{primary vertex position $-160~\millimeter < z_{\mathrm{pv}} < 120~\millimeter$;}\\
 & \multicolumn{3}{l}{secondary vertex $|\eta|<2.4$, $|z|<400~\millimeter$ and $r>20~\millimeter$;}\\
 & \multicolumn{3}{l}{number of tracks associated with the secondary vertex is two.}\\
\multicolumn{4}{l}{\textbf{In-flight decay veto}}\\
 & \multicolumn{3}{l}{\Kshort veto: $|m_{\SV}(\pi\pi)-m_{\Kshort}| > 50~\MeV$;}\\
 & \multicolumn{3}{l}{photon conversion veto: $m_{\SV}(ee) > 100~\MeV$;}\\
 & \multicolumn{3}{l}{$\Lambda$ veto: $|m_{\SV}(p\pi)-m_{\Lambda}| > 15~\MeV$.}\\
\multicolumn{4}{l}{\textbf{Fake rejection}}\\
 & \multicolumn{3}{l}{tracks associated with secondary vertex: $\pt>300~\MeV$;}\\
 & \multicolumn{3}{l}{secondary vertex $\chi^{2}/N_{\mathrm{dof}}<4.5$.}\\
\hline
\hline
\multicolumn{4}{c}{Photon Conversions}\\
\hline
\multicolumn{4}{l}{\textbf{Requirement on primary vertices}}\\
 & \multicolumn{3}{l}{at least 15 tracks are associated with the primary vertex;}\\
 & \multicolumn{3}{l}{pile-up veto.}\\
\multicolumn{4}{l}{\textbf{Acceptance}}\\
 & \multicolumn{3}{l}{primary vertex position $-160~\millimeter < z_{\mathrm{pv}} < 120~\millimeter$;}\\
 & \multicolumn{3}{l}{tracks associated with secondary vertex: $\pt>250~\MeV$ and $N_{\mathrm{SCT}}\geq4$;}\\
 & \multicolumn{3}{l}{conversion $\ptgamma>1~\GeV$ and $|\etagamma|<1.5$.}\\
\multicolumn{4}{l}{\textbf{Quality selection criteria}}\\
 & conversion vertex $\chi^{2}/N_{\mathrm{dof}}<1.0$;\\
 & \multicolumn{3}{l}{the photon trajectory must point back to the primary vertex to within 15 \millimeter in the longitudinal plane and within 4.5 \millimeter}\\
 & \multicolumn{3}{l}{in the transverse plane.}\\
\hline
\hline
\multicolumn{4}{c}{Track-Extension Efficiency}\\
\hline
\multicolumn{2}{l}{\textbf{Tracklet reconstruction}}  & \multicolumn{2}{l}{\textbf{Requirement on tracklets}} \\
 & $N_{\mathrm{pix}}^{\mathrm{hole}}\leq1$;      & & $\pt>500~\MeV$ and $|\eta|<2.5$;\\
 & at least three non-shared hits;               & & $|z_{\mathrm{vtx}}|<10~\millimeter$;\\
 & $\pt>50~\MeV$.                                 & & at least four pixel hits: $N_{\mathrm{pix}}\geq4$;\\
\multicolumn{2}{l}{\textbf{Requirement on primary vertices}}& & $|d_{0}|<2~\millimeter$ and $|z_{0}\sin\theta|<2~\millimeter$.\\
 & pile-up veto.                                 & \multicolumn{2}{l}{\textbf{Requirement on combined tracks}}\\
 &                                               & & $\pt>100~\MeV$ and $N_{\mathrm{SCT}}\geq 4$;\\
 &                                               & & at least one shared hit with the matched tracklet.\\
\hline
\hline
\multicolumn{4}{c}{Transverse Impact Parameter Resolution}\\
\hline
\multicolumn{4}{l}{\textbf{Requirement on tracks: the \emph{loose} requirement}}\\
 & $\pt>400~\MeV$ and $|\eta|<0.5$;\\
 & $N_{\mathrm{Si}}\geq 7$; $N_{\mathrm{Si}}^{\mathrm{sh}}\leq1$; $N_{\mathrm{Si}}^{\mathrm{hole}}\leq2$; $N_{\mathrm{pix}}^{\mathrm{hole}}\leq1$.\\
\multicolumn{4}{l}{\textbf{Requirement on primary vertices}}\\
 & at least 10 tracks associated with the primary vertex;\\
 & pile-up veto.\\
\hline
\hline
\end{tabular}

\label{tbl:selection_summary}
\end{table}

\subsubsection{Hadronic interactions}
\label{sec:reco_hadint}
Events are required to have exactly one primary vertex which has at least five tracks with $\pt> 400~\MeV$~and $|\eta|<2.5$~which satisfy the quality selection referred to as the \emph{loose-primary} condition~\cite{ATL-PHYS-PUB-2015-051}. The track selection criteria are summarised in Table~\ref{tbl:selection_summary}. Primary vertices are required to be contained in the approximately $\pm3\sigma$ range of the distribution of $pp$ interactions in $z$, namely $-160~\mathrm{mm}<z<120~\mathrm{mm}$. The centroid of the luminous region is near $z=-20~\mathrm{mm}$. Events which have \emph{pile-up} vertices are rejected. Secondary vertices are required to satisfy $|\eta|<2.4$ where the pseudorapidity is measured with respect to the primary vertex, $|z|<400$~\millimeter and $r>20~\millimeter$. In addition, the number of tracks associated with each secondary vertex is required to be exactly two so that the reconstruction efficiency can be compared to that of $\Kshort\rightarrow\pi^{+}\pi^{-}$~decays, as is discussed in Section~\ref{sec:hadInt_recoeff_corr}. This keeps approximately 90\% of the hadronic interaction vertex candidates.

To reject $\Kshort$, $\lambdaBaryon$ decays and photon conversion vertices, the following requirements are applied for vertices whose associated tracks have oppositely signed charges:
\begin{itemize}
\item~$\Kshort\rightarrow\pi^{+}\pi^{-}$ \emph{veto}: it is required that $|m_\SV(\pi\pi) - m_{\Kshort}| > 50~\MeV$, where $m_{\SV}(\pi\pi)$ is the so-called secondary-vertex invariant mass, calculated using the track parameters at the vertex. The $m_{\SV}(\pi\pi)$ value is calculated assuming pion masses for both tracks, and $m_{\Kshort}$ is the mass of \Kshort ($497.61~\MeV$).
\item~$\lambdaBaryon\rightarrow p\pi$ \emph{veto}: it is required that $\left|m_\SV(p\pi)-m_{\lambdaBaryon}\right| > 15~\MeV$, where $m_{\SV}(p\pi)$ is calculated assuming that the particle with larger $\pt$ is a proton or antiproton, and the other particle is assumed to be a pion, and $m_{\lambdaBaryon}$ is the mass of $\lambdaBaryon$ ($1115.68~\MeV$).
\item~\emph{Photon conversion veto}: the $m_\SV(\electron\electron) > 100~\MeV$ requirement is applied, where $m_\SV(\electron\electron)$ is calculated assuming electron masses for both tracks.
\end{itemize}

To reject combinatorial fake backgrounds, each track in the secondary vertex fit is required to satisfy $\pt > 300~\MeV$, and vertex $\chi^{2}/N_{\mathrm{dof}}$ must be less than 4.5. The secondary vertices satisfying all the above criteria are hereafter referred to as \emph{hadronic interaction candidates}.

\subsubsection{Photon conversions}
\label{sec:reco_conversion}

Events are required to contain exactly one primary vertex, reconstructed within $-160~\millimeter<z<120~\millimeter$, with at least 15 associated tracks. Events which contain any additional \emph{pile-up} vertices are rejected. Conversion vertex candidates reconstructed as described in Section~\ref{sec:vertex_reconstruction} are required to satisfy a number of quality criteria to reduce combinatorial backgrounds, as summarised in Table~\ref{tbl:selection_summary}. Conversion vertex candidates satisfying the following requirements are retained for further analysis:

\begin{itemize}
\item The reconstructed conversion vertex satisfies $\chi^{2}/N_{\mathrm{dof}} < 1$ and $r > 10\,\mathrm{mm}$.
\item Each of the tracks constituting the vertex must have $\pt>250~\MeV$ and at least four SCT hits.
\item The pseudorapidity of the converted photon candidate (calculated from the total momentum of the two tracks) must satisfy $|\etagamma|<1.5$.
\item The transverse momentum of the candidate converted photon must satisfy $\ptgamma > 1~\GeV$.
\item The absolute value of the impact parameter of the back-extrapolated photon trajectory with respect to the primary vertex must be less than $15\,\mathrm{mm}$ in the longitudinal plane and $4.5\,\mathrm{mm}$ in the transverse plane.
\end{itemize}

Conversion vertex candidates which satisfy these requirements are referred to hereafter as \emph{photon conversion candidates}. The region $|\etagamma|<1.5$ offers improved vertex position resolution in the radial direction (compared to the case of $|\etagamma|<2.5$). Furthermore, the simulated material between the pixel detector and SCT within $|\etagamma|<1.5$ (i.e. downstream of the radial region under study) is known to agree well with data, ensuring that the reconstruction efficiency of $\ee$ tracks is well described in the simulation. In simulated events, photon conversion candidates with a truth matched vertex are classified as true conversions while all other candidates are classified as background. The purity of photon conversion candidates is defined as the fraction of true conversions. The purity for vertices reconstructed within $22~\millimeter < r < 35\,\mathrm{mm}$ (the beam-pipe region) is around 80\%, which improves to over 95\% for $r > 35\,\mathrm{mm}$.

\subsubsection{Track-extension efficiency}
\label{sec:reco_sctext}

Tracklets are required to have at least four pixel hits, and to satisfy $\pt> 500~\MeV$ and $|\eta|< 2.5$. The requirement on the number of pixel hits is imposed to suppress the contribution from fake tracklets, and the $\pt$ requirement to suppress the contamination from non-primary charged particles and weakly decaying hadrons. In order to reduce the variation in track trajectories associated with a single value of $\eta$ arising from the variation of $z_{\mathrm{vtx}}$, a requirement of $|z_{\mathrm{vtx}}| < 10~\millimeter$ is imposed.
To further reject non-primary charged particles, the transverse and longitudinal impact parameters of tracklets are required to satisfy $|d_{0}| < 2~\millimeter$ and $|z_{0}\sin\theta|<2~\millimeter$. After applying these requirements, the fraction of non-primary charged particles in the tracklet sample is approximately 3\%. A summary of the tracklet selection is given in Table~\ref{tbl:selection_summary}. Combined tracks are required to have at least four SCT hits. A tracklet is classified as \emph{matched} if the tracklet and a selected combined track share at least one common pixel hit.

\subsubsection{Transverse impact parameter resolution}
Events are required to have a primary vertex with at least $10$ tracks. Events with pile-up vertices are rejected. Only tracks within the range $0.4~\GeV<\pt<10~\GeV$ and $|\eta|<0.5$ which satisfy the \emph{loose} criteria (see Table~\ref{tbl:selection_summary}) are used in the analysis.

\section{Characterisation of material in data and MC simulation}
\label{sec:results_qualitative}
\begin{figure}[t!]
\begin{center}
\subfigure[]{
  \label{fig:hadInt_data_map_XY_zoom}
  \includegraphics[width=0.47\textwidth]{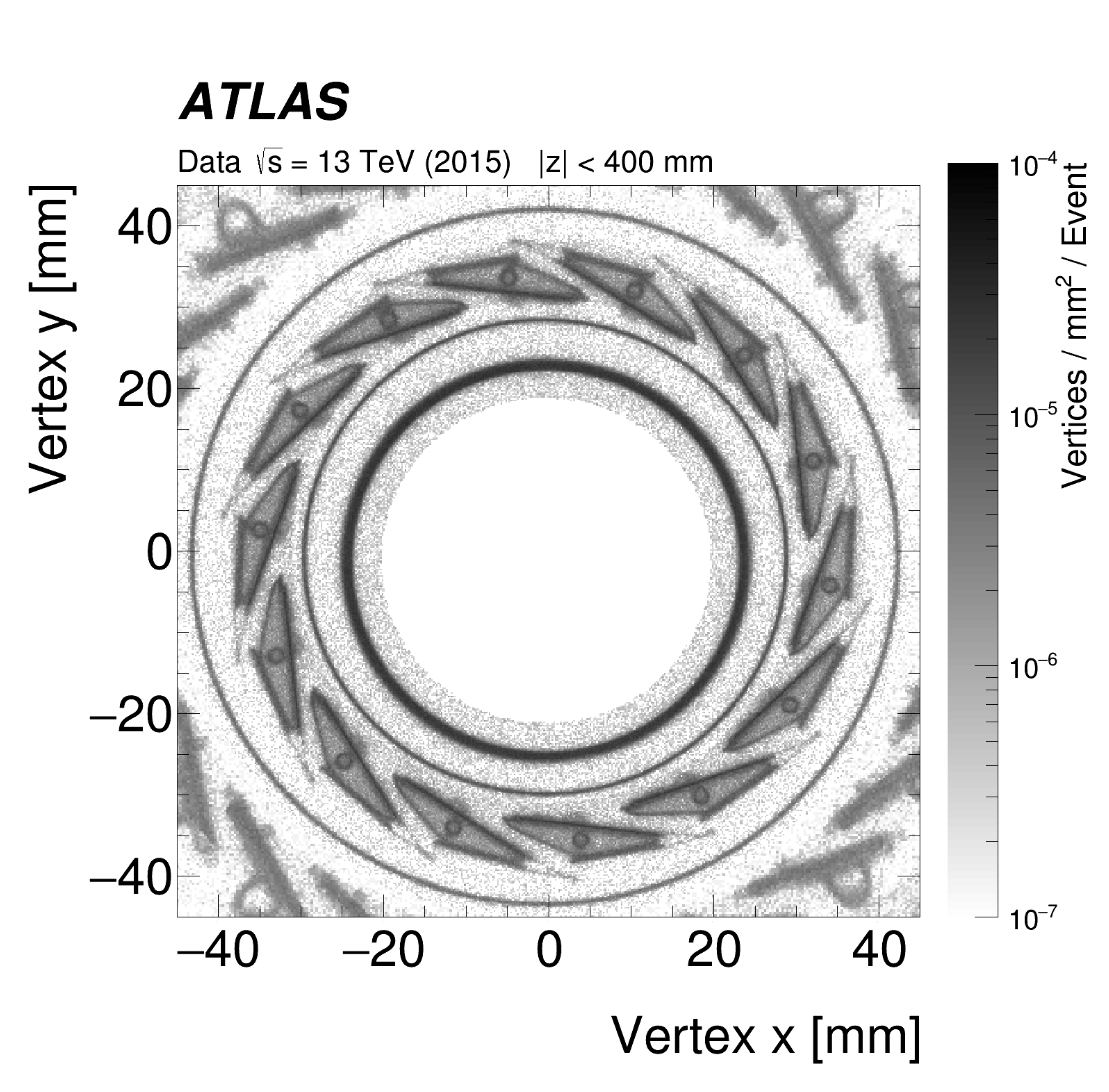}
}
\subfigure[]{
  \label{fig:hadInt_py8_map_XY_zoom}
  \includegraphics[width=0.47\textwidth]{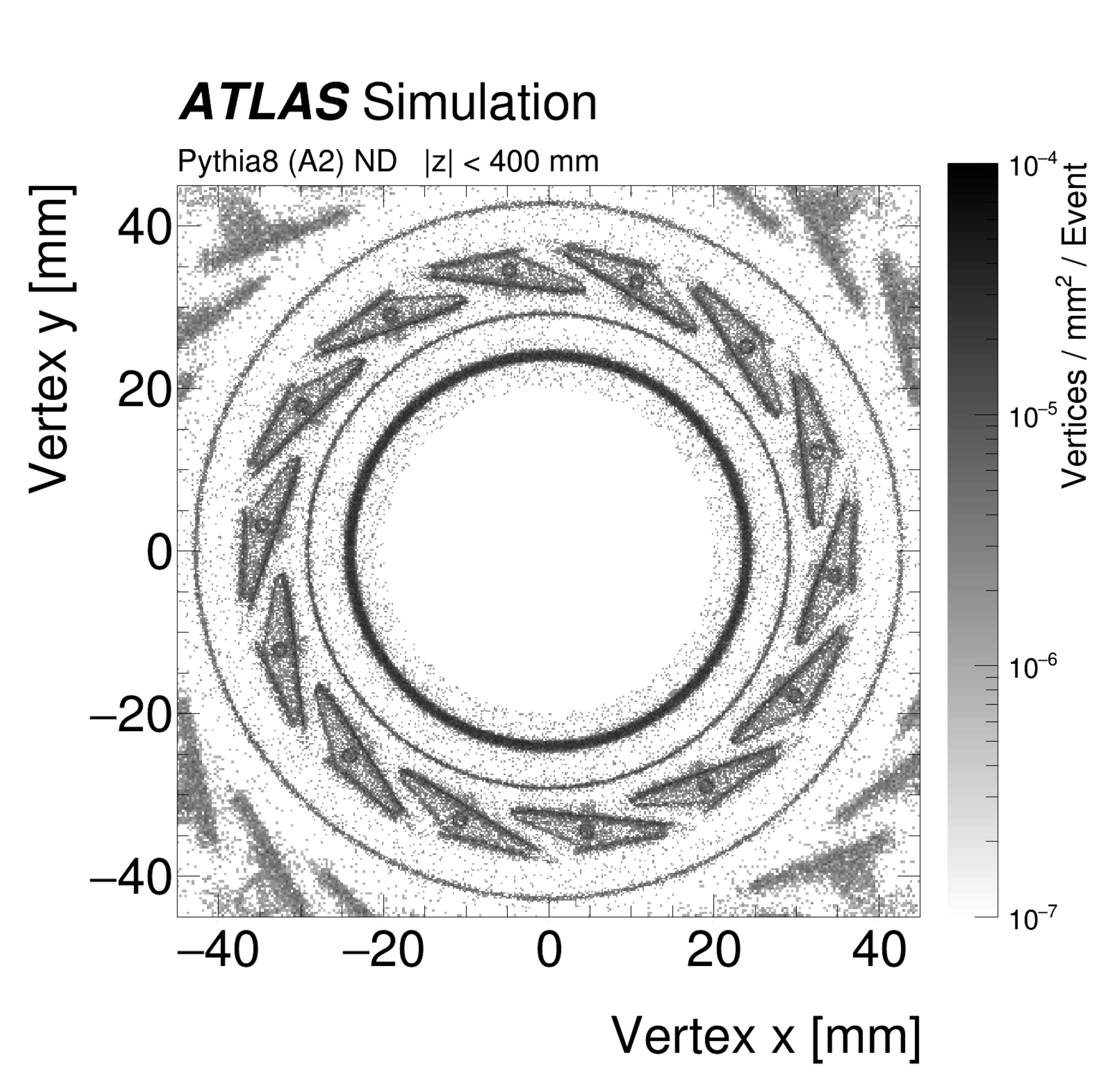}
}
\subfigure[]{
  \label{fig:hadInt_data_map_XY}
  \includegraphics[width=0.47\textwidth]{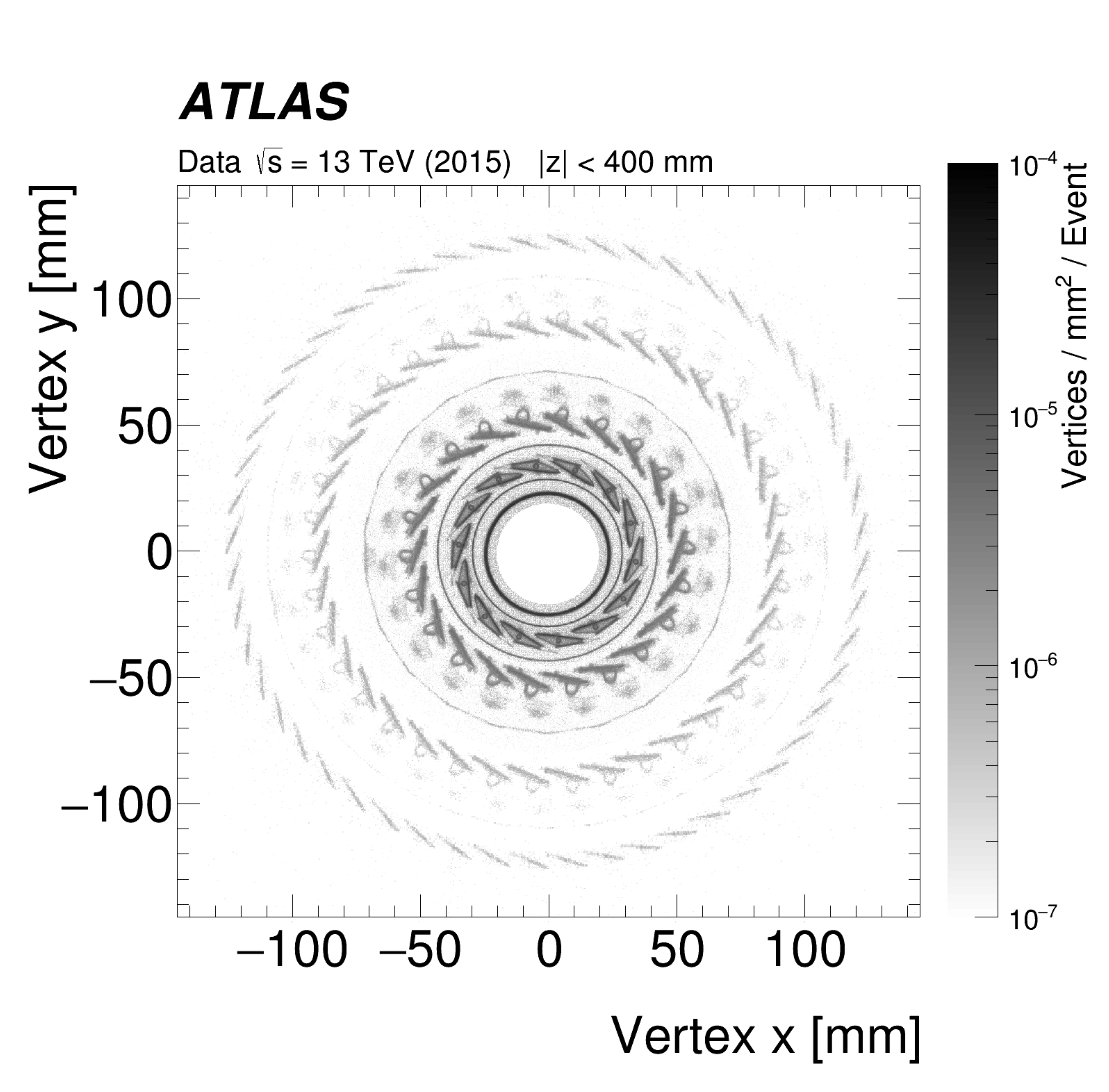}
}
\subfigure[]{
  \label{fig:hadInt_py8_map_XY}
  \includegraphics[width=0.47\textwidth]{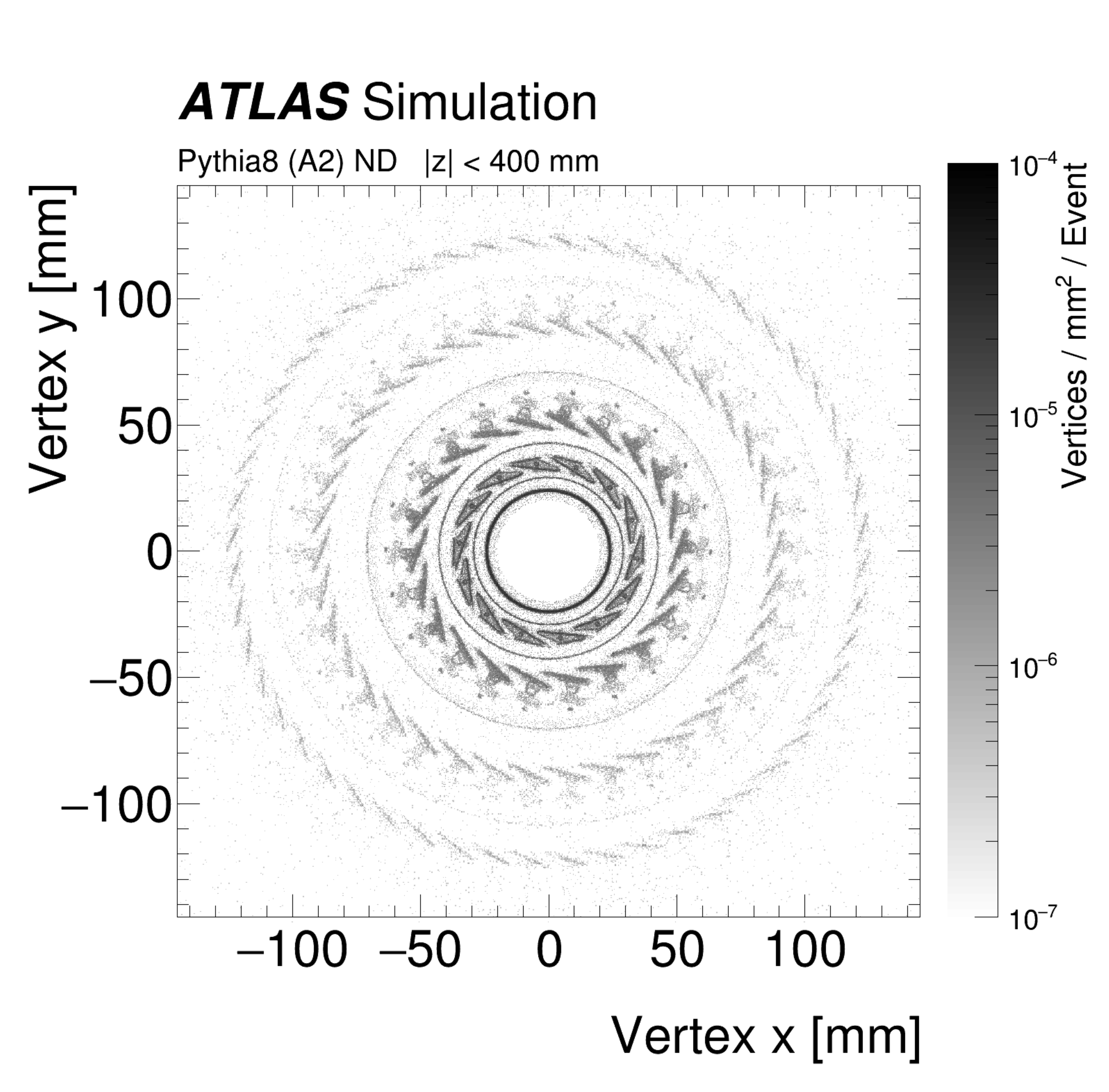}
}
\caption{Distribution of hadronic-interaction vertex candidates in $|\eta|<2.4$ and $|z|<400~\millimeter$ for data and the  \textsc{Pythia~8} MC simulation with the \emph{updated} geometry model. \subref{fig:hadInt_data_map_XY_zoom}, \subref{fig:hadInt_py8_map_XY_zoom} The $x$--$y$ view zooming-in to the beam pipe, IPT, IBL staves and IST, and \subref{fig:hadInt_data_map_XY}, \subref{fig:hadInt_py8_map_XY} of the pixel detector. Some differences between the data and the \textsc{Pythia~8} MC simulation, observed at the position of some of the cooling pipes in the next-to-innermost layer (PIX1), are due to mis-modelling of the coolant fluids, as discussed in Ref.~\cite{PERF-2011-08}.}
\label{fig:hadInt:map_XYs}
\end{center}
\end{figure}

This section reviews the qualitative aspects of the comparison between data and simulation.
The distributions of hadronic interaction vertices are shown in  Figures \ref{fig:hadInt:map_XYs} to \ref{fig:hadInt:compare_R}
while the distributions for photon conversion vertices are shown in Figure \ref{fig:conv:qual}.
The track-extension efficiency is shown as a function of both tracklet $\pt$ and $\eta$ in Figure \ref{fig:SCTExt_DataEposPythia_Diff}.

Reconstruction of hadronic interaction vertices enables a detailed visual inspection of the material distribution due to its superb position resolution. Figure~\ref{fig:hadInt:map_XYs} shows the distribution of vertices for hadronic interaction candidates in the $x$--$y$ plane for the data and the \textsc{Pythia~8} MC simulation for the \emph{updated} simulation. The qualitative features of the two distributions indicate that the geometry model description is generally accurate.

\subsection{Radial and pseudorapidity regions}
For the hadronic interaction and photon conversion analyses, the measurable ID volumes are divided into several groups by radii, which are referred to hereafter as \emph{radial regions}. Table~\ref{tbl:hadInt:section_definition} lists the radial regions. The boundaries are chosen to classify distinct barrel layers of the ID. Two regions, referred to as Gap1 (between PIX2 and PIX3) and Gap2 (between PIX3 and PSF), are also introduced as the regions with low purity of hadronic interactions in order to control the background yield of hadronic interaction signals. The gap regions and the other regions overlap so that the number of vertices in the gap regions is increased while also significantly reducing the number of hadronic interactions. For the photon conversion analysis, the regions of IPT, IBL and IST are combined into one region and denoted by ``IBL'' since the method does not have good enough resolution to differentiate between these components. For the track-extension efficiency study, the $\eta$-range is binned with a bin width of 0.1 for $1.5<|\eta|<2.5$.

\subsection{Radial position offset}
\label{sec:overview_offset}
\begin{figure}[t!]
\begin{center}
\subfigure[]{
  \label{fig:bp_prof}
  \includegraphics[width=0.462\textwidth]{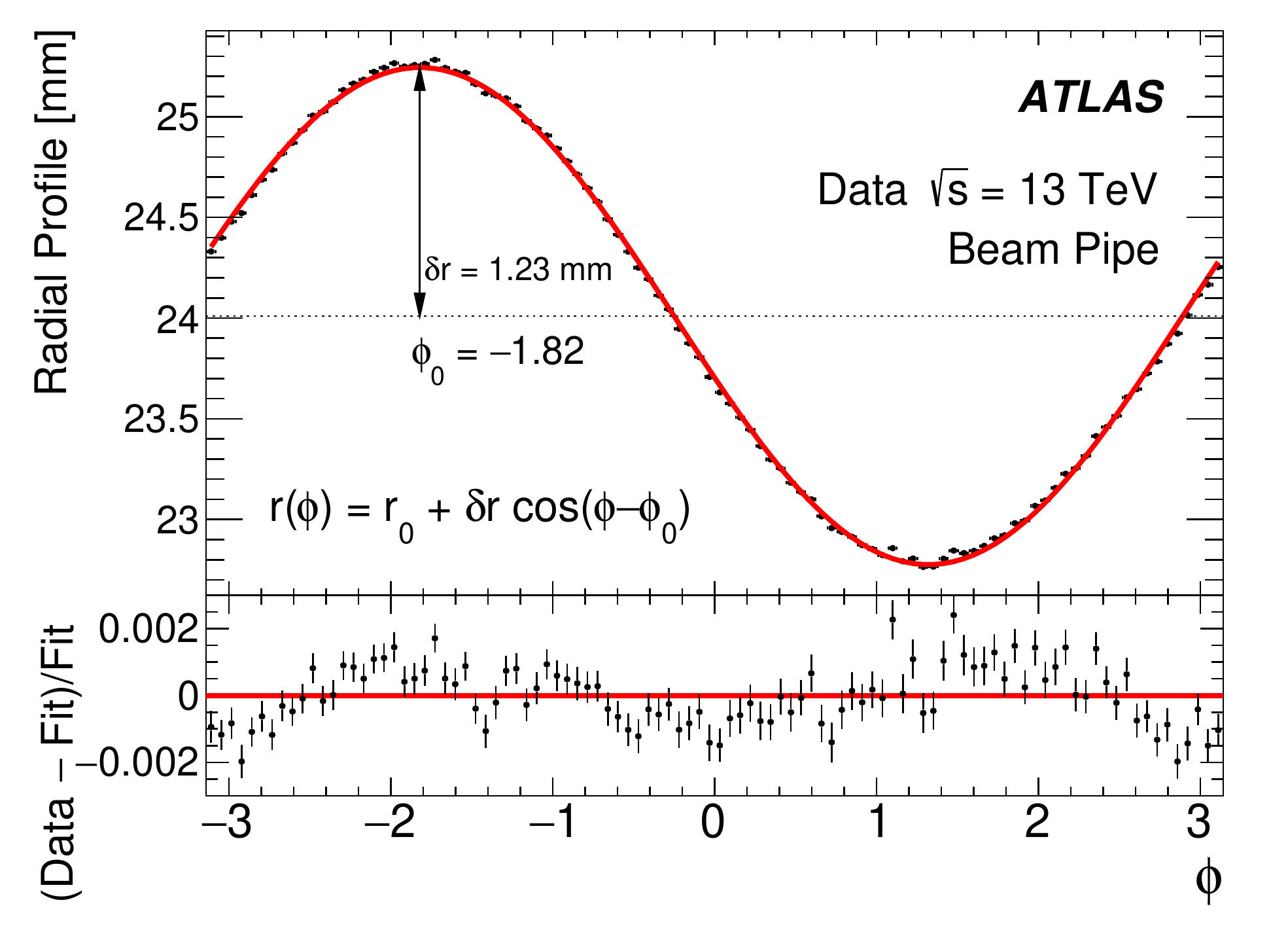}
}
\subfigure[]{
  \label{fig:compare_R_corr_twoMCs_IBL_2}
  \includegraphics[width=0.455\textwidth]{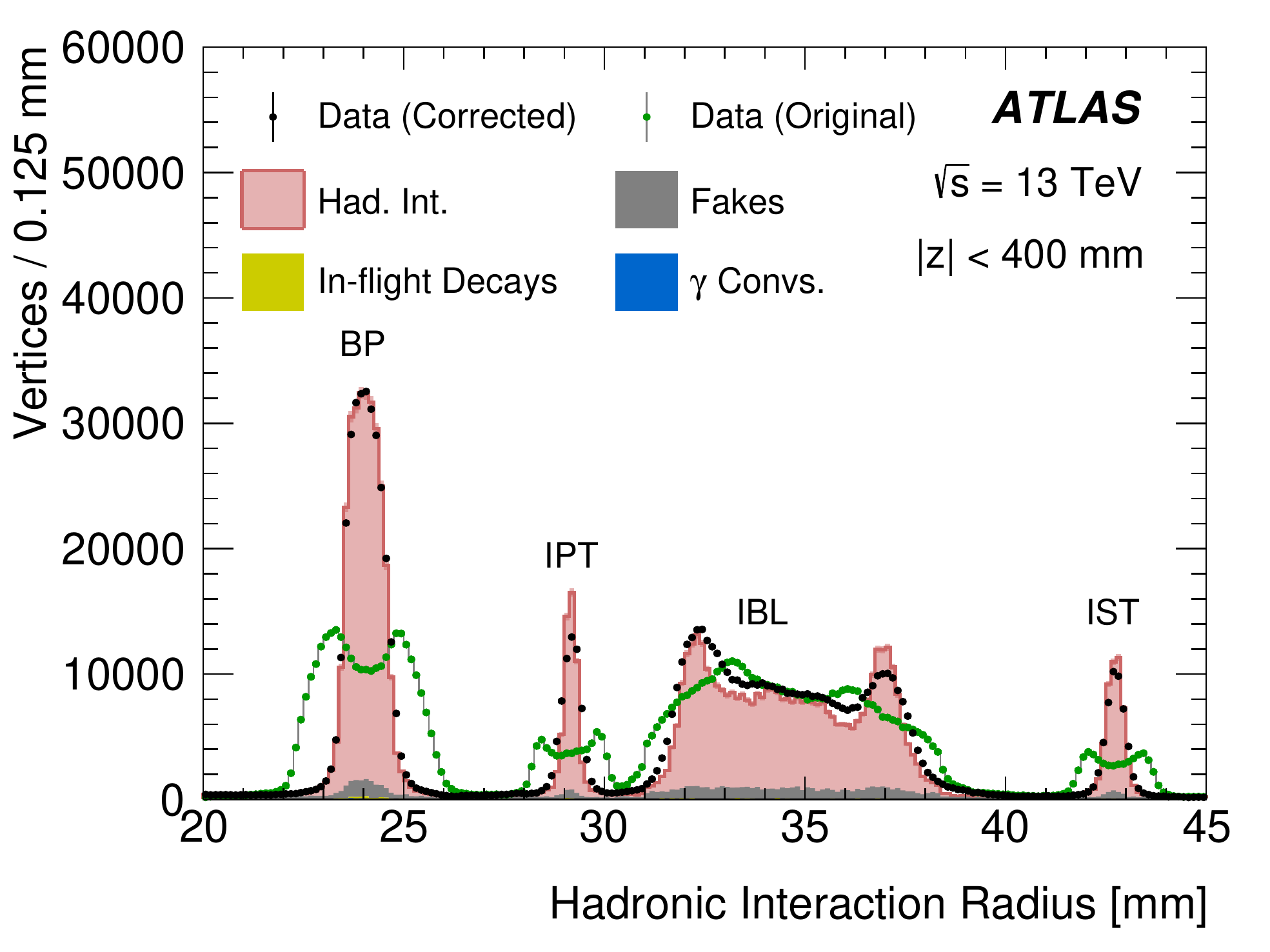}
}
\caption{\subref{fig:bp_prof} The $r$--$\phi$ profile of hadronic interaction candidates at the beam pipe, fitted with a sinusoidal curve to determine the shift of the pipe from the origin of the ATLAS coordinate system in the $x$--$y$ plane. In the ratio plot in the bottom panel, a small sinusoidal deviation in data from the fit is observed. This may be reflect a slight misalignment of the beam pipe with respect to the $z$-axis, but this does not affect the result of the analysis. \subref{fig:compare_R_corr_twoMCs_IBL_2} Comparison of the radial distribution of hadronic interaction candidates to the \textsc{Epos} \emph{updated} geometry model before and after the radial offset correction to the data for each barrel layer within $20~\millimeter<r<45~\millimeter$.}
\label{fig:hadInt_bp_prof}
\end{center}
\end{figure}

In data, the axis of each cylindrical layer of the beam pipe, IBL, pixel barrel layers and other support tubes has an offset perpendicular to the $z$-axis from the origin of the ATLAS coordinate system due to the placement precision. Figure~\ref{fig:bp_prof} shows the sinusoidal profile of the average radial position of the beam-pipe material as a function of $\phi$. Similar offsets were observed in a previous analysis~\cite{PERF-2011-08,PERF-2015-06}.
The offset of each layer is estimated by fitting a sinusoidal curve to the $r$--$\phi$ profile. The obtained size of the offset varies by layers within the range of around $0.3$~mm to $1.2$~mm. The radial distribution of hadronic interaction candidates is compared to the MC simulation in Figure~\ref{fig:compare_R_corr_twoMCs_IBL_2} both with and without the application of the radial position corrections.

\subsection{Beam pipe}
\label{sec:overview_beampipe}
\begin{figure}[htbp]

\begin{center}

  \label{fig:hadInt_bp_dist_reweighted}
  \includegraphics[width=0.45\textwidth]{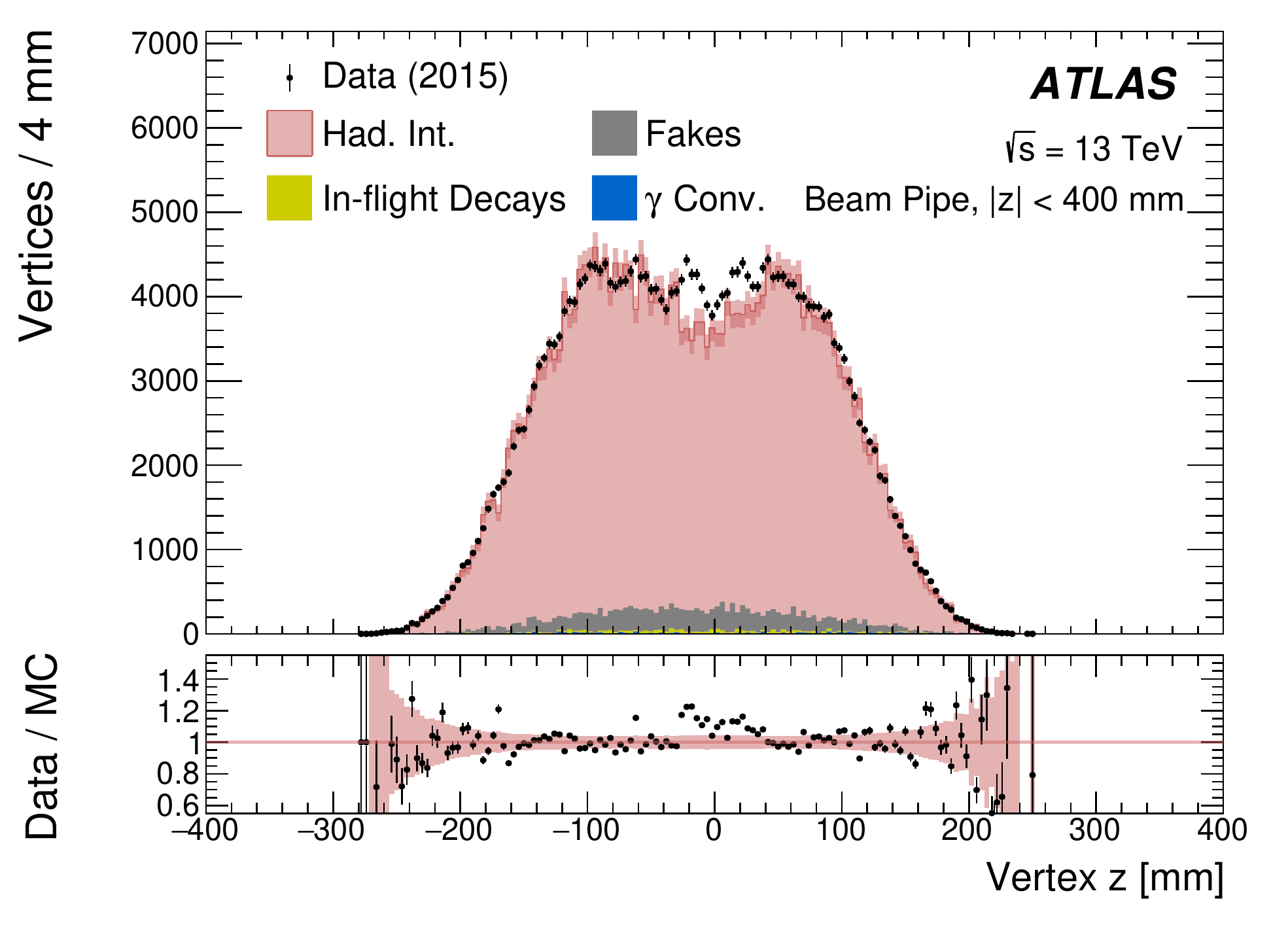}

\caption{Comparison of the $z$-distribution of the rate of hadronic interaction candidates at the beam pipe in data and \textsc{Epos} MC simulation with the \emph{updated} geometry model. The band shown in the bottom panel indicates statistical uncertainty of the MC simulation.}
\label{fig:hadInt_bp_dist}
\end{center}

\begin{center}
\subfigure[$-400~\millimeter < z < -40~\millimeter$]{
  \label{fig:}
  \includegraphics[width=0.31\textwidth]{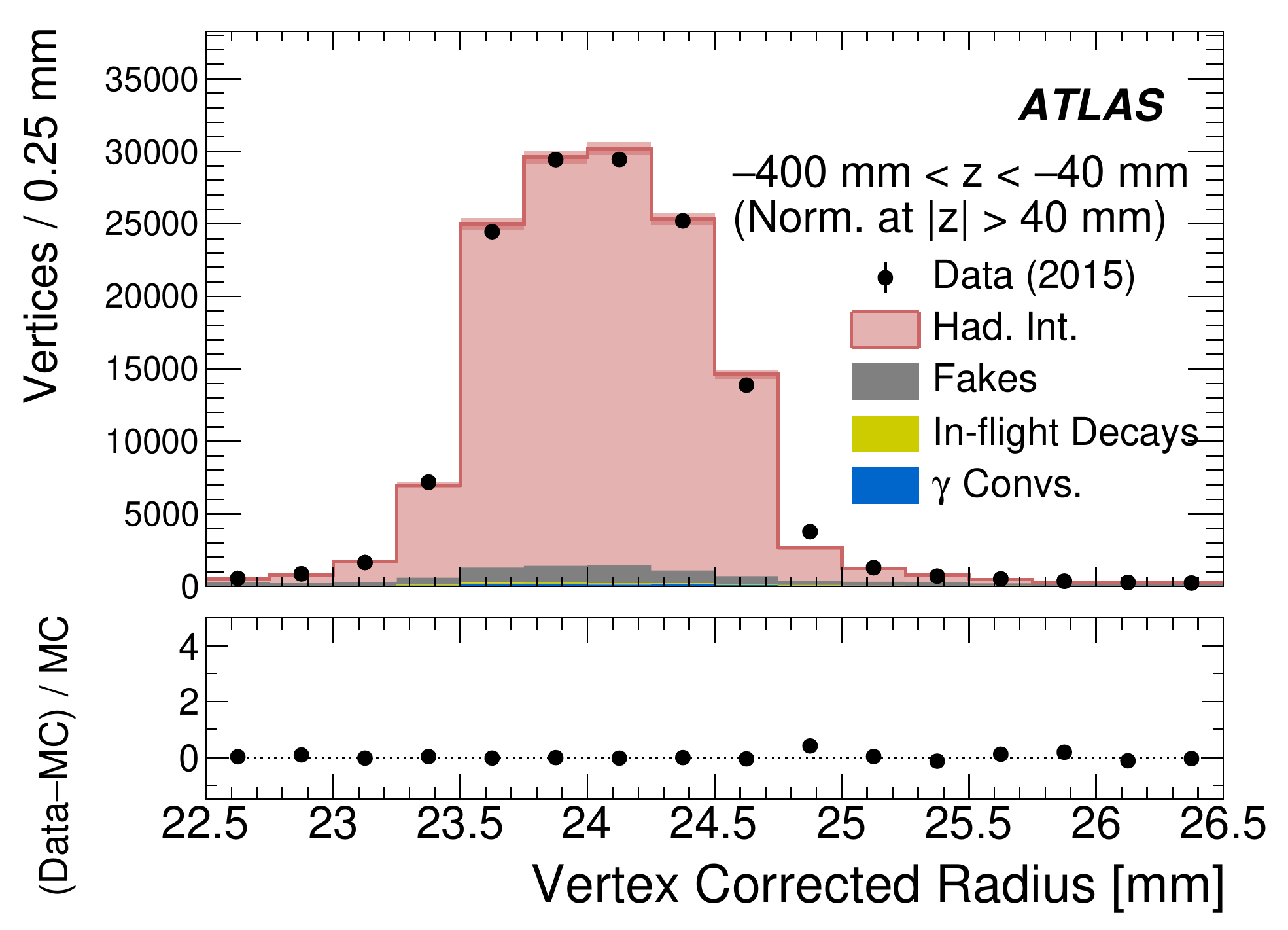}
}
\subfigure[$|z| < 40~\millimeter$]{
  \label{fig:}
  \includegraphics[width=0.31\textwidth]{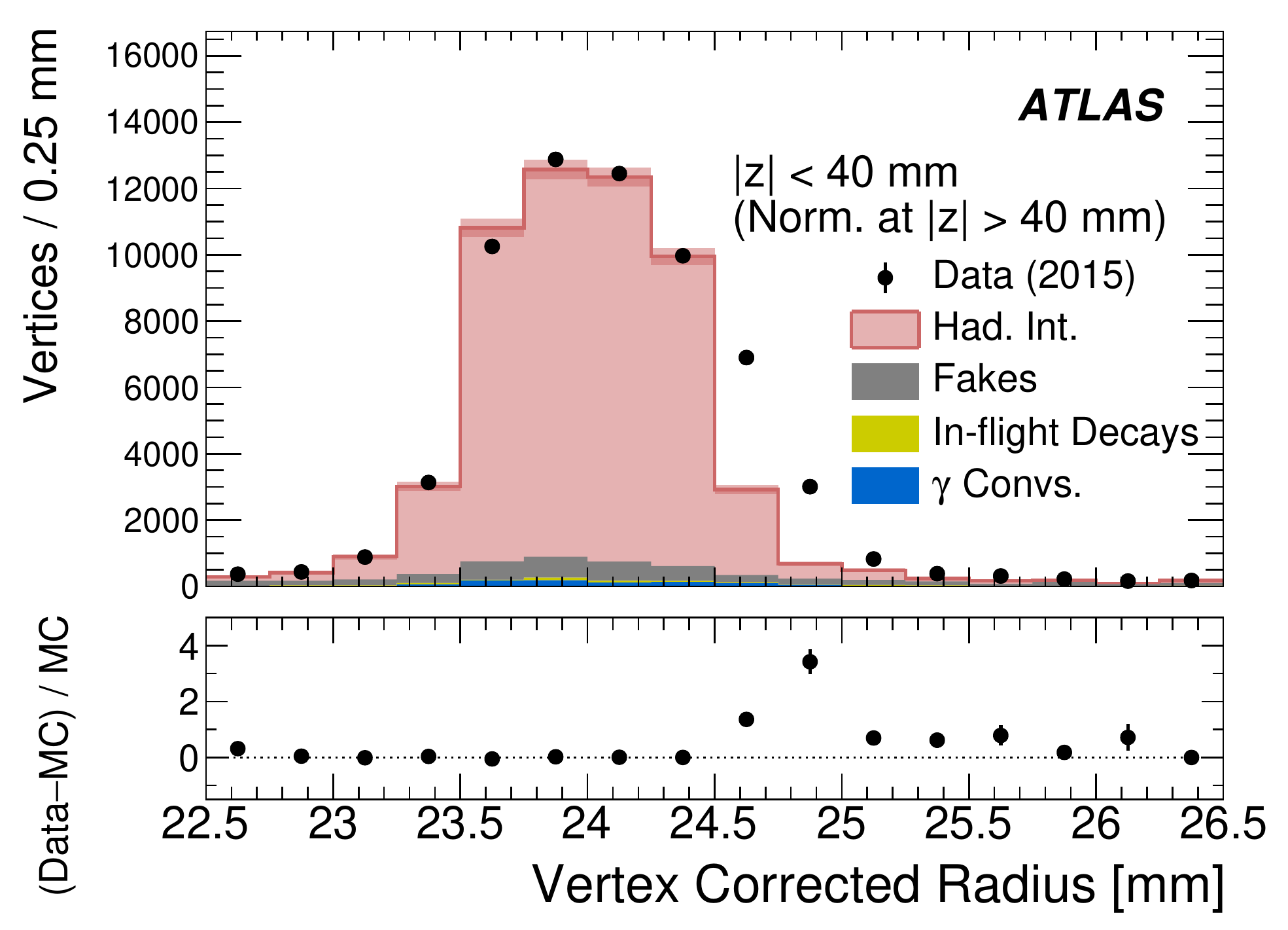}
}
\subfigure[$40~\millimeter < z < 400~\millimeter$]{
  \label{fig:}
  \includegraphics[width=0.31\textwidth]{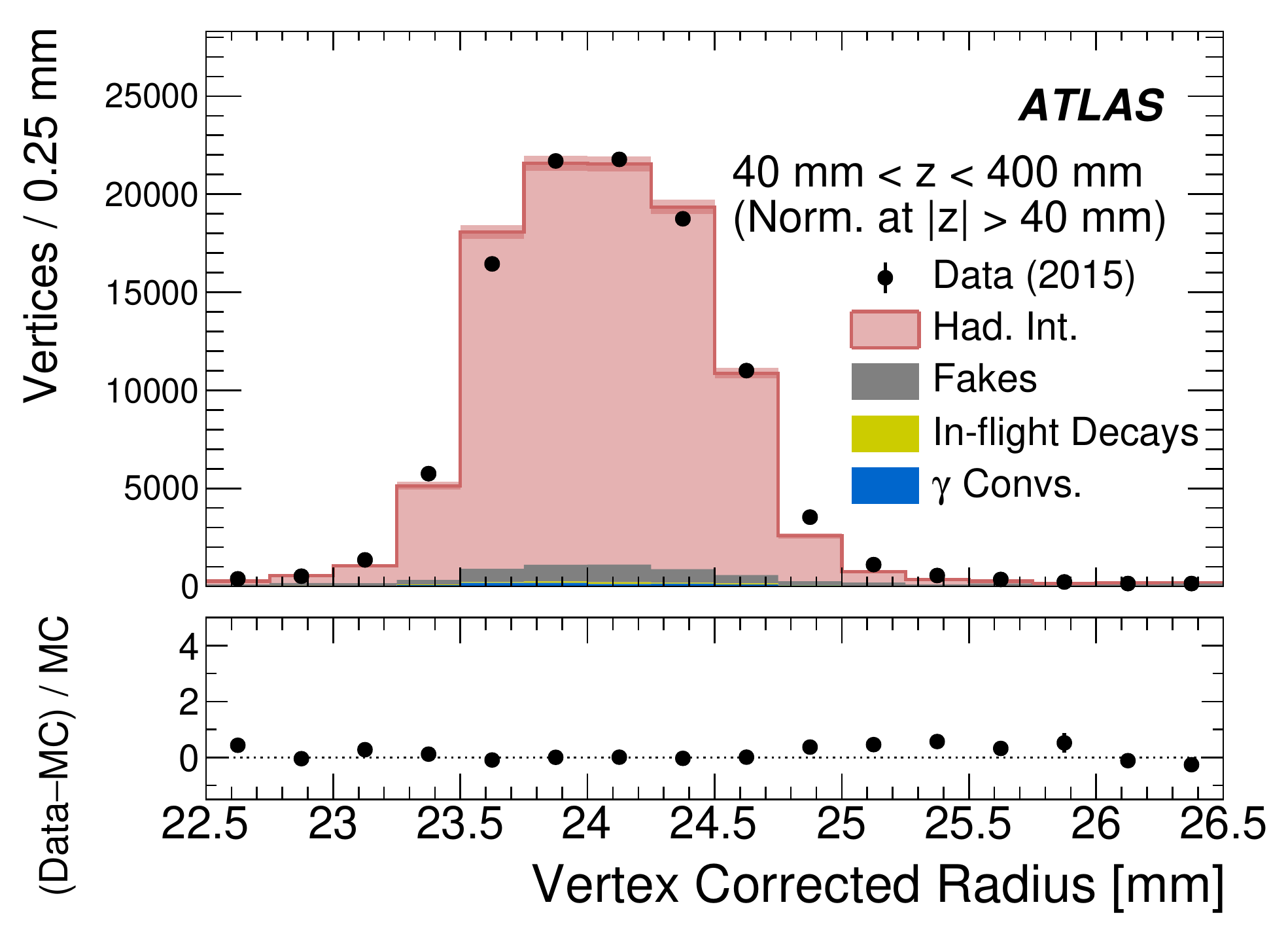}
}
\caption{Comparison between data and simulation of the $r$-distribution of the hadronic interaction candidates at the beam pipe ($22.5~\millimeter<r<26.5~\millimeter$) in different $z$ sections. The MC simulation is normalised to the data using the rate at $|z|>40~\millimeter$. An excess is observed at the outer surface of the beam pipe for $|z|<40~\millimeter$.}
\label{fig:hadInt:beampipe_radial_compare}
\end{center}
\end{figure}

The acceptance of the hadronic interaction reconstruction is such that interactions within the beam pipe are only reconstructed within the range $|z|<250~\millimeter$, as shown in Figure~\ref{fig:hadInt_bp_dist}. The description of the geometry model is generally good, but an excess of candidates is observed in data at the centremost part of the beam pipe within $|z|<40~\millimeter$.
The radial distributions of the beam pipe in different $z$-ranges are shown in Figure~\ref{fig:hadInt:beampipe_radial_compare} normalised to the rate in the beam pipe at $|z|>40~\millimeter$.
While the radial distribution is well described for $|z|>40~\millimeter$, there is a significant excess within $|z|<40~\millimeter$, which appears to be localised to the outer surface of the beam pipe.
The excess is 12\% of the rate at $|z|<40~\millimeter$, corresponding to approximately $N_{X_{0}}= 0.03$--$0.04\%$. This excess is also observed in the photon conversion case, as shown in Figure~\ref{fig:conv:bp_z} in the region $|z|<50\,\mathrm{mm}$.
Investigations of engineering records suggest that several 60-\micrometer-thick polyimide tape layers are missing in the simulated description of the beam pipe in the \emph{updated} geometry model.

\begin{table}[t!]
\caption{Definition of the radial regions used for comparing data to MC simulation. In the case of the photon conversion analysis, the IPT, IBL and IST regions are always considered together, due to the limited resolution of the approach. The corresponding $z$ region used for the data to MC simulation comparison is $|z|<400\,\millimeter$ for all of the radial regions listed.}
\centering
\small
\begin{tabular}{lcl}
\hline
\hline
Radial Region & Radial range [mm]& Description\\
\hline
BP      & 22.5--26.5   &beam pipe \\
IPT     & 28.5--30.0   &inner positioning tube \\
IBL     & 30.0--40.0   &IBL staves (for photon conversion: IPT+IBL+IST) \\
IST     & 41.5--45.0   &inner support tube  \\
PIX1    & 45.0--75.0   &first pixel barrel layer \\
PIX2    & ~83--110      &second pixel barrel layer \\
PIX3    & 118--145     &third pixel barrel layer \\
PSF     & 180--225     &pixel support frame\\
PST     & 225--240     &pixel support tube \\
SCT-ITE & 245--265     &SCT inner thermal enclosure \\
SCT1    & 276--320     &first SCT barrel layer \\
SCT2    & 347--390     &second SCT barrel layer \\
\hline
Gap1   & 73--83        &material gap between PIX1 and PIX2\\
Gap2   & 155--185      &material gap between PIX3 and PSF \\
\hline
\hline
\end{tabular}
\label{tbl:hadInt:section_definition}
\end{table}

\subsection{IBL and its support tubes}
For the IBL staves, the rate of hadronic interactions in the simulation with the \emph{original} geometry model is found to be significantly smaller than in the data around $r\simeq 32~\millimeter$, as shown in Figure~\ref{fig:compare_R_corr_twoMCs_IBL}. A corresponding deficit is observed for photon conversions. Investigations clarified that some surface-mounted components, e.g.~capacitors, located on the front-end chips of the IBL modules, are missing in the \emph{original} geometry model. As described in Section~\ref{sec:geo}, the \emph{updated} geometry model was created to resolve this issue; this gave significantly better agreement with the data. The description of the rate as a function of radius in $31<r<40~\mathrm{mm}$ is not perfect, but this is believed to be due to misalignment of each stave in the data compared to the nominal design. These effects produce a few-hundred $\micrometer$ of smearing, which could explain the difference between the data and the simulation.

The material composition of the IPT and IST is studied with hadronic interaction vertices. The nominal thickness of the IPT is 0.325~mm at $|z|<311$~mm. The observed thickness of the tube in terms of the FWHM (full width at half maximum) of the peak (at $r = 29$~mm in Figure \ref{fig:compare_R_corr_twoMCs_IBL}), divided by 2.35, is 0.55~mm for the data, while it is 0.34~mm for the MC simulation, a difference which is greater than the estimated radial resolution at the IPT radius (0.13~mm).
In the \emph{updated} geometry, the density of the IPT is scaled. The agreement in the observed number of hadronic interaction candidates in the IPT region compared between the data to the simulation with the \emph{updated} geometry is improved. For the IST, the rate in data is approximately 16\% smaller than in the \emph{original} geometry model, while the thickness in the data and simulation are similar. In the \emph{updated} geometry model, the density of the IST is also artificially scaled to give better agreement.

\subsection{Outer barrel layers}
\begin{figure}[t!]
\begin{center}
\subfigure[$20~\millimeter<r<75$~\millimeter]{
  \label{fig:compare_R_corr_twoMCs_IBL}
  \includegraphics[width=1.0\textwidth]{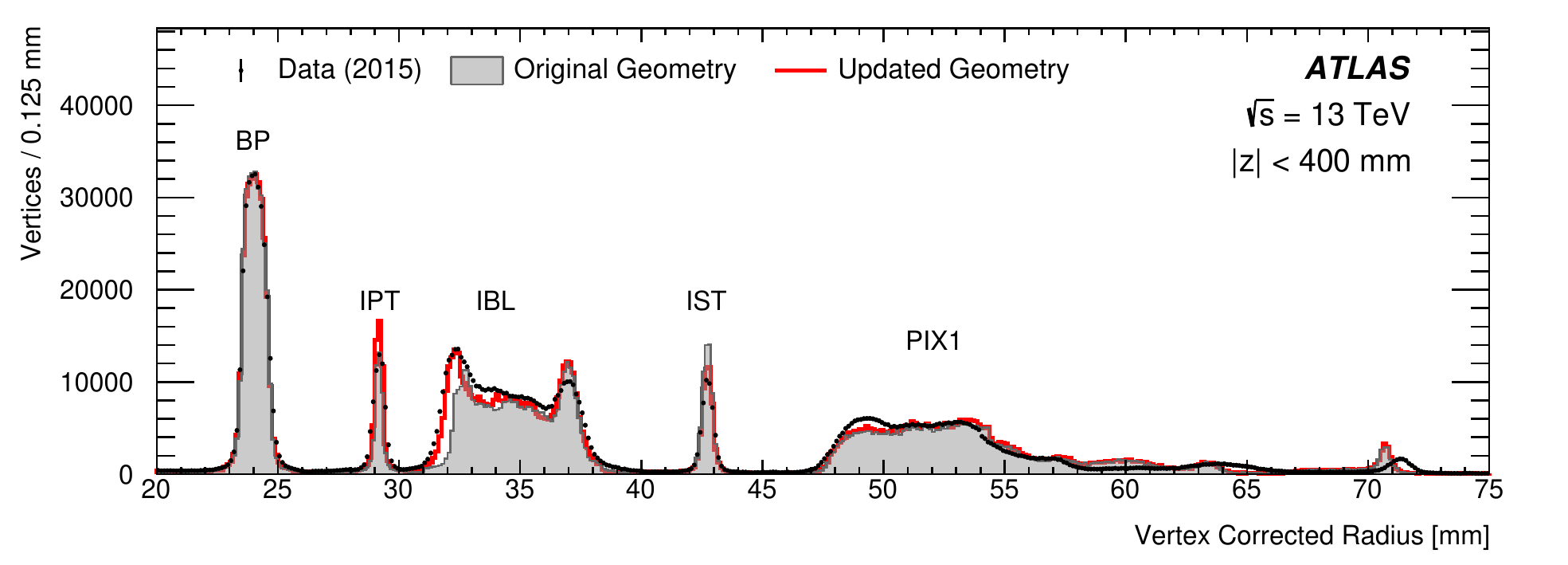}
}
\subfigure[$45$~\millimeter$<r<150$~\millimeter]{
  \label{fig:compare_R_corr_twoMCs_Pixel}
  \includegraphics[width=1.0\textwidth]{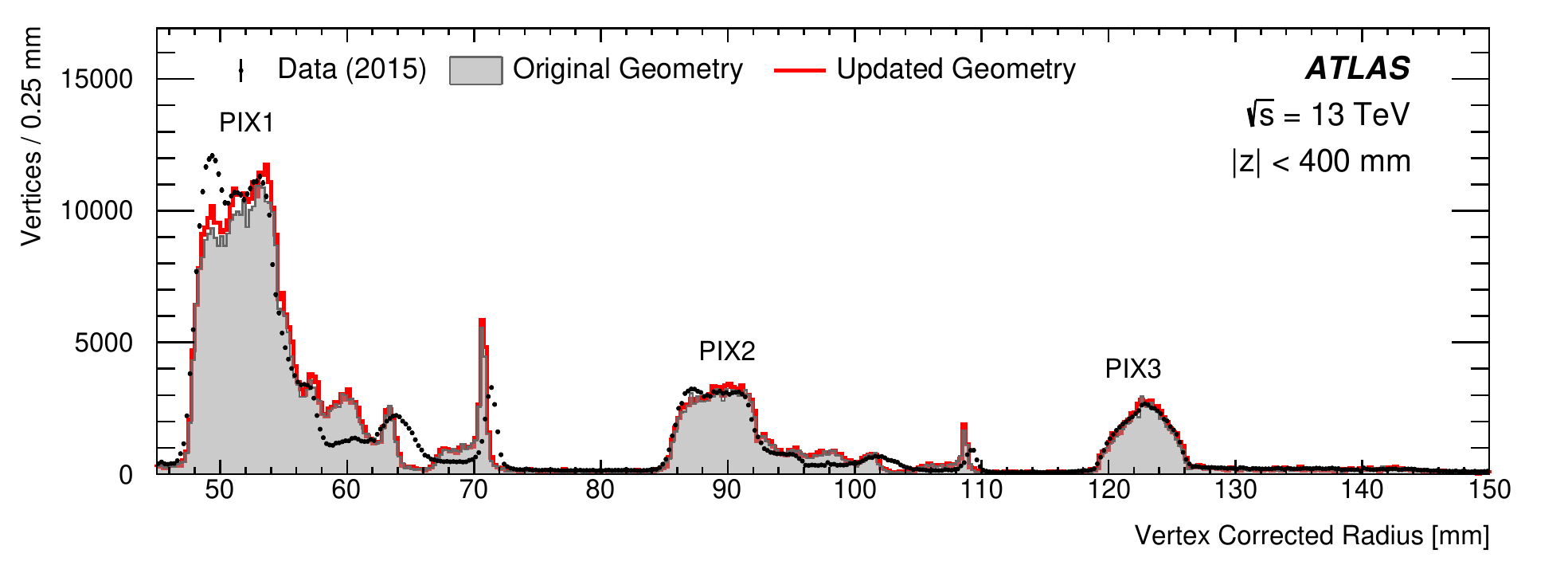}
}
\caption{Comparison of the radial distribution of hadronic interaction candidates between data and simulation (\emph{original} and \emph{updated} simulations) for \subref{fig:compare_R_corr_twoMCs_IBL} $20$~\millimeter$ < r < 75$ mm and \subref{fig:compare_R_corr_twoMCs_Pixel} $45$~\millimeter $< r <$ 150 mm.}
\label{fig:hadInt:compare_R}
\end{center}
\end{figure}

\begin{figure}[tbp]
\begin{center}
\subfigure[]{
  \label{fig:conv:bp_z}
  \includegraphics[width=0.45\textwidth]{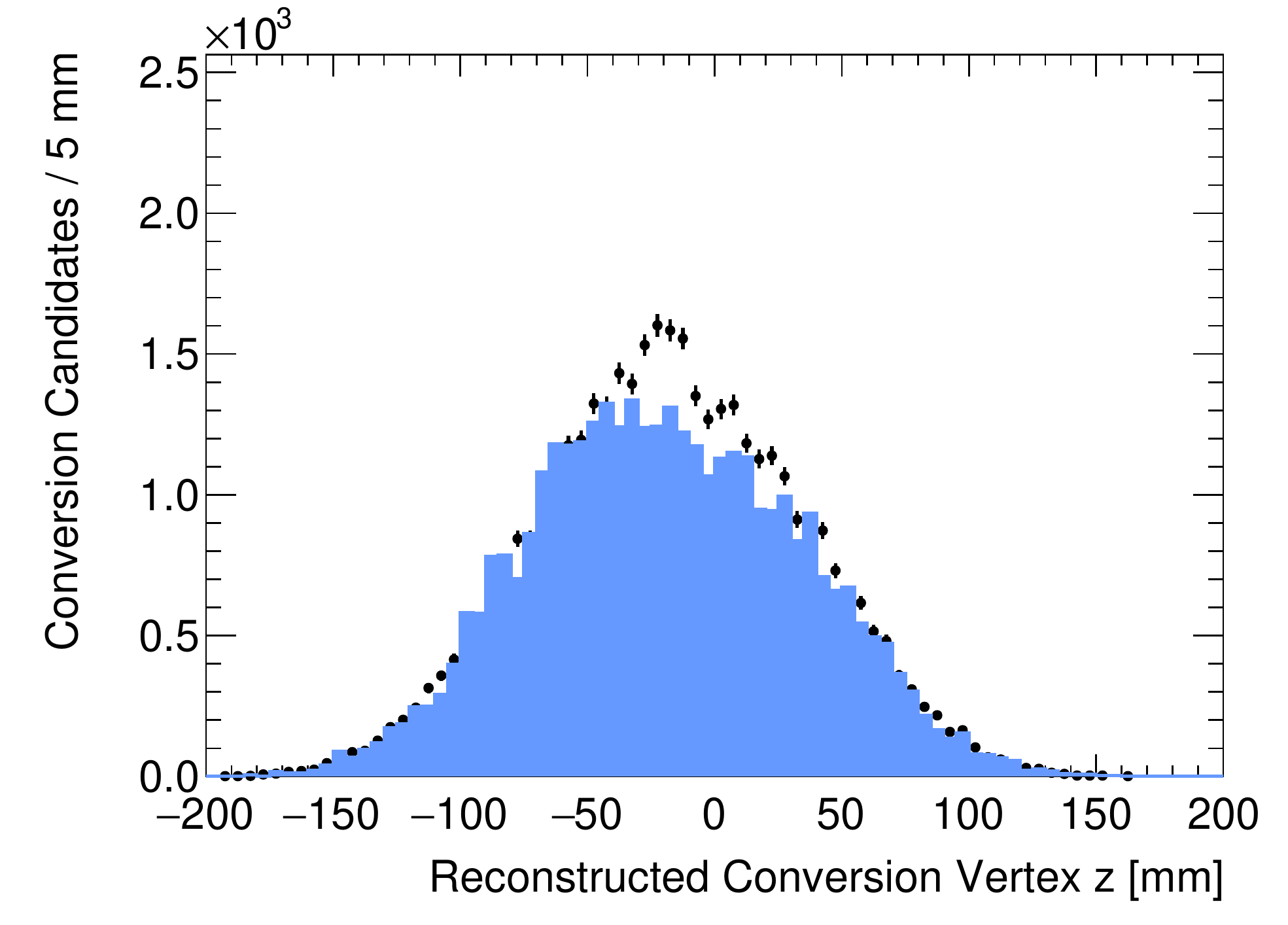}
}
\subfigure[]{
  \label{fig:conv:bp_r}
  \includegraphics[width=0.45\textwidth]{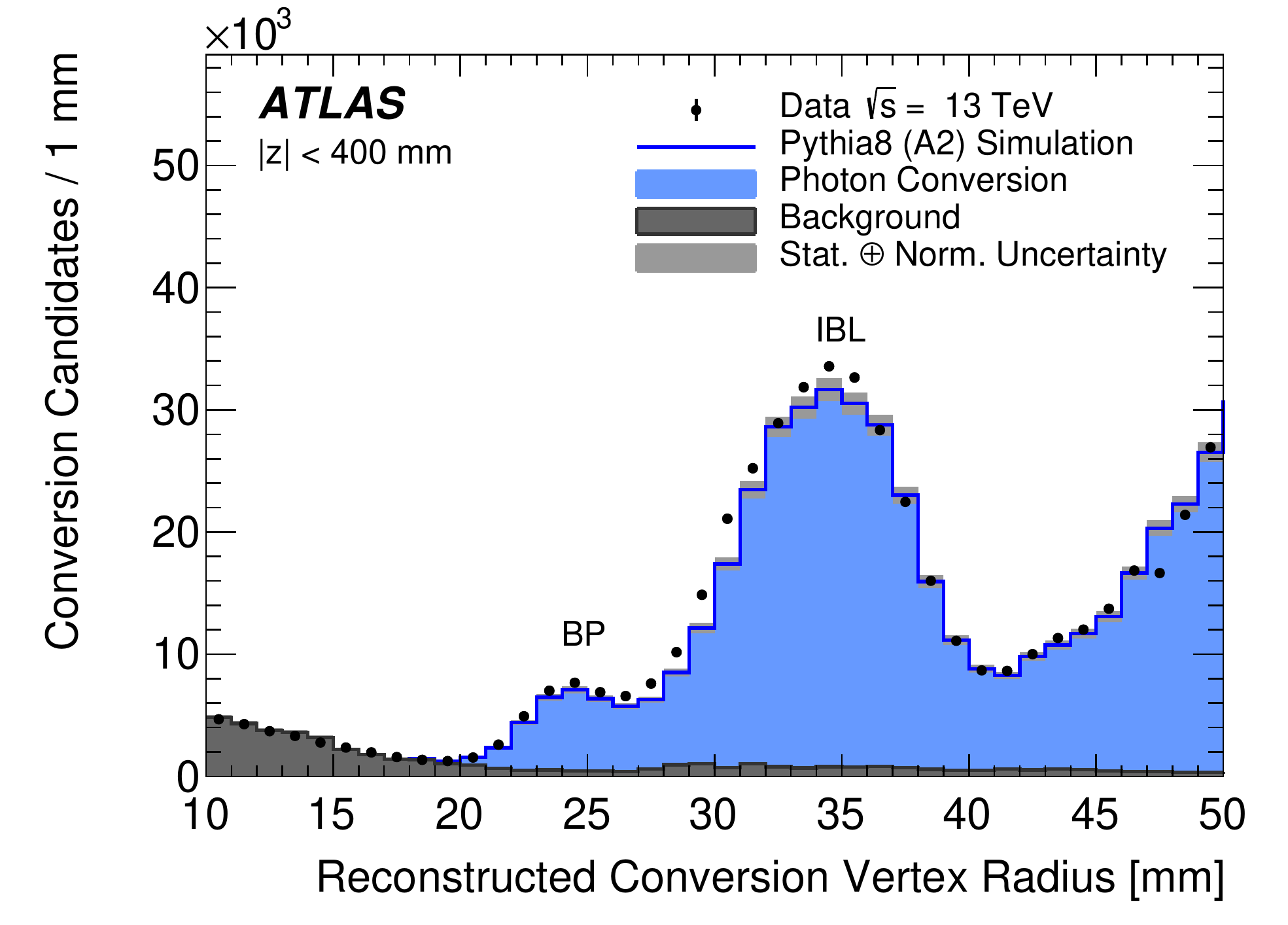}
}
\subfigure[]{
  \label{fig:conv:pix_r}
  \includegraphics[width=0.45\textwidth]{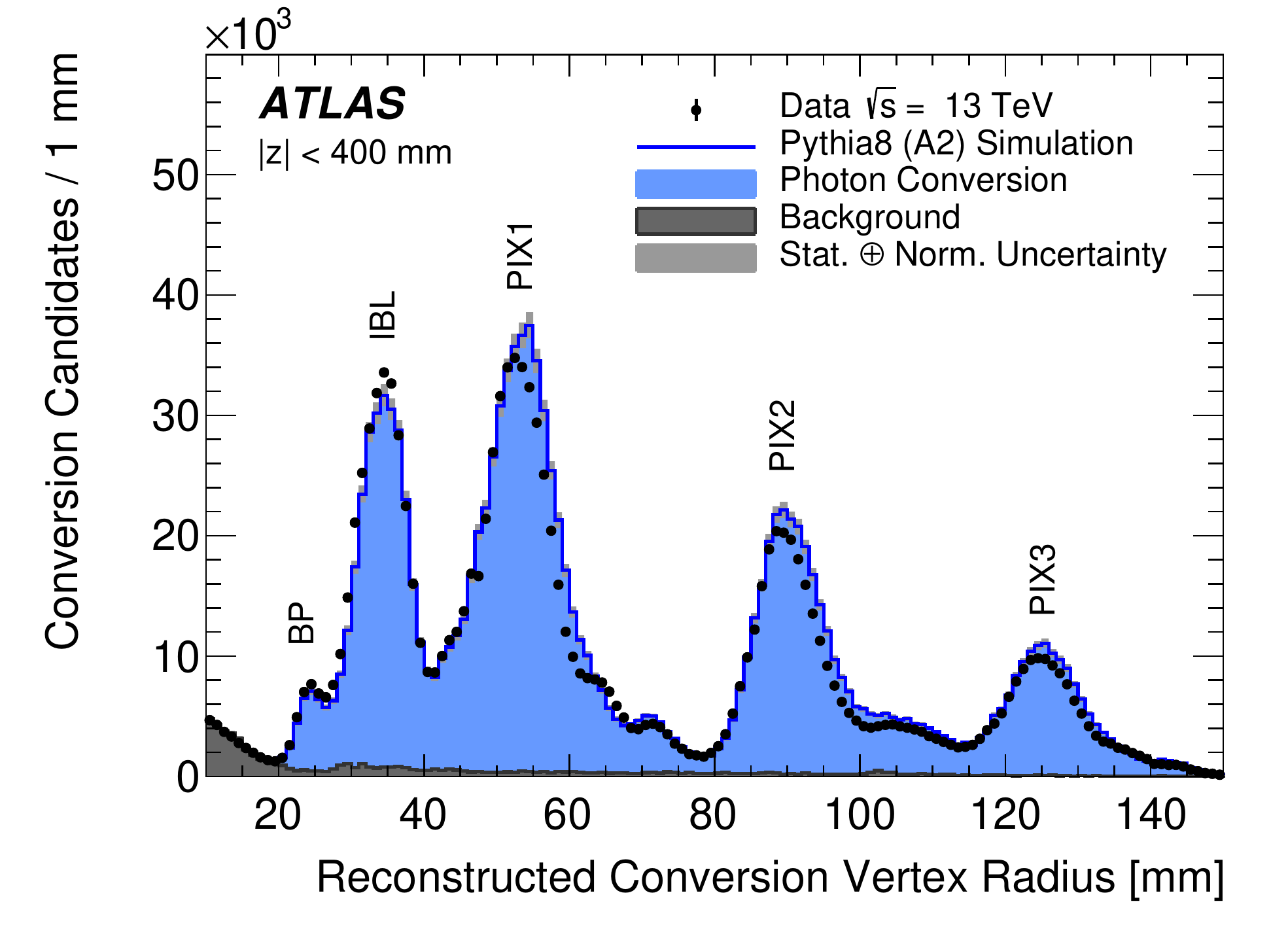}
}
\subfigure[]{
  \label{fig:conv:sct_r}
  \includegraphics[width=0.45\textwidth]{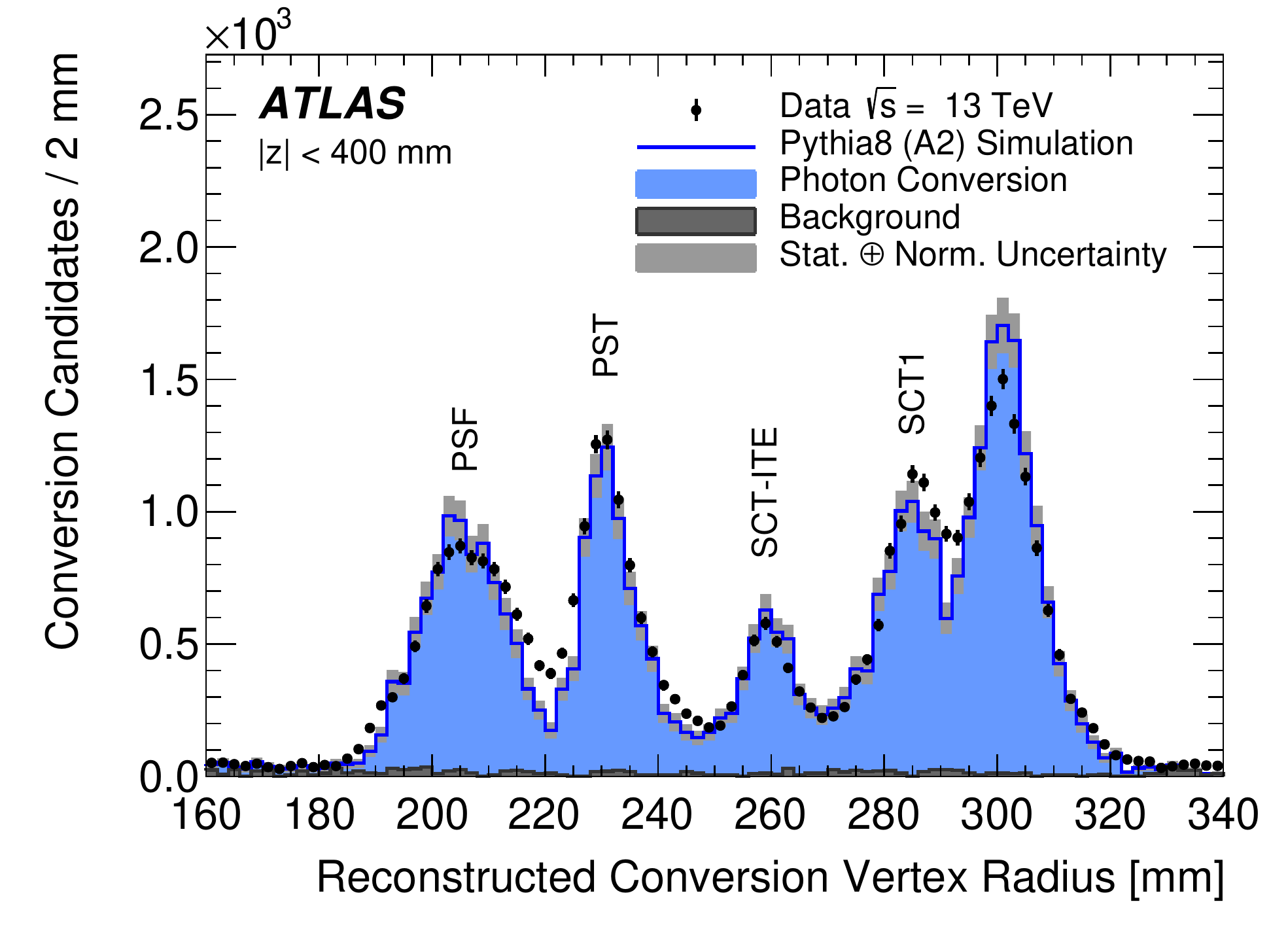}
}
\caption{Conversion vertex position distributions for \textsc{Pythia~8} simulation with the \emph{updated} geometry model compared to data, including \subref{fig:conv:bp_z} the conversion vertex $z$-position distribution in the beam-pipe radial region and the conversion vertex radial distributions in \subref{fig:conv:bp_r} the beam-pipe and IBL region, \subref{fig:conv:pix_r} region up to and including the third pixel layer and \subref{fig:conv:sct_r} region between the PSF second SCT layer.}
\label{fig:conv:qual}
\end{center}
\end{figure}

The pixel barrel layers were refurbished between Run~1 and Run~2, but the material composition is unchanged. The radial distribution of the outer barrel layers is shown in Figure~\ref{fig:compare_R_corr_twoMCs_Pixel} and Figure \ref{fig:conv:pix_r}. Due to careful investigations of Run~1 data, the distribution is reasonably well described by the MC simulation in all three layers. Nevertheless, some small deficit in the MC simulation observed around $r\simeq50~\millimeter$ and $r\simeq86~\millimeter$ in the hadronic interaction result may indicate that some components are missing from the simulated pixel modules. Furthermore, a discrepancy in the shape of the distribution is apparent in the region of the stave and cabling structures at $58~\millimeter<r<72~\millimeter$ and $96~\millimeter<r<112~\millimeter$. An excess in the MC simulation is also observed in the photon conversion measurements in this region (see Figure~\ref{fig:conv:pix_r}). The material composition of the PSF, PST and SCT barrel layers remains unchanged since Run~1. The radial distributions in this range are shown in Figures~\ref{fig:compare_R_SCT_Epos} and \ref{fig:conv:sct_r}, both of which exhibit good agreement. For hadronic interactions, the fraction of background vertices in this outer region is much larger, relative to the inner layers.

\subsection{Regions between pixel and SCT detectors}
\begin{figure}[t!]
\begin{center}
\subfigure[]{
  \label{fig:SCTExt_Eta_DataEposPythia_Diff}
  \includegraphics[width=0.45\textwidth]{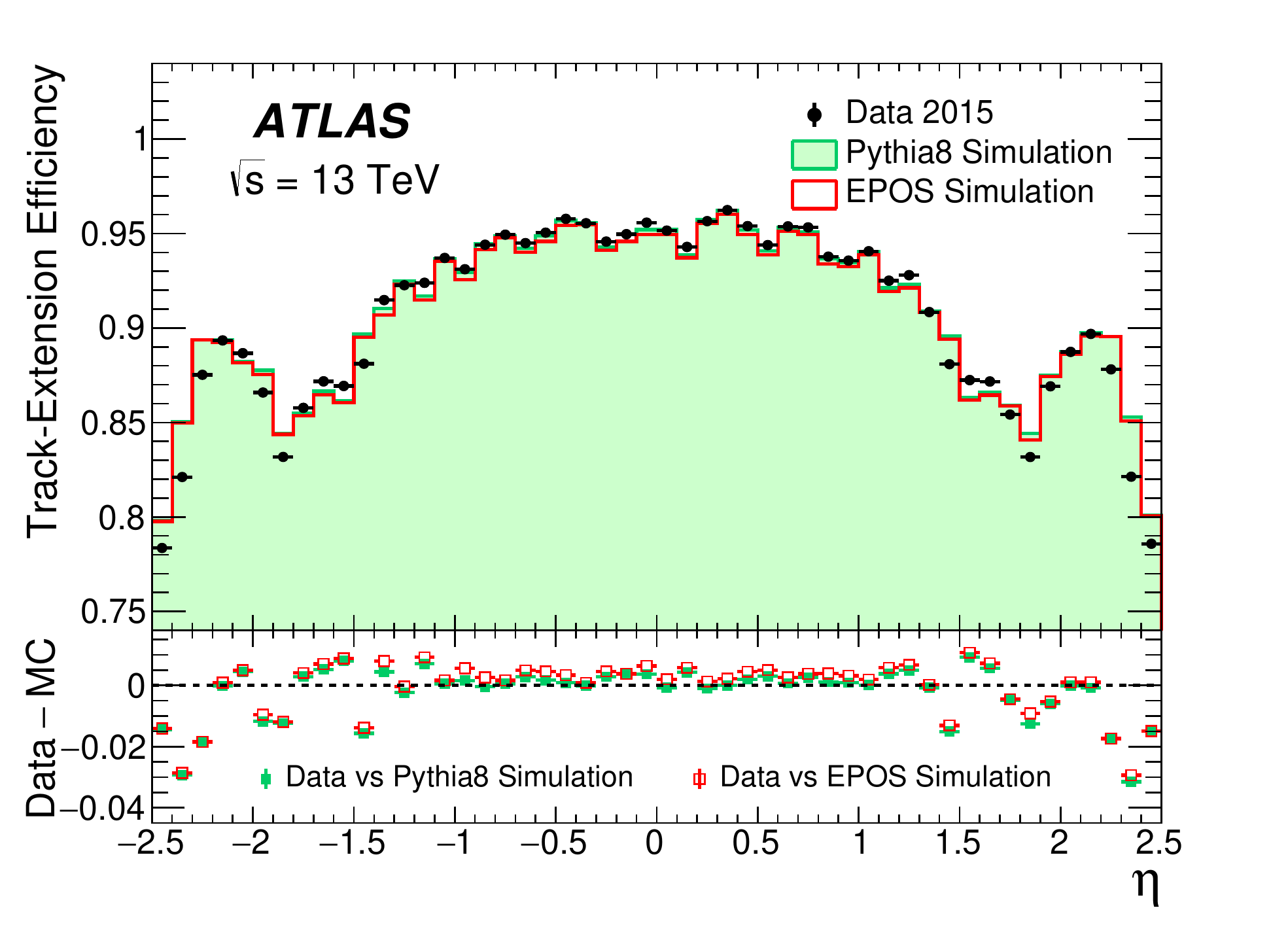}

}
\subfigure[]{
  \label{fig:SCTExt_Pt_DataEposPythia_Diff}
  \includegraphics[width=0.45\textwidth]{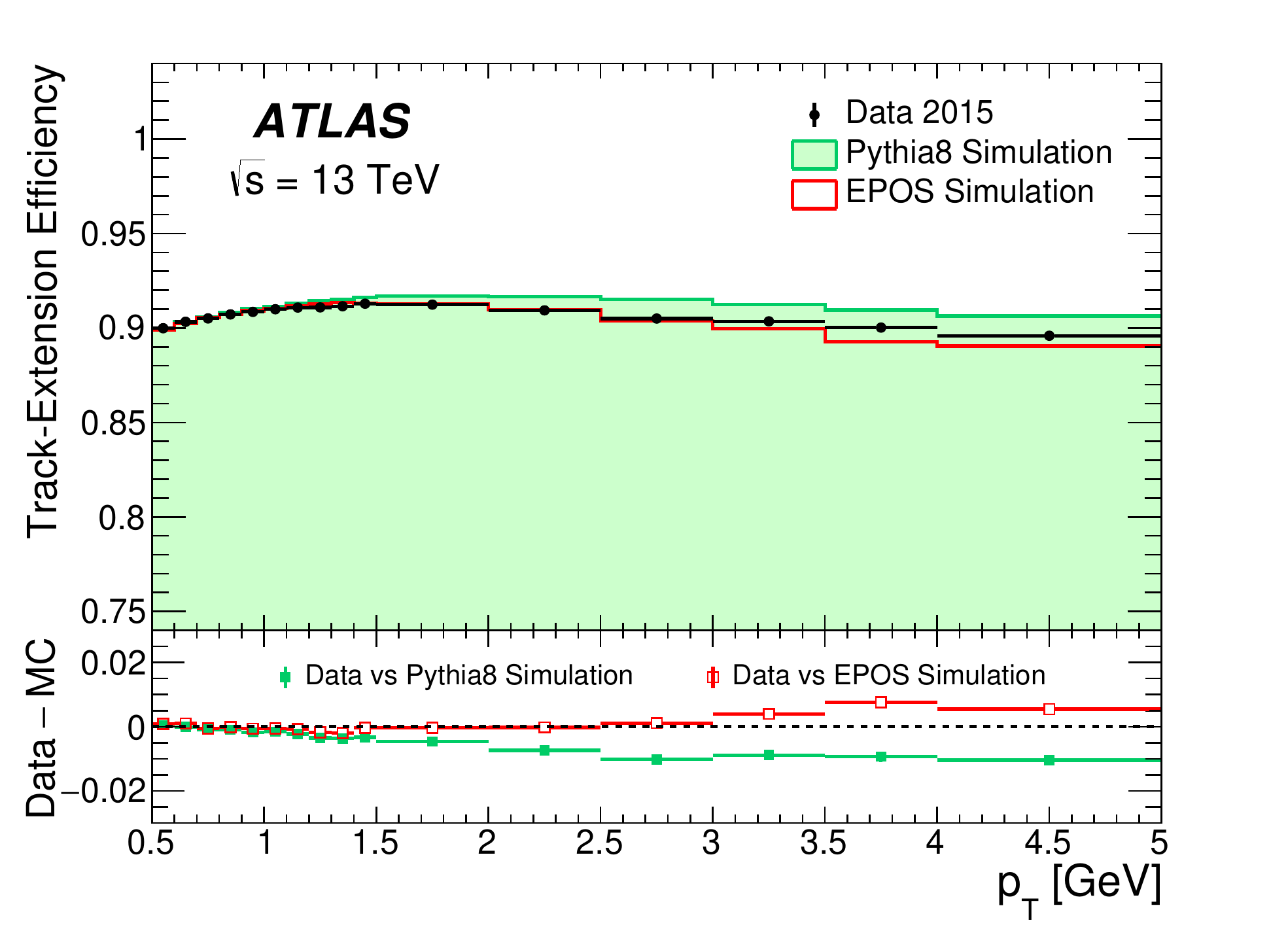}
}
\caption{Track-extension efficiency as a function of \subref{fig:SCTExt_Eta_DataEposPythia_Diff} $\eta$ and \subref{fig:SCTExt_Pt_DataEposPythia_Diff} $\pt$ of the tracklets in a comparison between data, \textsc{Pythia~8} and \textsc{Epos}.}
\label{fig:SCTExt_DataEposPythia_Diff}
\end{center}
\end{figure}

The track-extension efficiency $\SctExtEff(\eta)$, averaged over $\phi$, is shown in Figure \ref{fig:SCTExt_Eta_DataEposPythia_Diff}. The distribution is approximately constant around a value of 95\% within $|\eta|<0.5$, gradually falling towards a local minimum of around $83\%$ at $|\eta|\simeq1.9$. The efficiency recovers to around 90\% at $|\eta|\simeq 2.2$, and then falls again as $|\eta|$ increases further. This structure of $\SctExtEff(\eta)$ reflects the distribution of material as a function of $\eta$, as shown in Figure \ref{figures:matmap_etascan}. The MC simulation describes the overall structure of the  $\eta$ dependence, and there is good agreement in the central region of $|\eta|<1$. Nevertheless, discrepancies at the level of a few percent are observed in the forward region. Figure \ref{fig:SCTExt_Pt_DataEposPythia_Diff} shows the average track-extension efficiency as a function of $\pt$, integrated over $\eta$ and $\phi$. The $\pt$ dependence is also well described by the MC simulation, and the data points are between those of the two MC generators, \textsc{Pythia8} and \textsc{Epos}.

\section{Measurement of material in data and MC simulation}
\label{sec:analysis}

An assessment of the accuracy of the geometry model is performed through the comparison of the data and the MC simulation. In this section, the quantitative details of each material measurement are described.

\subsection{Hadronic interactions}
\label{sec:meas_hadint}
The ratio of the numbers of hadronic interaction vertices in data and \textsc{Epos} MC simulation using the \emph{updated} geometry model, referred to as the \emph{rate ratio}, is used as the primary measurement observable. Two comparisons are presented, the first measurement is referred to as the \emph{inclusive rate ratio}, $R^{\mathrm{incl}}_{i}$. It is determined for each radial region~$i$ listed in Table~\ref{tbl:hadInt:section_definition}, and defined as:

\begin{eqnarray*}
R_{i}^{\mathrm{incl}} = \frac{n_{i}^{\mathrm{data}}}{S_{\mathrm{BP}}\cdot c_{i}\cdot n_{i,{\mathrm{total}}}^{\mathrm{MC}}}~,
\end{eqnarray*}

where $n_{i}^{\mathrm{data}}$ represents the number of hadronic interaction candidates in the sample within the radial region~$i$. The term $S_{\mathrm{BP}}$ is the normalisation factor for the MC simulation common to all the radial regions and is derived from the rate observed at the beam pipe. The factor $c_{i}$ is the relative track reconstruction efficiency correction to the MC simulation for the radial region~$i$, which is estimated using \Kshort~samples. The number $n_{i,{\mathrm{total}}}^{\mathrm{MC}}$ is the sum of all hadronic interaction candidates, including true hadronic interactions, combinatorial fakes, in-flight decays and photon conversions. The rate of in-flight decay background vertices is scaled by an appropriate correction factor, following the approach in Ref.~\cite{PERF-2011-08}.

The second measurement is the \emph{background-subtracted rate ratio}, $R^{\mathrm{subtr}}_{i}$:

\begin{eqnarray*}
R_{i}^{\mathrm{subtr}} = \frac{n_{i}^{\mathrm{data}} - S_{\mathrm{BP}}\cdot c_{i}\cdot n_{i,{\mathrm{BG}}}^{\mathrm{MC}}}{S_{\mathrm{BP}}\cdot c_{i}\cdot n_{i,{\mathrm{had}}}^{\mathrm{MC}}}~,
\end{eqnarray*}

where $n_{i,{\mathrm{had}}}^{\mathrm{MC}}$ is the number of truth-matched hadronic interactions in the sample of candidate vertices, and $n_{i,{\mathrm{BG}}}^{\mathrm{MC}}$ represents the other candidate vertices. For in-flight decays, $n_{i,{\mathrm{BG}}}^{\mathrm{MC}}$ is corrected as discussed in Section~\ref{sec:hadInt_sys}.  Since $c_{i}$ is related to the track reconstruction efficiency, the same correction factor is applied to both the signal and the background processes. If the geometry model description is accurate, $R_{i}^{\mathrm{incl}}$ and $R_{i}^{\mathrm{subtr}}$ should be consistent with unity, while any deviation from unity, outside of the measurement uncertainty, may be associated with an inaccuracy in the material description.

Several corrections must be applied to both the data and simulation in order to compare the hadronic interaction rate in data with simulation in a given radial region. In this section, the corresponding systematic uncertainties are discussed. The values of systematic uncertainties are summarised in Table \ref{tbl:hadInt:result} in Section~\ref{sec:results}.

\subsubsection{Corrections}
\label{sec:hadInt_corr}
\paragraph{Radial position of barrel layers}
As discussed in Section~\ref{sec:overview_offset}, the barrel layers in data have a finite offset perpendicular to the $z$-axis. In order to compare radial distributions in data and simulation, the positions of secondary vertices in the data are corrected by the offset in the $x$--$y$ plane. Since the classification of the radial regions is unambiguous after the offset corrections, no systematic uncertainties are assigned to this correction.

\paragraph{Normalisation of rate at the beam pipe}
The material in the beam pipe is the part of the inner detector's material which is known with the greatest accuracy. Consequently, an \emph{in situ} rate normalisation using the beam pipe is applied in this study. The geometry model description of the beam pipe at $|z|>40~\millimeter$ is assumed to be accurate to within 1\% precision. The range $|z|<40~\millimeter$ is not used as a part of the reference material due to the observation of a deficit of material in the simulation corresponding to the polyimide tape, as described in Section~\ref{sec:overview_beampipe}.

\paragraph{Primary interaction reweighting}
In order to correct the primary-vertex $z$-distribution in the MC simulation, as well as the primary-particle flux, as a function of $\eta$, a reweighting correction is applied. The track multiplicity density, as a function of the primary-vertex $z$-position, $\pt$ and $\eta$ of the track, is calculated. The ratio of the spectra in the data and simulation is used as a weight for each secondary vertex. The $\pt$ of the primary particle that created the hadronic interaction vertex cannot be directly determined due to the possible production of undetected neutral particles in hadronic interactions. Instead, the reconstructed vertex's vectorial sum $\pt$ is used to parameterise the correction. 
The impact of primary-particle reweighting is found to change the data-to-MC simulation rate ratio by less than 1\%.

\begin{figure}[htbp]
\begin{center}
\subfigure[]{
  \label{fig:effratio}
  \includegraphics[width=0.47\textwidth]{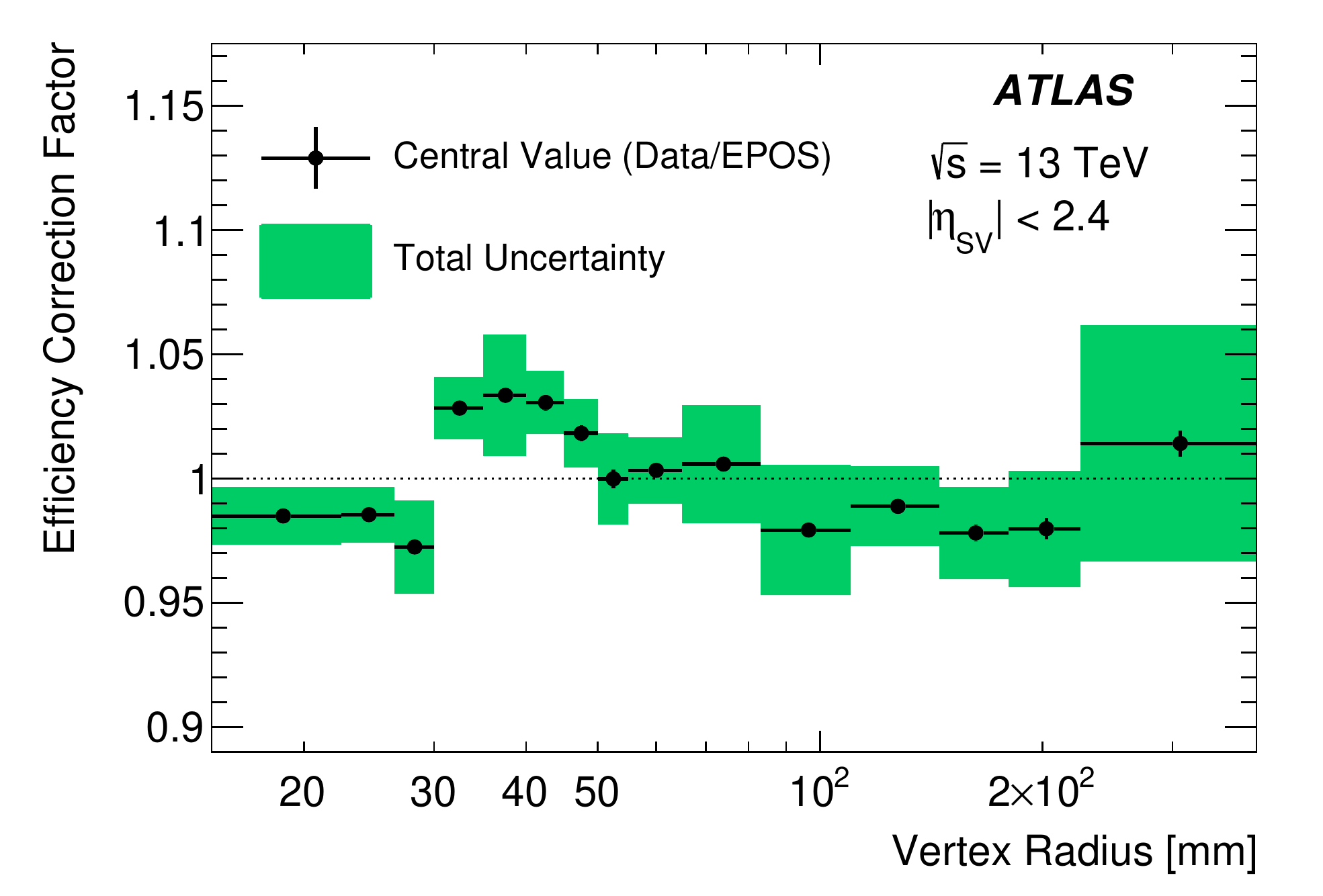}
}
\subfigure[]{
  \label{fig:compare_R_corr_distorted}
  \includegraphics[width=0.47\textwidth]{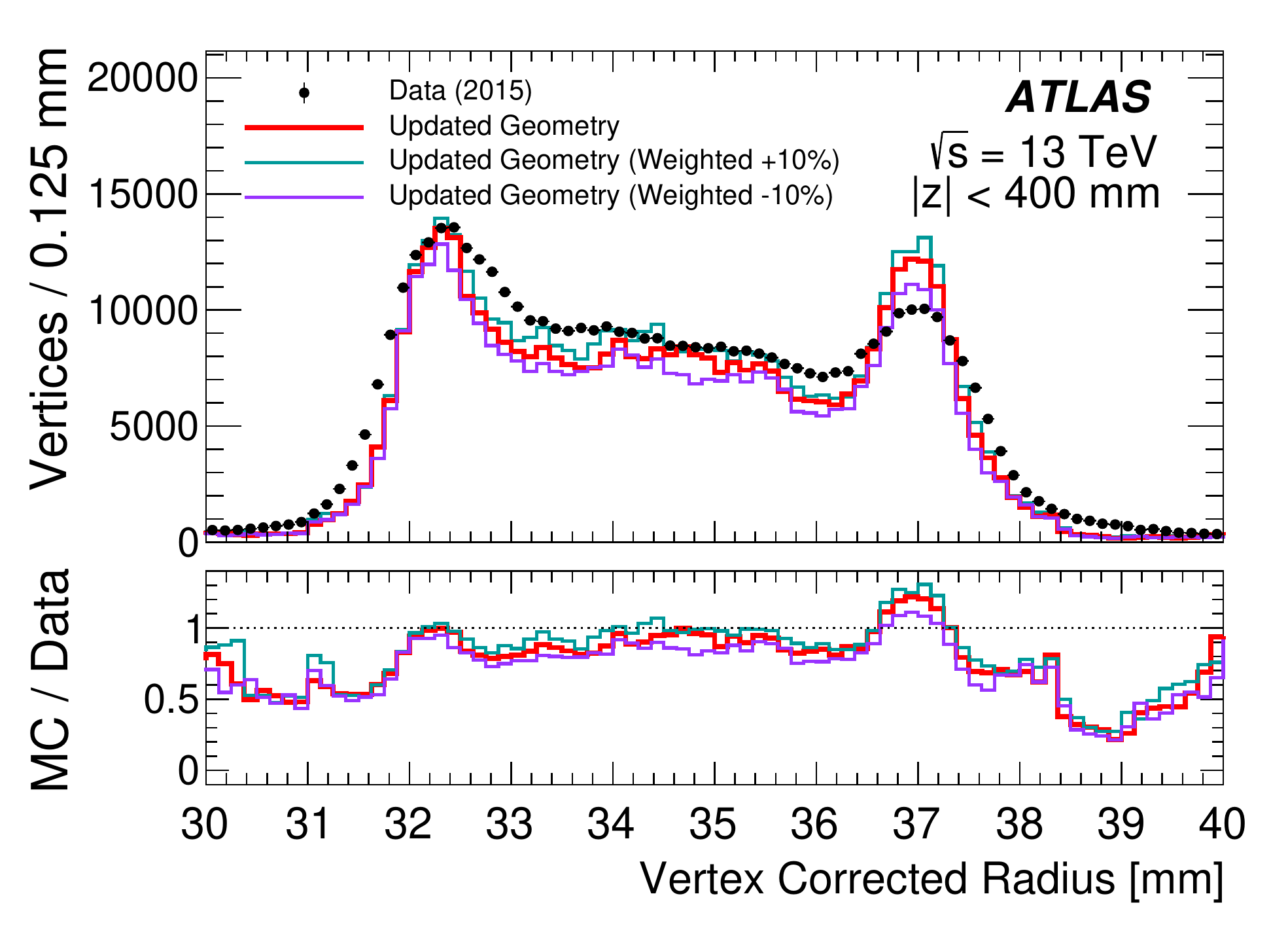}
}
\caption{\subref{fig:effratio} The estimated data-to-MC ratio of reconstruction efficiency and its uncertainty as a function of vertex radius. \subref{fig:compare_R_corr_distorted} Radial distribution of hadronic interaction candidates at the IBL region ($30~\millimeter<r<40~\millimeter$) for the data and the \textsc{Pythia~8} MC simulation with the \emph{updated} geometry model, together with the ``IBL $+10\%$'' and ``IBL $-10\%$'' distorted geometry samples listed in Table~\ref{tbl:mc_lists}.}
\label{fig:hadInt_bg_control}
\end{center}
\end{figure}

\paragraph{Reconstruction efficiency}
\label{sec:hadInt_recoeff_corr}
The reconstruction efficiency is assumed to be qualitatively well described in the MC simulation~\cite{STDM-2015-02}. The correction to the reconstruction efficiency in the MC simulation as a function of vertex radius is estimated using \Kshort~decays as a control sample. A sample of \Kshort~candidates is obtained using the same selection criteria as used for hadronic interaction candidates with an inverted $\Kshort$ veto requirement. The rate of \Kshort~for a given bin of \Kshort-system $\pt$ is obtained by fitting the invariant mass spectrum around the \Kshort~mass. For the fitting, the sum of a double Gaussian function (for the signal) constrained to have a common mean and a linear function (for the background) is used. The integral of the double Gaussian component is used to deduce the background subtracted-$\Kshort$ rate. The MC simulation rate is then re-weighted for each \Kshort-system $\pt$ bin to fit to the data before comparing the rate as a function of vertex radius. Reweighting performed as a function of the \Kshort-system $\pt$ is considered to give a more accurate normalisation than simply reweighting the total rate, due to the correlation between \Kshort system-$\pt$ and the decay vertex position due to the lifetime of the \Kshort meson. The ratio of the data rate to the MC simulation rate at a given radius after reweighting, 0.97--1.03 depending on radius (see Figure \ref{fig:effratio}), is considered as an estimate of the correction factor to be applied to the vertex reconstruction efficiency for the hadronic interaction candidates in the MC simulation.

\begin{figure}[tbp]
\begin{center}
\subfigure[]{
  \label{fig:compare_h1_sumpT_BP_all}
  \includegraphics[width=0.47\textwidth]{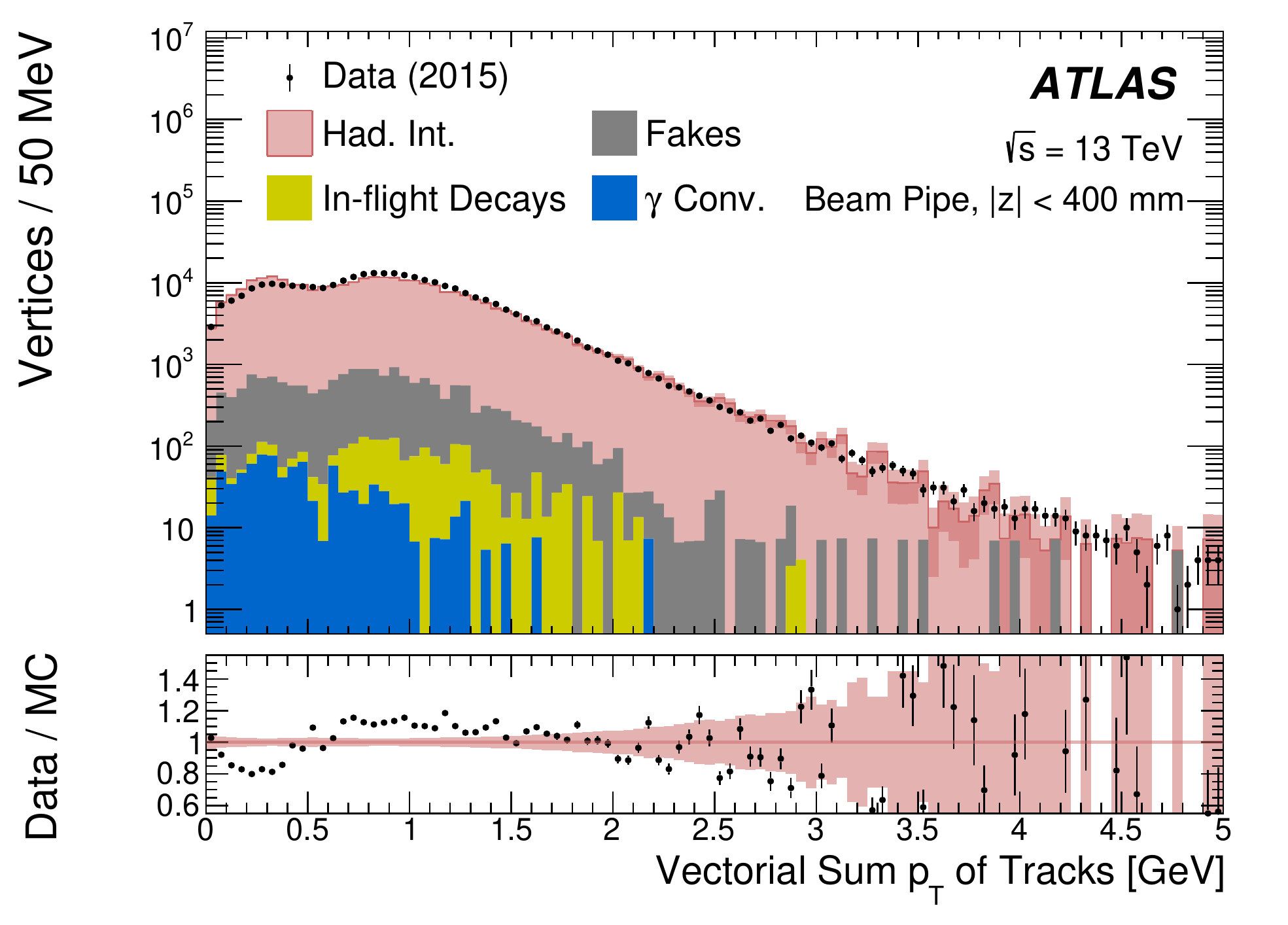}
}
\subfigure[]{
  \label{fig:compare_h1_sumpT_BLayer_all}
  \includegraphics[width=0.47\textwidth]{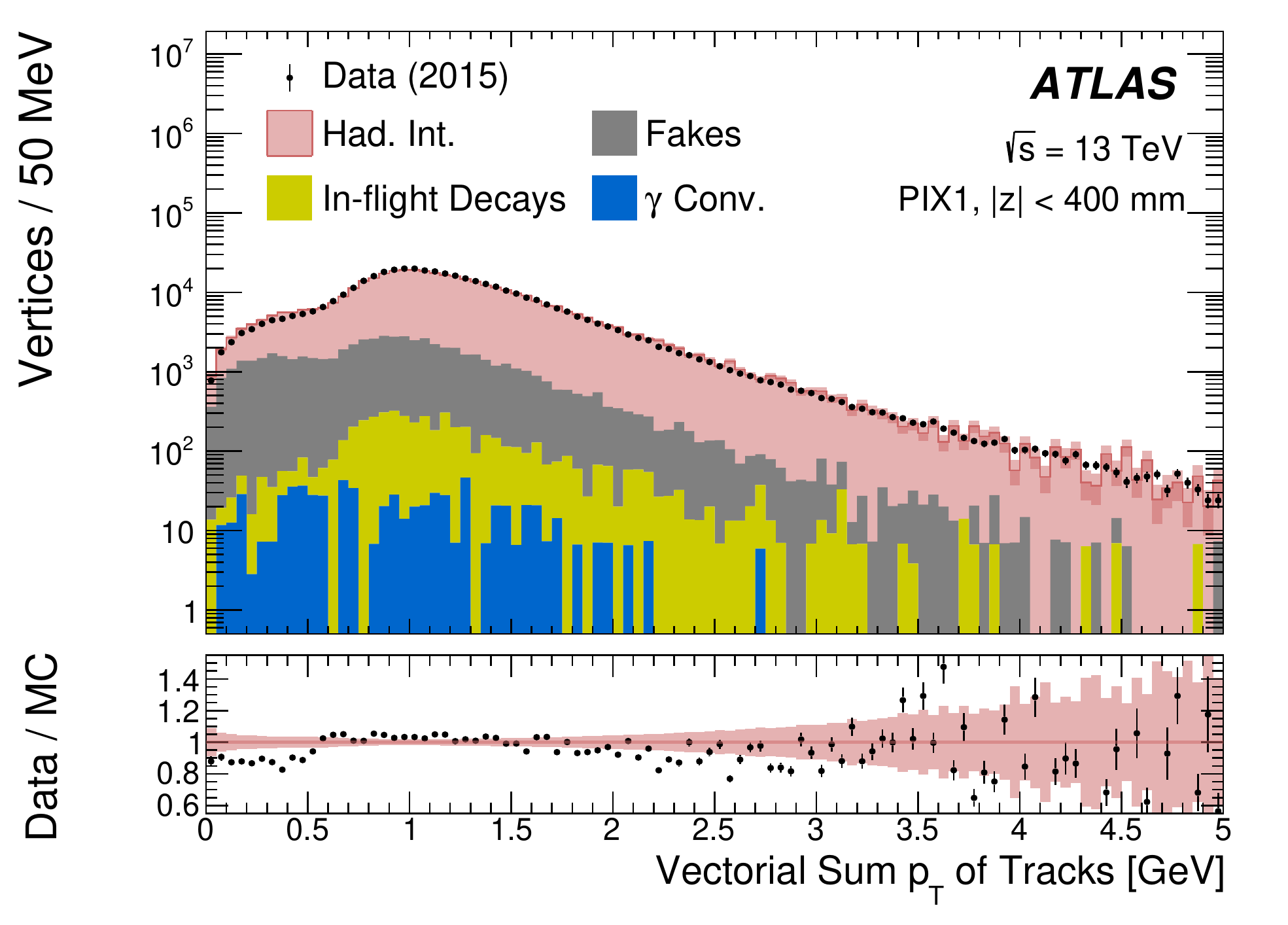}
}
\subfigure[]{
  \label{fig:compare_R_SCT_Epos}
  \includegraphics[width=0.47\textwidth]{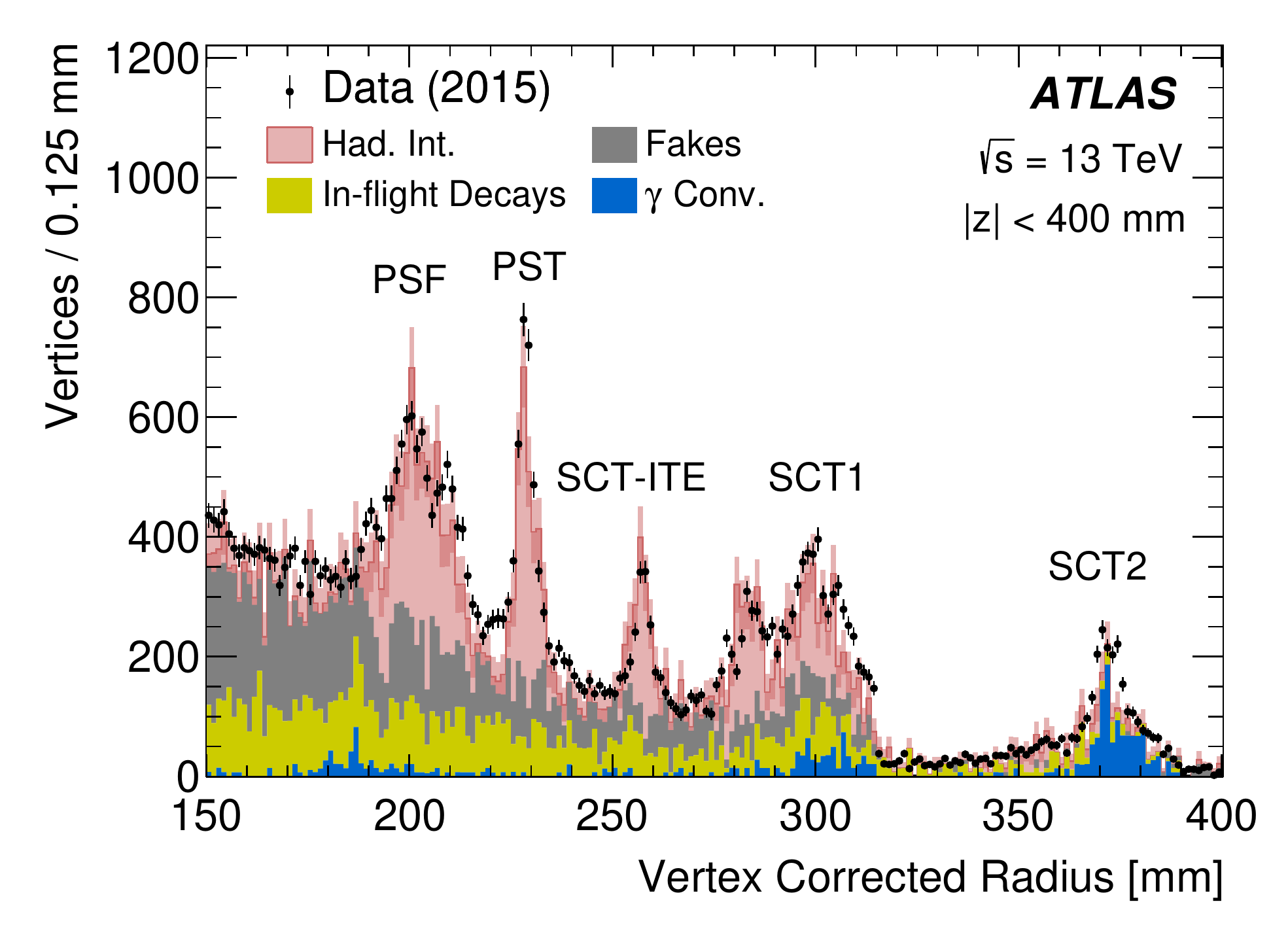}
}
\subfigure[]{
  \label{fig:compare_h1_minOpAng_gap1_all_Epos}
  \includegraphics[width=0.47\textwidth]{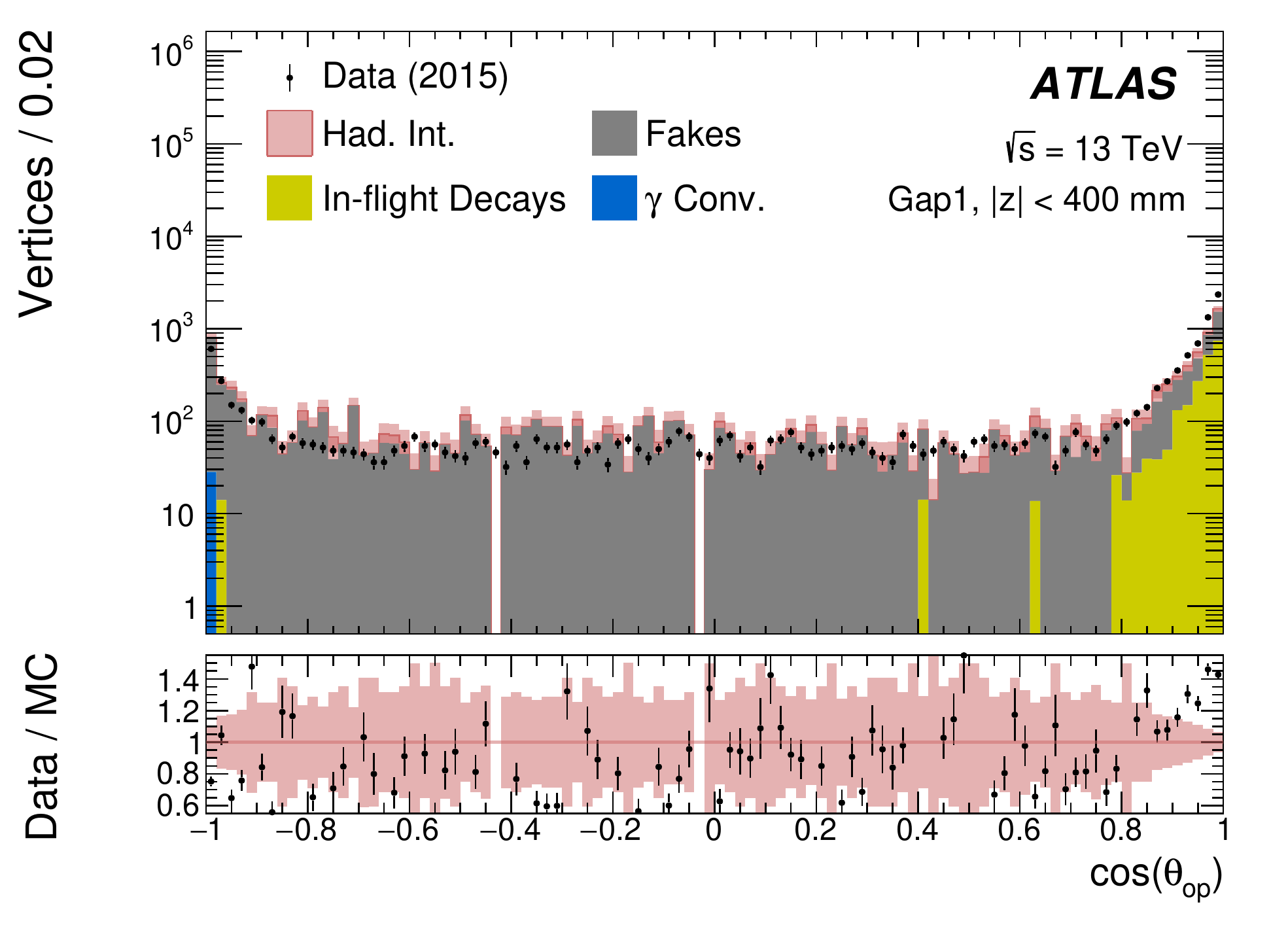}
}

\caption{Distribution of the vertex vectorial sum of \pt in hadronic interaction candidates \subref{fig:compare_h1_sumpT_BP_all} at the beam pipe in $22.5~\millimeter<r<26.5~\millimeter$, and \subref{fig:compare_h1_sumpT_BLayer_all} at the innermost pre-existing pixel layer (PIX1) in $45~\millimeter<r<75~\millimeter$. \subref{fig:compare_R_SCT_Epos} Radial distribution of hadronic interaction candidates in $150~\millimeter<r<400~\millimeter$. Background rates are not weighted for the \textsc{Epos} MC simulation. \subref{fig:compare_h1_minOpAng_gap1_all_Epos} Distribution of the cosine of the opening angle between two tracks in the laboratory frame $\cos(\theta_{\mathrm{op}})$ for hadronic interaction candidates within the material gap at $73~\millimeter<r<83~\millimeter$ (Gap1) where fake vertices and in-flight decays are enhanced. Background rates are not weighted for the \textsc{Epos} MC simulation. The band shown in \subref{fig:compare_h1_sumpT_BP_all}, \subref{fig:compare_h1_sumpT_BLayer_all} and \subref{fig:compare_h1_minOpAng_gap1_all_Epos} indicates the statistical uncertainty of the MC simulation.}
\label{fig:hadInt_bg_control}
\end{center}
\end{figure}

\subsubsection{Description of systematic uncertainty estimation}
\label{sec:hadInt_sys}
\paragraph{Physics modelling of hadronic interactions}
Modelling of hadronic interactions in the \textsc{Geant4} simulation is a source of uncertainty in the MC simulation rate, since the acceptance and efficiency of the secondary vertex reconstruction depend on the hadronic interaction kinematics. The model used in the simulation, \texttt{FTFP\_BERT}, is found to describe the kinematic properties of hadronic interactions fairly well, and no correction is applied to the obtained rate. However, it is also observed that the description is not totally accurate, and some differences are visible in particular at smaller vertex radii. Figures~\ref{fig:compare_h1_sumpT_BP_all} and \ref{fig:compare_h1_sumpT_BLayer_all} show vectorial sum of $\pt$ of the tracks associated with the vertex of hadronic interaction candidates at the beam pipe and at the first layer of the pre-existing pixel detector (PIX1) respectively. The description of the distribution of various kinematic variables is found to be generally better at outer radii than at the IBL. The fact that agreement between MC simulation and data for various kinematic distributions is better at outer radii is related to the acceptance of the track reconstruction. At outer radii, the angular phase space is more collimated due to the track reconstruction acceptance, so the kinematic distribution is less dependent on the detailed modelling of the angular distribution of outgoing particles from hadronic interactions.

In order to assess the systematic uncertainty of the hadronic interaction rate associated with the modelling of hadronic interactions for \texttt{FTFP\_BERT}, a data-driven approach is taken by varying the kinematic selection criteria. Four variables (the cosine of the opening angle between two tracks in the laboratory frame $\cos(\theta_{\mathrm{op}})$, vertex vectorial sum of $\pt$, leading-track $\pt$, and sub-leading-track $\pt$) are considered in order to assess the level of agreement between the data and simulation. The degree of agreement is evaluated by comparing the data and simulation rates over the entire spectrum to the integral over the two 50\% quantiles of the distribution, where the common quantile threshold for both data and MC simulation is calculated based on the data distribution. The simulation rate is renormalised at the beam-pipe radius (for $|z|>40~\mathrm{mm}$) for each selection. For the beam pipe, the variation of the data-to-MC simulation rate ratio before renormalisation is taken. The maximum difference amongst the four kinematic variables is taken as the systematic uncertainty in the physics modelling of the data-to-MC simulation rate ratio in the given radial region. Such a variation is evaluated for the inclusive rate ratio and the background-subtracted rate ratio separately. The estimated uncertainty is 5--18\% depending on the radial region.

\paragraph{Background estimation}
The purity of hadronic interactions in the sample of hadronic interaction candidates decreases as a function of radius, as presented in Table~\ref{tbl:hadInt:purity}. The major background components are combinatorial fake vertices for smaller radii up to around the pixel support tube. For the SCT region, contamination from in-flight decays and photon conversions is significant, as shown in Figure~\ref{fig:compare_R_SCT_Epos}. There is a small fraction of hadronic interactions in Gap1 (10\%) and Gap2 (5\%), potentially reflecting the migration of vertices from the nearby material regions and interactions with gases.

\begin{table}[t!]
\begin{scriptsize}
\centering
\begin{tabular}{c|cccccccccccc|cc}
\hline
\hline
Radial region & BP & IPT & IBL & IST & PIX1 & PIX2 & PIX3 & PSF & PST & SCT-ITE & SCT1 & SCT2 & Gap1 & Gap2\\
Purity [\%]   & 94 & 90  & 90  & 87  & 87   & 85   & 72  & 43   & 64  & 58      & 49   & 23   & 10   & 5\\
\hline
\hline
\end{tabular}
\caption{Estimated purity of hadronic interactions for the candidate vertices in each radial region in the \textsc{Epos} MC simulation. Purity values are given before correcting the background scale factors. For the definition of radial regions, see Table \ref{tbl:hadInt:section_definition}.}
\label{tbl:hadInt:purity}
\end{scriptsize}
\end{table}

The uncertainty in the rate of the combinatorial fake vertices is estimated using the rate within the two material gap regions: $73~\millimeter<r<83~\millimeter$ (Gap1) and $155~\millimeter<r<185~\millimeter$ (Gap2). A pure sample of fake vertices may be found within the region $\cos(\theta_{\mathrm{op}})<0.8$, as shown in Figure~\ref{fig:compare_h1_minOpAng_gap1_all_Epos}. It is confirmed that the rate in the MC simulation agrees with the data. No additional corrections are applied to the rate in the MC simulation for these components. The difference between the rate of combinatorial fake vertices in \textsc{Epos} and \textsc{Pythia~8} simulations for each radial region is taken as an estimate of uncertainty in the scale factors.

A difference in the total rate of hadronic interaction candidates is observed in the material gap regions mentioned above. The differences is associated with the rate of in-flight decays. To correct for this, a scale factor for this background is calculated to bring the rate into agreement with the data within each gap region. The estimated scale factor is 1.12 for Gap1 and 1.35 for Gap2. Since Gap2 is closer to the region where the contamination from in-flight decays is significant, the scale factor obtained at Gap2 is taken as the central value, and the difference between the scale factors obtained within the two gap regions is taken as the uncertainty in the scaling of in-flight decays. This uncertainty is dominant in regions beyond the outermost pixel layer, where the background contamination is most significant.

\paragraph{Primary particle flux}
The hadronic interaction vertex rate depends on the flux and species of primary hadrons and has an associated uncertainty. This uncertainty is partially suppressed by the \emph{in situ} normalisation using the rate at the beam pipe. The residual effect of this uncertainty after the normalisation is estimated by taking the relative difference in the rate of hadronic interaction candidates between the \textsc{Epos} and \textsc{Pythia~8} simulation. The size of this systematic uncertainty is found to be negligible (less than 1\%) compared to other uncertainties.

\paragraph{Reconstruction efficiency}
The dominant uncertainty in the correction factor for the reconstruction efficiency is found to be the spectrum of primary hadrons, and its size is estimated by taking the difference in the efficiency between the \textsc{Epos} and \textsc{Pythia~8} simulations. The uncertainty in the reweighting of the reconstruction efficiency is estimated from an alternative reweighting derived as a function of both the \Kshort-system $\pt$ and $\eta$. The variation is relatively small compared to the event generator dependence. The size and uncertainty of the correction is 1--6\% depending on vertex radii, as shown in Figure~\ref{fig:effratio}.

\paragraph{Measurement closure}
It is expected that the rate of reconstructed vertices in a given local region is primarily proportional to the amount of the material in the same local region. Any effects which deviate from this proportionality should be taken into account as another systematic uncertainty, referred to as \emph{closure}. The closure of the measurement is tested by using MC samples with distorted geometry models and comparing the measured rates to the predictions. Any deviation of a measurement from the predicted effect of a material change, larger than the MC simulation's statistical uncertainty, is considered as a systematic uncertainty. Figure~\ref{fig:compare_R_corr_distorted} illustrates the variation of the vertex rate in the IBL region ($30~\millimeter<r<40~\mathrm{mm}$) using the IBL $+10\%$ and IBL $-10\%$ samples. Only variations with a size greater than twice the statistical uncertainties are considered, and for each radial region, the maximum difference among all different samples is taken as the systematic uncertainty for non-closure of the measurement. This uncertainty is found to be 2--4\% depending on the vertex radius.

\subsection{Photon conversions}
\label{sec:meas_conv}

Due to the relatively poor resolution of the radial position of the reconstructed conversion vertices, the accuracy of the simulated description of the material is assessed in nine radial regions of the ID using a dedicated template fitting procedure. The nine radial regions, as shown in Table\,\ref{tbl:hadInt:section_definition}, are defined at the truth level of the simulation, based on the true radial position of the conversion vertex.

The fitting procedure consists of a simultaneous binned maximum-likelihood template fit of the reconstructed conversion radius distribution in two regions of the longitudinal position of the conversion vertex: $|z| < 50\,\mathrm{mm}$ (referred to as the \emph{inner} region) and $50~\millimeter < |z| < 400\,\mathrm{mm}$ (referred to as the \emph{outer} region). These two regions are treated separately, as the \emph{outer} region is used to normalise the simulated photon flux to the data at the beam-pipe radius, independently of the \emph{inner} region, which exhibits an excess of conversion candidates in the beam-pipe region. The \emph{updated} geometry and the \textsc{Pythia~8} MC generator are used to derive all templates.

\begin{figure}[t!]
\begin{center}

\subfigure[]{
\label{fig:ConvFit:InnerPIX}
\includegraphics[width=0.45\textwidth]{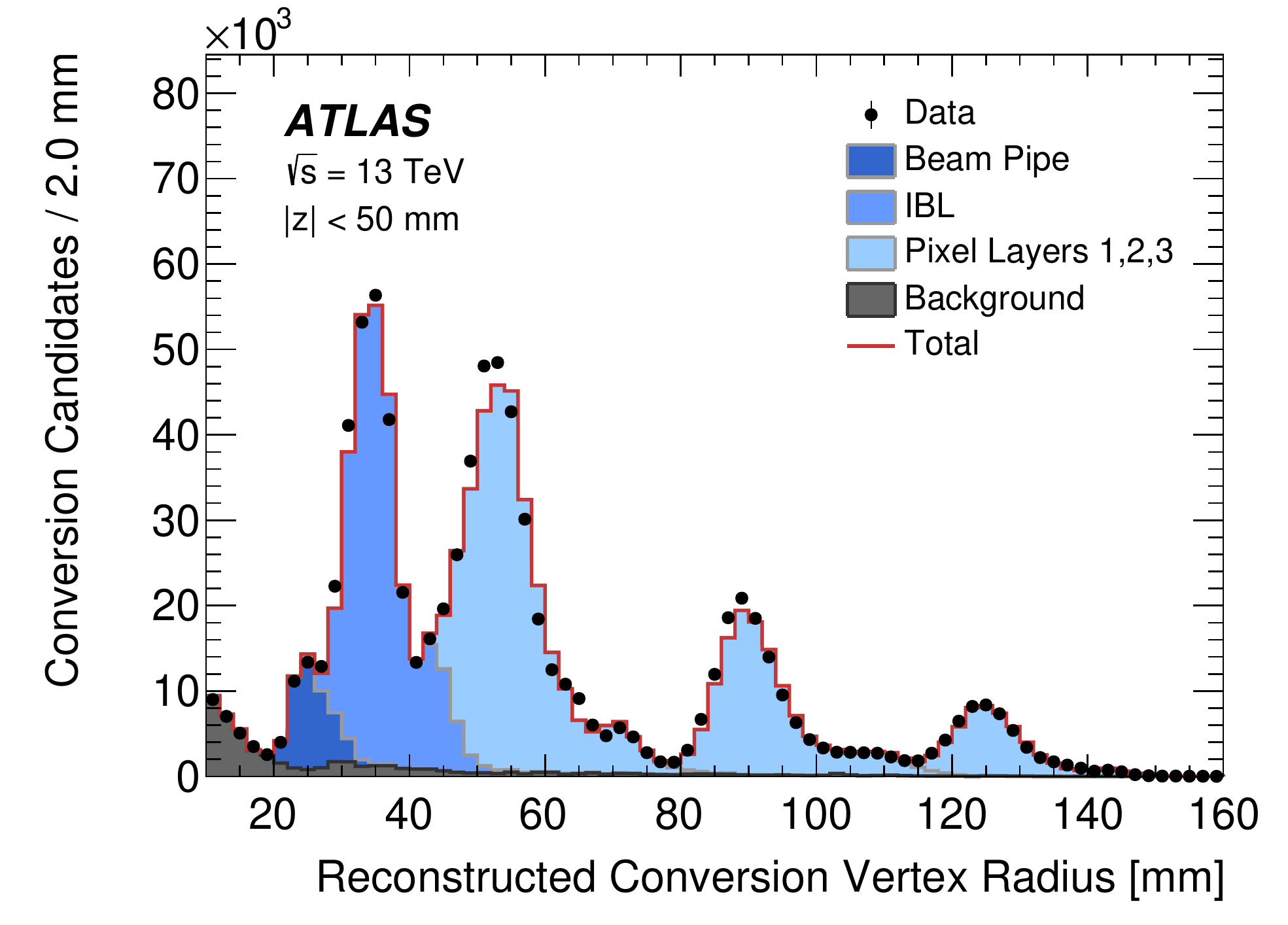}
}
\subfigure[]{
\label{fig:ConvFit:OuterPIX}
\includegraphics[width=0.45\textwidth]{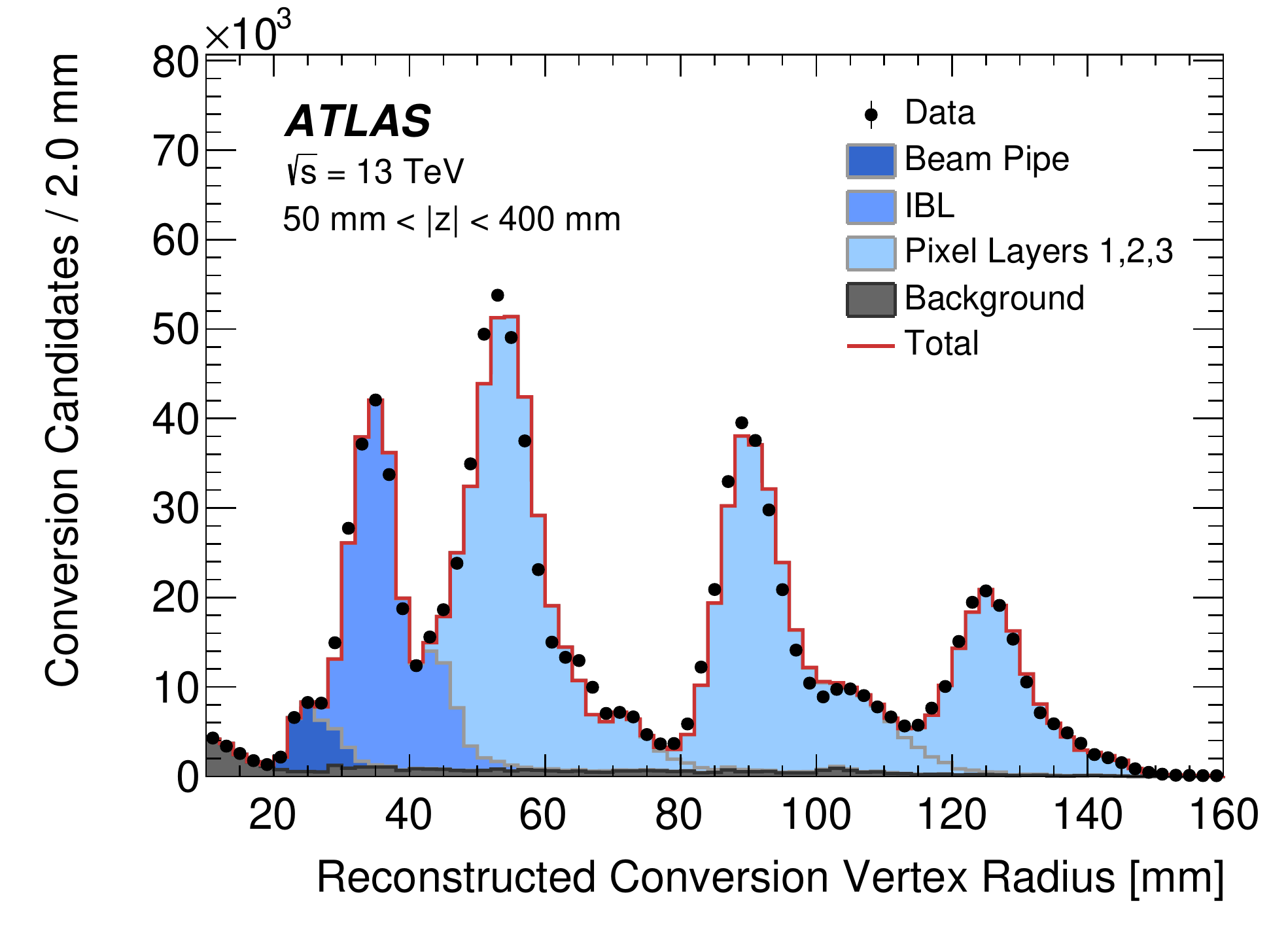}
}

\subfigure[]{
\label{fig:ConvFit:InnerSCT}
\includegraphics[width=0.45\textwidth]{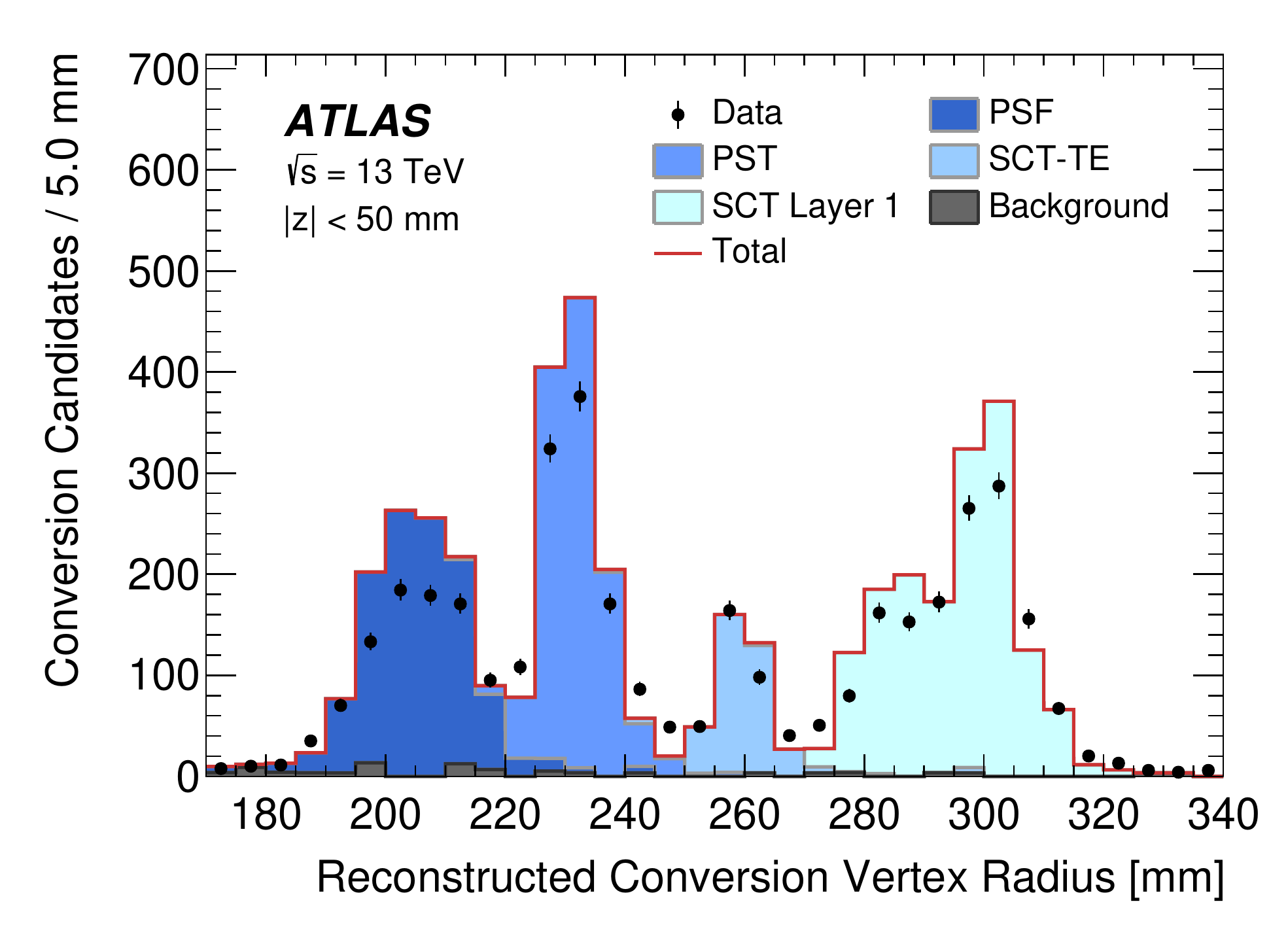}
}
\subfigure[]{
\label{fig:ConvFit:OuterSCT}
\includegraphics[width=0.45\textwidth]{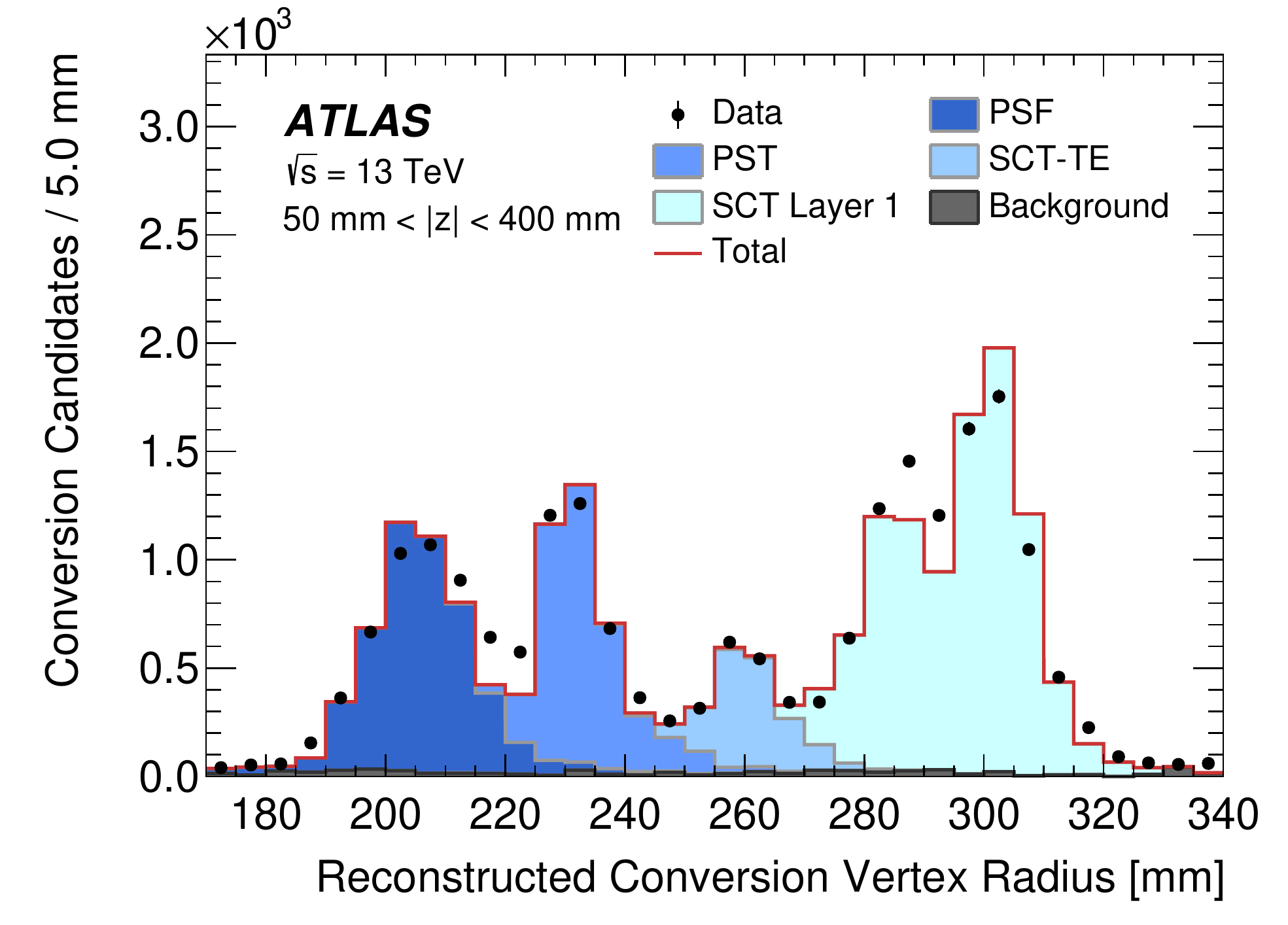}
}

\caption{The result of the binned maximum-likelihood fit to the data sample, described in Section~\ref{sec:meas_conv}. The \emph{inner} pixel and SCT regions are shown in \subref{fig:ConvFit:InnerPIX} and \subref{fig:ConvFit:InnerSCT}, respectively, while the \emph{outer} pixel and SCT regions are shown in \subref{fig:ConvFit:OuterPIX} and \subref{fig:ConvFit:OuterSCT}, respectively.}
\label{fig:ConvFit}
\end{center}
\end{figure}

Nine individual probability distribution functions (PDFs), based on the radial regions, are derived from the reconstructed conversion radius distributions of the simulated samples, where the reconstructed conversion candidate is matched to a true photon conversion. An additional template, derived from reconstructed conversion candidates which are not matched to true photon conversions, is derived to describe the background (fake) contribution to the reconstructed conversion radius distribution. The rate of photon conversion candidates associated with each individual PDF (including the fake conversion background contribution), denoted by the index $i$, is defined by:

\begin{equation*}
n_{i}^{\mathrm{data}} = R_{i}\cdot n_{i}^{\mathrm{MC}}\cdot S\,,
\end{equation*}

where $n_{i}^{\mathrm{data}}$ denotes the expected number of conversions modelled by the PDF,~~$n_{i}^{\mathrm{MC}}$ denotes the raw number of photon conversion candidates  from which the template is built, $R_{i}$ is an individual layer scale factor and $S$ is a scale factor common to all of the templates (including the background template). The parameter $S$ is determined from the background-subtracted ratio (assuming the purity determined from the simulation sample) of the number of reconstructed conversions in data to the number in simulation in the region $20~\mathrm{mm} < r < 25\,\mathrm{mm}$ of the \emph{outer} $z$ region alone. This parameter, $S$, effectively scales the photon flux of the model to match that observed in data. The parameters of interest, $R_{i}$, are determined by the fitting procedure. They represent the ``best fit'' scale factors required to adjust the number of reconstructed conversions associated with each radial region in the model (after accounting for the global flux normalisation) to match the number of photon conversion candidates observed in data. The $R_{i}$ are common to the \emph{inner} and \emph{outer} regions in $z$, with the exception of the beam-pipe region background templates, where each of them has an independent $R$ parameter. The $R$ parameter associated with the beam-pipe radial region, for the \emph{outer} $z$ region, is fixed to unity to facilitate the photon flux normalisation, while the corresponding $R$ for the \emph{inner} $z$ region is freely determined in the fit. 
The result of the fitting is displayed in Figure~\ref{fig:ConvFit} for both the \emph{inner} and \emph{outer} regions. The values of systematic uncertainties are summarised in Table \ref{tbl:conv:result} in Section~\ref{sec:results}. The background contribution is obtained directly in the fit to data and systematic uncertainties in the modelling of the background beyond those discussed later in this section are considered negligible. The fit result in the PSF and PST radial regions for $|z| < 50\,\mathrm{mm}$, shown in Figure~\ref{fig:ConvFit:InnerSCT}, exhibits a local excess in the MC simulation of around 20\% with respect to the data. This effect is due to material structures localised within $|z| < 40\,\mathrm{mm}$ which induce a conversion rate, relative to the region $40\,\mathrm{mm} < |z| < 400\,\mathrm{mm}$, which is larger in the simulation than that observed in data. Such a local effect cannot be accommodated by the fit model since only a single $R_{i}$ parameter is considered for the PSF and PST across the full $|z|$ region studied.

\subsubsection{Corrections}
Several corrections are applied to the simulation in order to reliably compare the photon conversion rates in data and simulation. The corrections described are applied to the photon-conversion candidates used to build the template distributions described in Section~\ref{sec:meas_conv}.

\paragraph{Primary interaction reweighting}
Weights are assigned to photon-conversion candidates reconstructed within the simulation samples to account for small differences between the characteristics of events in data and simulation. These weights are constructed such that the primary vertex $z$ position and track multiplicity distributions in the simulation samples match the corresponding distributions in the data sample.

\paragraph{Radial position of barrel layers}
As discussed in Section~\ref{sec:overview_offset}, the axis of each cylindrical layer of the beam pipe, IBL, pixel barrel layers and other support tubes in data has a non-zero offset from the origin perpendicular to the $z$-axis of the ATLAS coordinate system. This effect is not present in the simulation samples. The conversion vertices reconstructed in the simulation are offset by a small additive correction. The corrections are derived from the shifts observed in the hadronic interactions analysis, described in Section~\ref{sec:overview_offset}. The magnitude of the correction is determined based on the true conversion vertex position and applied to the reconstructed conversion vertex position.

\subsubsection{Systematic uncertainties}
\paragraph{Photon flux normalisation}

Several sources of systematic uncertainty in the measured values of the $R_{i}$ parameters are considered in the following paragraphs. The uncertainty in the scale parameter $S$, described in Section\,\ref{sec:meas_conv}, is estimated to be $2.6\%$ and is associated with the statistical uncertainty in the number of photon conversion candidates reconstructed in the data and simulation, which are used to determine the value of the parameter $S$. The fit is repeated with the parameter $S$ varied by $\pm1\sigma$, and the average change in the $R_{i}$ parameters is used to estimate the uncertainty associated with the simulated photon flux normalisation. This uncertainty, statistical in nature, is summed in quadrature for each $R_{i}$, with the statistical uncertainty returned by the fitting procedure used to derive an overall statistical uncertainty in the measured values of the $R_{i}$.

\paragraph{Modelling of primary photon flux}

The photon conversion reconstruction efficiency and the photon conversion probability depend upon the kinematic properties of the primary photon flux. In particular, the photon conversion reconstruction efficiency depends strongly on the transverse momentum of the incident photon. In order to estimate the systematic uncertainty associated with the modelling of the primary photon flux by the \textsc{Pythia~8} generator, the analysis is repeated with the \textsc{Epos} simulation sample in place of data. This is motivated by the different $\ptgamma$ and $\etagamma$ spectra predicted by the two MC generators. The average change in the $R_{i}$ parameters outside of the statistical uncertainty is used to estimate the systematic uncertainty associated with the modelling of the primary photon flux by the \textsc{Pythia~8} generator.

\paragraph{Simulated description of photon conversion reconstruction efficiency}
The reconstruction efficiency directly affects the number of reconstructed photon conversions. Limitations in the accuracy of the ATLAS detector simulation can lead to differences in the photon conversion reconstruction efficiency between data and simulation. Potential differences in the behaviour of the photon conversion reconstruction efficiency between data and simulation would manifest themselves as changes in the relative number of conversions reconstructed in the two samples as the selection criteria are varied. To estimate the effects of these potential differences, the fit procedure is repeated for each of the following variations in the selection criteria described in Section~\ref{sec:reco_conversion}:

\begin{itemize}
\item the reconstructed conversion vertex satisfies: $\chi^{2}/N_{\mathrm{dof}} < 0.33$, $0.33 < \chi^{2}/N_{\mathrm{dof}} < 0.66$ or $0.66 < \chi^{2}/N_{\mathrm{dof}} < 1.00$;
\item the reconstructed photon conversion $\ptgamma$ satisfies: $0.5~\GeV < \ptgamma < 1~\GeV$, $1~\GeV < \ptgamma < 1.5~\GeV$ or $\ptgamma > 1.5~\GeV$.
\end{itemize}

These variations are chosen as they induce large changes in the photon conversion reconstruction efficiency, as a function of vertex position, with respect to the nominal selection.

The standard deviation of the variations in the $R_{i}$ obtained from the ensemble of six alternative fits from the nominal value is used to estimate the systematic uncertainty in the measured value of each $R_{i}$ associated with the simulated description of photon conversion reconstruction efficiency. The statistical contribution to the estimate of this uncertainty (associated with splitting the samples into subsets) is expected to be less than $1\%$ and is neglected.

\paragraph{Measurement closure}

To validate the performance of the method described in Section~\ref{sec:meas_conv}, a number of tests were performed in which data was replaced with simulated samples generated by the \textsc{Pythia~8} generator but simulated with modified detector geometries, shown in Table~\ref{tbl:mc_lists}. The average change in the $R_{i}$ parameters from their expected values, outside of the statistical uncertainty, is used to estimate the systematic uncertainty associated with any residual bias in the fitting procedure. This approach leads to the assignment of a systematic uncertainty of $3\%$, common to all $R_{i}$.

\subsection{Track-extension efficiency}
\label{sec:meas_sctext}

The difference in the number of nuclear interaction lengths between the real detector and the geometry model, $\varDelta N_{\lambda_{I}}^{\mathrm{Data-MC}}$,
results in a difference in the track-extension efficiency between data and simulation, $\varDelta\SctExtEff^{\mathrm{Data-MC}}(\eta)$.
This relation can be expressed as:

\begin{eqnarray}
\varDelta\SctExtEff^{\mathrm{Data-MC}}(\eta) \simeq -K(\eta) \cdot \varDelta N_{\lambda_{I}}^{\mathrm{Data-MC}}(\eta),
\label{eq:sctexteff_correction_factor}
\end{eqnarray}

where $K(\eta)$ is a scale factor.
The sensitivity of this method is proportional to the amount of material along the track's path but it is unable to identify accurately the radial position of the material.
The factor $K(\eta)$ accounts for the algorithmic reconstruction efficiency of the combined track as well as the fact that the tracklets arise from not only a sample of stable hadrons, but also contain contributions from weakly decaying hadrons and fake tracks.  It is necessary to establish the appropriate value of $K(\eta)$ to calculate the difference in the material.

The fraction of particles that remain on average after travelling through $N_{\lambda_{I}}$ of material is given by:
\begin{equation*}
f(N_{\lambda_{I}}) = {\mathrm e}^{-N_{\lambda_{I}}}.
\label{eqn:lambda}
\end{equation*}

Assuming the only loss is from interactions with the material and considering all the material located between the pixel and SCT detectors, then $f(N_{\lambda_{I}})=\SctExtEff$.
Thus, in the presence of any additional passive material ($\varDelta N_{\lambda_{I}}$) sufficiently thin ($\ll 1$),
the difference in track-extension efficiency between the nominal (\emph{nom}) geometry and in the modified (\emph{mod}) geometry
can be expressed as:
\begin{eqnarray}
\varDelta\SctExtEff^{\mathrm{mod-nom}} &=& f(N_{\lambda_{I,0}}+\varDelta N_{\lambda_{I}}) - f(N_{\lambda_{I,0}})\nonumber\\
&\simeq& -f(N_{\lambda_{I,0}})\cdot \varDelta N_{\lambda}\nonumber\\
&=& -\SctExtEff^{\mathrm{nom}}\cdot\varDelta N_{\lambda} ,
\label{eq:sctexteff_exp}
\end{eqnarray}

where $N_{\lambda_{I,0}}$ is the nominal material.
By comparing Eq. (\ref{eq:sctexteff_correction_factor}) with Eq. (\ref{eq:sctexteff_exp}), it is clear that $K(\eta) \approx \SctExtEff^{\mathrm{nom}}$.
This means that the change in material is approximatively equal to the normalised difference in track-extension efficiency between the two geometries (\NormSctExtEff):

\begin{equation}
\varDelta{N}^{\mathrm{mod-nom}}_{\lambda_{I}} \approx - \NormSctExtEff.
\label{eqn:NormExtEff}
\end{equation}

To verify the value of $K(\eta)$, a set of MC simulation samples where the material in the pixel service region is modified was created (see Table \ref{tbl:mc_lists}). The relation is shown in Figure \ref{fig:NormExtVSMaterial}, and any deviation from the expected dependence is taken as a systematic uncertainty. The values of systematic uncertainties are summarised in Table \ref{tab:SCTEffFinalResults} in Section~\ref{sec:results}.

\subsubsection{Systematic uncertainties}
\paragraph{Particle composition}

As the tracklets used in the measurement of $\SctExtEff$ originate from a variety of particles, the final  $\SctExtEff$ is actually the weighted sum of the $\SctExtEff$ for all particles:
\begin{eqnarray*}
\SctExtEff = \sum_{i = \mathrm{species} } { f_{i}\,\SctExtEff^{i}}~,
\end{eqnarray*}
where $f_i$ is the fraction of reconstructed tracklets associated with a particular particle species and  $\SctExtEff^{i}$ is its associated track-extension efficiency.

The probability that a particle interacts with matter, and hence its $\SctExtEff$, depends on the species of the particle.  In addition, in-flight decays of short-lived charged hadrons, e.g.~weakly decaying strange baryons, represent a source of irreducible background to the $\SctExtEff$ measurement as they exhibit an experimental signature identical to stable particles (e.g. pions) interacting with matter. Both of these effects give rise to a dependence of the $\SctExtEff$ measurement on the particle composition in the simulation.

Considerable differences in the predicted rate of particles between various event generators are observed. For example, the predicted cross-section of weakly decaying strange baryons in \textsc{Epos} is twice that of \textsc{Pythia~8}~\cite{STDM-2015-02}. As such, the impact of the particle composition on the measurement of $\SctExtEff$ needs to be evaluated in data or estimated from simulation.

Both the particle composition and the material interaction probability vary as a function of hadron $\pt$ and $\eta$. If there were a perfect description of the particle composition and material interactions in the simulation, no difference in the relative change of the $\SctExtEff$ between data and simulation would be observed when varying the $\pt$ range used to perform the measurement. However, notable deficiencies in the modelling are present in all MC generators, such as the fraction of short-lived charged hadrons as a function $\pt$ and thus the distribution of their decay length. To estimate the impact of these potential discrepancies on the measurement, the relative change in $\SctExtEff$ evaluated in data with respect to simulation (\textsc{Pythia~8}) is measured in four regions of $\pt$ within the range $0.5~\GeV<\pt<5~\GeV$. The maximum variation from the inclusive value is used to estimate a systematic uncertainty of 0.50\%. Furthermore, the difference in $\SctExtEff(\eta)$, integrated over $\pt$, between the \textsc{Epos} and \textsc{Pythia~8} simulations, as shown in Figure~\ref{fig:SCTExt_Eta_DataEposPythia_Diff}, reaches a maximum of 0.21\% and it is treated as a systematic uncertainty. To encompass both of these effects, a systematic uncertainty of $0.54\%$, the sum in quadrature of 0.50\% and 0.21\%, is applied in each $\eta$ bin.

\paragraph{Fake tracklets}
Fake tracklets are another source of bias in the measurement of $\SctExtEff$. In all simulated samples, the fraction of fake tracklets is less than 0.3\%. To estimate the uncertainty in $\SctExtEff$ associated with fake tracklets, a variation of $\pm50\%$ of the fake tracklet rate is considered, as recommended in Ref.~\cite{ATL-PHYS-PUB-2015-051}, and the corresponding variation of $\SctExtEff(\eta)$ is assigned as an uncertainty for each $\eta$ bin.

\paragraph{Calibration procedure}

\begin{figure}
\begin{center}
  \includegraphics[width=0.65\textwidth]{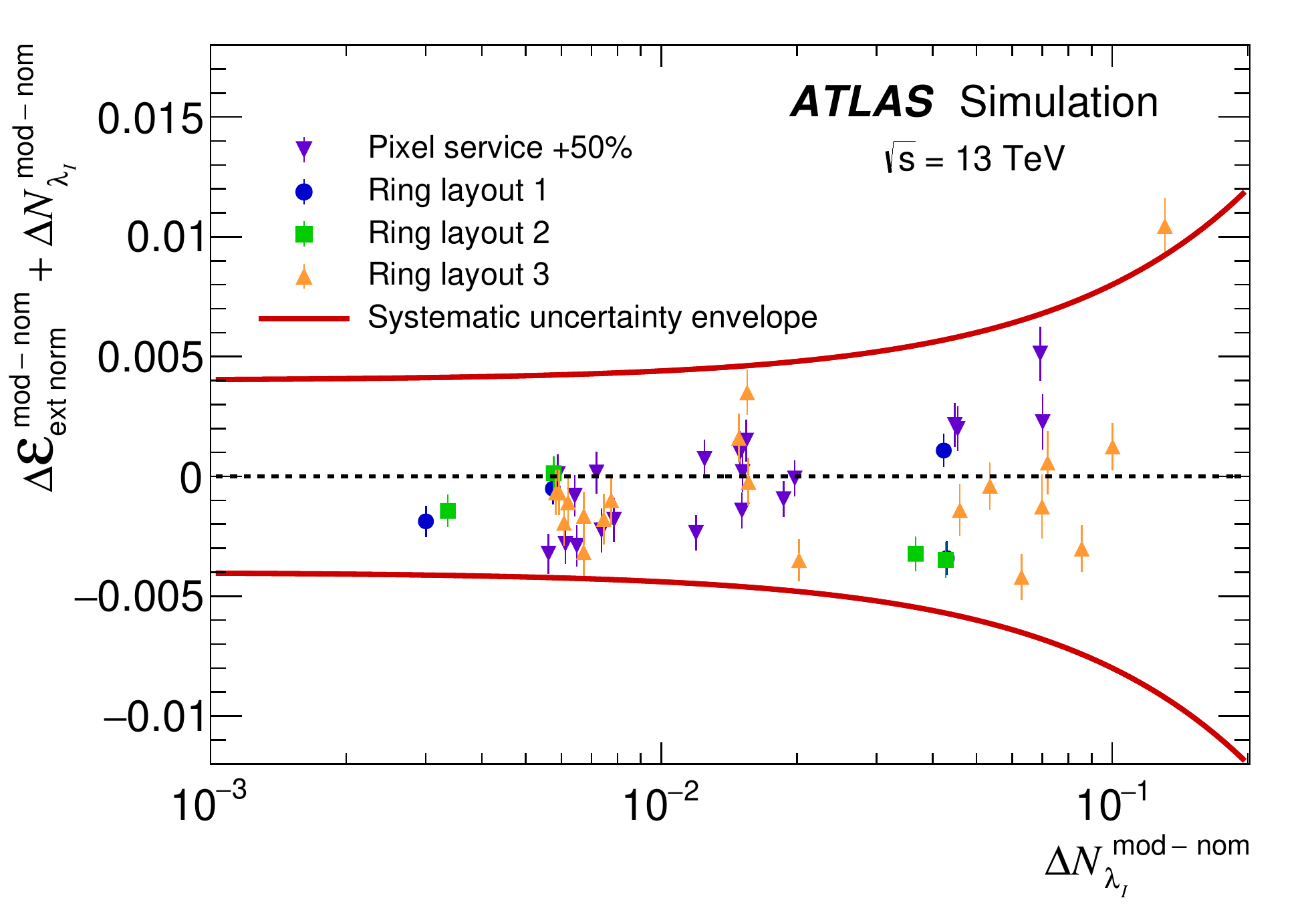}
\caption{The sum of \NormSctExtEff (normalised track-extension efficiency in the nominal geometry and in the modified geometries) and $\varDelta{N}^{\mathrm{mod-nom}}_{\lambda_{I}}$ (number of nuclear interaction lengths in the nominal geometry and in the modified geometries).
A modified geometry results in different values of $\varDelta{N}^{\mathrm{mod-nom}}_{\lambda_{I}}$ over several $\eta$ bins.
The solid red lines are envelopes described by the curves $\pm (0.004 + 0.04 \cdot \varDelta{N}^{\mathrm{mod-nom}}_{\lambda_{I}})$.}
\label{fig:NormExtVSMaterial}
\end{center}
\end{figure}
The amount of material associated with nuclear interactions for each $\eta$ bin, $N_{\lambda_{I}}(\eta)$, is established from the difference between the \emph{original} geometry model and the modified geometry models, namely the sample with the density of the material in the pixel service region scaled up by 50\% (pixel service + 50\%) and the three samples with rings of passive material added between the pixel and the SCT detectors (ring layout 1, 2 and 3).

For each geometry model, $N_{\lambda_{I}}$ is calculated by integrating the material along the path of a virtual neutral non-interacting particle, referred to as a \emph{geantino} in the \textsc{Geant4} simulation. The difference between the two geometry models is taken as the contribution of the weighted material in the pixel service region.

For the geantino-based calibration to be accurate, the path travelled by the geantinos must match that of the particles used in the measurement.  As such, the distribution of geantino production locations in $z$ should match the distribution of $z_{\mathrm{vtx}}$ observed in data.
Thus, the $z$-distribution of the geantinos is re-weighted to match the $z_{\mathrm{vtx}}$-distribution observed in the data.

The track-extension efficiency is affected by the radial position and orientation of material in the detector.
The dependence on the location and orientation of the material can be simply explained: if the missing material is located nearer to the first layer of the SCT there is a higher probability that one of the secondary particles arising from hadronic interactions produces a hit in the SCT compatible with the tracklet and thus considered as an extension.
This artificial increase of the track-extension efficiency is highly suppressed by requiring four SCT hits (which correspond to hits on at least two layers of the SCT detector) on the combined track.
As described in Section~\ref{sec:geo}, a series of detector geometries were created in which an additional ring of passive material was added to the detector at different radii covering $2.2 < |\eta| < 2.3$.
Simulated samples were created based on these geometries and the track-extension efficiency per unit of material was calculated.
The variation in this quantity is shown in Figure \ref{fig:NormExtVSMaterial}. This variation is constrained by an envelope described by the equation $\pm (0.004 + 0.04 \cdot \varDelta{N}^{\mathrm{mod-nom}}_{\lambda_{I}})$.
The value calculated as $0.004 + 0.04 \cdot \varDelta{N}^{\mathrm{Data-MC}}_{\lambda_{I}}$ is taken as an additional systematic uncertainty in the final results.

\paragraph{Other sources}
Many other aspects which may potentially contribute to the uncertainty were investigated and found to be negligible ($<0.01\%$). These include the differences in the physics list in the \textsc{Geant4} simulation, hit efficiency in the pixel and SCT detectors, passive material between the first and second layer of the SCT detector, and residual misalignment between the pixel and SCT detectors.
Furthermore, adjusting the requirement on the minimum number of hits shared between the tracklet and selected combined track was also found to have a negligible effect on the final measurement.

\begin{figure}[t]
\begin{center}
\includegraphics[width=0.7\textwidth]{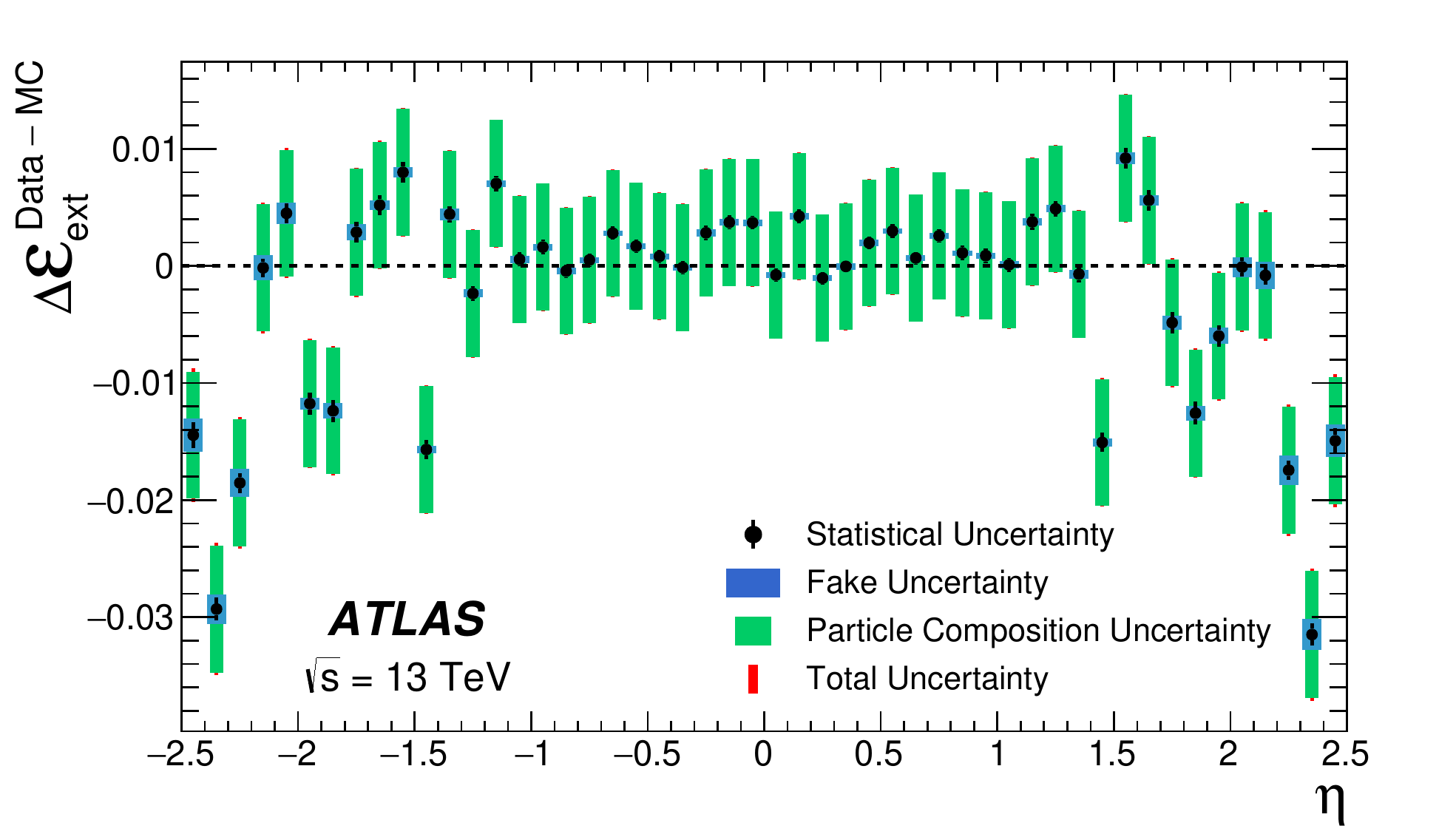}
\caption{The difference between the track-extension efficiency measured in data and in simulation, $\varDelta\SctExtEff^{\mathrm{Data}-\mathrm{MC}}(\eta)$, is shown together with the uncertainties. The total uncertainty includes the uncertainty from fake tracklets, the uncertainty from the particle composition and $\pt$-dependence, and the statistical uncertainty; these are all summed in quadrature. The particle composition uncertainty overwhelmingly dominates the total uncertainty and thus the vertical extent of its uncertainty band is very close to that of the total uncertainty.    }
\label{fig:DeltaSCTExt_alt}
\end{center}
\end{figure}

The difference in $\SctExtEff$ between the data and the \textsc{Pythia~8} simulation as a function of $\eta$, $\varDelta\SctExtEff^{\mathrm{Data}-\mathrm{MC}}(\eta)$, together with the associated statistical and systematic uncertainties is shown in Figure~\ref{fig:DeltaSCTExt_alt}.
This difference is translated to an amount of material using Eq.~(\ref{eqn:NormExtEff}) and is shown in Figure~\ref{fig:DeltaMaterial_Lambda0_alt}.
It can be observed that the main sources of systematic uncertainty are the uncertainties related to particle composition,
which have an impact on the track-extension efficiency, and the uncertainties arising from the calibration procedure, which affect directly the
final measurements of $\varDelta N_{\lambda_{I}}^{\mathrm{Data-MC}}$.

\subsection{Transverse impact parameter resolution}
\label{sec:meas_d0}
In this analysis, the transverse impact parameter, $d_{0}$, of a track is calculated with respect to the primary vertex position of the event. The \emph{visible} $d_{0}$-resolution, $\sigma_{d_{0}}^{\mathrm{vis}}$, is calculated for each $\pt$--$\eta$ slice by fitting a Gaussian function to the core part of the $d_{0}$-distribution. However, $\sigma_{d_{0}}^{\mathrm{vis}}$ is smeared by the position resolution of the primary vertex reconstruction, $\sigma_{\mathrm{PV}}$. In order to remove this effect, an iterative unfolding is applied. At the $i$-th iteration, the $d_{0}$ is varied by a factor $\hat{\sigma}_{d_{0}}^{(i)}/(\hat{\sigma}_{d_{0}}^{(i)}\oplus\sigma_{\mathrm{PV}})$, where $\hat{\sigma}_{d_{0}}^{(i)}$ is the estimator of $\sigma_{d_{0}}$ in the $i$-th iteration. For the first iteration, $\sigma_{d_{0}}^{\mathrm{vis}}$ is used as the estimator. The iteration is repeated typically two or three times before $\hat{\sigma}_{d_{0}}^{(i)}$ converges upon a stable value.
A full description of the methodology and validation is found in Ref.~\cite{ATLAS-CONF-2010-070}. For the MC simulation, the true primary vertex position is used and the unfolding is not applied. The self-consistency of the method was explicitly checked with MC simulation.

\section{Results and discussion}
\label{sec:results}

\begin{figure}[t!]
\begin{center}
\subfigure[]{
\label{fig:hadInt:result}
\includegraphics[width=0.47\textwidth]{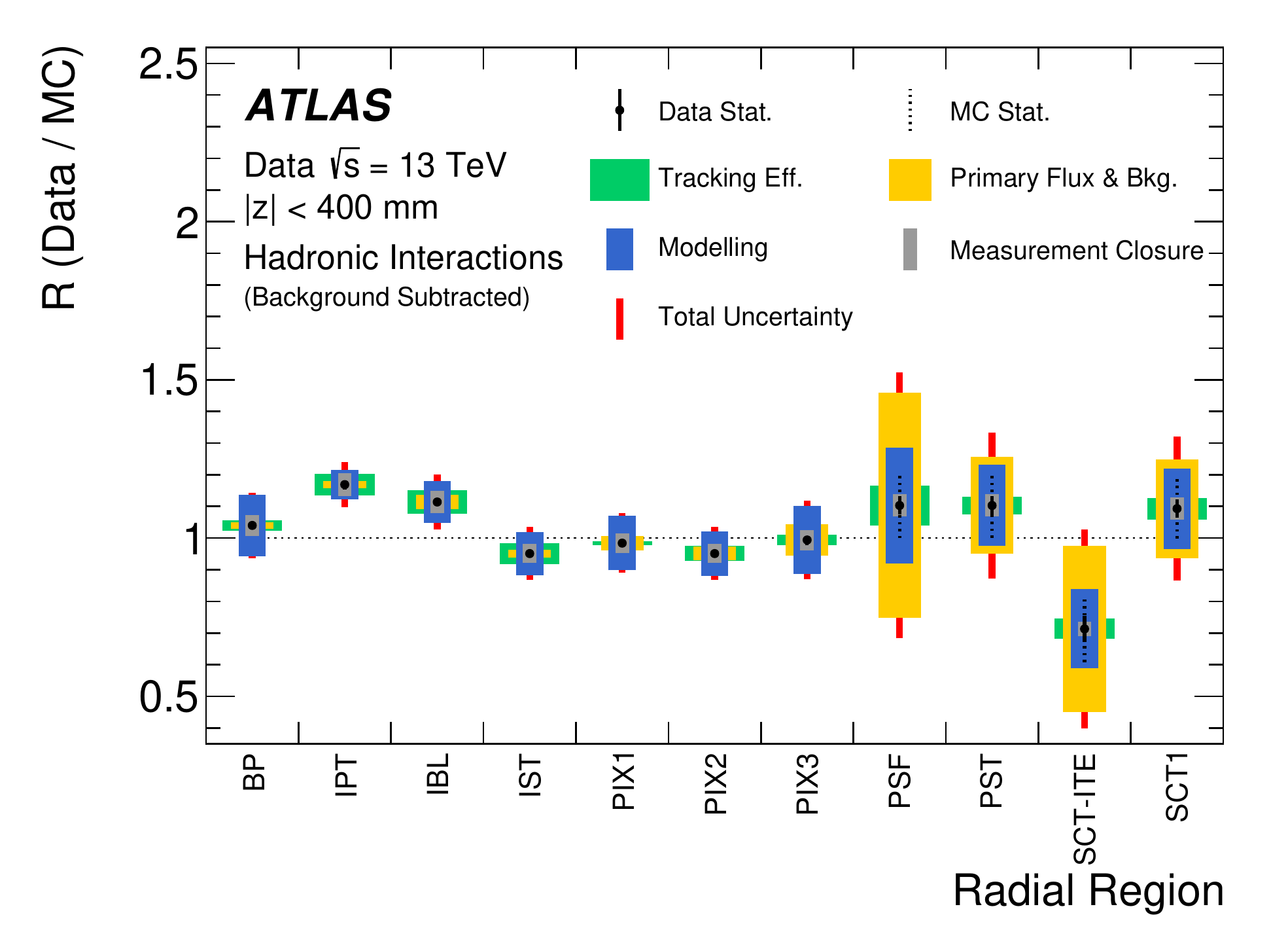}
}
\subfigure[]{
\label{fig:Conv:SummaryPlot}
\includegraphics[width=0.47\textwidth]{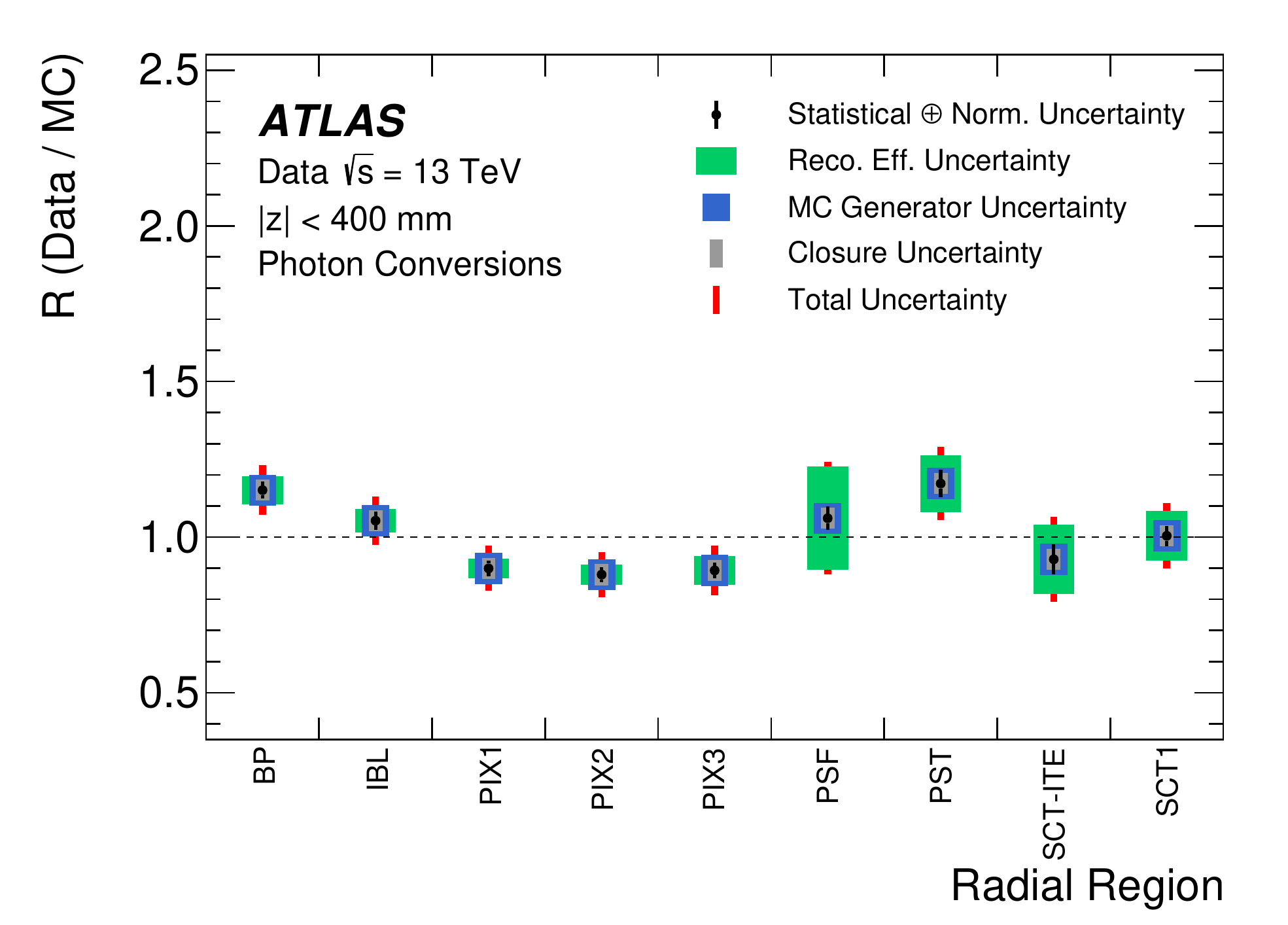}

}
\caption{Comparison of the rate ratio, denoted $R$, between data and MC simulation for \subref{fig:hadInt:result} hadronic interactions  and \subref{fig:Conv:SummaryPlot} photon conversions using the \emph{updated} geometry model. The hadronic interaction results are shown after background subtraction. The systematic uncertainties for each radial section are also shown. The components of the uncertainty in \subref{fig:hadInt:result} and \subref{fig:Conv:SummaryPlot} are the same as those listed in Tables~\ref{tbl:hadInt:result}~and~\ref{tbl:conv:result}, respectively.}
\end{center}

\begin{center}
\includegraphics[width=1.0\textwidth]{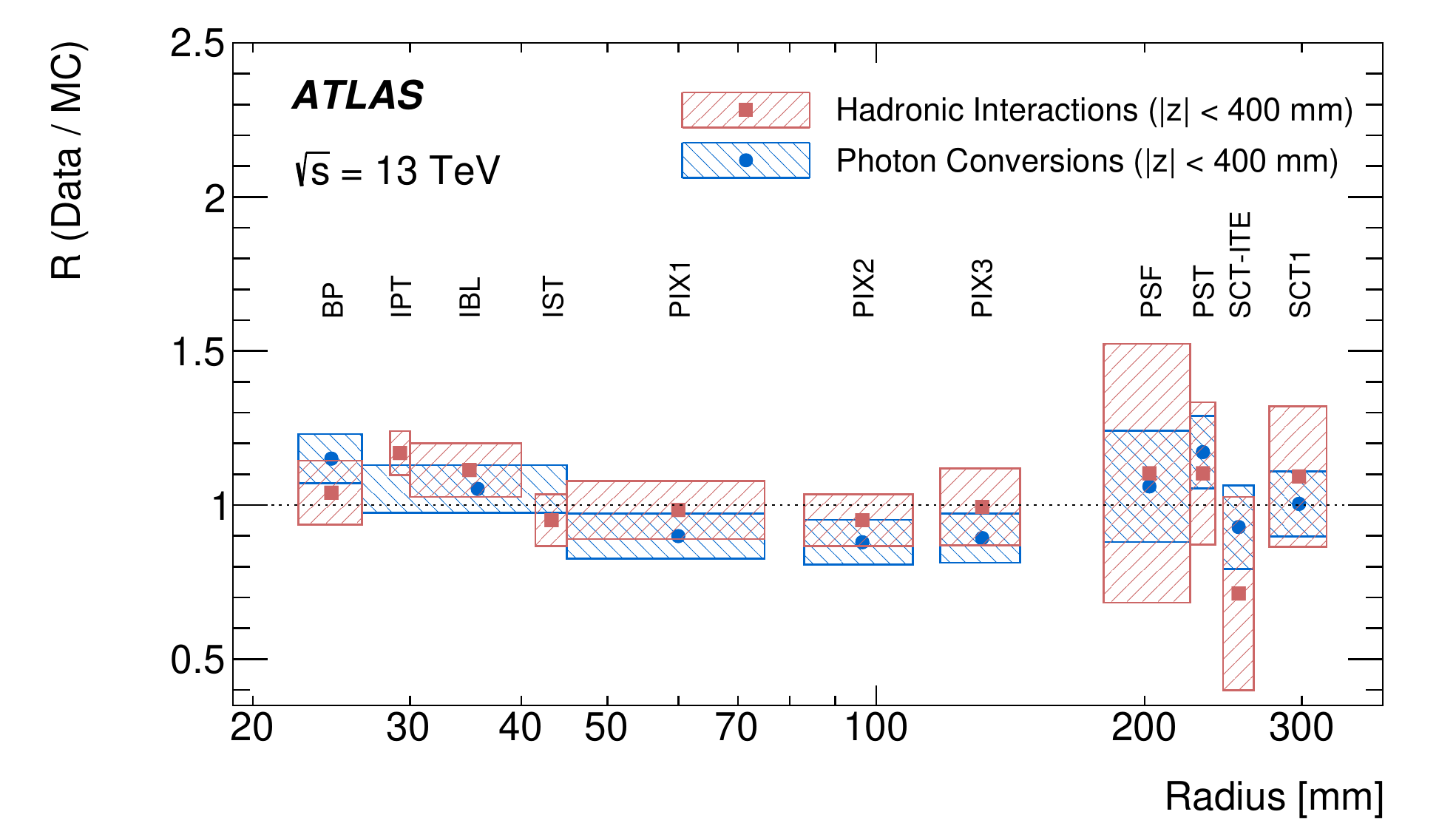}
\caption{Comparison of the rate ratio, denoted $R$, between data and MC simulation, for hadronic interactions and photon conversions as a function of radius. The horizontal range of each marker represents the radial range of vertices used in each measurement, while the vertical range represents the total uncertainty.}
\label{fig:BothSummaryPlot}
\end{center}
\end{figure}

Results of the rate ratio measurements using hadronic interaction vertices for 11 radial regions are presented in Table~\ref{tbl:hadInt:result} and Figure~\ref{fig:hadInt:result}. With the exception of the SCT-ITE, the background subtracted rate ratio measurements remain within $1.00\pm0.17$ for all of the radial sections, spanning the cylindrical region $r < 320\,\millimeter$ and $|z|< 400\,\millimeter$. The total uncertainty is dominated by systematic uncertainties. For the radial region up to and including the IST, the major source of systematic uncertainty is physics modelling of hadronic interactions. At larger radii, the background uncertainty becomes significant as the purity of the hadronic interaction decreases. The total measurement uncertainty for the background-subtracted comparison is estimated to be 7--13\% for the inner radial regions from the beam pipe up to the PIX3, and 22--42\% for outer radial regions from the PSF to the SCT1. The large variation in the size of the uncertainty between the radial regions arises mainly from variations in the purity of the reconstructed hadronic interaction candidates. The uncertainty in the background rate is enhanced at the PSF and outer layers. The uncertainty of the physics modelling is smaller in the IPT--PIX2 regions relative to the beam pipe, reflecting the large correlation in the values of the rate ratio resulting from changes associated with the selection criteria used in the uncertainty evaluation. The beam-pipe data/MC ratio is not unity due to the presence of the excess within $|z|<40~\millimeter$, which is excluded from the normalisation. The results obtained from the background-inclusive rate ratio, $R_{i}^{\mathrm{incl}}$, for layers unchanged since Run~1 (PIX1--PIX3 and SCT1) are consistent with the previous analyses presented in Refs.~\cite{PERF-2011-08,PERF-2015-06}.

Figure~\ref{fig:Conv:SummaryPlot} shows the measurement of the photon conversion rate in data with respect to simulation, as also shown in Table \ref{tbl:conv:result}. The values of the $R$ parameters remain within $1.00\pm0.17$ for all of the radial sections, spanning the cylindrical region $r < 320\,\millimeter$ and $|z|< 400\,\millimeter$.
Good agreement between data and simulation is observed for the IBL region and the first SCT layer. An excess in the observed conversion rate in data with respect to simulation of around 10--15\% is observed in the beam-pipe region. The cause of this excess, also observed with hadronic interactions, is interpreted as a localised region of material missing from the \emph{updated} geometry. The largest deviations in the measured $R$ parameters from unity are observed in all three original pixel layers, which exhibit a systematic deficit in the conversion rate of 10--12\% compared to that predicted by the simulation.

Differences in the material content of one detector layer between data and simulation would also affect the relative hadronic interaction and photon conversion rates observed at all downstream layers due to a modification of the hadron/photon flux incident on all downstream layers. This effect was studied and found to affect the measurements at a level far below the systematic uncertainties associated with both the hadronic interaction and photon conversion measurements and no explicit corrections are applied. Figure~\ref{fig:BothSummaryPlot} shows both the hadronic interaction and photon conversion measurements. While sensitive to slightly different properties of the ID material (the nuclear interaction and radiation lengths), the two measurements are compatible.

\begin{table}[t!]

\caption{Hadronic interaction rate ratio of data with respect the \textsc{Epos} MC simulation using the \emph{updated} geometry model for different radial sections. \emph{Syst.(model)} is the uncertainty of the physics modelling of hadronic interactions, \emph{Syst.(flux \& bkg.)} is the primary particle flux uncertainty and the uncertainty of the fakes and decays backgrounds, \emph{Syst.(eff.)} is the systematic uncertainty of track reconstruction efficiency, and \emph{Syst.(closure)} is the uncertainty of the closure of the measurement. The total uncertainty is calculated from the sum in quadrature of the statistical and systematic uncertainties.}
\scriptsize
\centering
\begin{tabular}{l|c|cc|cccc|c}
\hline
\hline
\multicolumn{9}{c}{Hadronic interaction inclusive rate ratio: $R_{i}^{\mathrm{incl}}$}\\
\hline
Radial region & Value & Stat.(data) & Stat.(MC) & Syst.(model) & Sys(flux \& bkg.) & Syst.(eff.) & Syst.(closure) & Total uncertainty\\
\hline
BP      & 1.04 & $ \pm 0.00 $  & $ \pm 0.01 $  & $ \pm 0.09 $  & $ \pm 0.01 $  & $ \pm 0.02 $  & $ \pm 0.03 $  & $ \pm 0.10 $  \\
IPT     & 1.16 & $ \pm 0.01 $  & $ \pm 0.01 $  & $ \pm 0.05 $  & $ \pm 0.01 $  & $ \pm 0.03 $  & $ \pm 0.04 $  & $ \pm 0.07 $  \\
IBL     & 1.10 & $ \pm 0.00 $  & $ \pm 0.01 $  & $ \pm 0.07 $  & $ \pm 0.02 $  & $ \pm 0.03 $  & $ \pm 0.04 $  & $ \pm 0.09 $  \\
IST     & 0.96 & $ \pm 0.01 $  & $ \pm 0.01 $  & $ \pm 0.07 $  & $ \pm 0.01 $  & $ \pm 0.03 $  & $ \pm 0.03 $  & $ \pm 0.08 $  \\
PIX1    & 0.99 & $ \pm 0.00 $  & $ \pm 0.01 $  & $ \pm 0.08 $  & $ \pm 0.02 $  & $ \pm 0.01 $  & $ \pm 0.03 $  & $ \pm 0.09 $  \\
PIX2    & 0.96 & $ \pm 0.00 $  & $ \pm 0.01 $  & $ \pm 0.07 $  & $ \pm 0.02 $  & $ \pm 0.02 $  & $ \pm 0.03 $  & $ \pm 0.08 $  \\
PIX3    & 1.00 & $ \pm 0.00 $  & $ \pm 0.01 $  & $ \pm 0.10 $  & $ \pm 0.03 $  & $ \pm 0.01 $  & $ \pm 0.03 $  & $ \pm 0.11 $  \\
PSF     & 1.03 & $ \pm 0.01 $  & $ \pm 0.03 $  & $ \pm 0.12 $  & $ \pm 0.11 $  & $ \pm 0.02 $  & $ \pm 0.03 $  & $ \pm 0.17 $  \\
PST     & 1.06 & $ \pm 0.02 $  & $ \pm 0.05 $  & $ \pm 0.14 $  & $ \pm 0.09 $  & $ \pm 0.02 $  & $ \pm 0.03 $  & $ \pm 0.17 $  \\
SCT-ITE & 0.89 & $ \pm 0.02 $  & $ \pm 0.05 $  & $ \pm 0.07 $  & $ \pm 0.09 $  & $ \pm 0.01 $  & $ \pm 0.03 $  & $ \pm 0.13 $  \\
SCT1    & 1.04 & $ \pm 0.01 $  & $ \pm 0.04 $  & $ \pm 0.11 $  & $ \pm 0.07 $  & $ \pm 0.02 $  & $ \pm 0.03 $  & $ \pm 0.14 $  \\
\hline
\hline
\multicolumn{9}{c}{Hadronic interaction background-subtracted rate ratio: $R_{i}^{\mathrm{subtr}}$}\\
\hline
Radial region & Value & Stat.(data) & Stat.(MC) & Syst.(model) & Syst.(flux \& bkg.) & Syst.(eff.) & Syst.(closure) & Total uncertainty\\
\hline
BP      & 1.04 & $ \pm 0.00 $  & $ \pm 0.01 $  & $ \pm 0.10 $  & $ \pm 0.01 $  & $ \pm 0.02 $  & $ \pm 0.03 $  & $ \pm 0.10 $  \\
IPT     & 1.17 & $ \pm 0.01 $  & $ \pm 0.02 $  & $ \pm 0.05 $  & $ \pm 0.01 $  & $ \pm 0.03 $  & $ \pm 0.04 $  & $ \pm 0.07 $  \\
IBL     & 1.11 & $ \pm 0.00 $  & $ \pm 0.01 $  & $ \pm 0.07 $  & $ \pm 0.02 $  & $ \pm 0.04 $  & $ \pm 0.04 $  & $ \pm 0.09 $  \\
IST     & 0.95 & $ \pm 0.01 $  & $ \pm 0.01 $  & $ \pm 0.07 $  & $ \pm 0.01 $  & $ \pm 0.03 $  & $ \pm 0.03 $  & $ \pm 0.08 $  \\
PIX1    & 0.98 & $ \pm 0.00 $  & $ \pm 0.01 $  & $ \pm 0.09 $  & $ \pm 0.02 $  & $ \pm 0.01 $  & $ \pm 0.03 $  & $ \pm 0.09 $  \\
PIX2    & 0.95 & $ \pm 0.01 $  & $ \pm 0.01 $  & $ \pm 0.07 $  & $ \pm 0.02 $  & $ \pm 0.02 $  & $ \pm 0.03 $  & $ \pm 0.08 $  \\
PIX3    & 1.00 & $ \pm 0.01 $  & $ \pm 0.02 $  & $ \pm 0.11 $  & $ \pm 0.05 $  & $ \pm 0.02 $  & $ \pm 0.03 $  & $ \pm 0.13 $  \\
PSF     & 1.10 & $ \pm 0.03 $  & $ \pm 0.10 $  & $ \pm 0.18 $  & $ \pm 0.36 $  & $ \pm 0.06 $  & $ \pm 0.04 $  & $ \pm 0.42 $  \\
PST     & 1.10 & $ \pm 0.03 $  & $ \pm 0.10 $  & $ \pm 0.13 $  & $ \pm 0.15 $  & $ \pm 0.03 $  & $ \pm 0.04 $  & $ \pm 0.23 $  \\
SCT-ITE & 0.71 & $ \pm 0.04 $  & $ \pm 0.11 $  & $ \pm 0.13 $  & $ \pm 0.27 $  & $ \pm 0.03 $  & $ \pm 0.02 $  & $ \pm 0.31 $  \\
SCT1    & 1.09 & $ \pm 0.03 $  & $ \pm 0.09 $  & $ \pm 0.13 $  & $ \pm 0.16 $  & $ \pm 0.03 $  & $ \pm 0.04 $  & $ \pm 0.22 $  \\
\hline
\hline
\end{tabular}
\label{tbl:hadInt:result}

\caption{Photon conversion rate ratio and associated uncertainties in data measured with respect to simulation with the \emph{updated} geometry. Measurements are presented in nine radial regions of the detector in the cylindrical region $r < 325\,\mathrm{mm}$ and $|z|< 400\,\mathrm{mm}$. \emph{Stat.(data \& MC)} is the statistical and normalisation uncertainty, \emph{Syst.(eff.)} is the systematic uncertainty associated with the reconstruction efficiency, \emph{Syst.(MC gen.)} is the systematic uncertainty associated with the choice of MC Generator and \emph{Syst.(closure)} is the systematic uncertainty associated with the closure of the measurement. The total uncertainty is calculated from the sum in quadrature of the statistical and systematic uncertainties.}

\scriptsize
\centering
\begin{tabular}{l|c|c|ccc|c|c}

\hline
\hline
\multicolumn{8}{c}{Photon conversion rate ratio: $R_{i}$}\\
\hline
Radial region & Value & Stat.(data \& MC) & Syst.(MC gen.) & Syst.(eff.) & Syst.(closure) & Syst.(total) & Total uncertainty \\\hline
BP      & 1.15 & $\pm0.03$ & $\pm0.05$ & $\pm0.04$  & $\pm0.03$ & $\pm0.08$ & $\pm0.08$ \\
IBL     & 1.05 & $\pm0.03$ & $\pm0.05$ & $\pm0.04$  & $\pm0.03$ & $\pm0.07$ & $\pm0.08$ \\
PIX1    & 0.90 & $\pm0.02$ & $\pm0.05$ & $\pm0.03$  & $\pm0.03$ & $\pm0.07$ & $\pm0.07$ \\
PIX2    & 0.88 & $\pm0.02$ & $\pm0.05$ & $\pm0.03$  & $\pm0.03$ & $\pm0.07$ & $\pm0.07$ \\
PIX3    & 0.89 & $\pm0.02$ & $\pm0.05$ & $\pm0.05$  & $\pm0.03$ & $\pm0.08$ & $\pm0.08$ \\
PSF     & 1.06 & $\pm0.04$ & $\pm0.05$ & $\pm0.17$  & $\pm0.03$ & $\pm0.18$ & $\pm0.18$ \\
PST     & 1.17 & $\pm0.04$ & $\pm0.05$ & $\pm0.09$  & $\pm0.03$ & $\pm0.11$ & $\pm0.12$ \\
SCT-ITE & 0.93 & $\pm0.05$ & $\pm0.05$ & $\pm0.11$  & $\pm0.03$ & $\pm0.13$ & $\pm0.14$ \\
SCT1    & 1.00 & $\pm0.03$ & $\pm0.05$ & $\pm0.08$  & $\pm0.03$ & $\pm0.10$ & $\pm0.11$ \\
\hline
\hline
\end{tabular}
\label{tbl:conv:result}
\end{table}%

\begin{figure}[t!]
\begin{center}
\includegraphics[width=0.7\textwidth]{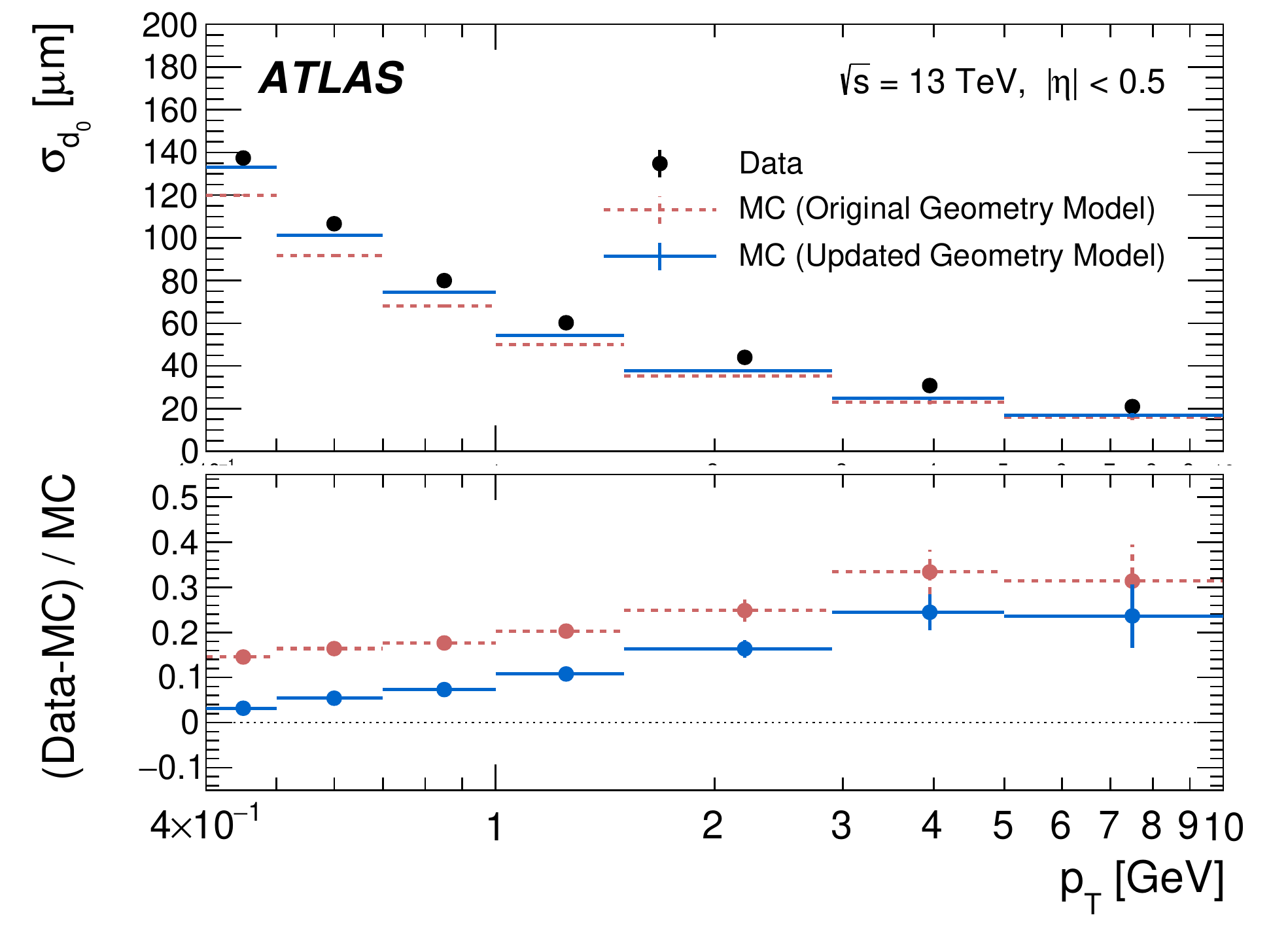}
\caption{Unfolded transverse impact parameter resolution measured in data as a function of $\pt$, compared to the simulation using the \emph{original} and \emph{updated} geometry models. Uncertainties are only statistical.}
\label{fig:d0reso_result}
\end{center}
\end{figure}

As a further qualitative cross-check, the transverse impact parameter resolution in the centremost barrel region in $|\eta|<0.5$ is compared between data and the simulation using different geometry models. Figure~\ref{fig:d0reso_result} shows a comparison of the \emph{original} and \emph{updated} geometry models to the data: the \emph{updated} geometry model provides better agreement with the data at $\pt<1~\GeV$, where multiple scattering by material dominates. The deviation of the data from the simulation at $\pt>1~\GeV$ is believed to be related to effects other than multiple scattering, e.g.~detector misalignment. This gives an independent qualitative cross-check of the validity of the \emph{updated} geometry model in the barrel region.

\begin{figure}[t!]
\begin{center}
\includegraphics[width=1.0\textwidth]{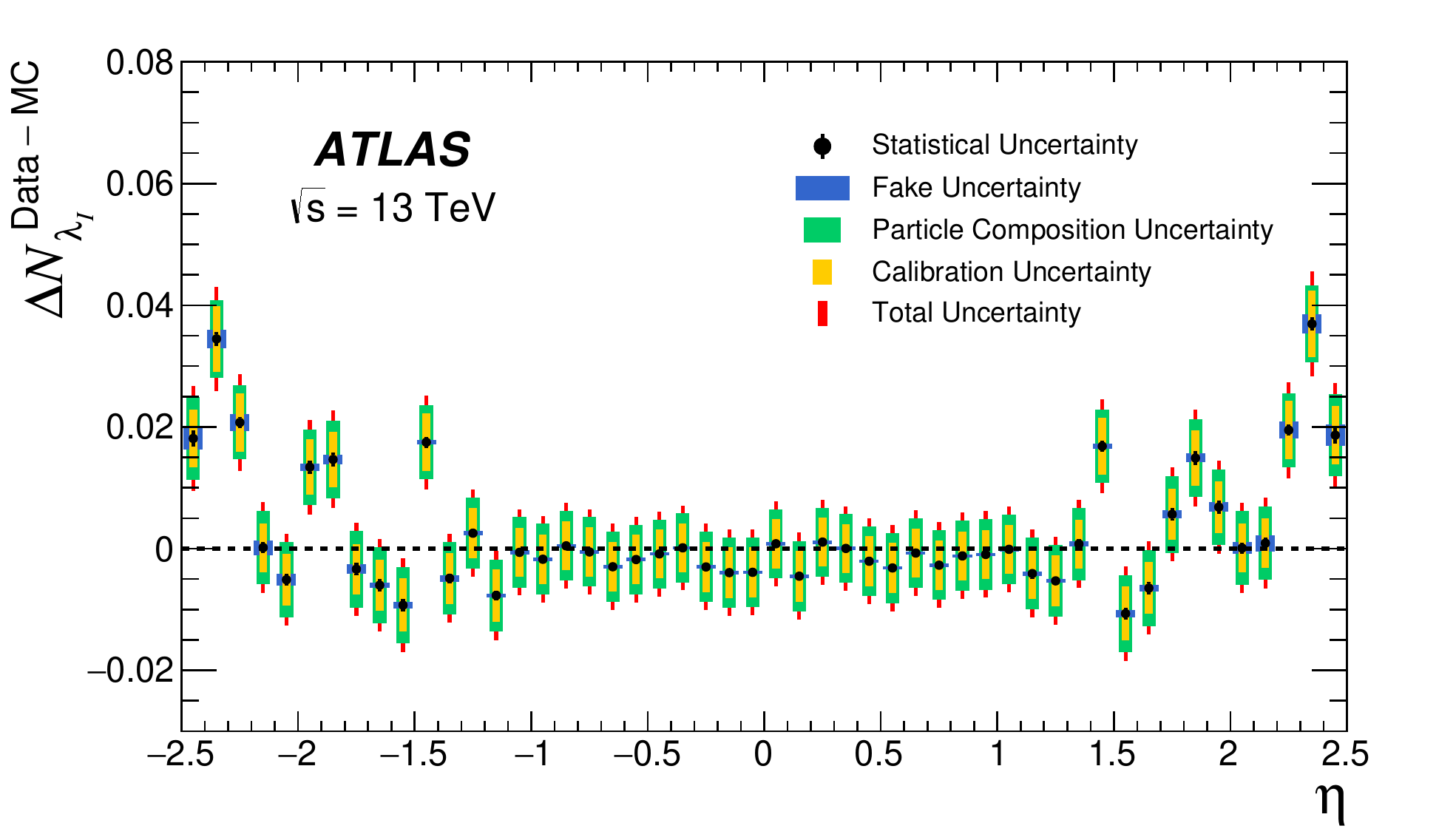}
\caption{The excess amount of material between the pixel and SCT detector associated with nuclear interactions in data, $\varDelta N_{\lambda_{I}}^{\mathrm{Data-MC}}$, based on the track-extension efficiency measurement. The uncertainties shown include the uncertainty from fake pixel-tracklets, the uncertainty from the particle composition and $\pt$-dependence, the uncertainty from the material location and the statistical uncertainties.}
\label{fig:DeltaMaterial_Lambda0_alt}
\end{center}
\end{figure}

The results of the track extension efficiency method, shown in Figure~\ref{fig:DeltaMaterial_Lambda0_alt}, exhibit, within the uncertainties, good agreement between data and simulation in the pseudorapidity region $|\eta|<1$. The geometry model of the pixel services in this region was highly optimised for Run~1 and no major changes occurred between Run~1 and Run~2. For the forward region, $\varDelta N_{\lambda_{I}}^{\mathrm{Data}-\mathrm{MC}}(\eta)$ is greater than zero for $1.4<|\eta|<1.5$, $1.8<|\eta|<2.0$ and $2.3<|\eta|<2.5$. This indicates some missing material in the corresponding regions of the geometry model. The maximum of $\varDelta N_{\lambda_{I}}^{\mathrm{Data}-\mathrm{MC}}$ of $(3.7\pm0.9)\%$ is observed at $2.3 < |\eta| < 2.4$, as can be seen in Table~\ref{tab:SCTEffFinalResults}. This corresponds to approximately 10\% more material in the pixel service region at the corresponding location in $\eta$.

\begin{table}[t!]
\caption{Excess amount of material associated with nuclear interactions in data compared to simulation, $\varDelta N_{\lambda_{I}}^{\mathrm{Data-MC}}$, derived from the track-extension efficiency as a function of $\eta$  together with the uncertainties. \emph{Syst.(particle comp.)} is the systematic uncertainty related to particle composition, \emph{Syst.(fake)} is the uncertainty of the fake rate and \emph{Syst.(calibration)} is the uncertainty associated with the calibration procedure.
The total uncertainty is the sum in quadrature of the statistical and systematic components. The measurements are labelled by their bin centre in $\eta$.}
\scriptsize
\begin{center}
\begin{tabular}{r|r|cccc|c}
\hline
\hline
\multicolumn{1}{c|}{\multirow{2}{*}{$\eta$}} & \multicolumn{6}{c}{Excess amount of material in data: $\varDelta N_{\lambda_{I}}^{\mathrm{Data-MC}}$ [\%]} \\
\cline{2-7}
\multicolumn{1}{c|}{} & Value & Stat. & Syst.(particle comp.) & Syst.(fake) & Syst.(calibration) & Total uncertainty\\
\hline

$-2.45$ & $ 1.81$ & $\pm 0.14$  & $\pm 0.68$ & $\pm 0.18$ & $\pm 0.47$ & $\pm 0.86$\\
$-2.35$ & $ 3.45$ & $\pm 0.11$  & $\pm 0.64$ & $\pm 0.15$ & $\pm 0.54$ & $\pm 0.85$\\
$-2.25$ & $ 2.07$ & $\pm 0.09$  & $\pm 0.61$ & $\pm 0.14$ & $\pm 0.48$ & $\pm 0.79$\\
$-2.15$ & $ 0.02$ & $\pm 0.09$  & $\pm 0.61$ & $\pm 0.12$ & $\pm 0.40$ & $\pm 0.74$\\
$-2.05$ & $-0.51$ & $\pm 0.10$  & $\pm 0.61$ & $\pm 0.10$ & $\pm 0.42$ & $\pm 0.76$\\
$-1.95$ & $ 1.34$ & $\pm 0.11$  & $\pm 0.62$ & $\pm 0.06$ & $\pm 0.45$ & $\pm 0.78$\\
$-1.85$ & $ 1.47$ & $\pm 0.11$  & $\pm 0.64$ & $\pm 0.08$ & $\pm 0.46$ & $\pm 0.80$\\
$-1.75$ & $-0.34$ & $\pm 0.10$  & $\pm 0.63$ & $\pm 0.08$ & $\pm 0.41$ & $\pm 0.77$\\
$-1.65$ & $-0.60$ & $\pm 0.10$  & $\pm 0.63$ & $\pm 0.05$ & $\pm 0.42$ & $\pm 0.76$\\
$-1.55$ & $-0.93$ & $\pm 0.10$  & $\pm 0.63$ & $\pm 0.05$ & $\pm 0.44$ & $\pm 0.77$\\
$-1.45$ & $ 1.75$ & $\pm 0.09$  & $\pm 0.60$ & $\pm 0.04$ & $\pm 0.47$ & $\pm 0.77$\\
$-1.35$ & $-0.49$ & $\pm 0.08$  & $\pm 0.60$ & $\pm 0.05$ & $\pm 0.42$ & $\pm 0.73$\\
$-1.25$ & $ 0.25$ & $\pm 0.07$  & $\pm 0.59$ & $\pm 0.04$ & $\pm 0.41$ & $\pm 0.72$\\
$-1.15$ & $-0.77$ & $\pm 0.07$  & $\pm 0.59$ & $\pm 0.03$ & $\pm 0.43$ & $\pm 0.74$\\
$-1.05$ & $-0.06$ & $\pm 0.06$  & $\pm 0.58$ & $\pm 0.03$ & $\pm 0.40$ & $\pm 0.71$\\
$-0.95$ & $-0.17$ & $\pm 0.07$  & $\pm 0.58$ & $\pm 0.03$ & $\pm 0.41$ & $\pm 0.71$\\
$-0.85$ & $ 0.05$ & $\pm 0.06$  & $\pm 0.57$ & $\pm 0.03$ & $\pm 0.40$ & $\pm 0.70$\\
$-0.75$ & $-0.05$ & $\pm 0.06$  & $\pm 0.57$ & $\pm 0.03$ & $\pm 0.40$ & $\pm 0.70$\\
$-0.65$ & $-0.30$ & $\pm 0.06$  & $\pm 0.58$ & $\pm 0.02$ & $\pm 0.41$ & $\pm 0.71$\\
$-0.55$ & $-0.18$ & $\pm 0.06$  & $\pm 0.57$ & $\pm 0.03$ & $\pm 0.41$ & $\pm 0.70$\\
$-0.45$ & $-0.09$ & $\pm 0.05$  & $\pm 0.57$ & $\pm 0.03$ & $\pm 0.40$ & $\pm 0.70$\\
$-0.35$ & $ 0.01$ & $\pm 0.06$  & $\pm 0.57$ & $\pm 0.03$ & $\pm 0.40$ & $\pm 0.70$\\
$-0.25$ & $-0.30$ & $\pm 0.07$  & $\pm 0.57$ & $\pm 0.03$ & $\pm 0.41$ & $\pm 0.71$\\
$-0.15$ & $-0.39$ & $\pm 0.06$  & $\pm 0.57$ & $\pm 0.03$ & $\pm 0.42$ & $\pm 0.71$\\
$-0.05$ & $-0.39$ & $\pm 0.06$  & $\pm 0.57$ & $\pm 0.03$ & $\pm 0.42$ & $\pm 0.71$\\
$0.05$  & $ 0.08$ & $\pm 0.06$  & $\pm 0.57$ & $\pm 0.03$ & $\pm 0.40$ & $\pm 0.70$\\
$0.15$  & $-0.45$ & $\pm 0.07$  & $\pm 0.58$ & $\pm 0.03$ & $\pm 0.42$ & $\pm 0.72$\\
$0.25$  & $ 0.11$ & $\pm 0.06$  & $\pm 0.57$ & $\pm 0.03$ & $\pm 0.40$ & $\pm 0.70$\\
$0.35$  & $-0.00$ & $\pm 0.05$  & $\pm 0.56$ & $\pm 0.03$ & $\pm 0.40$ & $\pm 0.69$\\
$0.45$  & $-0.21$ & $\pm 0.06$  & $\pm 0.57$ & $\pm 0.03$ & $\pm 0.41$ & $\pm 0.70$\\
$0.55$  & $-0.32$ & $\pm 0.06$  & $\pm 0.58$ & $\pm 0.03$ & $\pm 0.41$ & $\pm 0.71$\\
$0.65$  & $-0.07$ & $\pm 0.06$  & $\pm 0.57$ & $\pm 0.03$ & $\pm 0.40$ & $\pm 0.70$\\
$0.75$  & $-0.27$ & $\pm 0.06$  & $\pm 0.57$ & $\pm 0.03$ & $\pm 0.41$ & $\pm 0.71$\\
$0.85$  & $-0.12$ & $\pm 0.07$  & $\pm 0.58$ & $\pm 0.02$ & $\pm 0.40$ & $\pm 0.71$\\
$0.95$  & $-0.09$ & $\pm 0.07$  & $\pm 0.58$ & $\pm 0.03$ & $\pm 0.40$ & $\pm 0.71$\\
$1.05$  & $-0.01$ & $\pm 0.06$  & $\pm 0.58$ & $\pm 0.04$ & $\pm 0.40$ & $\pm 0.71$\\
$1.15$  & $-0.41$ & $\pm 0.07$  & $\pm 0.59$ & $\pm 0.04$ & $\pm 0.42$ & $\pm 0.73$\\
$1.25$  & $-0.53$ & $\pm 0.07$  & $\pm 0.59$ & $\pm 0.03$ & $\pm 0.42$ & $\pm 0.73$\\
$1.35$  & $ 0.08$ & $\pm 0.08$  & $\pm 0.60$ & $\pm 0.04$ & $\pm 0.40$ & $\pm 0.73$\\
$1.45$  & $ 1.68$ & $\pm 0.09$  & $\pm 0.60$ & $\pm 0.04$ & $\pm 0.47$ & $\pm 0.77$\\
$1.55$  & $-1.07$ & $\pm 0.10$  & $\pm 0.63$ & $\pm 0.06$ & $\pm 0.44$ & $\pm 0.78$\\
$1.65$  & $-0.65$ & $\pm 0.10$  & $\pm 0.63$ & $\pm 0.05$ & $\pm 0.43$ & $\pm 0.77$\\
$1.75$  & $ 0.56$ & $\pm 0.10$  & $\pm 0.63$ & $\pm 0.07$ & $\pm 0.42$ & $\pm 0.77$\\
$1.85$  & $ 1.49$ & $\pm 0.11$  & $\pm 0.64$ & $\pm 0.07$ & $\pm 0.46$ & $\pm 0.80$\\
$1.95$  & $ 0.68$ & $\pm 0.11$  & $\pm 0.62$ & $\pm 0.08$ & $\pm 0.43$ & $\pm 0.76$\\
$2.05$  & $ 0.01$ & $\pm 0.09$  & $\pm 0.61$ & $\pm 0.10$ & $\pm 0.40$ & $\pm 0.74$\\
$2.15$  & $ 0.09$ & $\pm 0.09$  & $\pm 0.60$ & $\pm 0.13$ & $\pm 0.40$ & $\pm 0.74$\\
$2.25$  & $ 1.95$ & $\pm 0.09$  & $\pm 0.61$ & $\pm 0.14$ & $\pm 0.48$ & $\pm 0.79$\\
$2.35$  & $ 3.69$ & $\pm 0.11$  & $\pm 0.64$ & $\pm 0.16$ & $\pm 0.55$ & $\pm 0.86$\\
$2.45$  & $ 1.87$ & $\pm 0.14$  & $\pm 0.68$ & $\pm 0.18$ & $\pm 0.47$ & $\pm 0.86$\\

\hline
\hline
\end{tabular}
\end{center}
\label{tab:SCTEffFinalResults}
\end{table}%

\clearpage
\section{Conclusion}
\label{sec:conclusion}

A good description of the distribution of material in the inner detector is crucial for understanding the performance of track reconstruction within ATLAS. Three complementary techniques, hadronic interaction and photon conversion vertex reconstruction together with an estimation using the track-extension efficiency, are applied to measure the inner detector's material using around $2.0\,\inb$ of low-luminosity $\rts=13~\TeV$ $pp$ collisions at the LHC.
While the first two methods probe the barrel region of the inner detector, in particular the new detector components installed in Run~2 (the beam pipe, the IBL and the supporting tubes of IPT and IST), the track-extension efficiency method has sensitivity also in the forward $\eta$ region of $1.0 < |\eta|<2.5$, in which most of the refurbished pixel services reside.

The description of the geometry model was examined in detail both in radial and longitudinal distributions of the rate of reconstructed hadronic interaction and photon conversion vertices. In the central barrel region, a significant amount of missing material in the IBL front-end electronics for the flex bus, surface-mounted devices on the front-end chips and the IPT and IST was identified in the \emph{original} geometry model (the geometry model used for ATLAS MC simulation in 2015). The \emph{updated} geometry model, which was created to resolve the above discrepancies, provides a much better description. The beam pipe is found to be very accurately described except the centremost region of $|z|<40~\millimeter$. The simulated material in the IBL within the \emph{updated} geometry model is found to be consistent with that in observed data, within the less than 10\% uncertainties of the hadronic interaction and conversion measurements. The existing pixel barrel layers are found to be described well, and the results from the analyses using the hadronic interactions and photon conversions agree within the systematic uncertainties. This result confirms the results of the previous hadronic interaction analysis using the Run~1 data set.

While sensitive to slightly different material properties and $z$ regions of the detector, both the hadronic interactions and photon conversions provide a consistent understanding of the barrel detector material. The \emph{updated} geometry model provides reasonable agreement with the data in the ratio of the rate measurements of hadronic interactions and photon conversions within the uncertainties of the measurements. The measured rates of photon conversions and hadronic interactions reconstructed in data are found to agree to within $7\%$--$18\%$ with those predicted by simulation, based on the \emph{updated} geometry model, out to the outer envelope of the pixel detector. This is also supported by a study of the transverse impact parameter resolution below $\pt=1~\GeV$, where the multiple scattering is dominant.

In the forward region, the material in the pixel service region is found to be underestimated in the geometry model by up to $\varDelta N_{\lambda_{I}}=(3.7\pm0.9)\%$ at some values of~$\eta$. This corresponds roughly to 10\% of the material in the pixel services in the corresponding regions.

The results of these studies have been taken into account in an improved description of the material in the ATLAS inner detector simulation, to be used in future analyses.

\section*{Acknowledgements}


We thank CERN for the very successful operation of the LHC, as well as the
support staff from our institutions without whom ATLAS could not be
operated efficiently.

We acknowledge the support of ANPCyT, Argentina; YerPhI, Armenia; ARC, Australia; BMWFW and FWF, Austria; ANAS, Azerbaijan; SSTC, Belarus; CNPq and FAPESP, Brazil; NSERC, NRC and CFI, Canada; CERN; CONICYT, Chile; CAS, MOST and NSFC, China; COLCIENCIAS, Colombia; MSMT CR, MPO CR and VSC CR, Czech Republic; DNRF and DNSRC, Denmark; IN2P3-CNRS, CEA-DSM/IRFU, France; SRNSF, Georgia; BMBF, HGF, and MPG, Germany; GSRT, Greece; RGC, Hong Kong SAR, China; ISF, I-CORE and Benoziyo Center, Israel; INFN, Italy; MEXT and JSPS, Japan; CNRST, Morocco; NWO, Netherlands; RCN, Norway; MNiSW and NCN, Poland; FCT, Portugal; MNE/IFA, Romania; MES of Russia and NRC KI, Russian Federation; JINR; MESTD, Serbia; MSSR, Slovakia; ARRS and MIZ\v{S}, Slovenia; DST/NRF, South Africa; MINECO, Spain; SRC and Wallenberg Foundation, Sweden; SERI, SNSF and Cantons of Bern and Geneva, Switzerland; MOST, Taiwan; TAEK, Turkey; STFC, United Kingdom; DOE and NSF, United States of America. In addition, individual groups and members have received support from BCKDF, the Canada Council, CANARIE, CRC, Compute Canada, FQRNT, and the Ontario Innovation Trust, Canada; EPLANET, ERC, ERDF, FP7, Horizon 2020 and Marie Sk{\l}odowska-Curie Actions, European Union; Investissements d'Avenir Labex and Idex, ANR, R{\'e}gion Auvergne and Fondation Partager le Savoir, France; DFG and AvH Foundation, Germany; Herakleitos, Thales and Aristeia programmes co-financed by EU-ESF and the Greek NSRF; BSF, GIF and Minerva, Israel; BRF, Norway; CERCA Programme Generalitat de Catalunya, Generalitat Valenciana, Spain; the Royal Society and Leverhulme Trust, United Kingdom.

The crucial computing support from all WLCG partners is acknowledged gratefully, in particular from CERN, the ATLAS Tier-1 facilities at TRIUMF (Canada), NDGF (Denmark, Norway, Sweden), CC-IN2P3 (France), KIT/GridKA (Germany), INFN-CNAF (Italy), NL-T1 (Netherlands), PIC (Spain), ASGC (Taiwan), RAL (UK) and BNL (USA), the Tier-2 facilities worldwide and large non-WLCG resource providers. Major contributors of computing resources are listed in Ref.~\cite{ATL-GEN-PUB-2016-002}.

\printbibliography
\newpage 
\begin{flushleft}
{\Large The ATLAS Collaboration}

\bigskip

M.~Aaboud$^\textrm{\scriptsize 137d}$,
G.~Aad$^\textrm{\scriptsize 88}$,
B.~Abbott$^\textrm{\scriptsize 115}$,
J.~Abdallah$^\textrm{\scriptsize 8}$,
O.~Abdinov$^\textrm{\scriptsize 12}$$^{,*}$,
B.~Abeloos$^\textrm{\scriptsize 119}$,
S.H.~Abidi$^\textrm{\scriptsize 161}$,
O.S.~AbouZeid$^\textrm{\scriptsize 139}$,
N.L.~Abraham$^\textrm{\scriptsize 151}$,
H.~Abramowicz$^\textrm{\scriptsize 155}$
% ...
\bigskip
\\
% ...
$^{8}$ Department of Physics, The University of Texas at Arlington, Arlington TX, United States of America\\
% ...
$^{12}$ Institute of Physics, Azerbaijan Academy of Sciences, Baku, Azerbaijan\\
% ...
$^{88}$ CPPM, Aix-Marseille Universit{\'e} and CNRS/IN2P3, Marseille, France\\
% ...
$^{115}$ Homer L. Dodge Department of Physics and Astronomy, University of Oklahoma, Norman OK, United States of America\\
% ...
$^{119}$ LAL, Univ. Paris-Sud, CNRS/IN2P3, Universit{\'e} Paris-Saclay, Orsay, France\\
% ...
$^{137}$ $^{(a)}$ Facult{\'e} des Sciences Ain Chock, R{\'e}seau Universitaire de Physique des Hautes Energies - Universit{\'e} Hassan II, Casablanca; $^{(b)}$ Centre National de l'Energie des Sciences Techniques Nucleaires, Rabat; $^{(c)}$ Facult{\'e} des Sciences Semlalia, Universit{\'e} Cadi Ayyad, LPHEA-Marrakech; $^{(d)}$ Facult{\'e} des Sciences, Universit{\'e} Mohamed Premier and LPTPM, Oujda; $^{(e)}$ Facult{\'e} des sciences, Universit{\'e} Mohammed V, Rabat, Morocco\\
% ...
$^{139}$ Santa Cruz Institute for Particle Physics, University of California Santa Cruz, Santa Cruz CA, United States of America\\
% ...
$^{151}$ Department of Physics and Astronomy, University of Sussex, Brighton, United Kingdom\\
% ...
$^{155}$ Raymond and Beverly Sackler School of Physics and Astronomy, Tel Aviv University, Tel Aviv, Israel\\
% ...
$^{161}$ Department of Physics, University of Toronto, Toronto ON, Canada\\
% ...
$^{a}$ Also at Department of Physics, King's College London, London, United Kingdom\\
$^{b}$ Also at Institute of Physics, Azerbaijan Academy of Sciences, Baku, Azerbaijan\\
$^{c}$ Also at Novosibirsk State University, Novosibirsk, Russia\\
$^{d}$ Also at TRIUMF, Vancouver BC, Canada\\
$^{e}$ Also at Department of Physics {\&} Astronomy, University of Louisville, Louisville, KY, United States of America\\
% ...
$^{*}$ Deceased
\end{flushleft}


\clearpage
\appendix
\part*{Auxiliary material}
\addcontentsline{toc}{part}{Auxiliary material}

\begin{figure}[htbp]
\begin{center}
\includegraphics[width=0.85\textwidth]{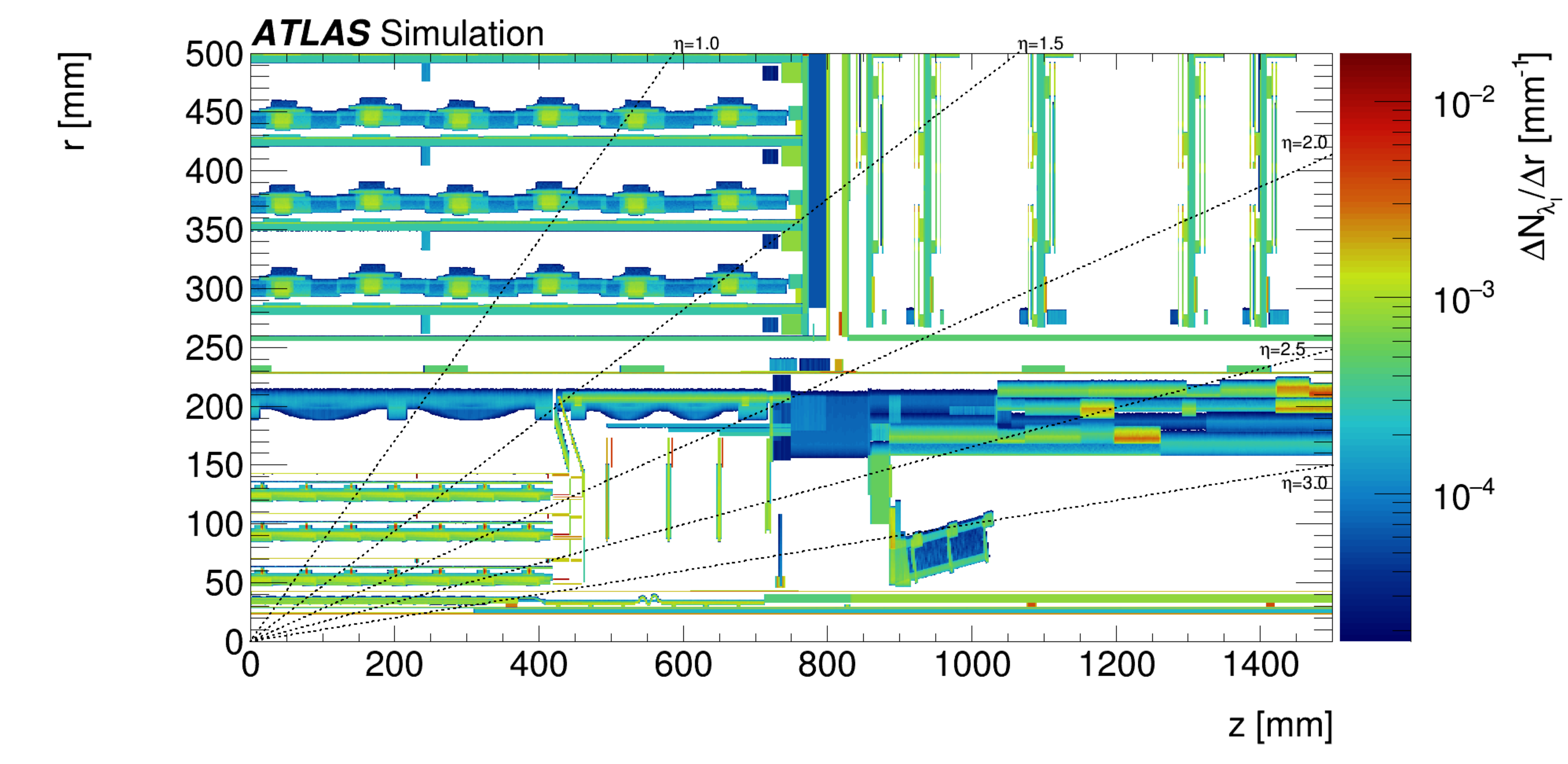}
\caption{The $r$--$z$ distribution of differential number of nuclear interaction length, $\varDelta N_{\lambda_{I}}/\varDelta r$, for the \emph{updated} geometry model of the quadrant of the inner detector barrel region of the pixel detector and the SCT.}
\label{fig:rzmap_Had}
\includegraphics[width=0.65\textwidth]{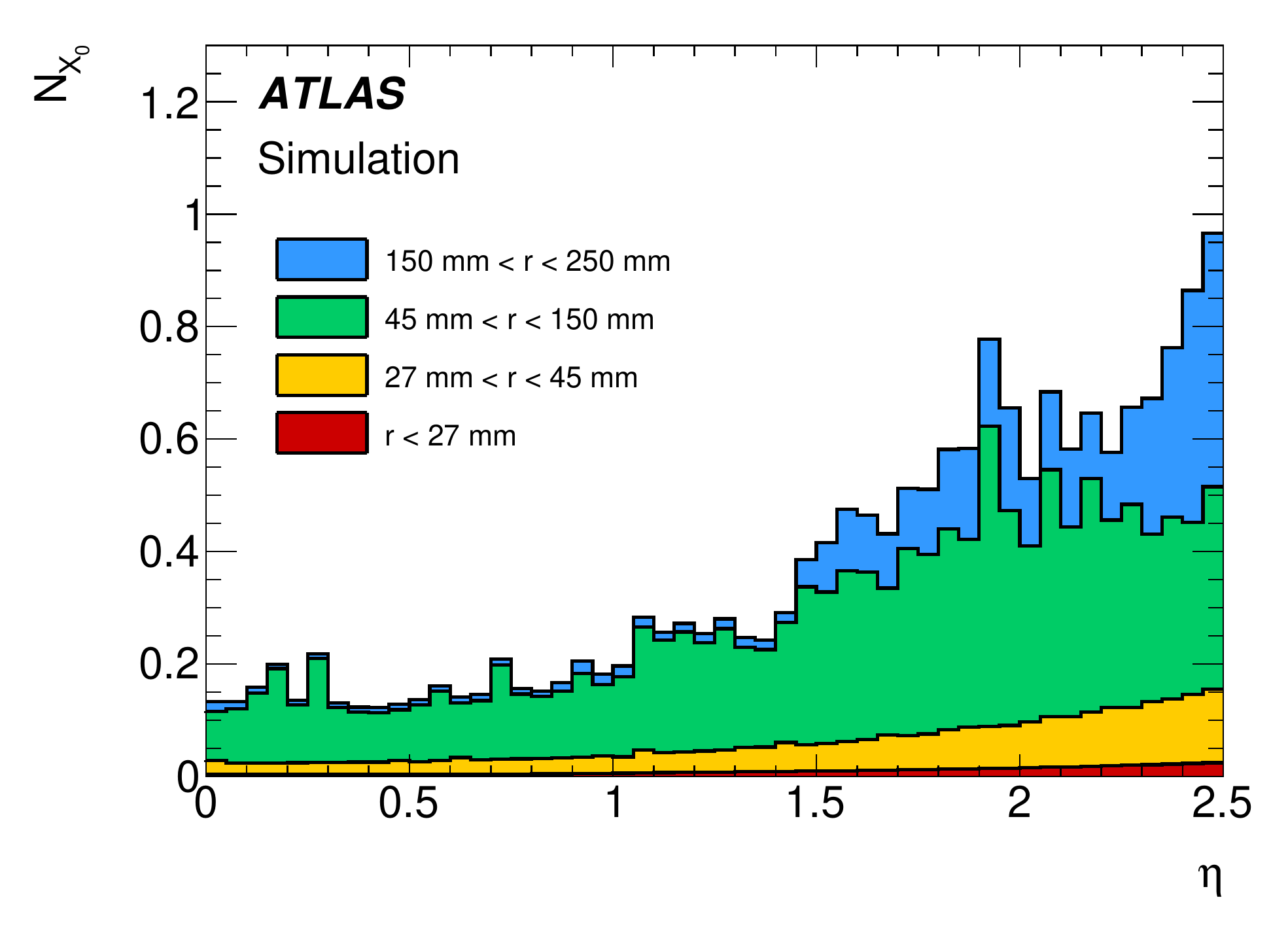}
\caption{The amount of material associated to electromagnetic interactions, $N_{X_{0}} = \int\!{\mathrm d}s\,X_{0}^{-1}$ as a function of $\eta$ in the positive $\eta$ range integrated up to $r=250~\millimeter$ of the \emph{updated} geometry model. Regions are categorised into $r<27~\millimeter$, $27~\millimeter<r<45~\millimeter$, $45~\millimeter<r<150~\millimeter$ and $150~\millimeter<r<250~\millimeter$, corresponding to the beam pipe, IBL, pixel barrel and pixel service region, respectively.}
\label{figures:matmap_etascan_X0}
\end{center}
\end{figure}

\begin{figure}[htbp]
\begin{center}
\subfigure[]{
  \label{fig:hadInt_data_map_RZ}
  \includegraphics[width=0.74\textwidth]{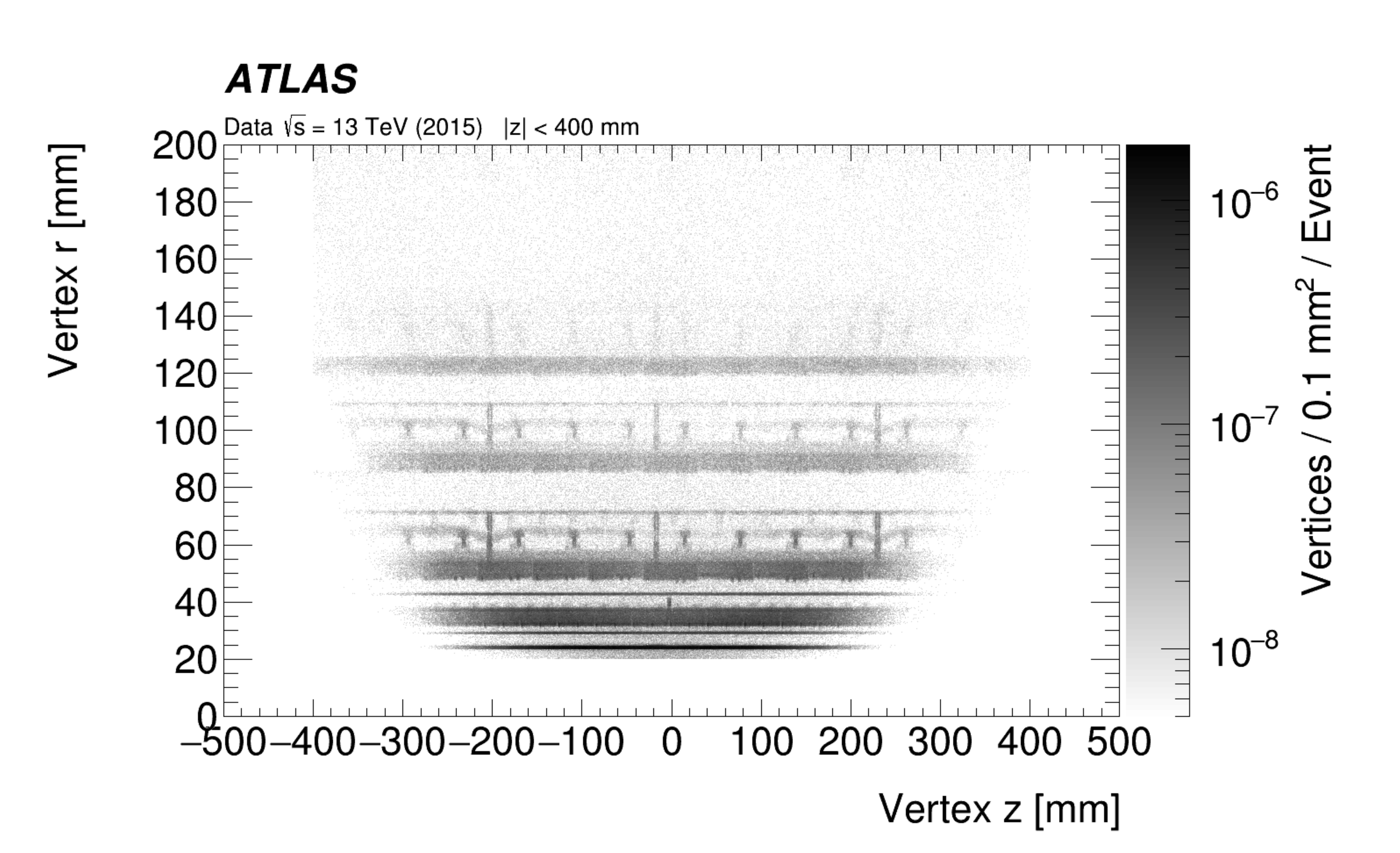}
}
\subfigure[]{
  \label{fig:hadInt_py8_map_RZ}
  \includegraphics[width=0.74\textwidth]{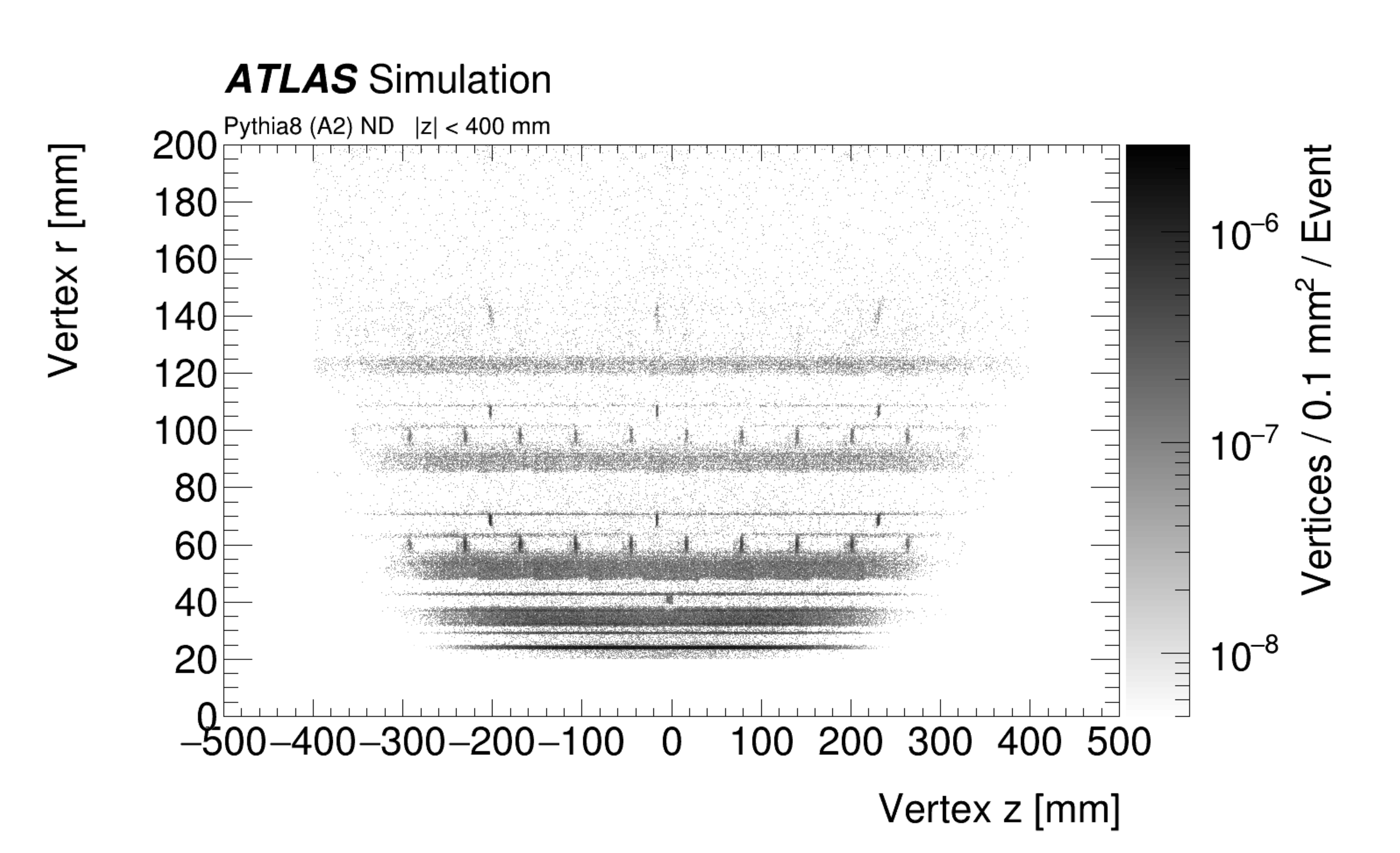}
}
\caption{Distribution of the hadronic-interaction vertex candidates within $|\eta|<2.4$ and $|z|<400~\mathrm{mm}$ in $r$--$z$ view for \subref{fig:hadInt_data_map_RZ} data and \subref{fig:hadInt_py8_map_RZ} \textsc{Pythia~8} simulation with the \emph{updated} geometry model. The integral path is taken in the radial direction for each $z$-position.
The vertex radius is corrected for the radial offset for the data.}
\label{fig:hadInt_map_RZ}
\end{center}
\end{figure}

\clearpage

\begin{figure}[htbp]
\begin{center}
\subfigure[]{
  \label{fig:hadInt_data_map_Rphi}
  \includegraphics[width=0.75\textwidth]{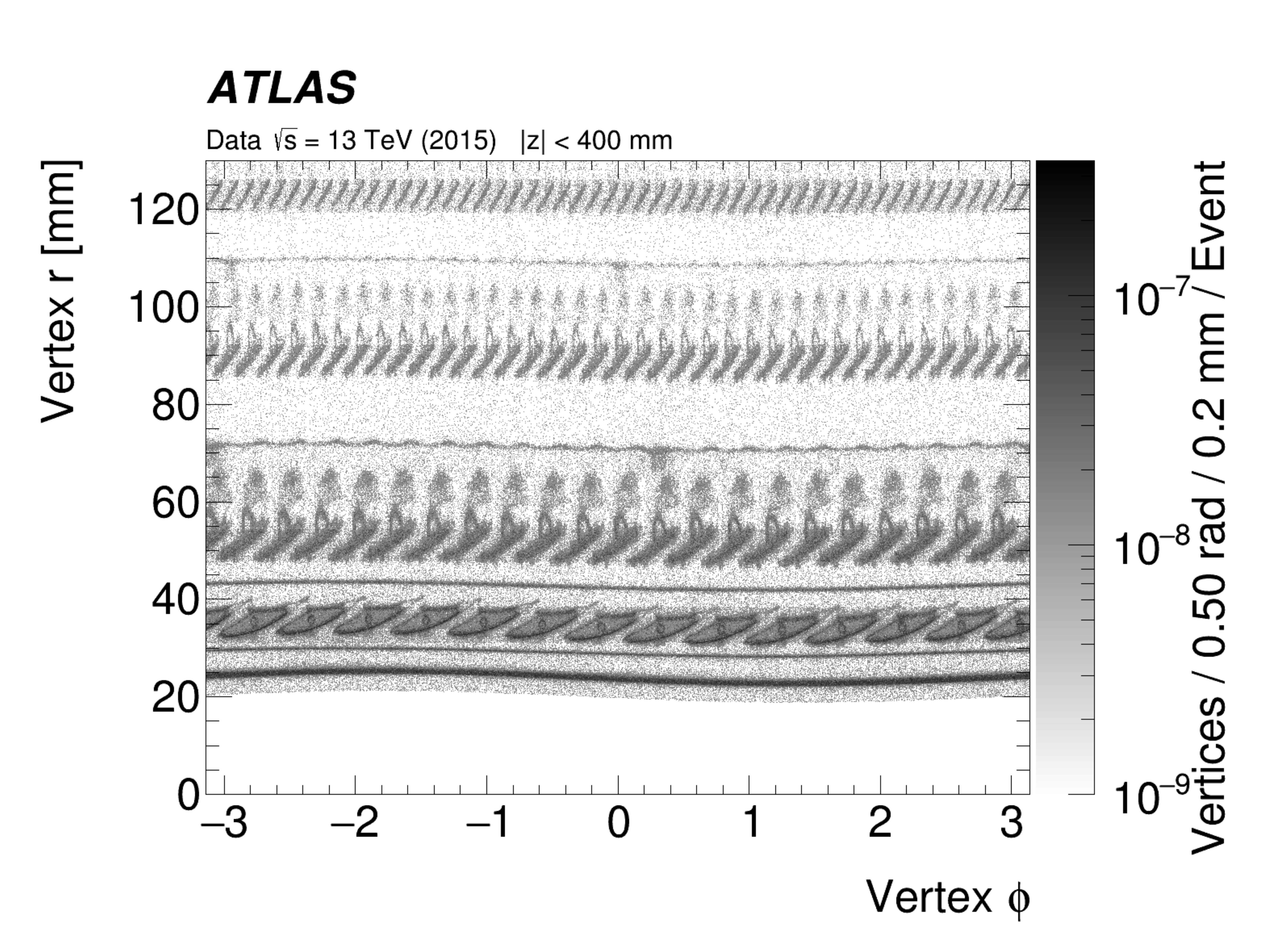}
}
\subfigure[]{
  \label{fig:hadInt_py8_map_Rphi}
  \includegraphics[width=0.75\textwidth]{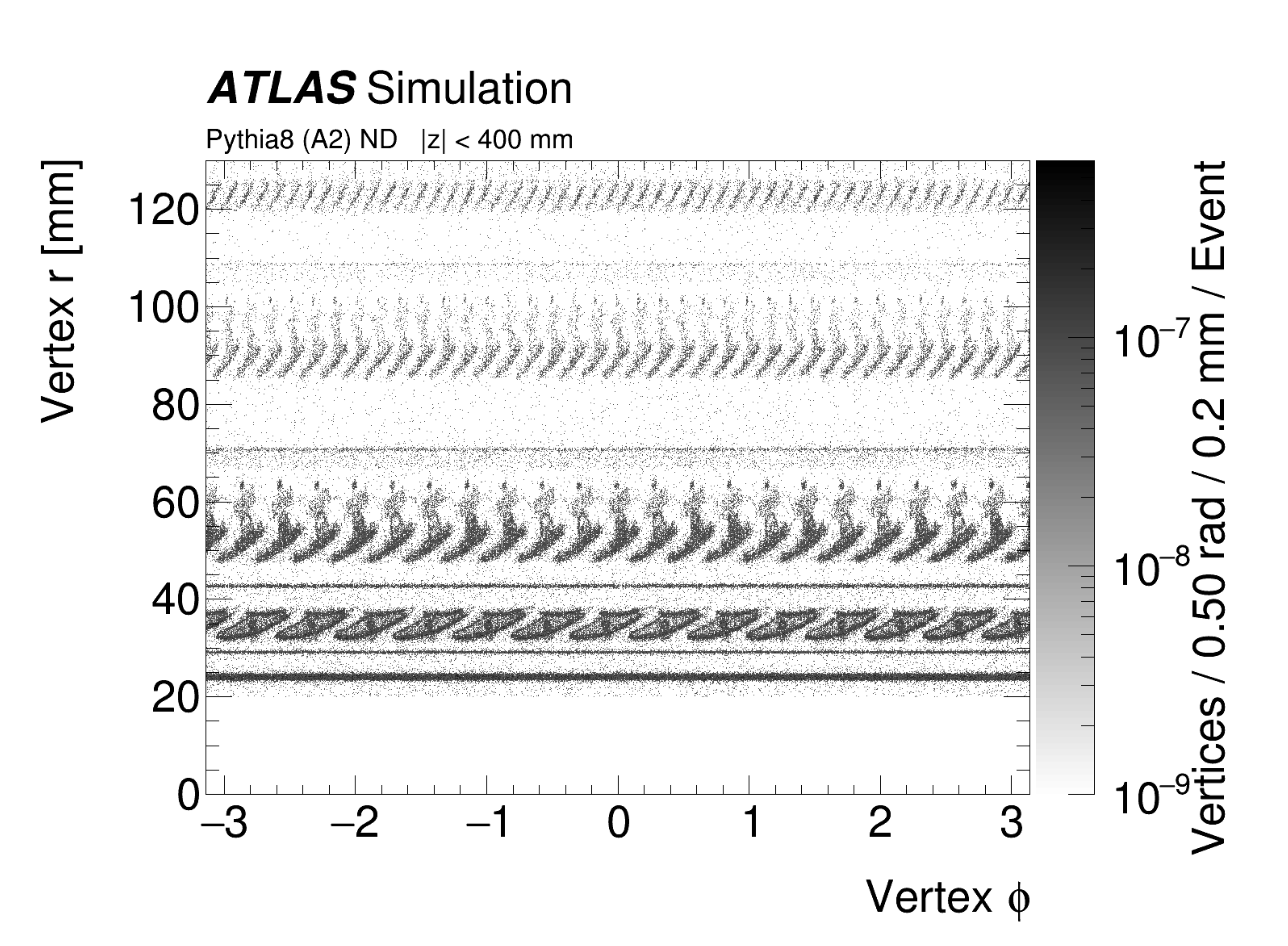}
}
\caption{Distribution of hadronic-interaction vertex candidates within $|\eta|<2.4$ and $|z|<400~\mathrm{mm}$, without radial offset correction in $r$--$\phi$ view for \subref{fig:hadInt_data_map_Rphi} data and \subref{fig:hadInt_py8_map_Rphi} \textsc{Pythia~8} simulation with the \emph{updated} geometry model. The vertex radius is not corrected for the radial offset for the data.}
\label{fig:hadInt_map_Rphi}
\end{center}
\end{figure}

\clearpage

\begin{figure}[htbp]
\begin{center}
\subfigure[]{
  \label{fig:compare_h1_phi_BP_all}
  \includegraphics[width=0.4\textwidth]{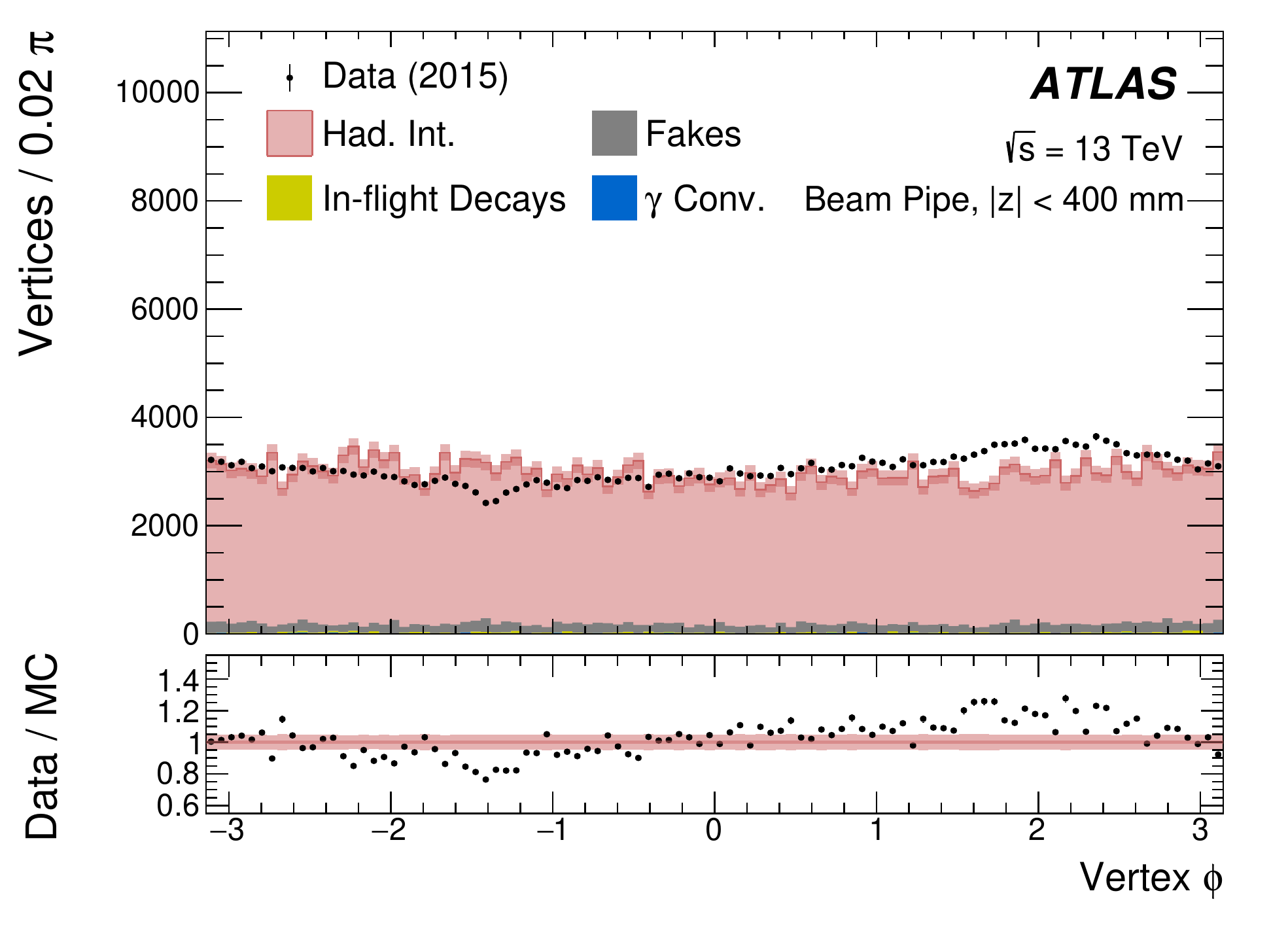}
}
\caption{Distribution of hadronic-interaction vertex candidates in $\phi$ at the beam pipe ($22.5~\millimeter<r<26.5~\millimeter$) in data compared to the \textsc{Epos} MC simulation sample for the \emph{updated} geometry. The band in the ratio plots in the bottom panel indicates statistical uncertainty of the MC simulation samples.}
\label{fig:compare_h1_BP}
\end{center}

\begin{center}
\subfigure[]{
  \label{fig:compare_h1_Z_IPT_all}
  \includegraphics[width=0.4\textwidth]{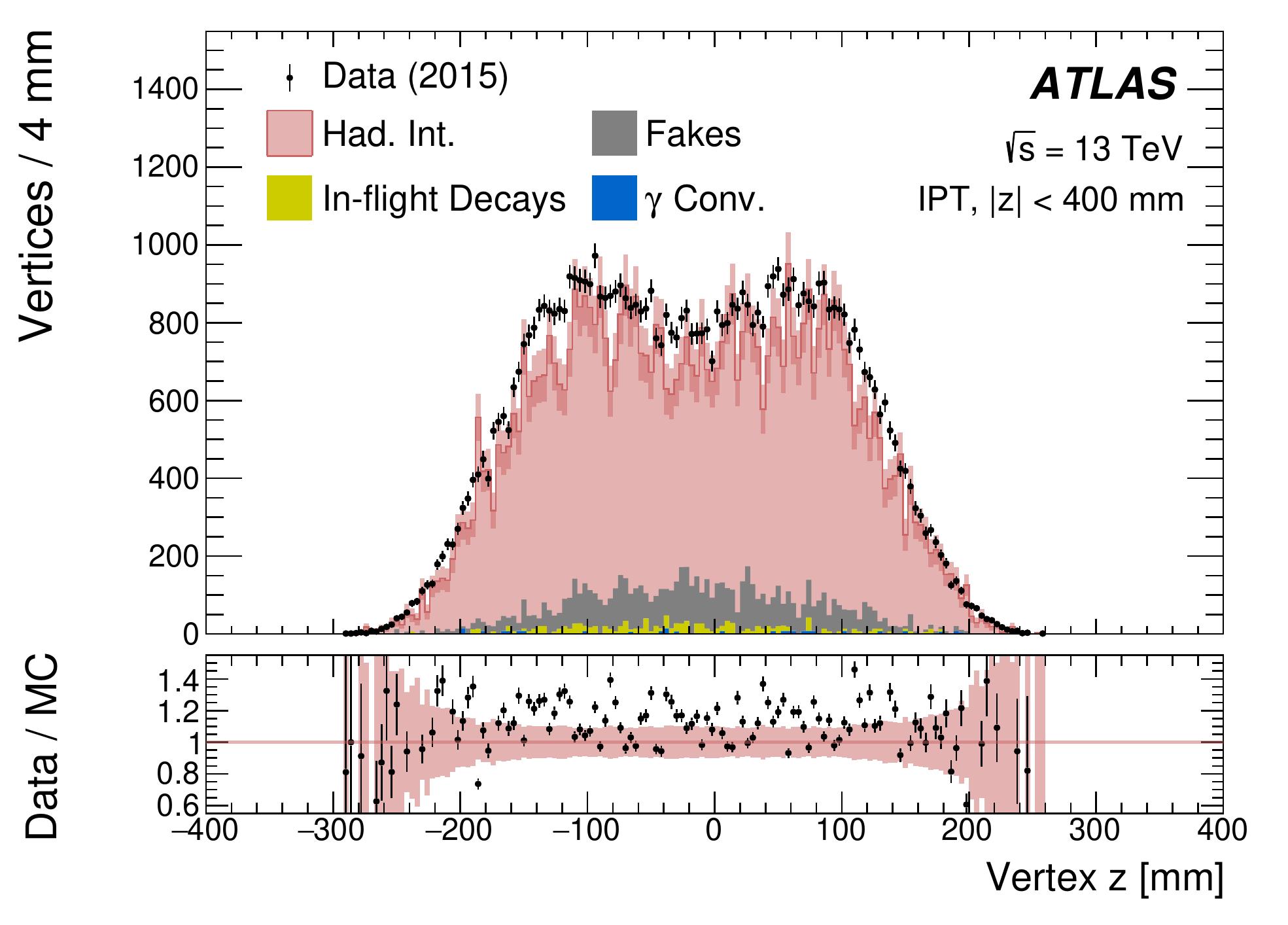}
}
\subfigure[]{
  \label{fig:compare_h1_phi_IPT_all}
  \includegraphics[width=0.4\textwidth]{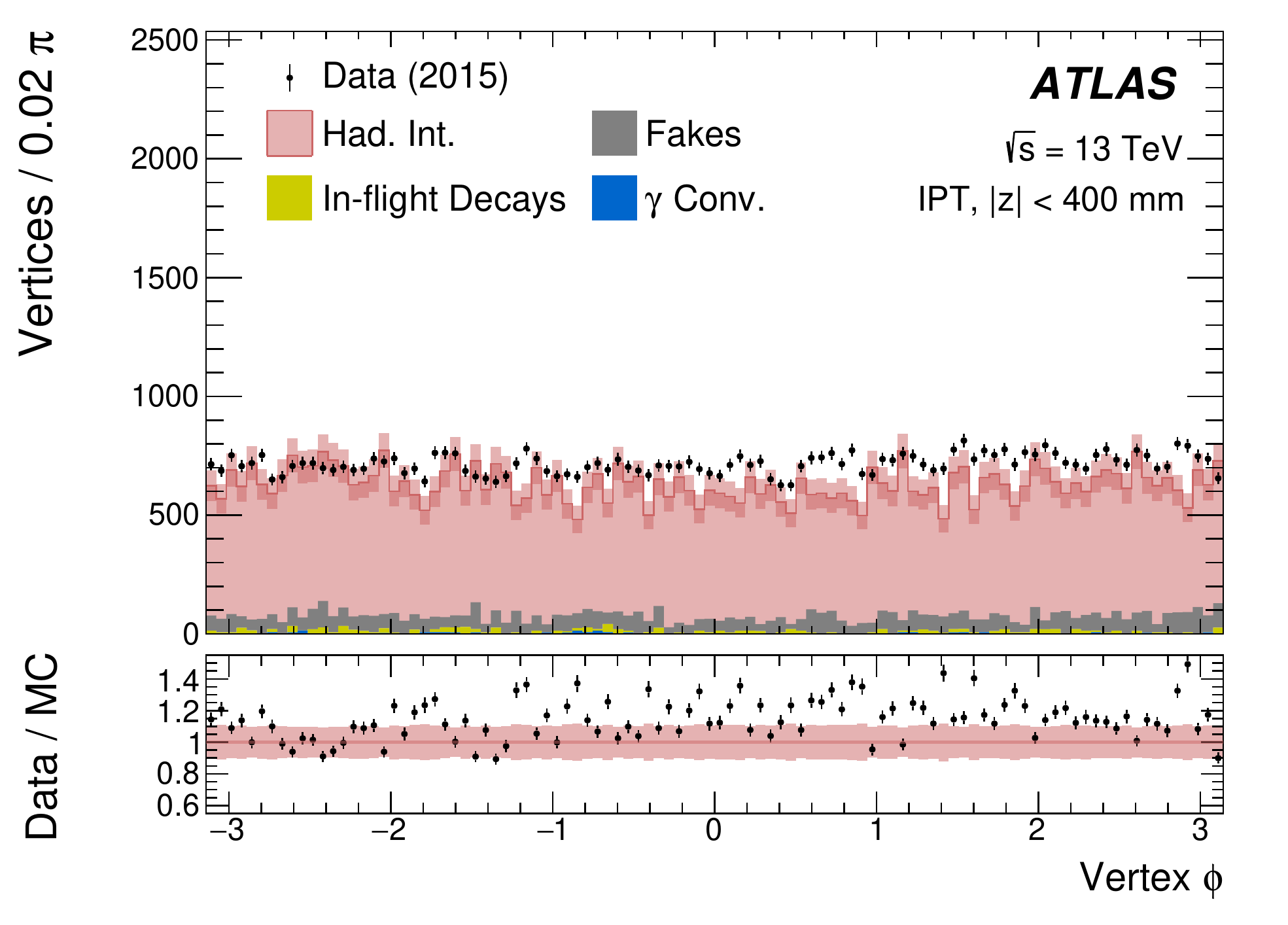}
}
\caption{Distribution of hadronic-interaction vertex candidates in $z$ and $\phi$ at the IPT ($28.5~\millimeter<r<30.0~\millimeter$) in data compared to the \textsc{Epos} MC simulation sample for the \emph{updated} geometry. The band in the ratio plots in the bottom panel indicates statistical uncertainty of the MC simulation samples.}
\label{fig:compare_h1_IPT}
\end{center}

\clearpage

\begin{center}
\subfigure[]{
  \label{fig:compare_h1_Z_IBL_all}
  \includegraphics[width=0.4\textwidth]{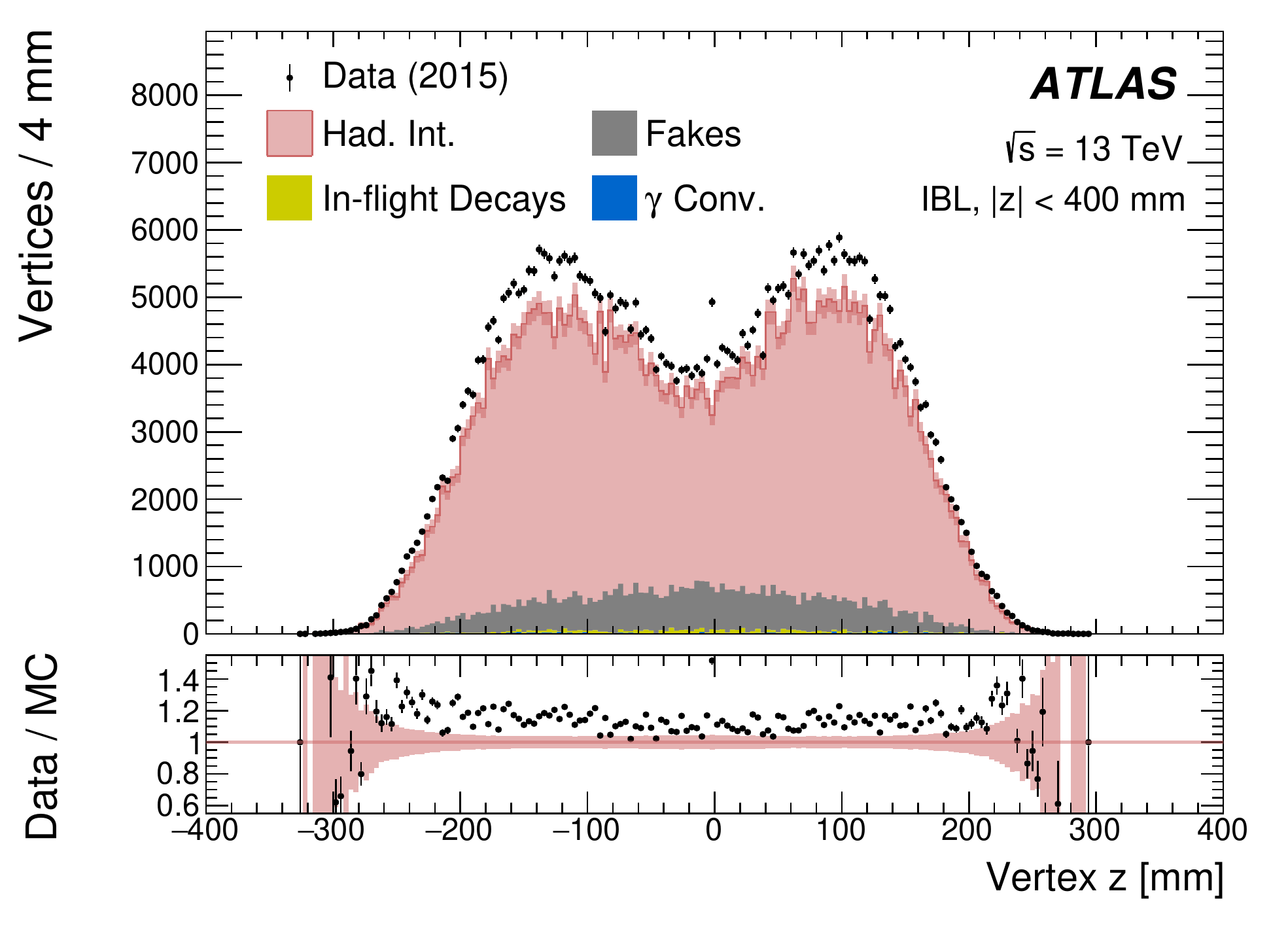}
}
\subfigure[]{
  \label{fig:compare_h1_phi_IBL_all}
  \includegraphics[width=0.4\textwidth]{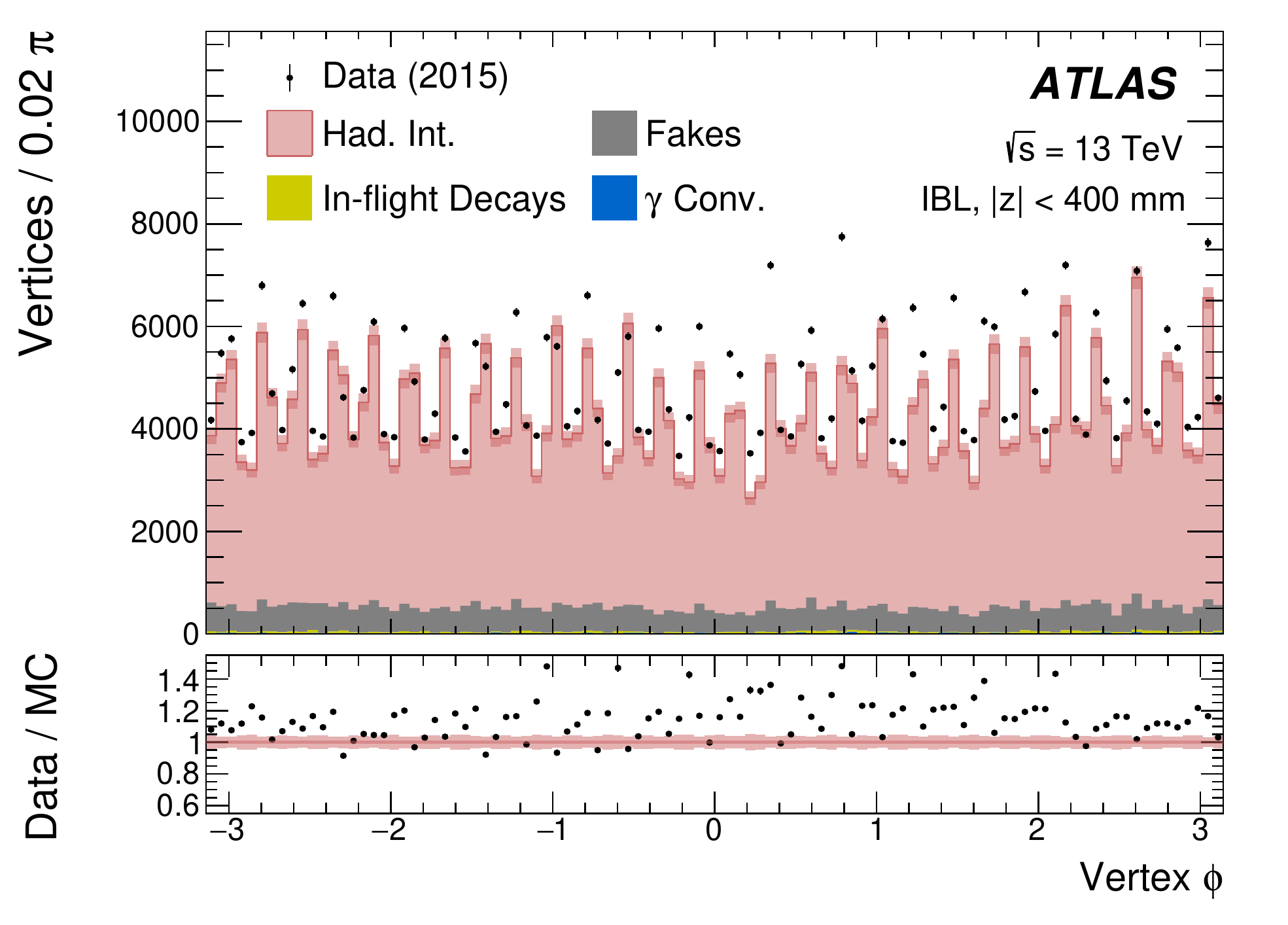}
}
\caption{Distribution of hadronic-interaction vertex candidates in $z$ and $\phi$ at the IBL ($30~\millimeter<r<40~\millimeter$) in data compared to the \textsc{Epos} MC simulation sample for the \emph{updated} geometry. The band in the ratio plots in the bottom panel indicates statistical uncertainty of the MC simulation samples.}
\label{fig:compare_h1_IPT}
\end{center}
\end{figure}

\clearpage

\begin{figure}[htbp]
\begin{center}
\subfigure[]{
  \label{fig:compare_h1_Z_IST_all}
  \includegraphics[width=0.4\textwidth]{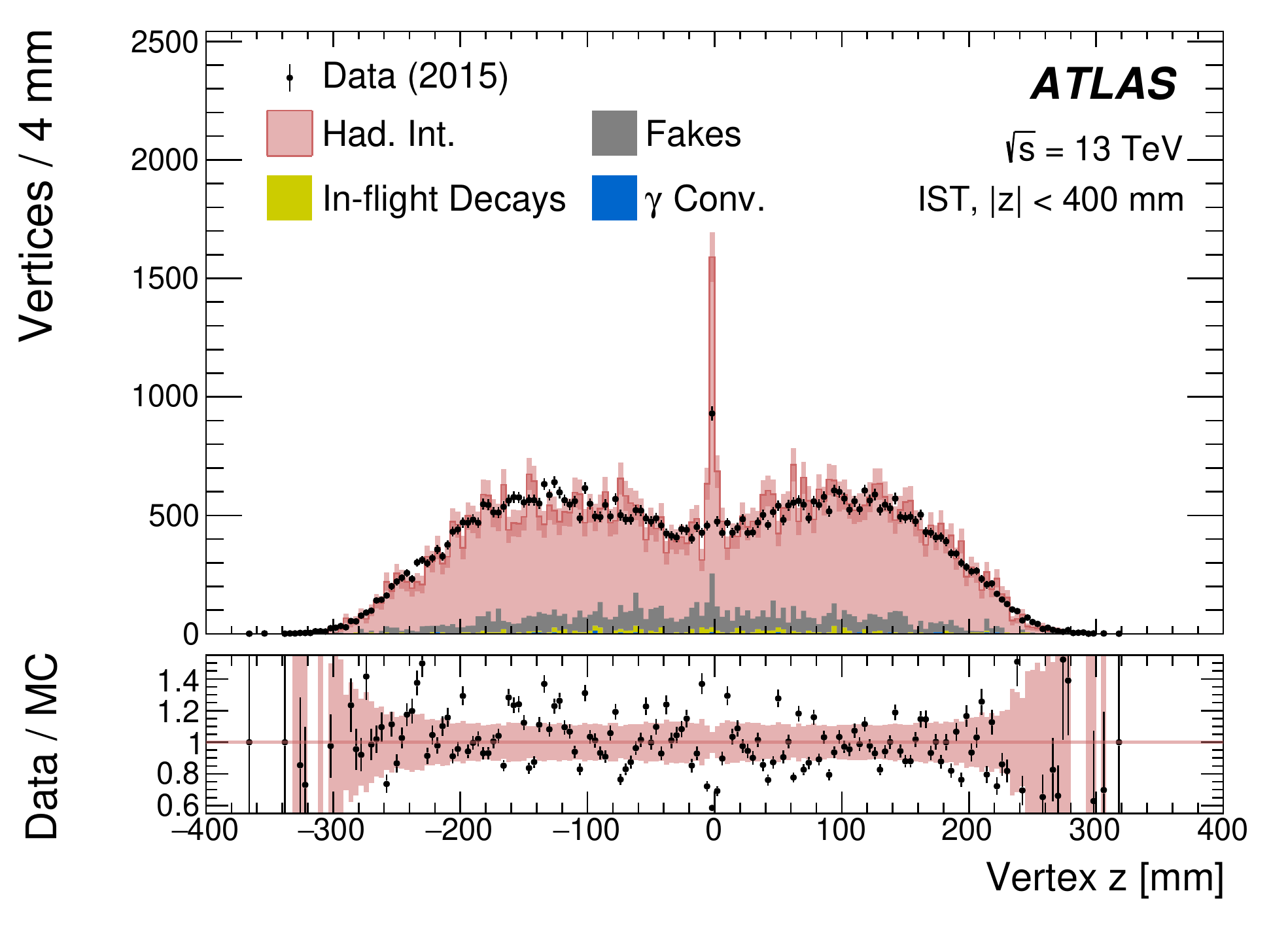}
}
\subfigure[]{
  \label{fig:compare_h1_phi_IST_all}
  \includegraphics[width=0.4\textwidth]{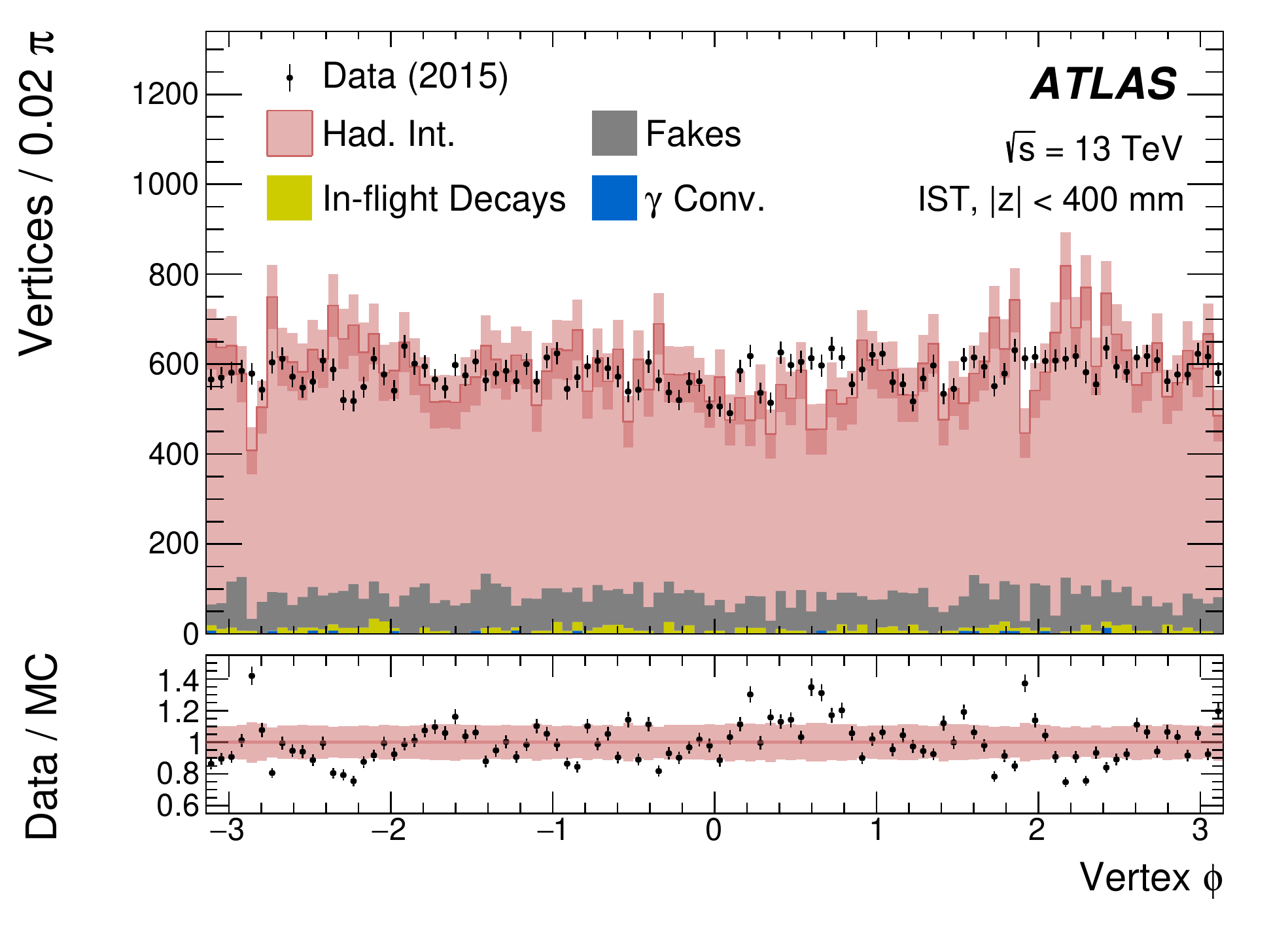}
}
\caption{Distribution of hadronic-interaction vertex candidates in $z$ and $\phi$ at the IST ($41.5~\millimeter<r<45~\millimeter$) in data compared to the \textsc{Epos} MC simulation sample for the \emph{updated} geometry. The band in the ratio plots in the bottom panel indicates statistical uncertainty of the MC simulation samples. The deficit of the data at $z=0~\mathrm{mm}$ is due to difference of classification of radial sections between data and the MC simulation reflecting the radial offset. The counterpart yield is observed in Figure~\ref{fig:compare_h1_Z_IBL_all}.}
\label{fig:compare_h1_IPT}
\end{center}

\begin{center}
\subfigure[]{
  \label{fig:compare_h1_Z_PIX1_all}
  \includegraphics[width=0.4\textwidth]{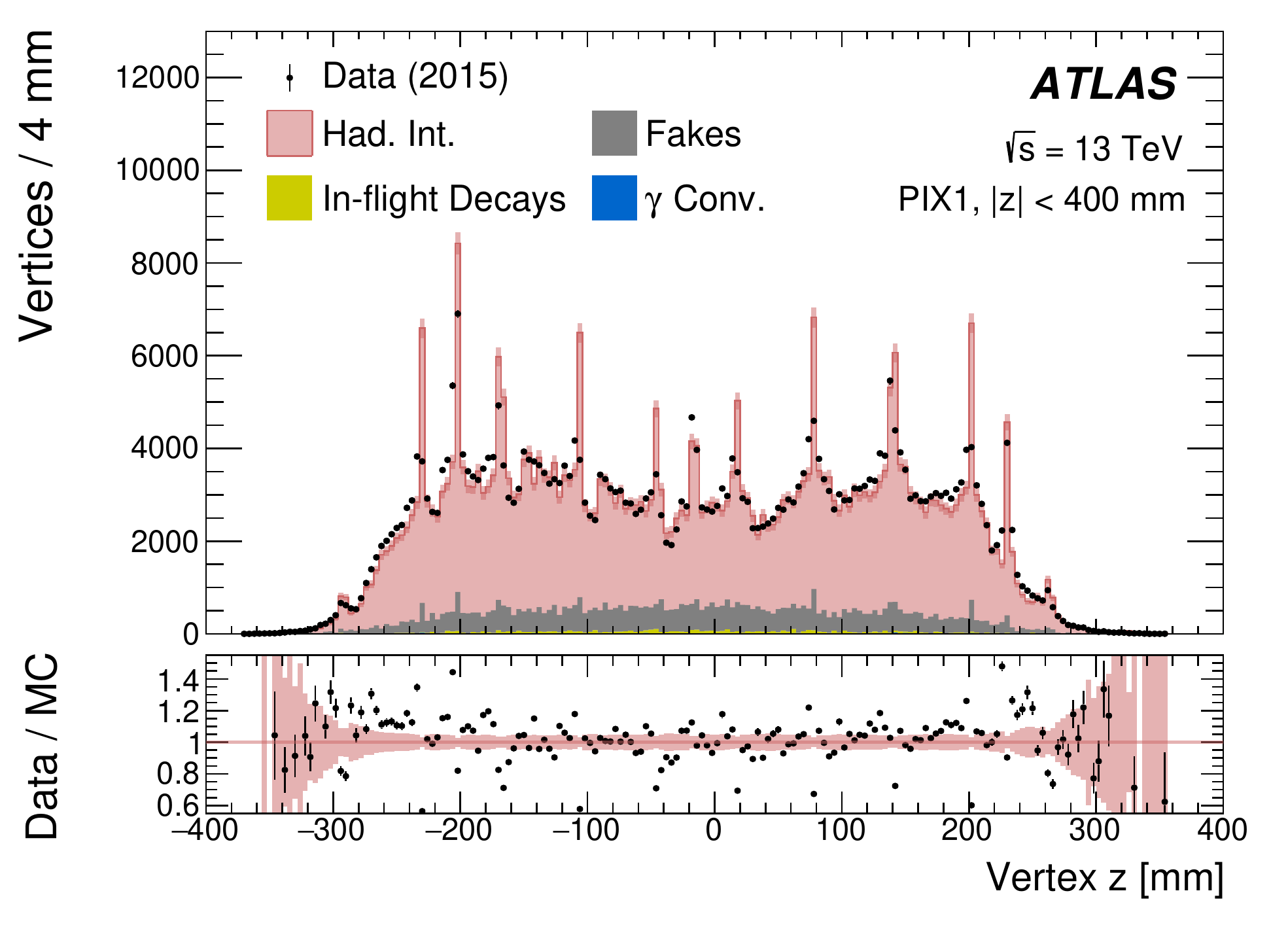}
}
\subfigure[]{
  \label{fig:compare_h1_phi_PIX1_all}
  \includegraphics[width=0.4\textwidth]{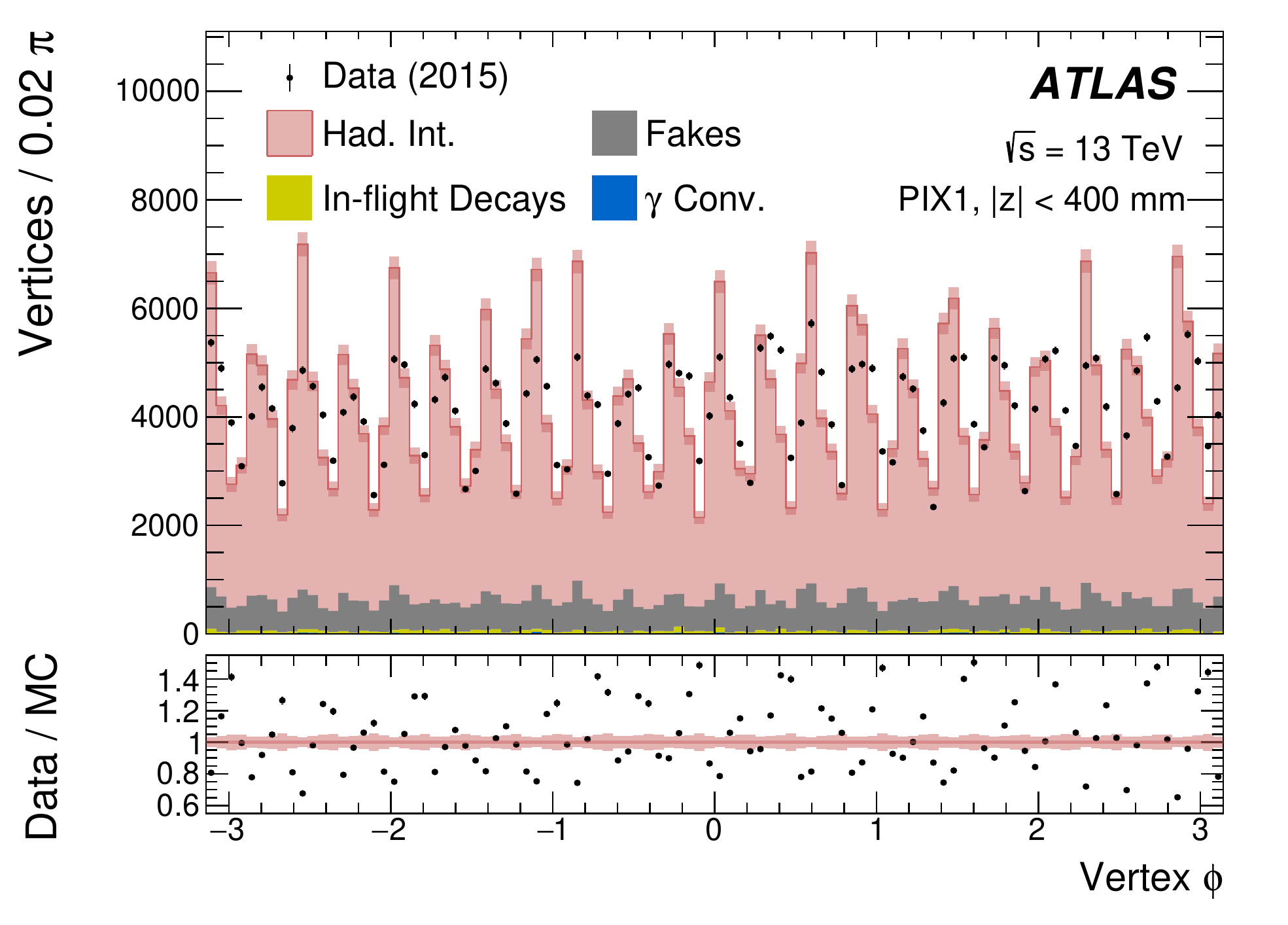}
}
\caption{Distribution of hadronic-interaction vertex candidates in $z$ and $\phi$ at the PIX1 ($45~\millimeter<r<75~\millimeter$) in data compared to the \textsc{Epos} MC simulation sample for the \emph{updated} geometry. The band in the ratio plots in the bottom panel indicates statistical uncertainty of the MC simulation samples.}
\label{fig:compare_h1_PIX1}
\end{center}

\begin{center}
\subfigure[]{
  \label{fig:compare_h1_Z_PIX2_all}
  \includegraphics[width=0.4\textwidth]{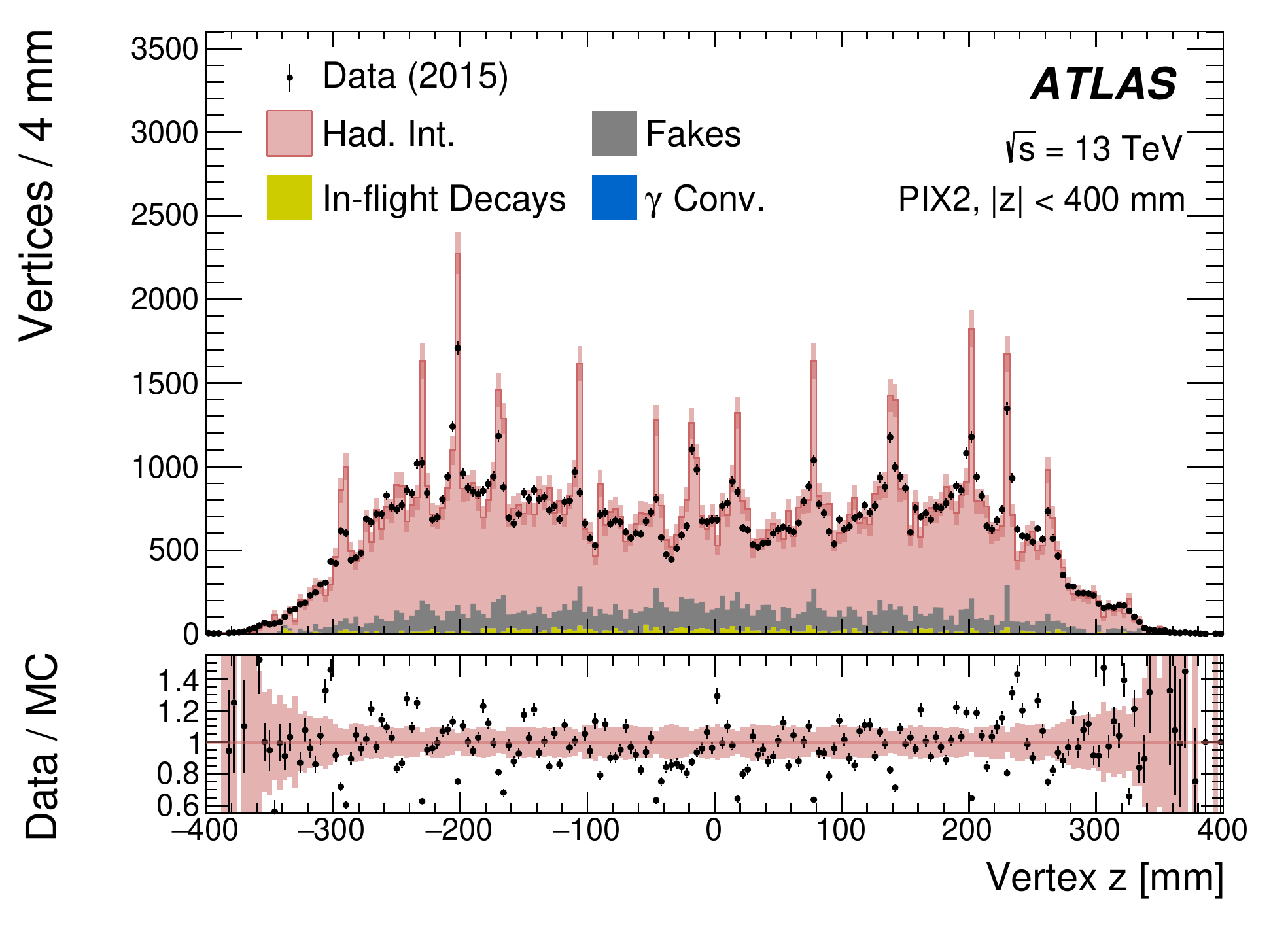}
}
\subfigure[]{
  \label{fig:compare_h1_phi_PIX2_all}
  \includegraphics[width=0.4\textwidth]{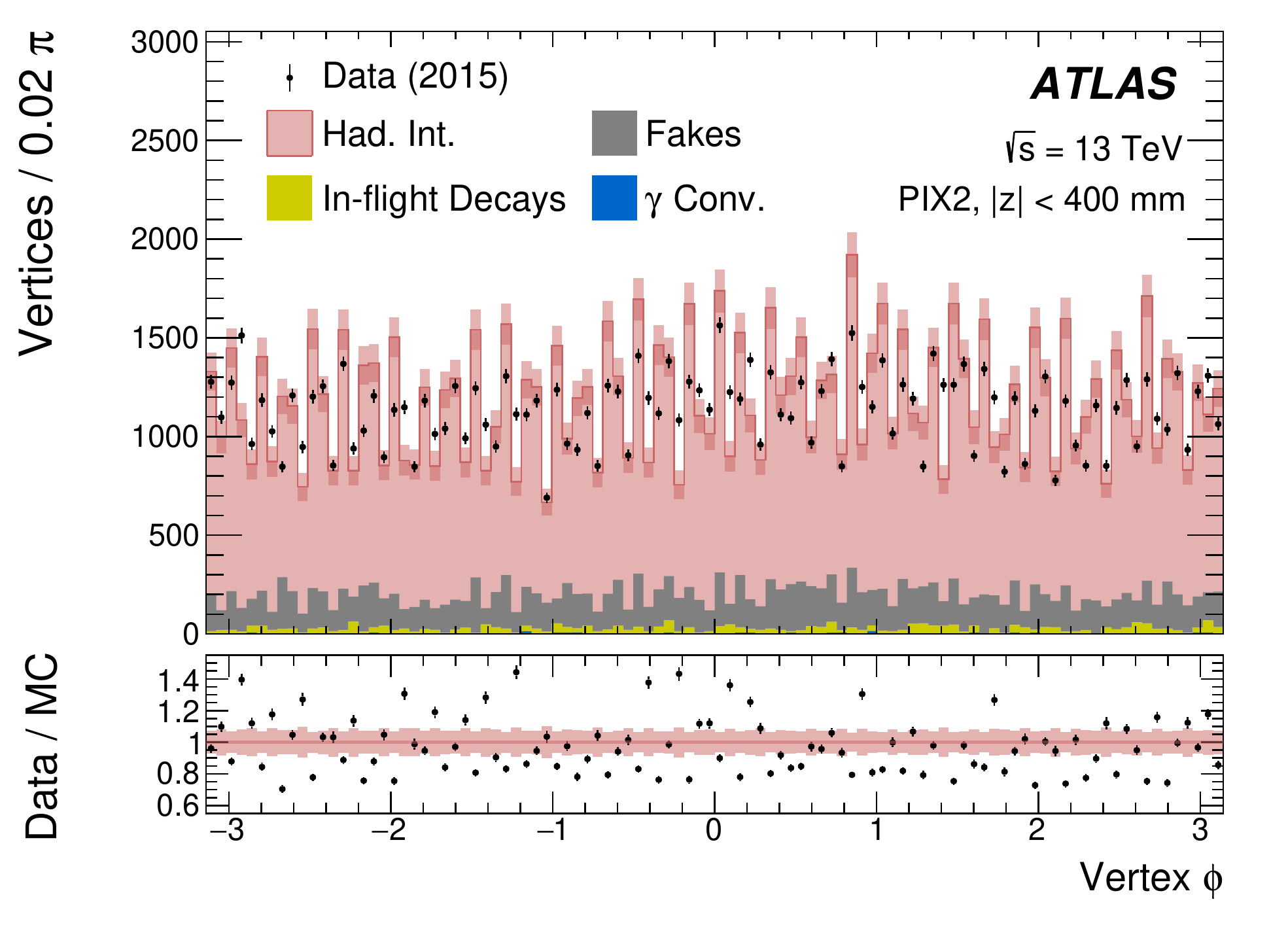}
}
\caption{Distribution of hadronic-interaction vertex candidates in $z$ and $\phi$ at the PIX2 ($83~\millimeter<r<110~\millimeter$) in data compared to the \textsc{Epos} MC simulation sample for the \emph{updated} geometry. The band in the ratio plots in the bottom panel indicates statistical uncertainty of the MC simulation samples.}
\label{fig:compare_h1_PIX2}
\end{center}
\end{figure}

\clearpage

\begin{figure}[htbp]
\begin{center}
\subfigure[]{
  \label{fig:compare_h1_Z_PIX3_all}
  \includegraphics[width=0.4\textwidth]{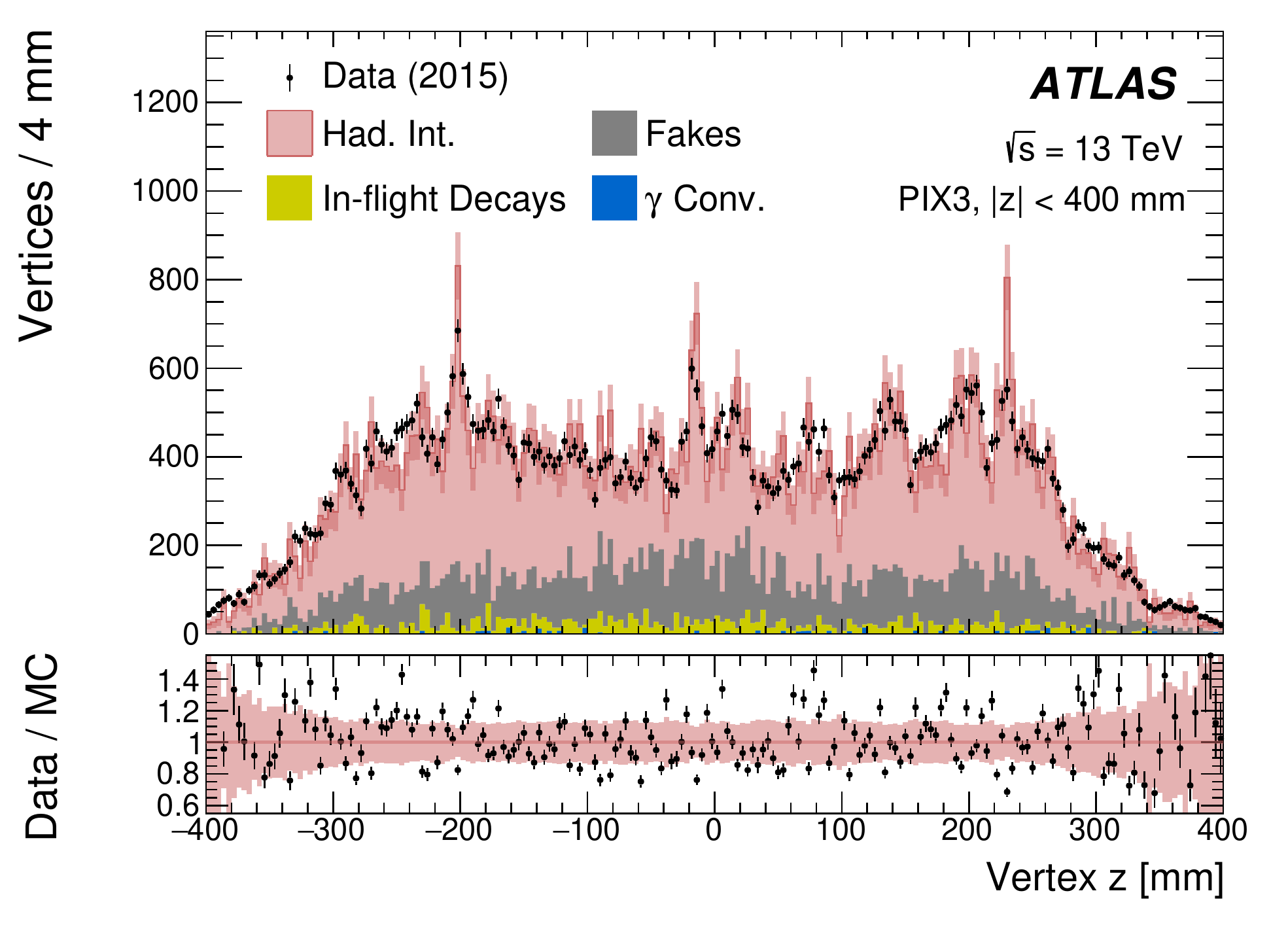}
}
\subfigure[]{
  \label{fig:compare_h1_phi_PIX3_all}
  \includegraphics[width=0.4\textwidth]{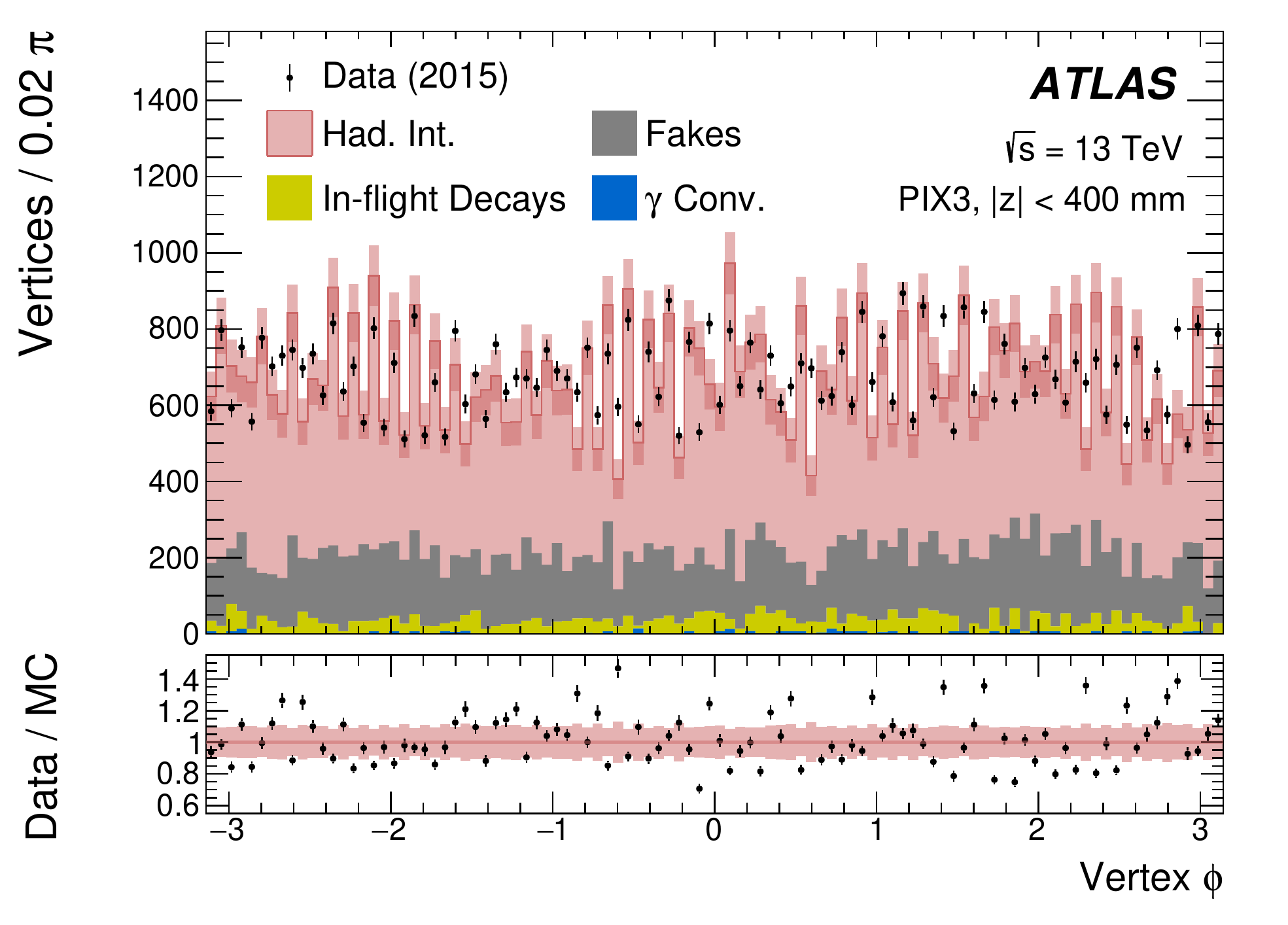}
}
\caption{Distribution of hadronic-interaction vertex candidates in $z$ and $\phi$ at the PIX3 ($118~\millimeter<r<145~\millimeter$) in data compared to the \textsc{Epos} MC simulation sample for the \emph{updated} geometry. The band in the ratio plots in the bottom panel indicates statistical uncertainty of the MC simulation samples.}
\label{fig:compare_h1_PIX3}
\end{center}

\begin{center}
\subfigure[]{
  \label{fig:compare_h1_Z_PST_all}
  \includegraphics[width=0.4\textwidth]{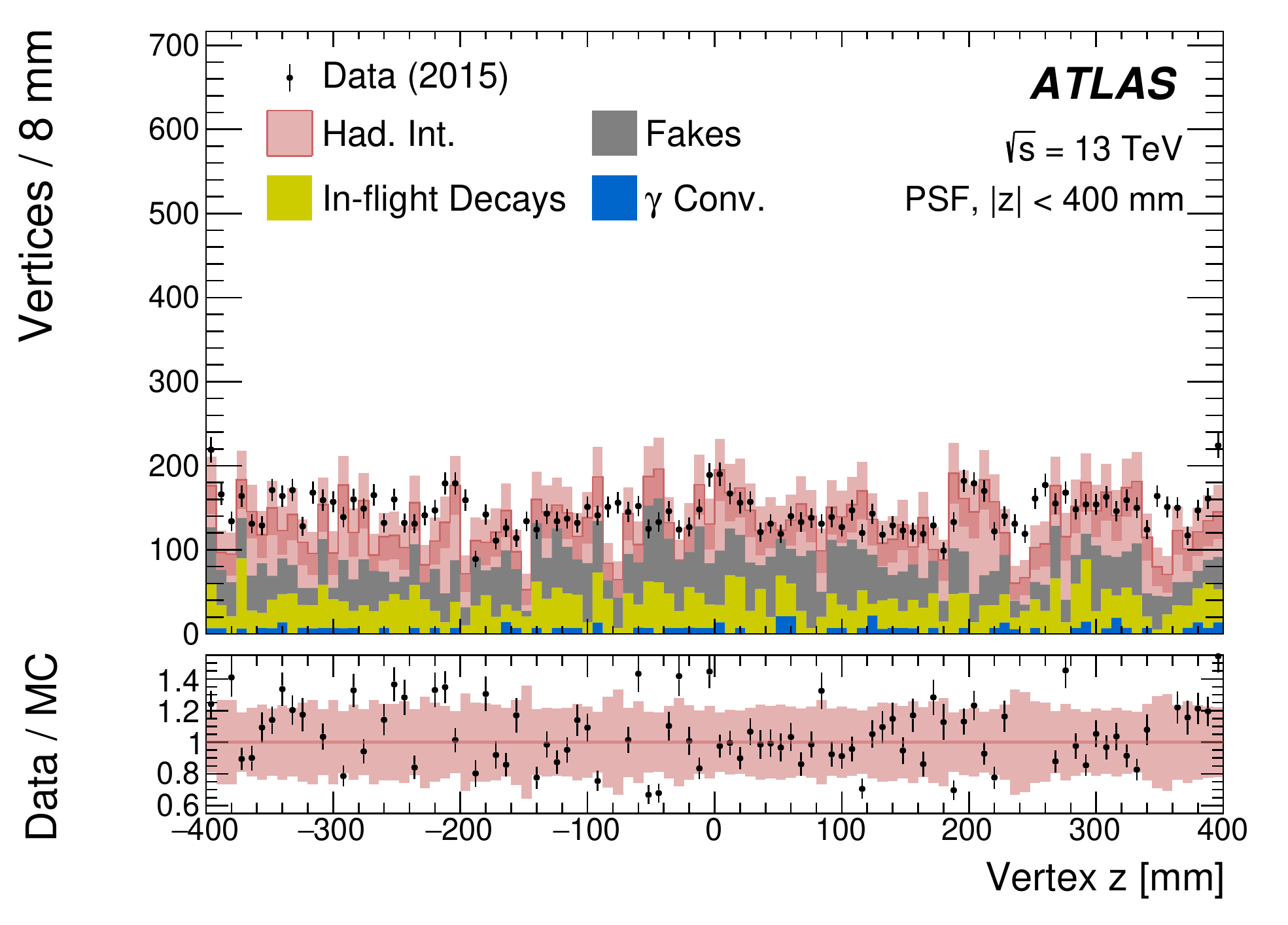}
}
\subfigure[]{
  \label{fig:compare_h1_phi_PST_all}
  \includegraphics[width=0.4\textwidth]{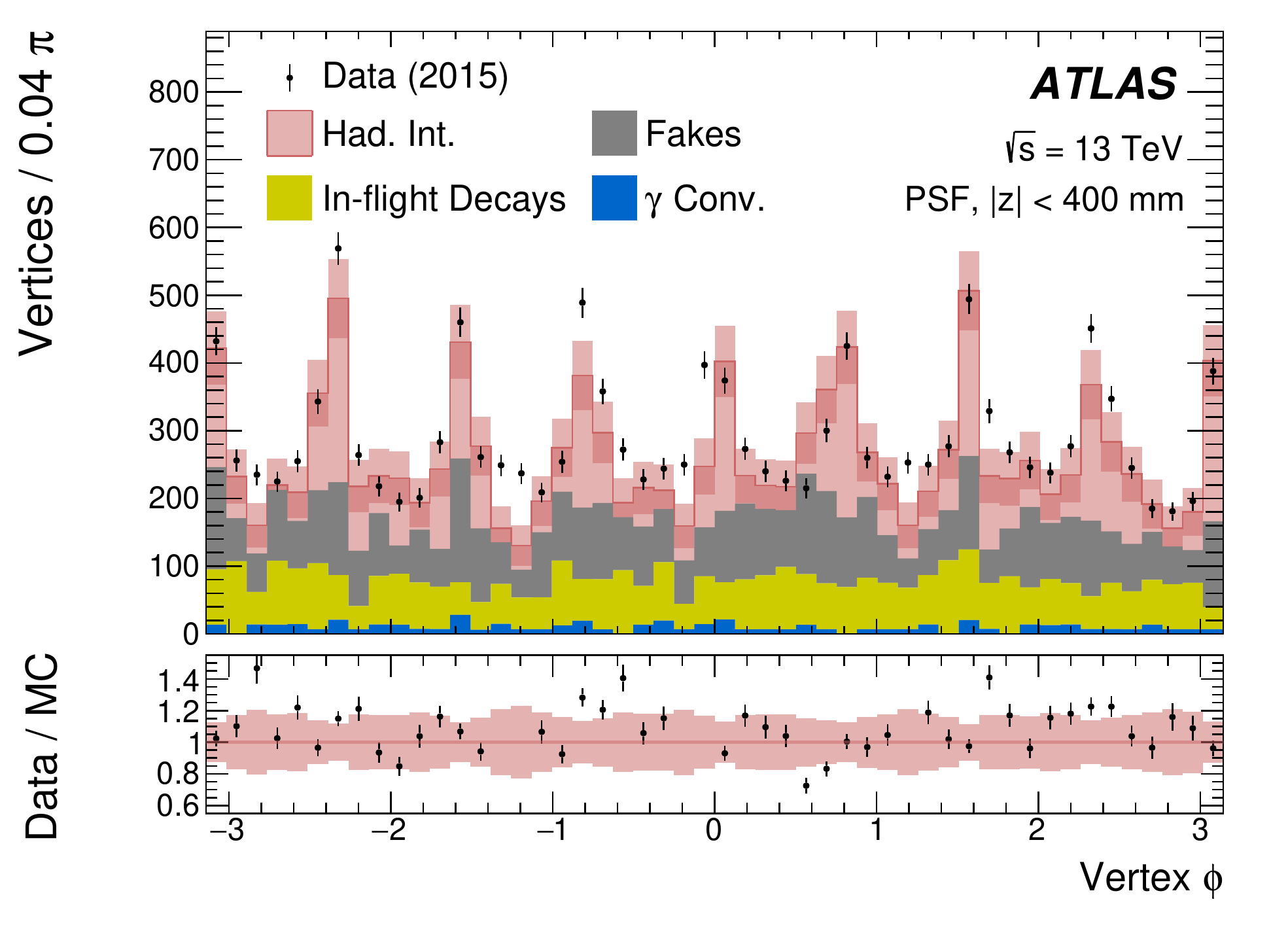}
}
\caption{Distribution of hadronic-interaction vertex candidates in $z$ and $\phi$ at the PSF ($180~\millimeter<r<225~\millimeter$) in data compared to the \textsc{Epos} MC simulation sample for the \emph{updated} geometry. The band in the ratio plots in the bottom panel indicates statistical uncertainty of the MC simulation samples.}
\label{fig:compare_h1_PST}
\end{center}

\begin{center}
\subfigure[]{
  \label{fig:compare_h1_Z_PST_all}
  \includegraphics[width=0.4\textwidth]{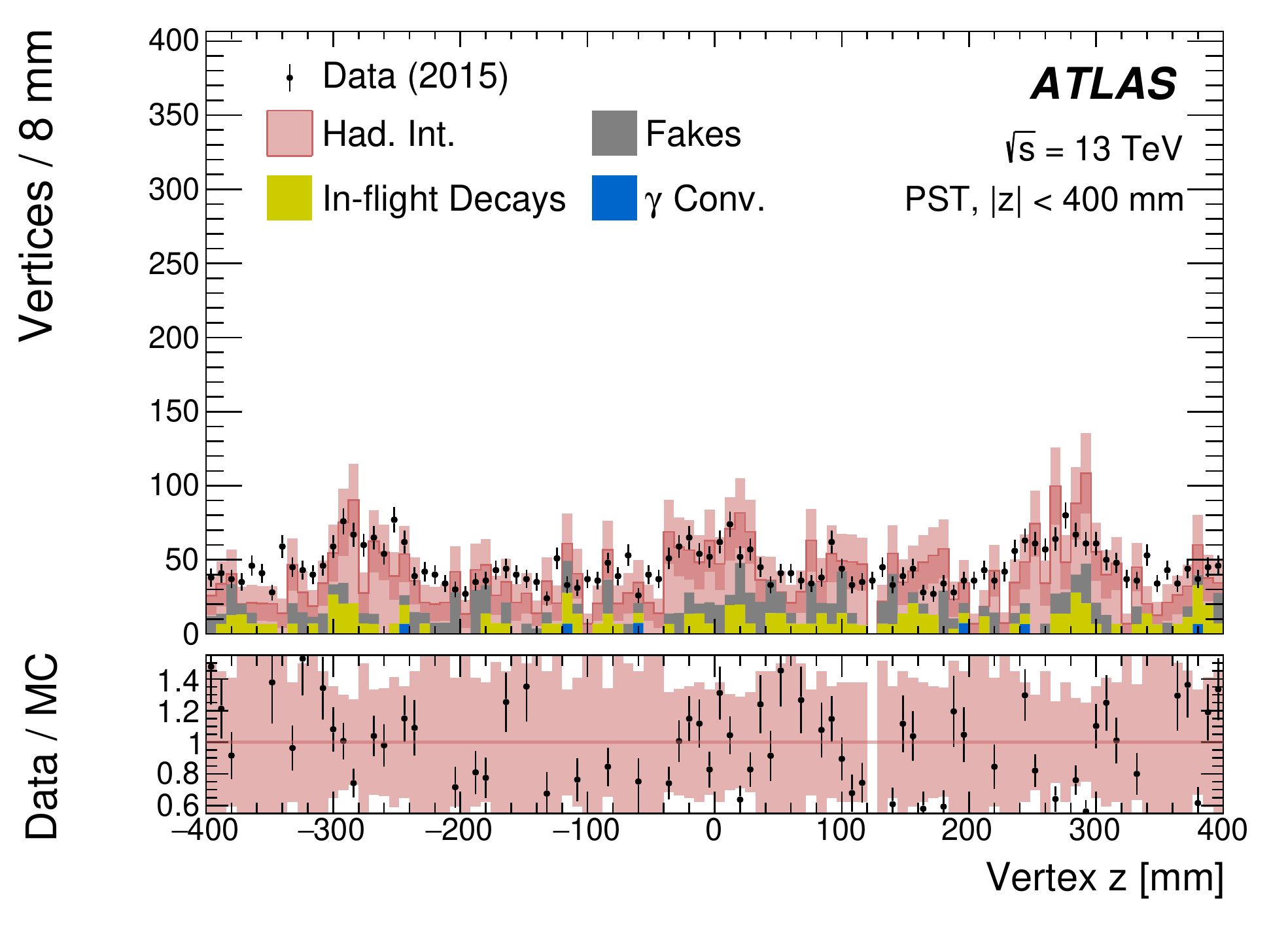}
}
\subfigure[]{
  \label{fig:compare_h1_phi_PST_all}
  \includegraphics[width=0.4\textwidth]{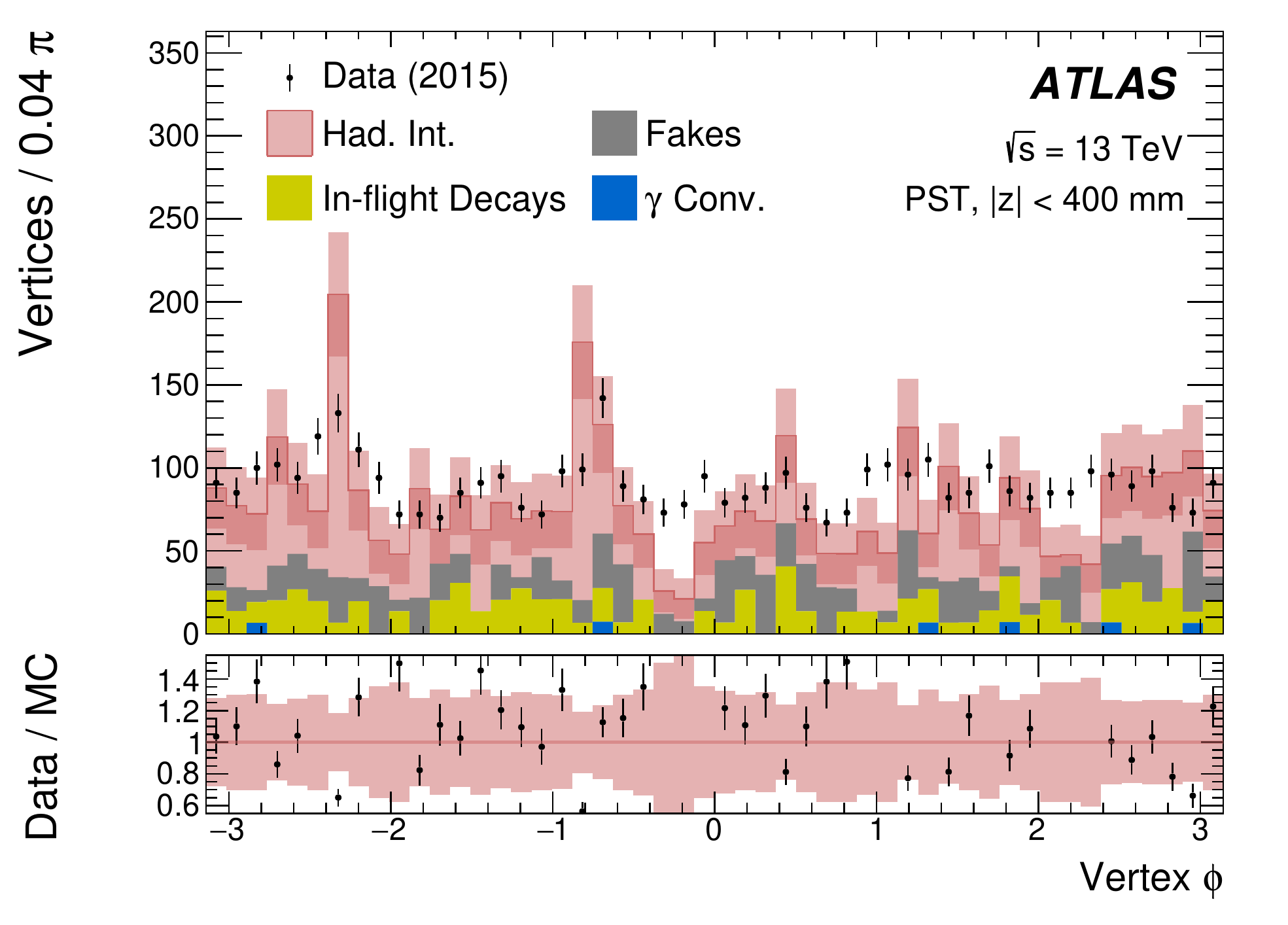}
}
\caption{Distribution of hadronic-interaction vertex candidates in $z$ and $\phi$ at the PST ($225~\millimeter<r<240~\millimeter$) in data compared to the \textsc{Epos} MC simulation sample for the \emph{updated} geometry. The band in the ratio plots in the bottom panel indicates statistical uncertainty of the MC simulation samples.}
\label{fig:compare_h1_PST}
\end{center}
\end{figure}

\clearpage

\begin{figure}[htbp]
\begin{center}
\subfigure[]{
  \label{fig:compare_h1_Z_PST_all}
  \includegraphics[width=0.4\textwidth]{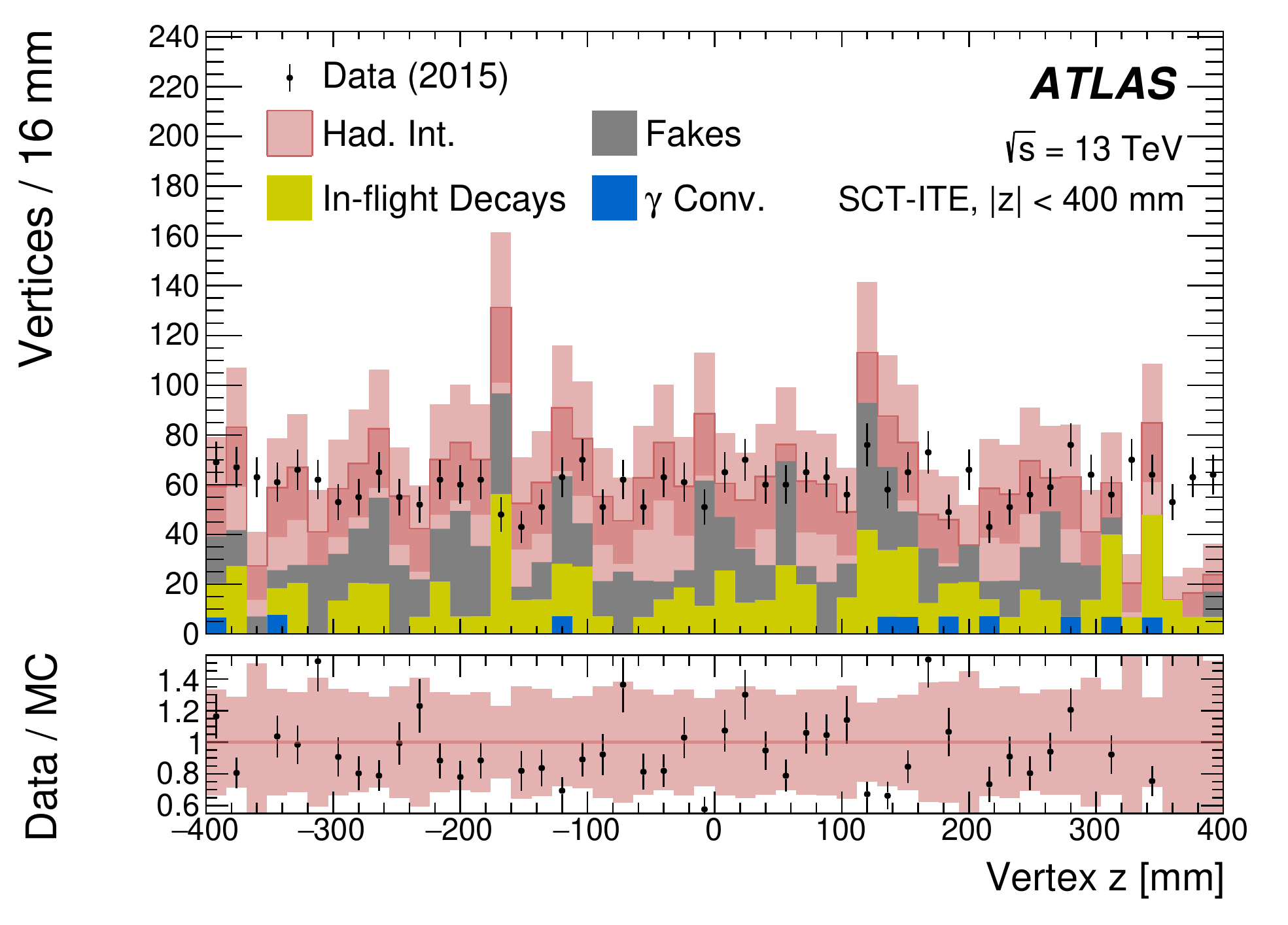}
}
\subfigure[]{
  \label{fig:compare_h1_phi_PST_all}
  \includegraphics[width=0.4\textwidth]{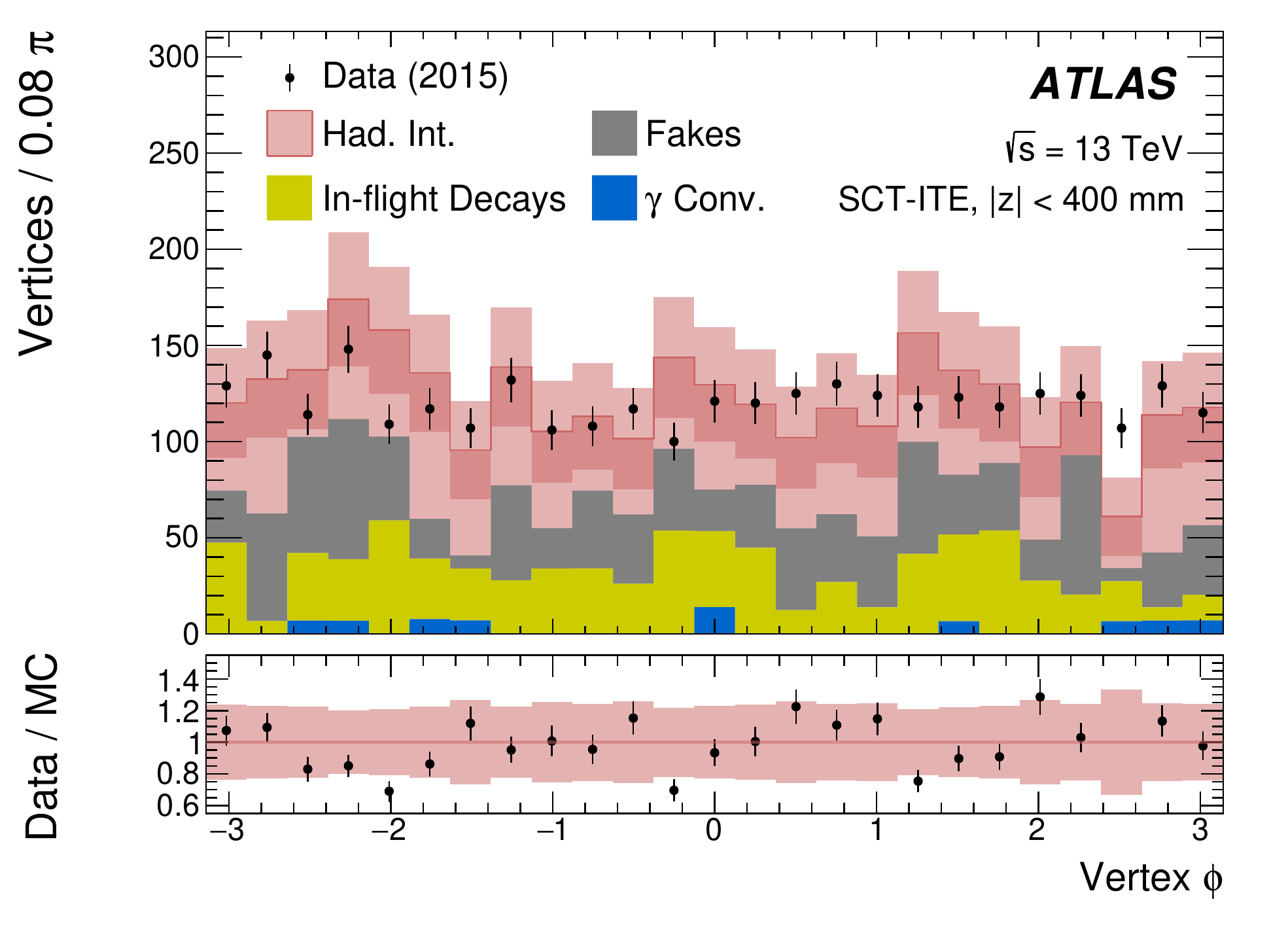}
}
\caption{Distribution of hadronic-interaction vertex candidates in $z$ and $\phi$ at the SCT-ITE ($245~\millimeter<r<265~\millimeter$) in data compared to the \textsc{Epos} MC simulation sample for the \emph{updated} geometry. The band in the ratio plots in the bottom panel indicates statistical uncertainty of the MC simulation samples.}
\label{fig:compare_h1_PST}
\end{center}

\begin{center}
\subfigure[]{
  \label{fig:compare_h1_Z_SCT1_all}
  \includegraphics[width=0.4\textwidth]{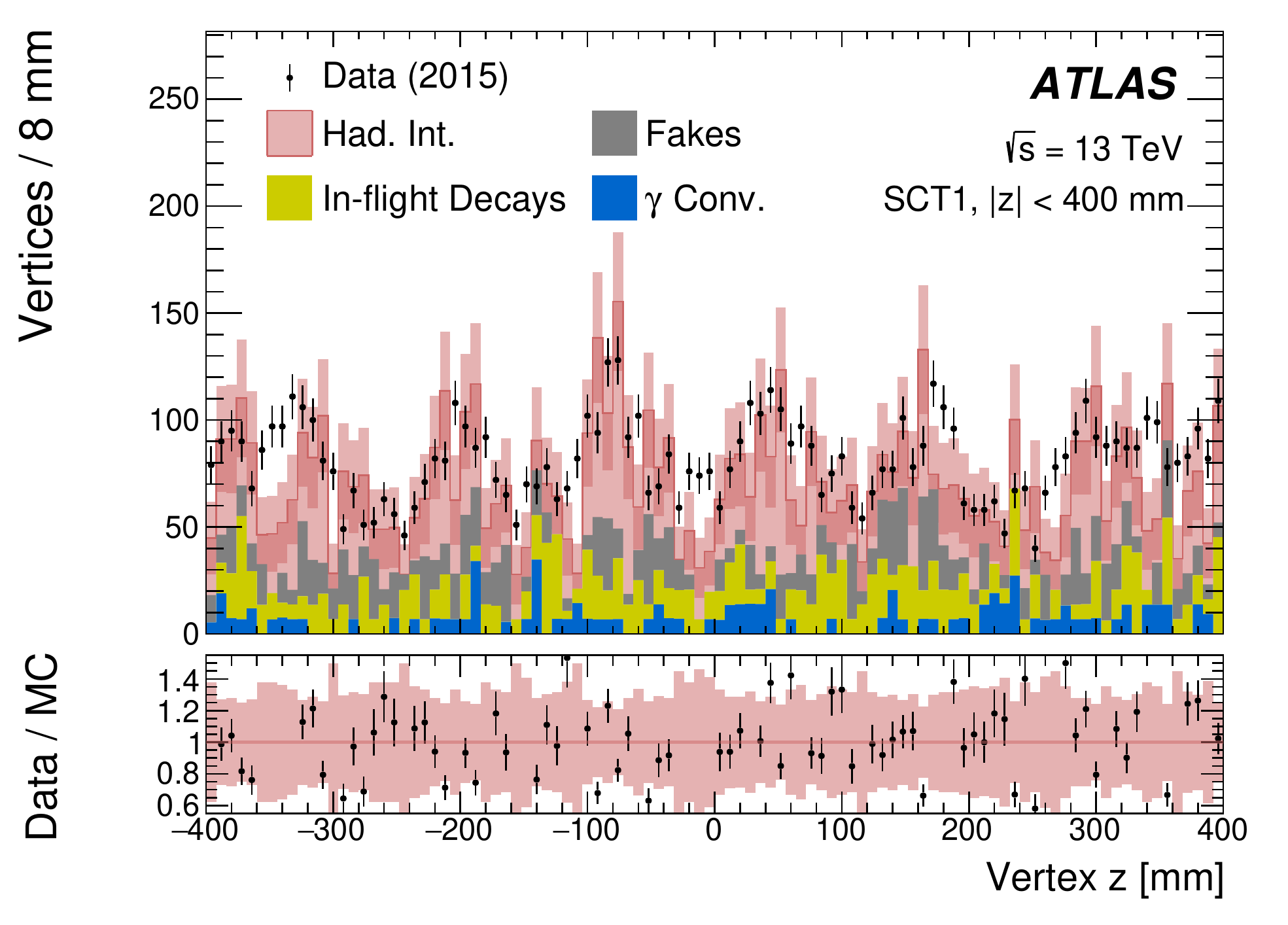}
}
\subfigure[]{
  \label{fig:compare_h1_phi_SCT1_all}
  \includegraphics[width=0.4\textwidth]{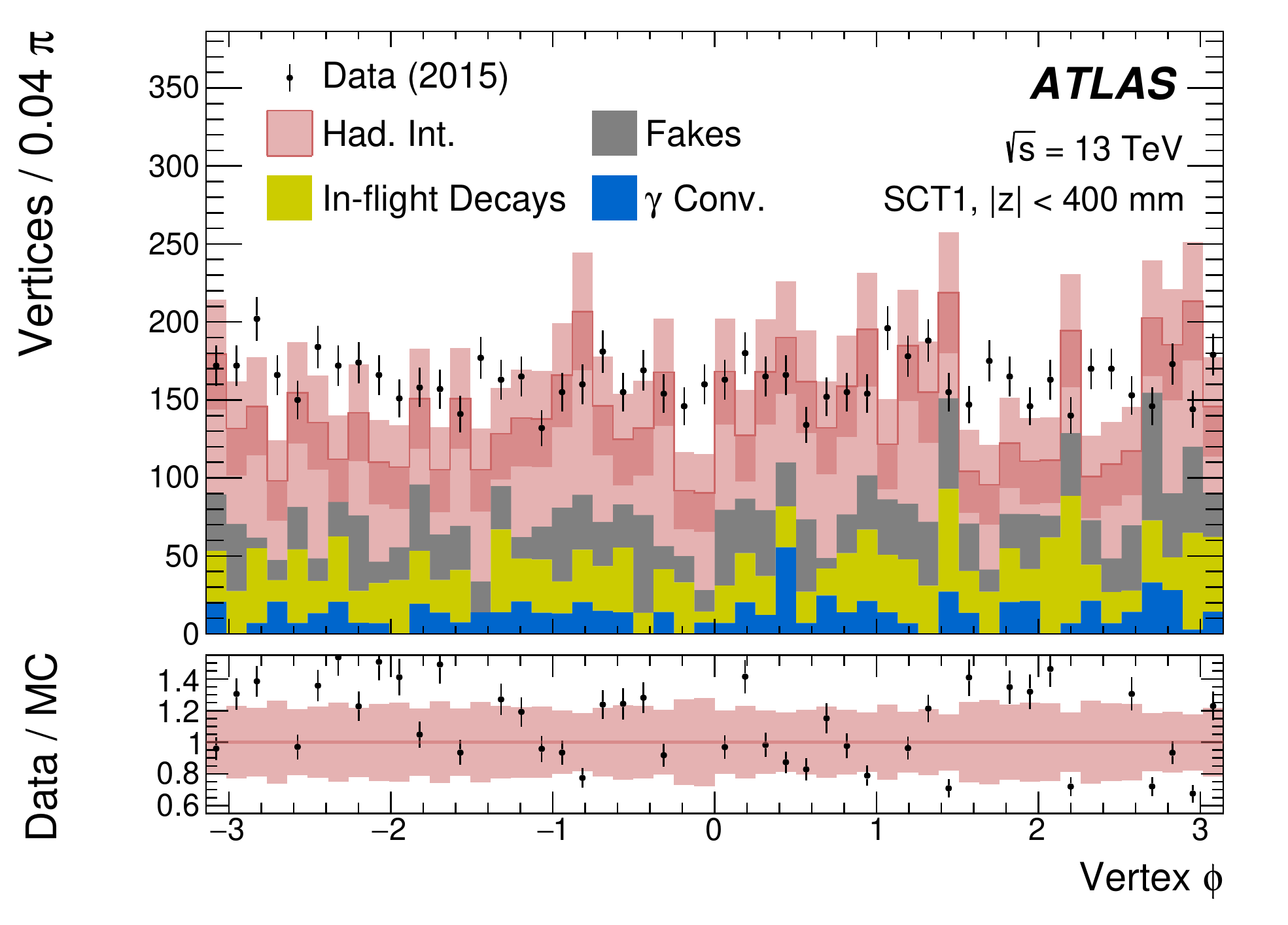}
}
\caption{Distribution of hadronic-interaction vertex candidates in $z$ and $\phi$ at the SCT1 ($276~\millimeter<r<320~\millimeter$) in data compared to the \textsc{Epos} MC simulation sample for the \emph{updated} geometry. The band in the ratio plots in the bottom panel indicates statistical uncertainty of the MC simulation samples.}
\label{fig:compare_h1_SCT1}
\end{center}

\begin{center}
\subfigure[]{
  \label{fig:compare_h1_Z_SCT2_all}
  \includegraphics[width=0.4\textwidth]{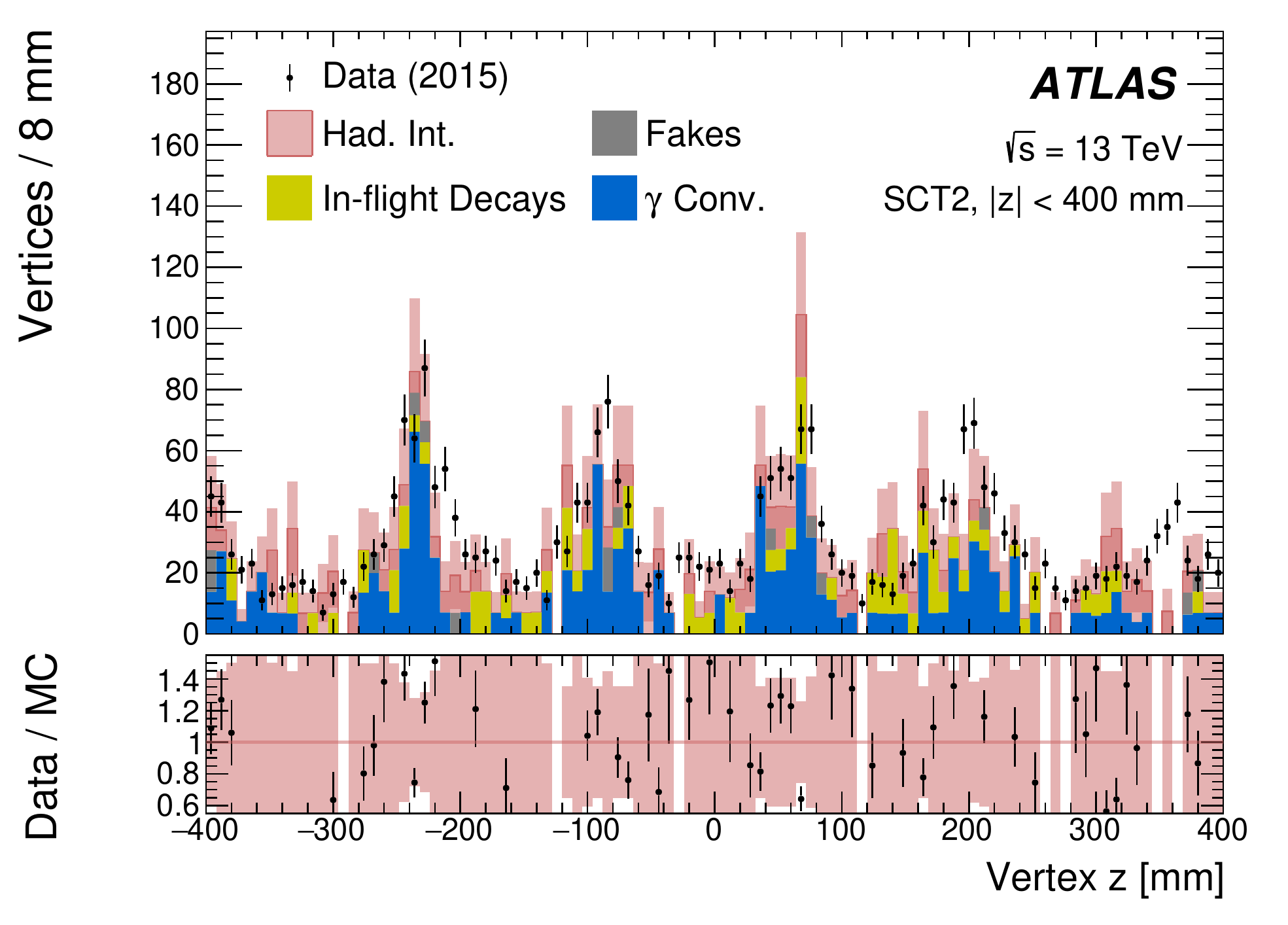}
}
\subfigure[]{
  \label{fig:compare_h1_phi_SCT2_all}
  \includegraphics[width=0.4\textwidth]{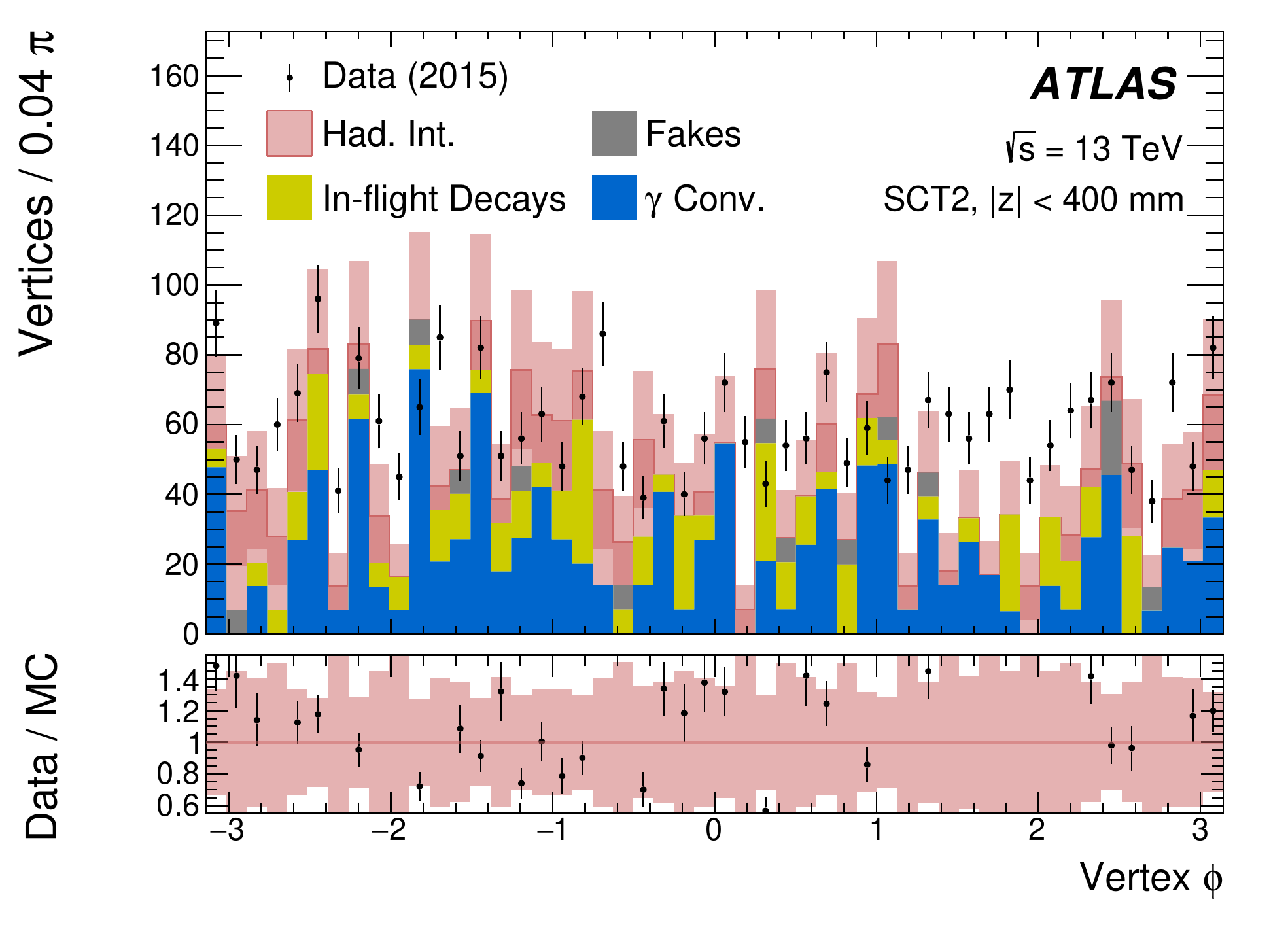}
}
\caption{Distribution of hadronic-interaction vertex candidates in $z$ and $\phi$ at the SCT2 ($347~\millimeter<r<390~\millimeter$) in data compared to the \textsc{Epos} MC simulation sample for the \emph{updated} geometry. The band in the ratio plots in the bottom panel indicates statistical uncertainty of the MC simulation samples.}
\label{fig:compare_h1_SCT2}
\end{center}
\end{figure}

\clearpage

\begin{figure}
\begin{center}
\includegraphics[width=0.45\textwidth]{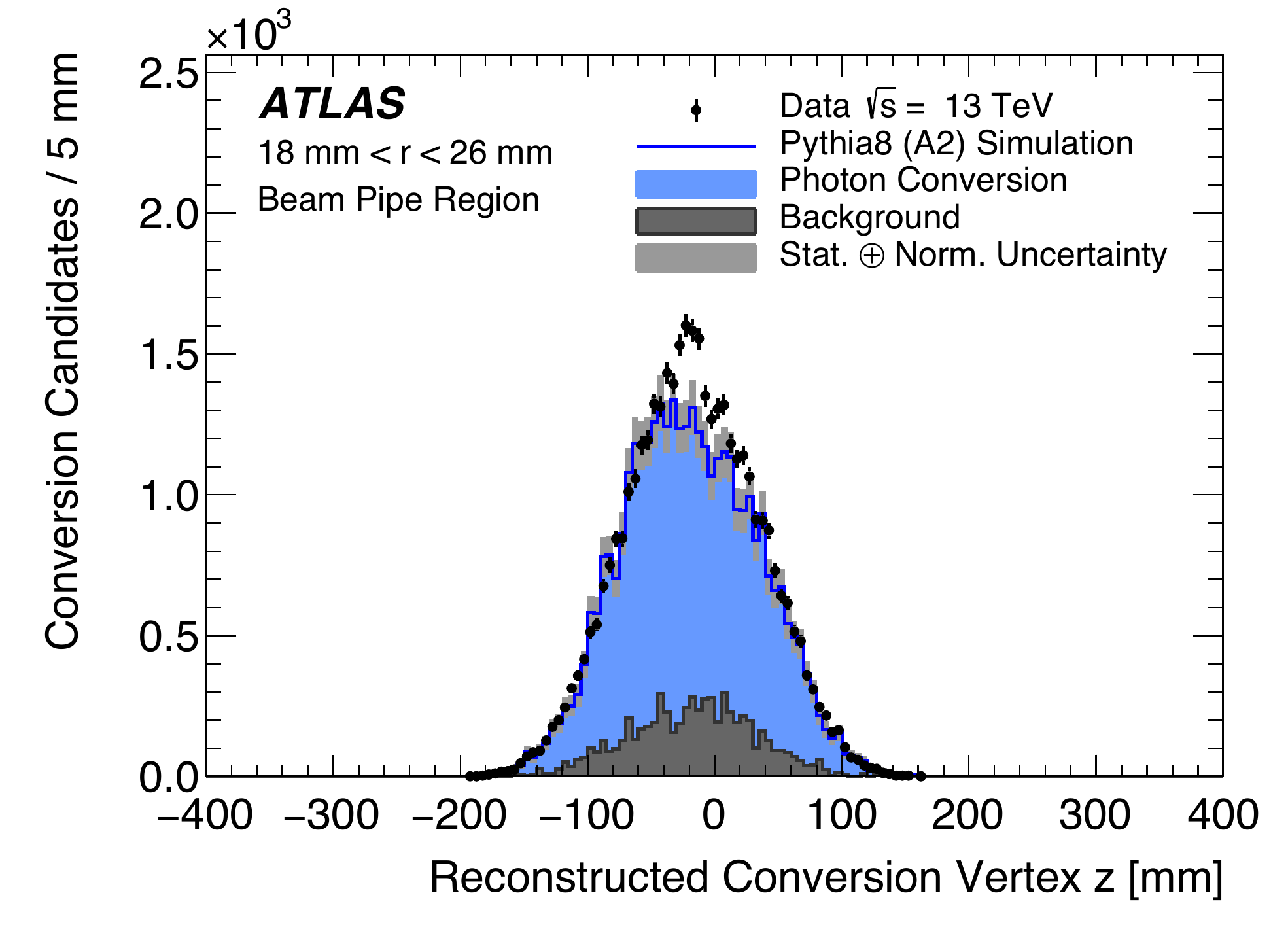}
\caption{Conversion vertex $z$-position distributions for \textsc{Pythia~8} simulation compared to data in the beam-pipe radial region.}
\label{figures:conv_aux1}
\end{center}
\end{figure}

\begin{figure}
\begin{center}
\includegraphics[width=0.45\textwidth]{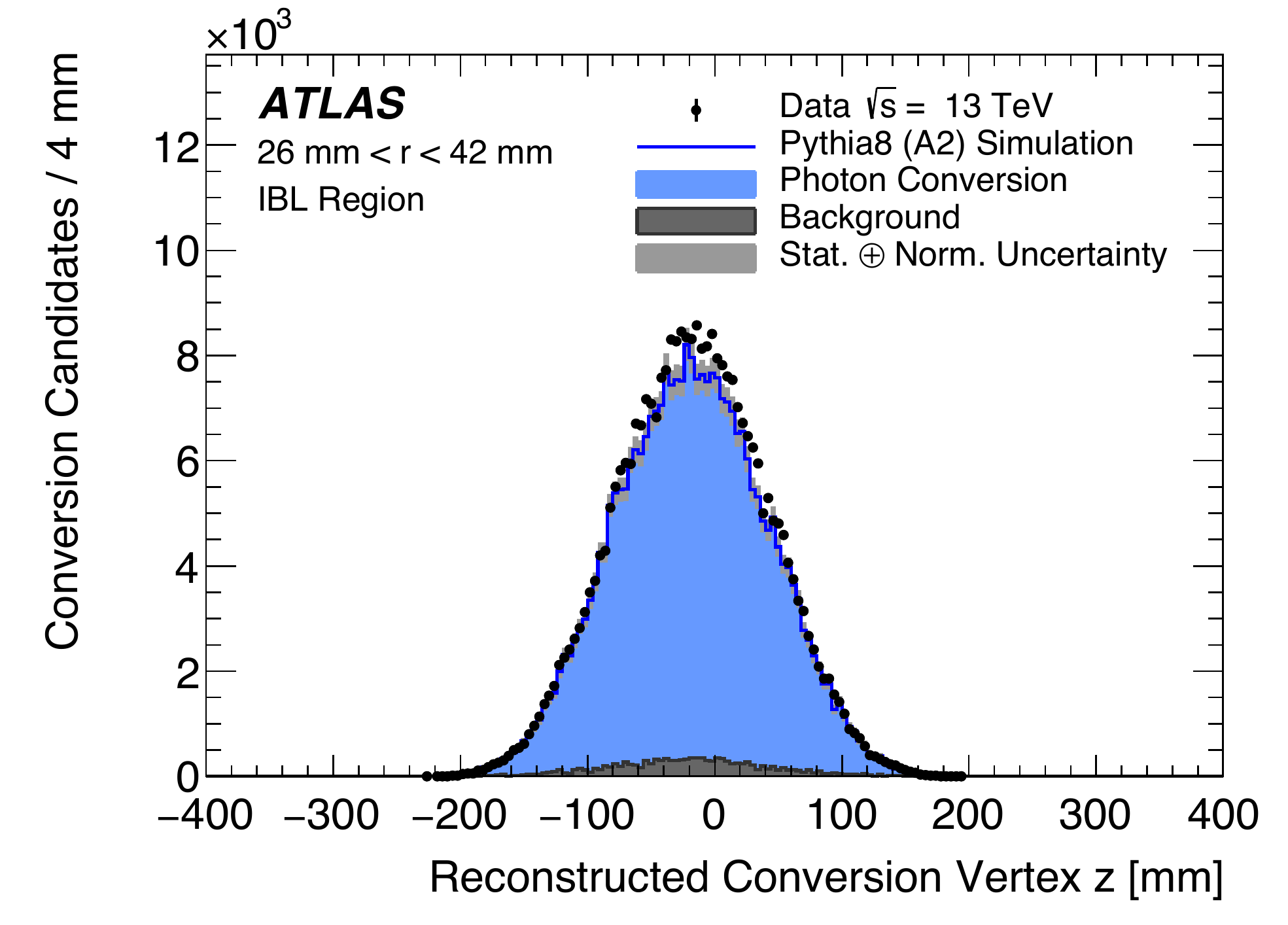}
\caption{Conversion vertex $z$-position distributions for \textsc{Pythia~8} simulation compared to data in the IBL radial region.}
\label{figures:conv_aux2}
\end{center}
\end{figure}

\begin{figure}
\begin{center}
\includegraphics[width=0.45\textwidth]{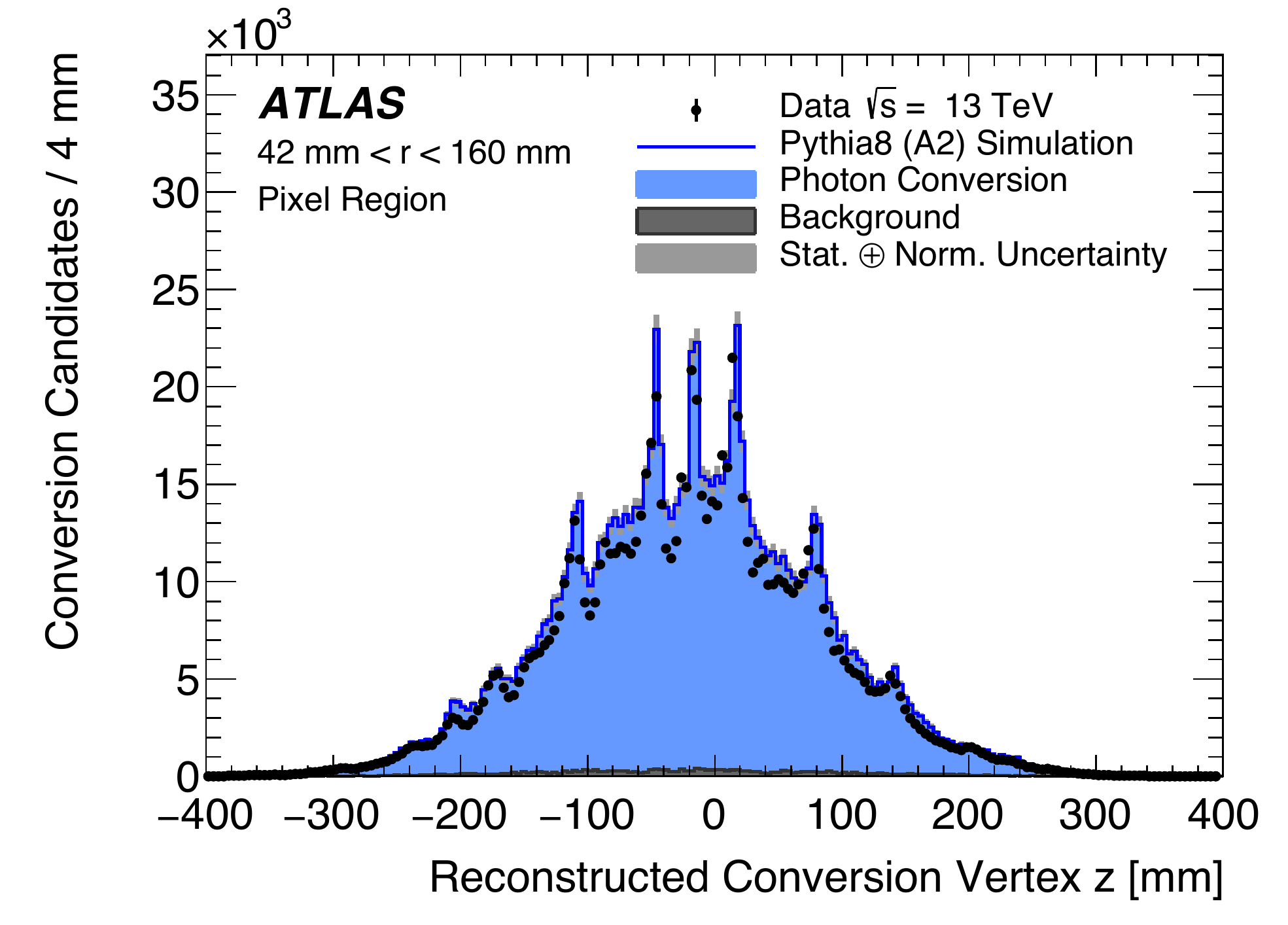}
\caption{Conversion vertex $z$-position distributions for \textsc{Pythia~8} simulation compared to data in the pixel radial region.}
\label{figures:conv_aux3}
\end{center}
\end{figure}

\begin{figure}
\begin{center}
\includegraphics[width=0.45\textwidth]{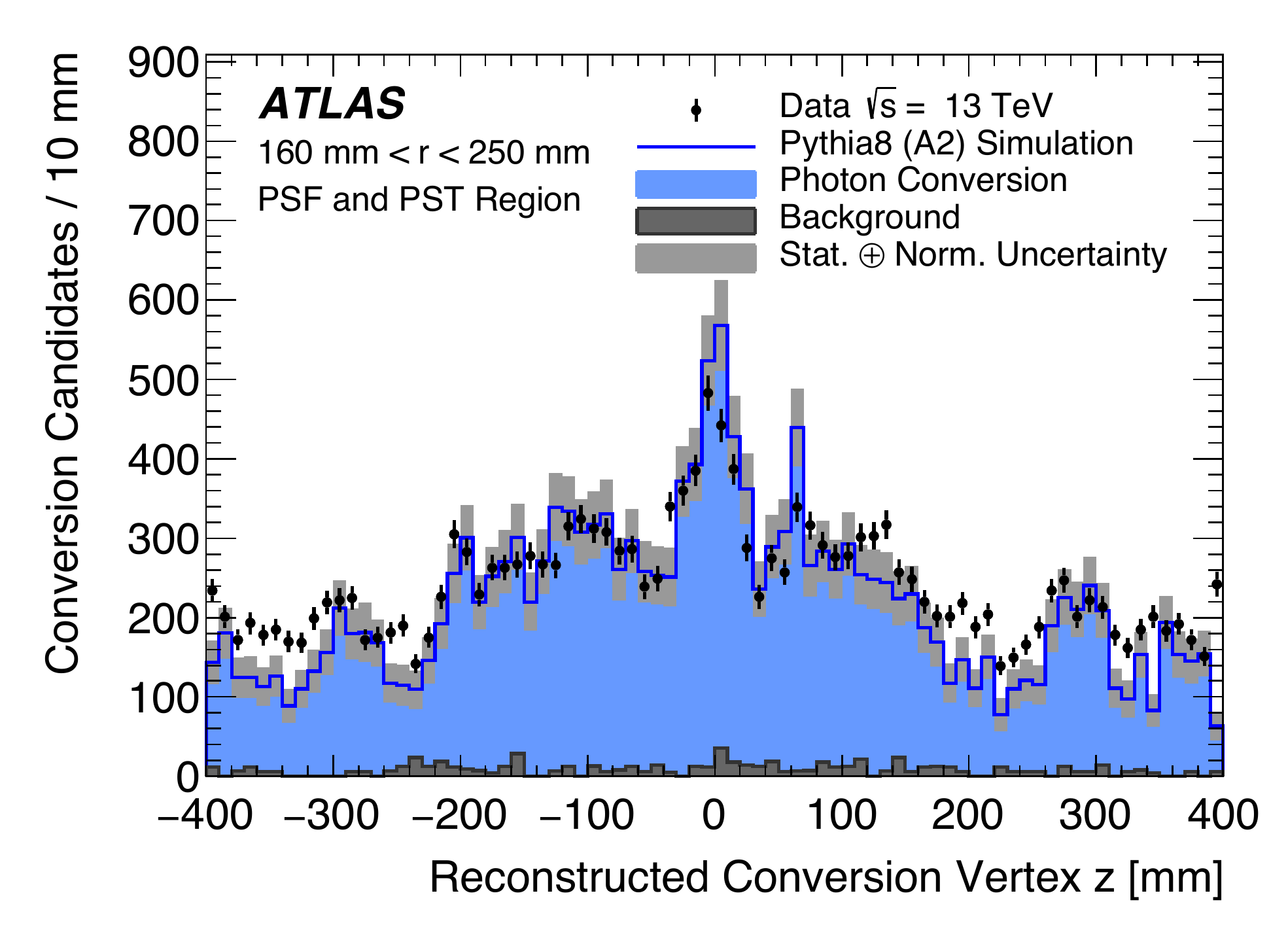}
\caption{Conversion vertex $z$-position distributions for \textsc{Pythia~8} simulation compared to data in the PSF and PST radial region.}
\label{figures:conv_aux4}
\end{center}
\end{figure}

\begin{figure}
\begin{center}
\includegraphics[width=0.45\textwidth]{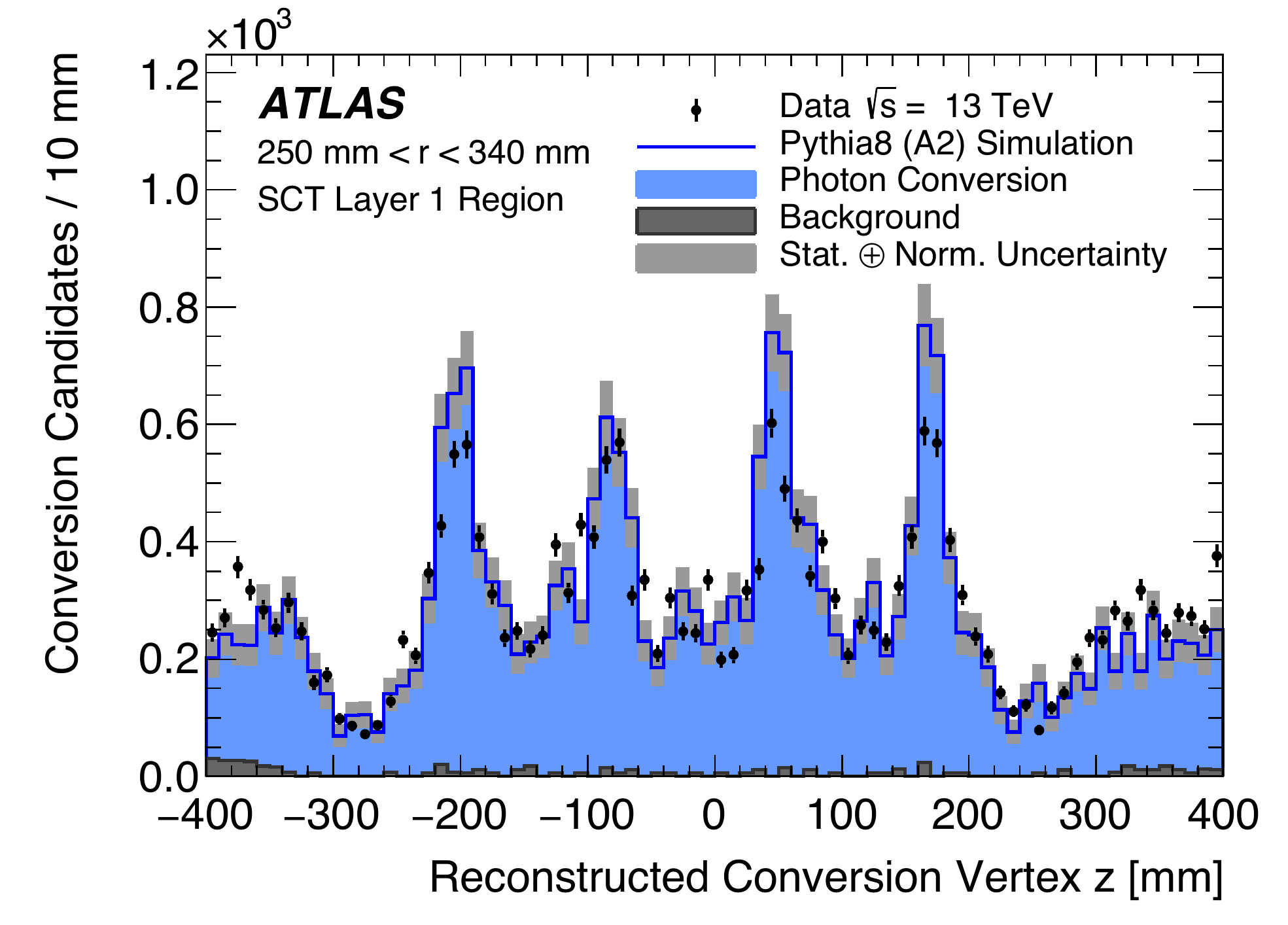}
\caption{Conversion vertex $z$-position distributions for \textsc{Pythia~8} simulation compared to data in the SCT Layer 1 radial region.}
\label{figures:conv_aux5}
\end{center}
\end{figure}

\begin{figure}
\begin{center}
\includegraphics[width=0.45\textwidth]{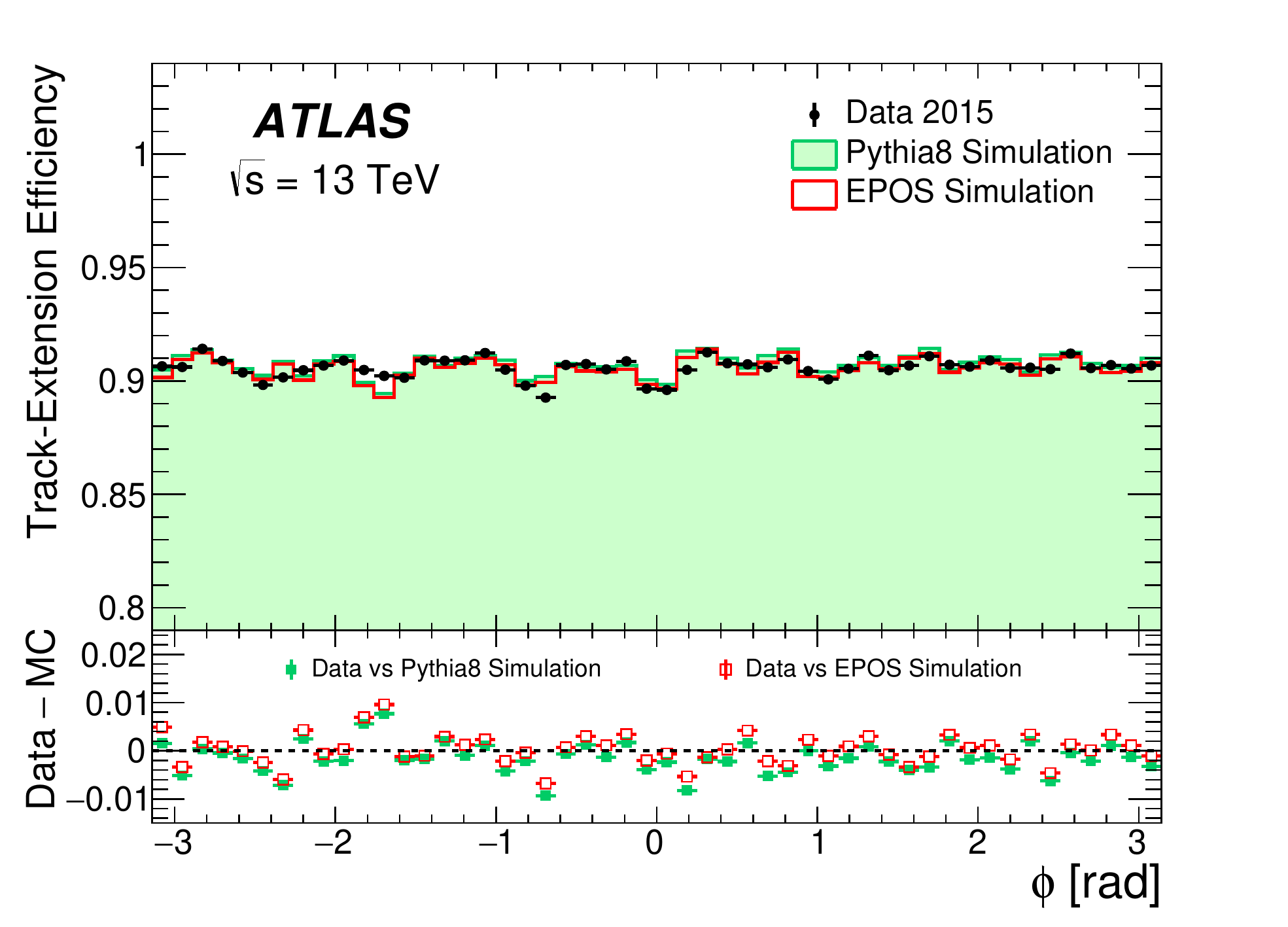}
\caption{Track-extension efficiency as a function of $\phi$ of the tracklets in a comparison between data, \textsc{Pythia~8} and \textsc{Epos}.}
\label{fig:SCTExt_DataEposPythia_Diff_Phi}
\end{center}
\end{figure}

\begin{figure}[tbp]
\begin{center}
\subfigure[]{
  \label{fig:SCTExt_EtaPhi_DataPythia_Diff}
  \includegraphics[width=0.45\textwidth]{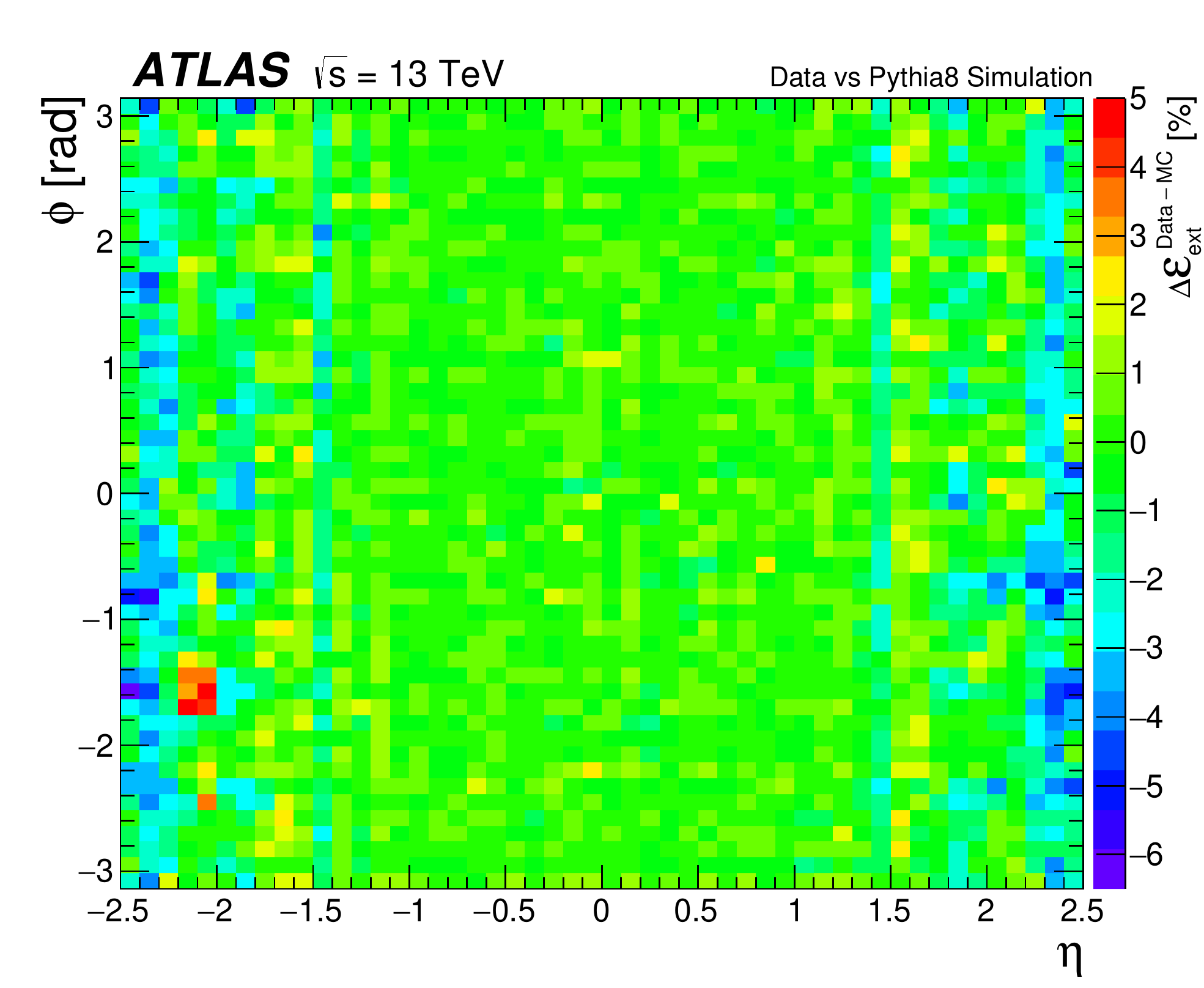}
}
\subfigure[]{
  \label{fig:SCTExt_EtaPhi_DataEpos_Diff}
  \includegraphics[width=0.45\textwidth]{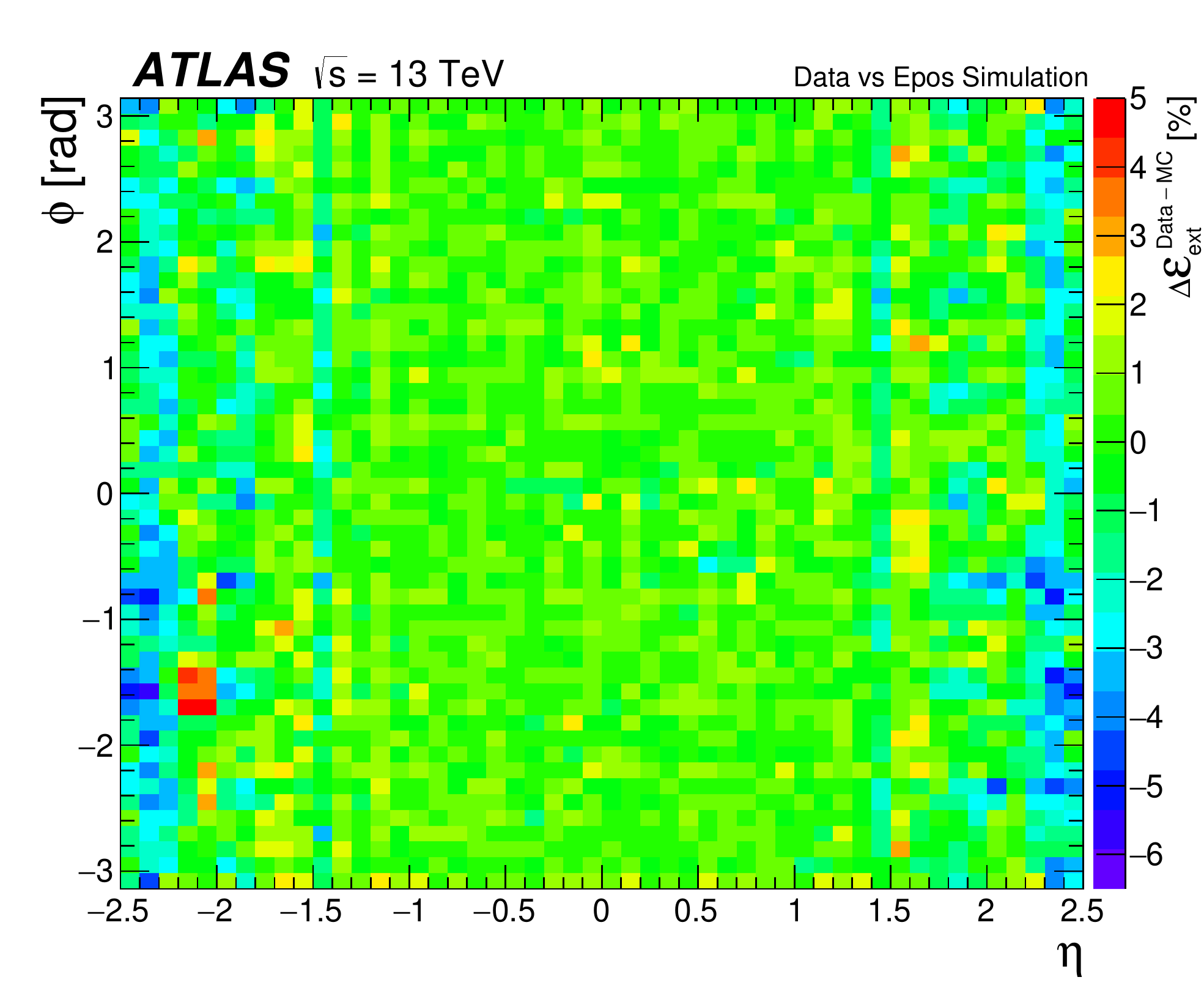}
}
\caption{Track-extension efficiency as a function of $\eta$ and $\phi$ of the tracklets in a comparison between \subref{fig:SCTExt_EtaPhi_DataPythia_Diff} data and \textsc{Pythia~8} and between \subref{fig:SCTExt_EtaPhi_DataEpos_Diff} data and \textsc{Epos}.}
\label{fig:SCTExt_DataEposPythia_Diff_EtaPhi}
\end{center}
\end{figure}

\begin{figure}
\begin{center}
\subfigure[]{
  \label{fig:SCTExt_Particles_Pythia_Eta}
  \includegraphics[width=0.47\textwidth]{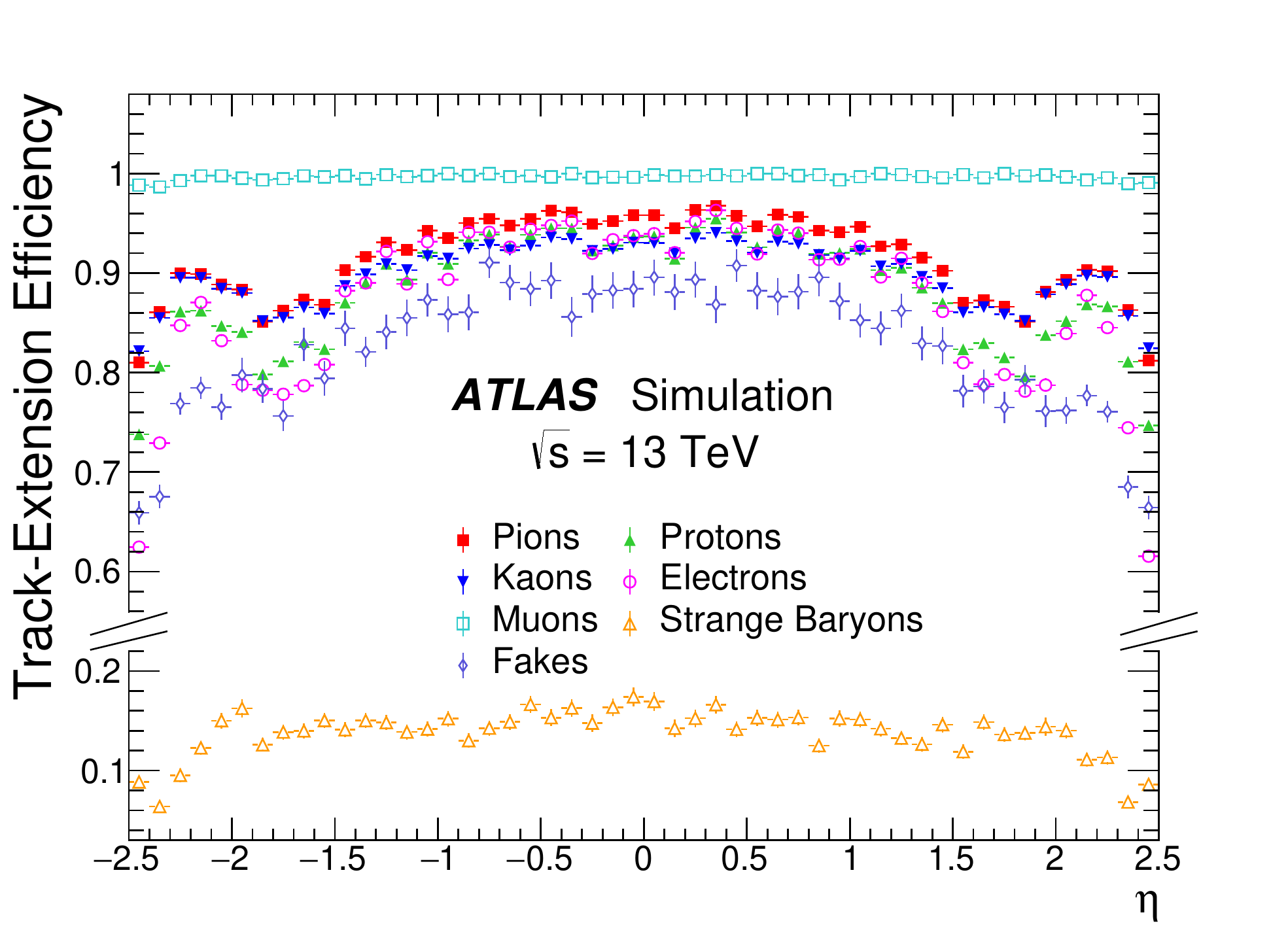}
}
\subfigure[]{
  \label{fig:SCTExt_Particles_Pythia_Pt}
  \includegraphics[width=0.47\textwidth]{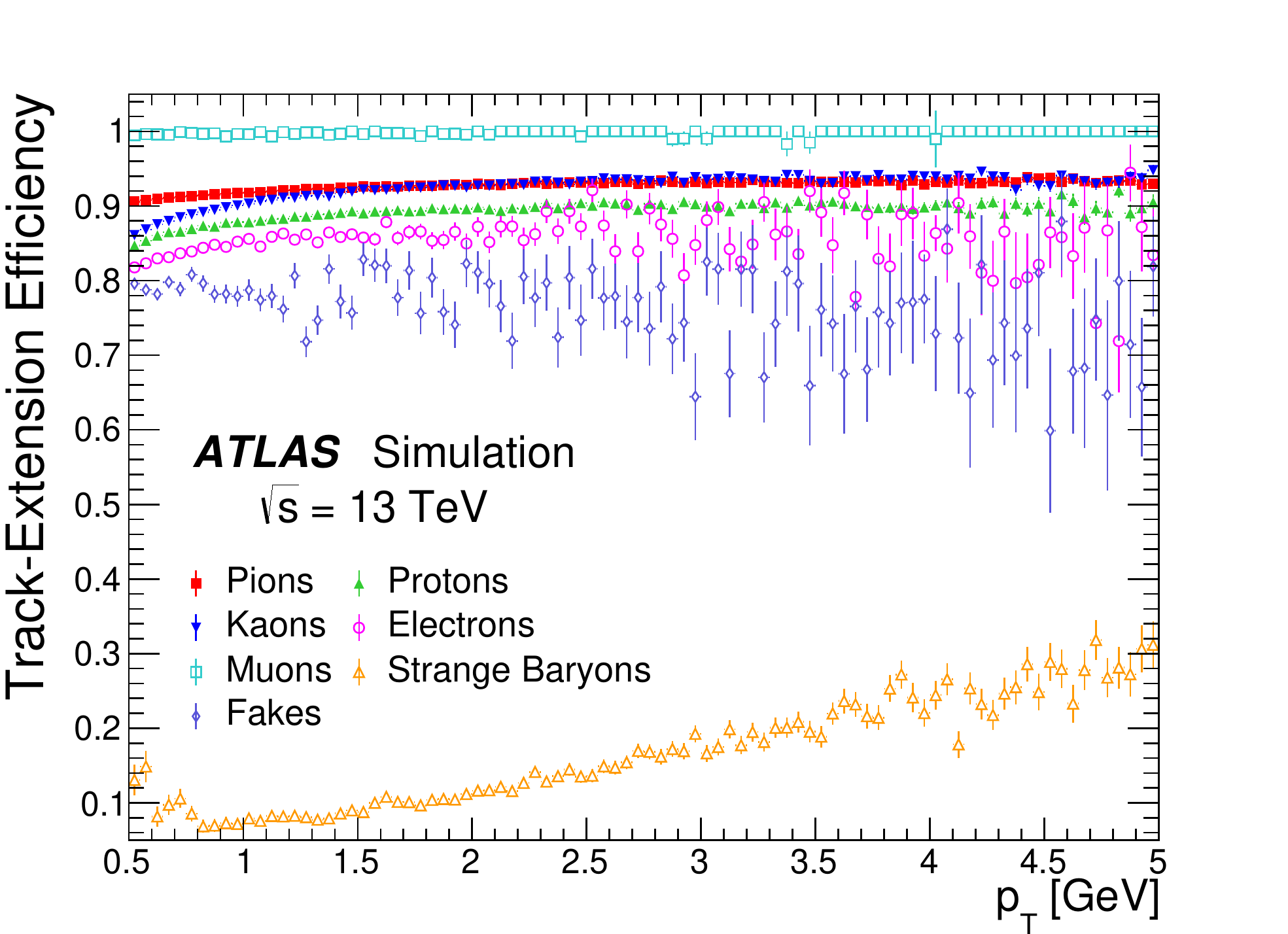}
}
\caption{Track-Extension efficiency exclusive for pions, protons, kaons, electrons, muons, weakly decaying strange baryons ($\Sigma^{\pm}$, $ \Xi^{-}$, $\Omega^{-}$ and their antiparticles) or fake tracklets as a function of \subref{fig:SCTExt_Particles_Pythia_Eta} $\eta$ and \subref{fig:SCTExt_Particles_Pythia_Pt} $\pt$ in simulation.}
\label{fig:SCTExt_Particles_Pythia}
\end{center}
\end{figure}

\begin{figure}
\begin{center}
\subfigure[]{
  \label{fig:TrackletFraction_Pythia_Eta}
  \includegraphics[width=0.47\textwidth]{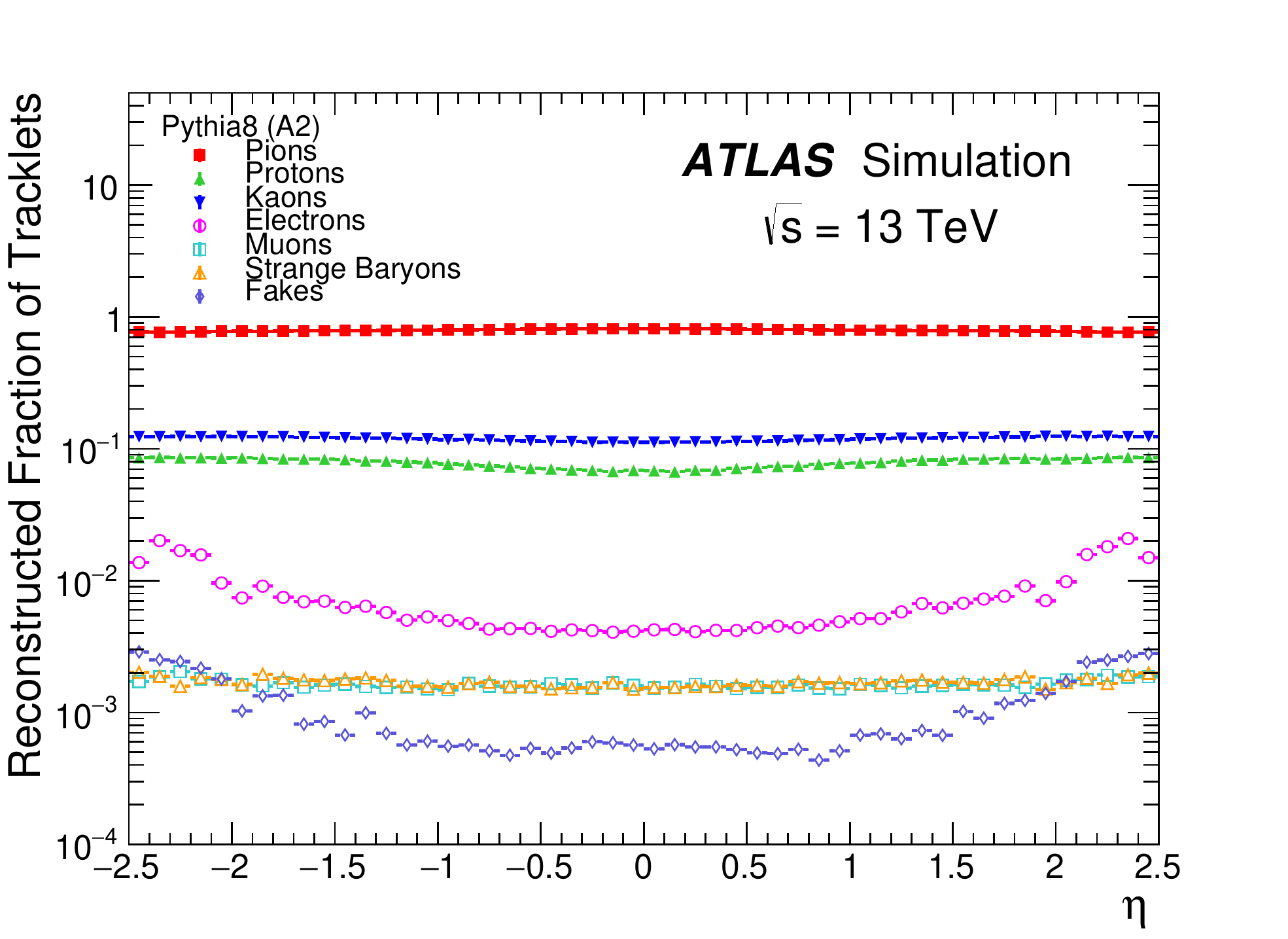}
}
\subfigure[]{
  \label{fig:TrackletFraction_Pythia_Pt}
  \includegraphics[width=0.47\textwidth]{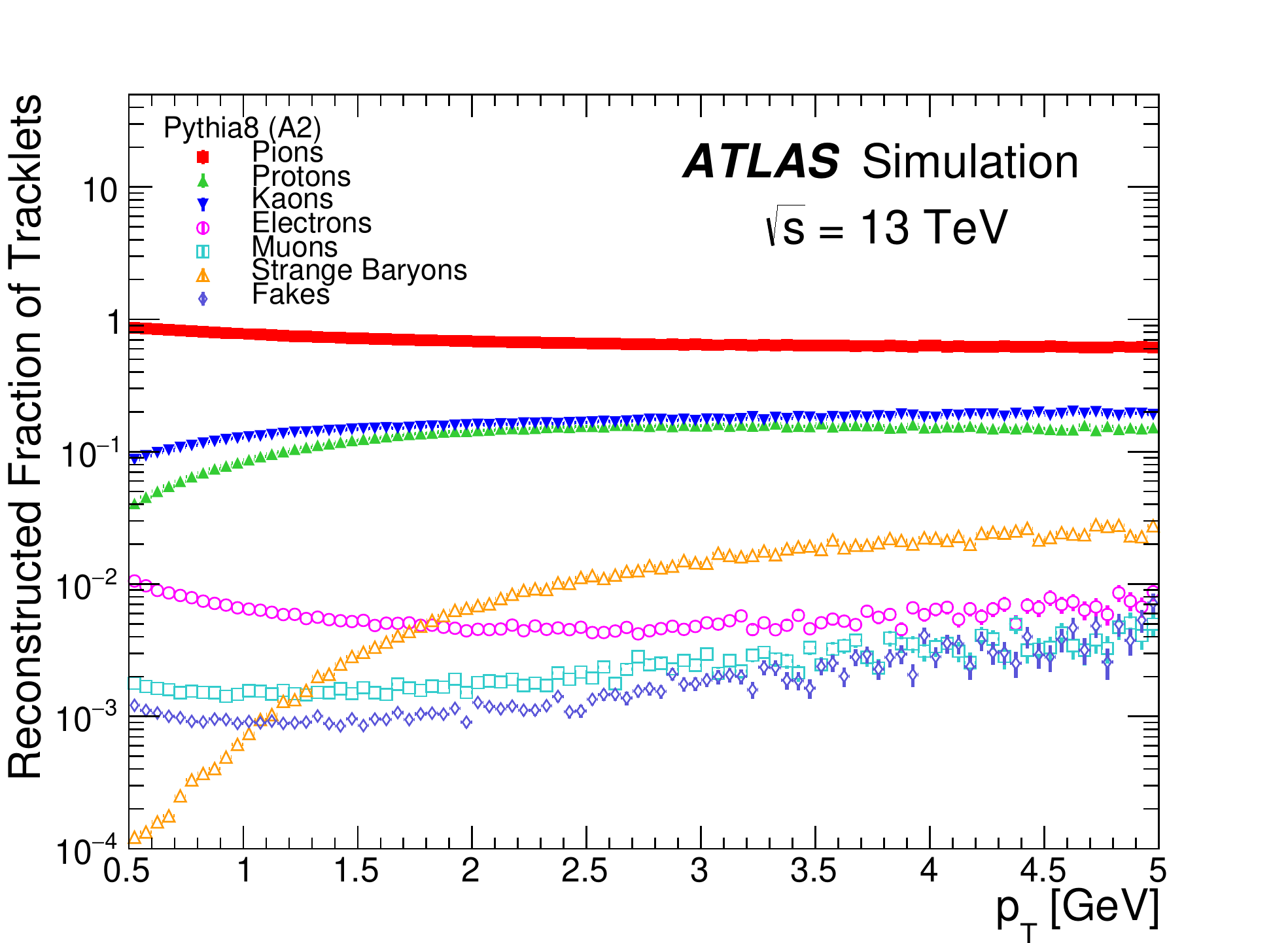}
}
\caption{Fraction of reconstucted tracklets associated with pions, protons, kaons, electrons, muons, weakly decaying strange baryons ($\Sigma^{\pm}$, $ \Xi^{-}$ and $\Omega^{-}$) or fake tracklets with respect to the total number of reconstructed tracklets as a function of \subref{fig:TrackletFraction_Pythia_Eta} $\eta$ and \subref{fig:TrackletFraction_Pythia_Pt} $\pt$ in the \textsc{Pythia~8} simulated sample.}
\label{fig:TrackletFraction_Pythia}
\end{center}

\begin{center}
\subfigure[]{
  \label{fig:TrackletFraction_Epos_Eta}
  \includegraphics[width=0.47\textwidth]{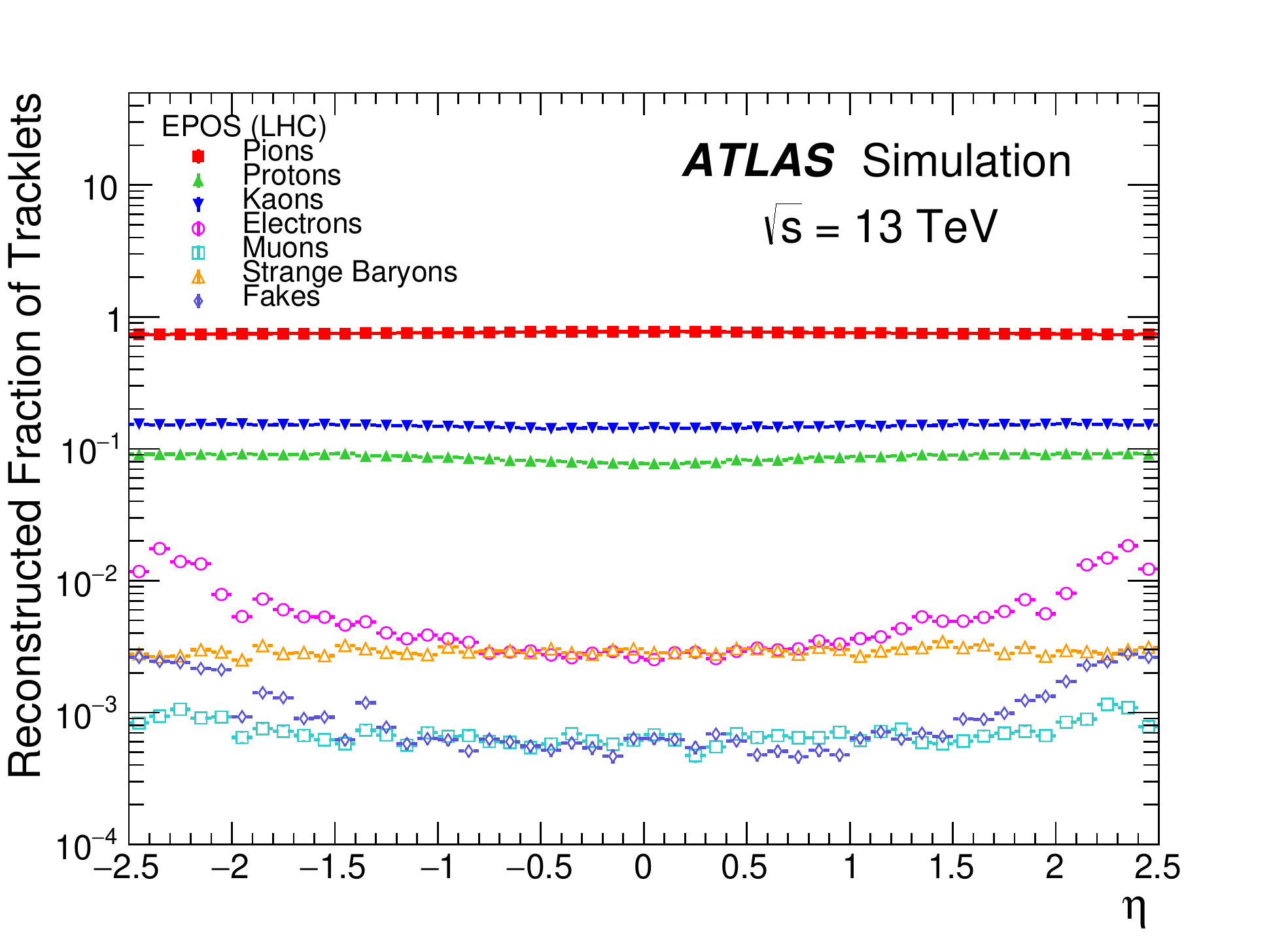}
}
\subfigure[]{
  \label{fig:TrackletFraction_Epos_Pt}
  \includegraphics[width=0.47\textwidth]{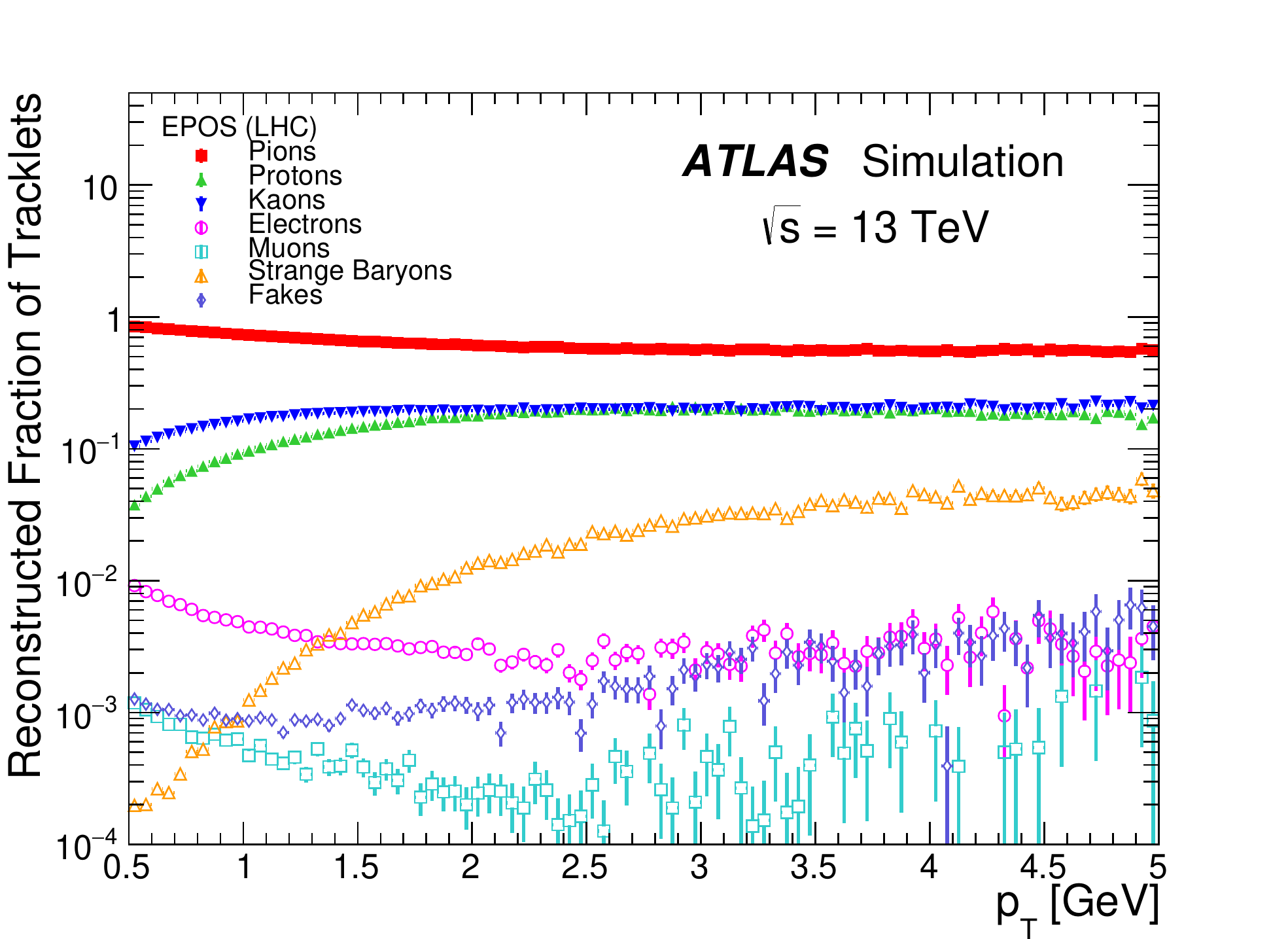}
}
\caption{Fraction of reconstucted tracklets associated with {pions}, {protons}, {kaons}, {electrons}, {muons}, {weakly decaying strange baryons} ($\Sigma^{\pm}$, $ \Xi^{-}$ and $\Omega^{-}$) or {fake tracklets} with respect to the total number of reconstructed tracklets as a function of \subref{fig:TrackletFraction_Epos_Eta} $\eta$ and \subref{fig:TrackletFraction_Epos_Pt} $\pt$ in the \textsc{Epos} simulated sample.}
\label{fig:TrackletFractio_Epos}
\end{center}
\end{figure}

\end{document}